\documentstyle[12pt,here,psfig,epsf]{report}

\begin{document}

\def\PH{\tilde\Psi_H}
\def\PHH{\tilde\Psi_{H_2}}

\newcommand{\dspexp}[1]{\mbox{$e$\raisebox{2ex}{$\displaystyle{#1}$}}}

\newcommand{\spandex}[1]{
   \mbox{\raisebox{#1}{\mbox{}}\raisebox{-#1}{\mbox{}}}
}

\newenvironment{encapfig}[1]{
  \begin{figure}[#1]\centering\begin{minipage}{144mm}
}{
  \end{minipage}\end{figure}
}

\newenvironment{encaptab}[1]{
  \begin{table}[#1]\centering\begin{minipage}{144mm}\begin{center}
}{
  \end{center}\end{minipage}\end{table}
}
\def\gp{g^\prime}
\def\gps{g^{\prime^2}}
\def\gpp{g^{\prime\prime}}
\def\gpps{g^{\prime\prime^2}}

\def\xw{x^{\mbox{}}_W}
\def\mQ{m^{\mbox{}}_{q^\prime}}
\def\mL{m^{\mbox{}}_{l^\prime}}
\def\mheavyl{m^{\mbox{}}_{L}}
\def\mheavyq{m^{\mbox{}}_{Q}}
\def\porbar{\!\!\!
   \raisebox{-0.35ex}{
      $\stackrel{\,\mbox{{\tiny  (}-{\tiny )}}}{p}$
   }\!\!
}
\def\tw{\theta^{\mbox{}}_W}
\def\nuheavyl{\nu_{L}^{\mbox{}}}
\def\mheavynu{m_{\nu_L^{\mbox{}}}^{\mbox{}}}
\def\missing{{p\!\!\!/\,}_{T}^{\mbox{}}}
\def\mw{m^{\mbox{}}_W}
\def\mz{m^{\mbox{}}_Z}
\def\ms{m^{\mbox{}}_S}
\def\mhi{m^{\mbox{}}_{H^0_i}}
\def\mhone{m^{\mbox{}}_{H^0_1}}
\def\ghone{\Gamma^{\mbox{}}_{H^0_1}}
\def\mhtwo{m^{\mbox{}}_{H^0_2}}
\def\ghtwo{\Gamma^{\mbox{}}_{H^0_2}}
\def\mhthree{m^{\mbox{}}_{H^0_3}}
\def\ghthree{\Gamma^{\mbox{}}_{H^0_3}}
\def\mzone{m^{\mbox{}}_{Z_1}}
\def\mztwo{m^{\mbox{}}_{Z_2}}
\def\gztwo{\Gamma^{\mbox{}}_{Z_2}}
\def\mpzero{m^{\mbox{}}_{P^0}}
\def\gpzero{\Gamma^{\mbox{}}_{P^0}}
\def\mhpm{m^{\mbox{}}_{H^\pm}}
\def\ghpm{\Gamma^{\mbox{}}_{H^\pm}}
\newcommand{\order}[1]{{\cal O}({#1})}
\newcommand{\timey}{$\times$}
\def\teva{\mbox{T{\sc evatron}}}

\def\R{{\rm R}}
\def\L{{\rm L}}

\def\bra{\langle}
\def\ket{\rangle}

\newcommand{\Doublet}[2]{
   \left(\begin{array}{@{}c@{}}{#1}\\{#2}\end{array}\right)
}
\newcommand{\Triplet}[3]{
   \left(\begin{array}{@{}c@{}}{#1}\\{#2}\\{#3}\end{array}\right)
}

\def\Rp{{\R^\prime}}
\def\Lp{{\L^\prime}}
\def\tRp{{\tilde\R^\prime}}
\def\tLp{{\tilde\L^\prime}}

   \def\NE{\nu_e}
   \def\Ep{e^\prime}
  \def\NpE{\nu_e^\prime}

  \def\NMU{\nu_\mu}
  \def\MUp{\mu^\prime}
 \def\NpMU{\nu_\mu^\prime}

 \def\NTAU{\nu_\tau}
 \def\TAUp{\tau^\prime}
\def\NpTAU{\nu_\tau^\prime}

   \def\UL{\left(\begin{array}{c}       u\\       d\end{array}\right)_\L}
   \def\EL{\left(\begin{array}{c}     \NE\\       e\end{array}\right)_\L}
  \def\EpL{\left(\begin{array}{c}    \NpE\\     \Ep\end{array}\right)_\L}
  \def\EpR{\left(\begin{array}{c}     \Ep\\    \NpE\end{array}\right)_\L^c}

   \def\CL{\left(\begin{array}{c}        c\\      s\end{array}\right)_\L}
  \def\MUL{\left(\begin{array}{c}     \NMU\\    \mu\end{array}\right)_\L}
 \def\MUpL{\left(\begin{array}{c}    \NpMU\\   \MUp\end{array}\right)_\L}
 \def\MUpR{\left(\begin{array}{c}     \MUp\\  \NpMU\end{array}\right)_\L^c}

   \def\TL{\left(\begin{array}{c}        t\\      b\end{array}\right)_\L}
 \def\TAUL{\left(\begin{array}{c}    \NTAU\\   \tau\end{array}\right)_\L}
\def\TAUpL{\left(\begin{array}{c}   \NpTAU\\  \TAUp\end{array}\right)_\L}
\def\TAUpR{\left(\begin{array}{c}    \TAUp\\ \NpTAU\end{array}\right)_\L^c}

 \def\UR{u_\L^c}
 \def\DR{d_\L^c}
\def\NER{\nu_{e_\L}^c}
 \def\ER{e_\L^c}
\def\DpL{d_\L^\prime}
\def\DpR{d_\L^{\prime c}}
\def\NppE{\nu_{e_\L}^{\prime\prime c}}
	   
  \def\CR{c_\L^c}
  \def\SR{s_\L^c}
\def\NMUR{\nu_{\mu_\L}^c}
 \def\MUR{\mu_\L^c}
 \def\SpL{s_\L^\prime}
 \def\SpR{s_\L^{\prime c}}
\def\NppMU{\nu_{\mu_\L}^{\prime\prime c}}
	   
   \def\TR{t_\L^c}
   \def\BR{b_\L^c}
\def\NTAUR{\nu_{\tau_\L}^c}
 \def\TAUR{\tau_\L^c}
  \def\BpL{b_\L^\prime}
  \def\BpR{b_\L^{\prime c}}
\def\NppTAU{\nu_{\tau_\L}^{\prime\prime c}}
	 
\newcommand{\SU}[2]{\mbox{${\rm SU}({#1})_{{\rm #2}}$}}
\newcommand{\U}[2]{\mbox{${\rm U}({#1})_{{\rm #2}}$}}

\newcommand\STM{\mbox{$\SU{3}{c}\otimes\SU{2}{L}\otimes\U{1}{Y}$}\ }

\newcommand{\Esix}{\mbox{${\rm E}_6$}\ }
\newcommand{\ExE}{\mbox{${\rm E}_8\otimes{\rm E}_8^\prime\;$}}

\newcommand{\ZE}{\mbox{${\rm Z}_{\rm E}\,$}}

\newcommand\downrighthookarrow{
  \mbox{
    \setlength{\unitlength}{1mm}
    \begin{picture}(6,7)
      \put(0,7){\line(0,-1){6}}
      \put(0,1){\vector(1,0){6}}
    \end{picture}
  }
}

\newcommand{\Super}[1]{\mbox{$\Phi_{{#1}}$}}

\newcommand{\sfrac}[2]{\mbox{\small$\frac{{#1}}{{#2}}$}}
\newcommand{\ffrac}[2]{\mbox{\footnotesize$\frac{{#1}}{{#2}}$}}
\newcommand{\tfrac}[2]{\mbox{\tiny$\frac{{#1}}{{#2}}$}}
\newcommand{\txtfrac}[2]{\mbox{$\textstyle\frac{{#1}}{{#2}}$}}

\def\approxle{\,\raisebox{-0.625ex}{$\stackrel{<}{\sim}$}\,}
\def\approxge{\,\raisebox{-0.625ex}{$\stackrel{>}{\sim}$}\,}

\def\eAB{\mbox{$\in$\raisebox{-1.0ex}{\scriptsize$AB$}}}
\def\real{\mbox{Re}\,}
\def\imag{\mbox{Im}\,}
\def\onetwo{\frac{1}{2}}
\def\onethree{\frac{1}{3}}
\def\twothree{\frac{2}{3}}
\def\onefour{\frac{1}{4}}

\newcommand\implies{\mbox{$\Longrightarrow$}}

\newcommand{\Lam}[1]{\mbox{$\lambda_{{#1}}$}}
\def\lrpartial{\raisebox{1.9ex}{
   \footnotesize$\leftrightarrow$}\!\!\!\!\!\partial
}

\newcommand{\mathrm}[1]{{\rm #1}}
\def\vdij{\delta{\vec{\mathrm{r}}}_{\mathrm{ij}}} 
\def\vdrg{\delta{\vec{\mathrm{r}}}_{\mathrm{rg}}} 
\def\vdgb{\delta{\vec{\mathrm{r}}}_{\mathrm{gb}}} 
\def\vdbr{\delta{\vec{\mathrm{r}}}_{\mathrm{br}}} 
\def\vn{{\vec{\mathrm{n}}}}                 
\def\vr{{\vec{\mathrm{r}}}}                 
\def\vri{{\vec{\mathrm{r}}}_i}              
\def\vrj{{\vec{\mathrm{r}}}_j}              
\def\vrij{{\vec{\mathrm{r}}}_{ij}}          
\def\vx{{\vec{\mathrm{x}}}}                 
\def\vrr{{\vec{\mathrm{r}}}_{\mathrm{r}}}   
\def\vrg{{\vec{\mathrm{r}}}_{\mathrm{g}}}   
\def\vrb{{\vec{\mathrm{r}}}_{\mathrm{b}}}   
\def\vrrg{{\vec{\mathrm{r}}}_{\mathrm{rg}}} 
\def\vrgb{{\vec{\mathrm{r}}}_{\mathrm{gb}}} 
\def\vrbr{{\vec{\mathrm{r}}}_{\mathrm{br}}} 
\def\rrg{{\mathrm{r}}_{\mathrm{rg}}}        
\def\rgb{{\mathrm{r}}_{\mathrm{gb}}}        
\def\rbr{{\mathrm{r}}_{\mathrm{br}}}        
\def\grad{\vec{\nabla}}                     
\def\chic{\mbox{\raisebox{1mm}{$\chi$}}_{\rm Correlation}}

\def\ggqzll{\raisebox{-1.693cm}{
\mbox{
\setlength{\unitlength}{1cm}
\begin{picture}(5.743,3.6)
 \put(-0.257,0.086){
  \mbox{\epsfxsize=6.0cm
   \epsffile{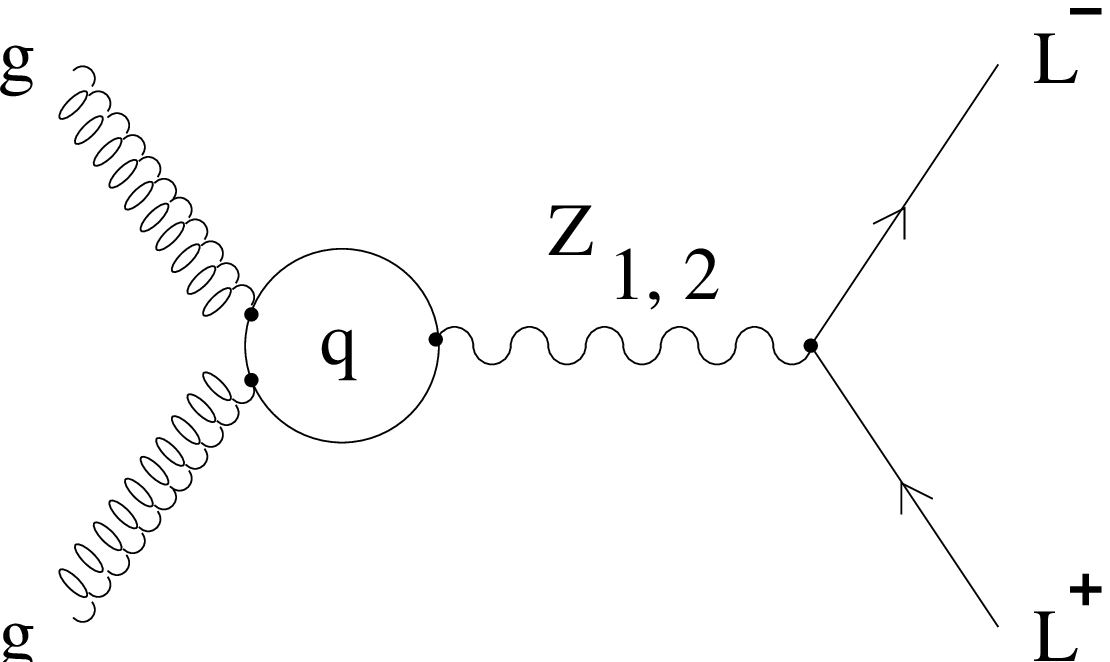}}
 }
\end{picture}
}}}

\def\ggqhpll{\raisebox{-1.693cm}{
\mbox{
\setlength{\unitlength}{1cm}
\begin{picture}(5.743,3.6)
 \put(-0.257,0.086){
  \mbox{\epsfxsize=6.0cm
   \epsffile{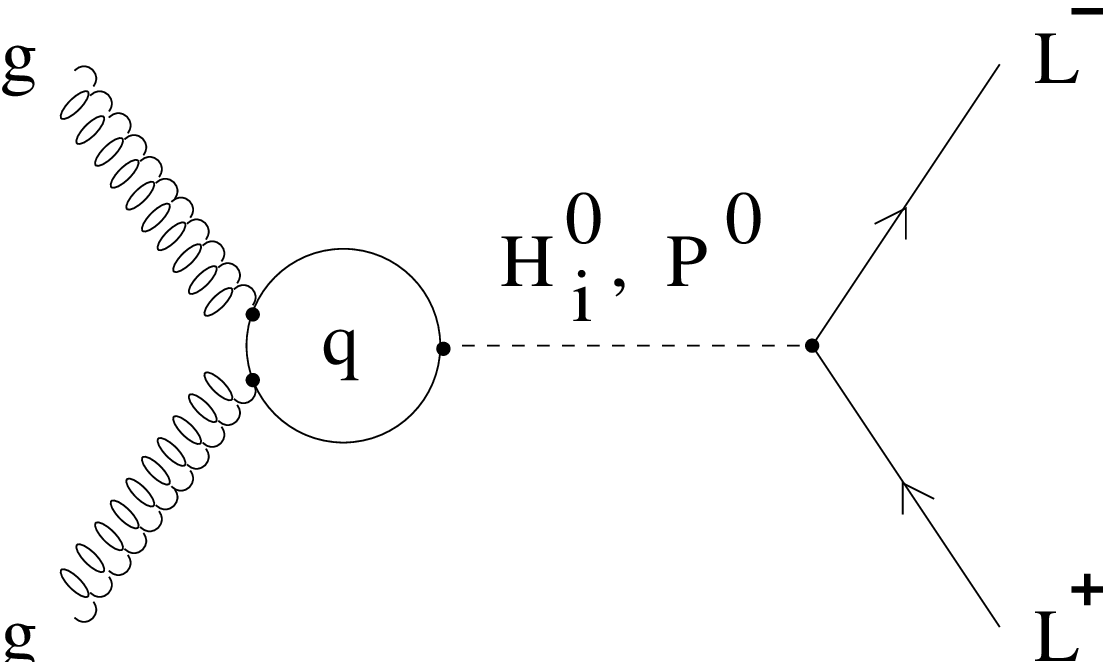}}
 }
\end{picture}
}}}

\def\ggqhll{\raisebox{-1.693cm}{
\mbox{
\setlength{\unitlength}{1cm}
\begin{picture}(5.743,3.6)
 \put(-0.257,0.086){
  \mbox{\epsfxsize=6.0cm
   \epsffile{ggqhll.eps}}
 }
\end{picture}
}}}

\def\ggsqhll{\raisebox{-1.693cm}{
\mbox{
\setlength{\unitlength}{1cm}
\begin{picture}(5.743,3.6)
 \put(-0.257,0.086){
  \mbox{\epsfxsize=6.0cm
   \epsffile{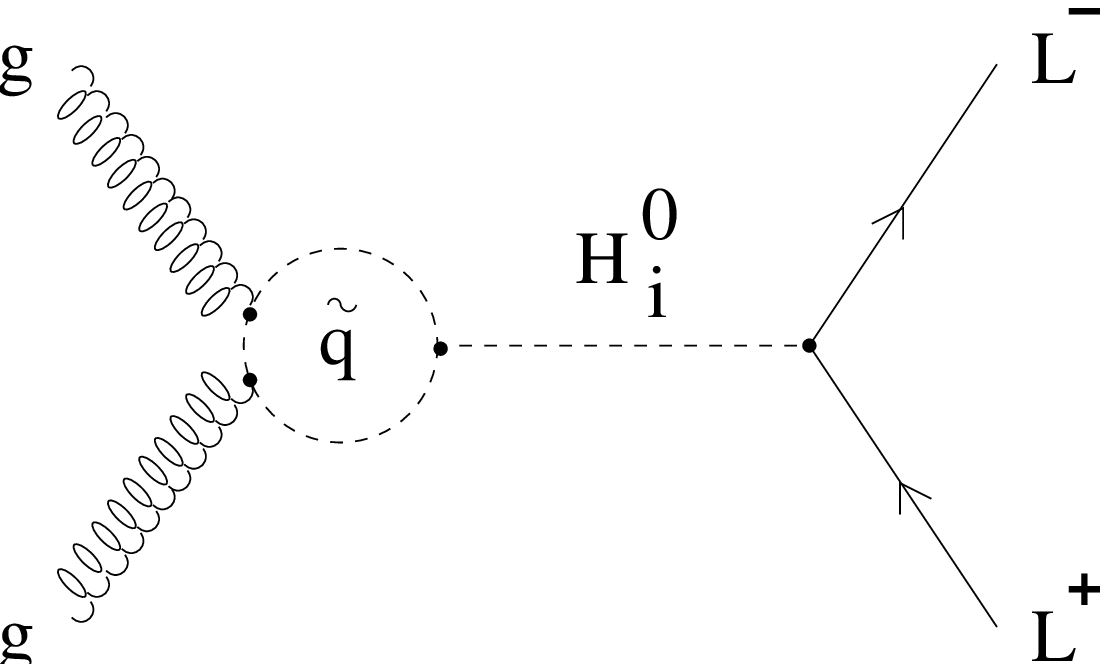}}
 }
\end{picture}
}}}

\def\ggvsqhll{\raisebox{-1.693cm}{
\mbox{
\setlength{\unitlength}{1cm}
\begin{picture}(5.743,3.6)
 \put(-0.257,0.086){
  \mbox{\epsfxsize=6.0cm
   \epsffile{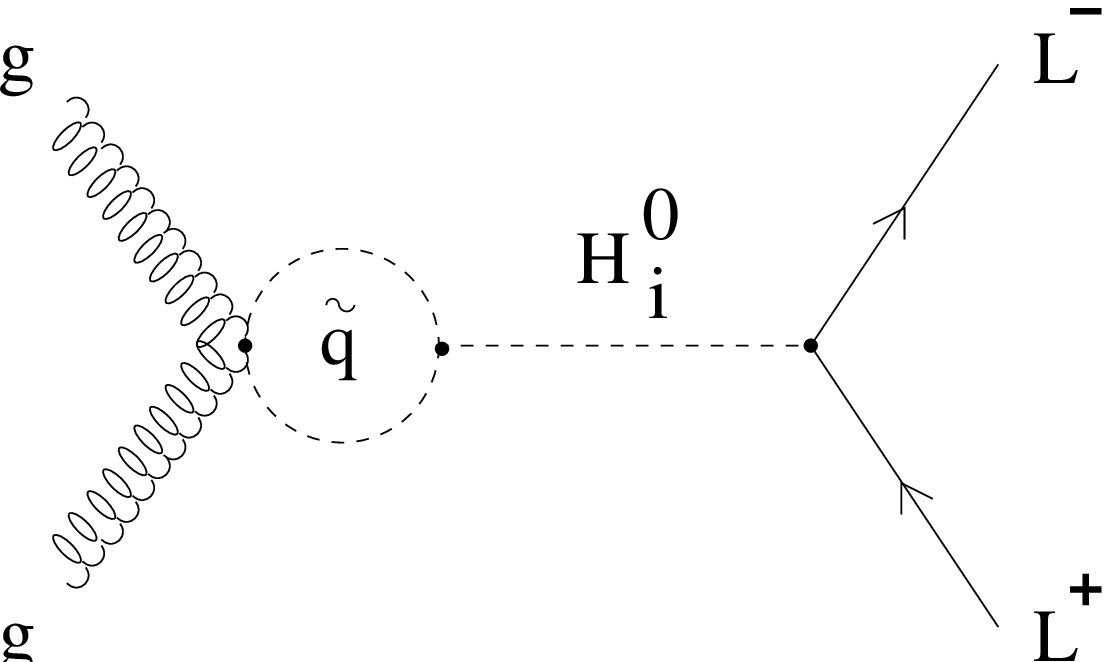}}
 }
\end{picture}
}}}

%


\newcommand{\gaugeff}[3]{\mbox{$\!\!$\raisebox{-1.65625cm}{
\mbox{
\setlength{\unitlength}{1cm}
\begin{picture}(3.25,3.5)
 \put(0.625,2.125){\mbox{${\displaystyle #1}$}}
 \put(3.0,0.125){\mbox{${\displaystyle #2}$}}
 \put(3.0,3.125){\mbox{${\displaystyle #3}$}}
 \put(-0.25,0.25){
  \mbox{\epsfxsize=3.0cm
   \epsffile{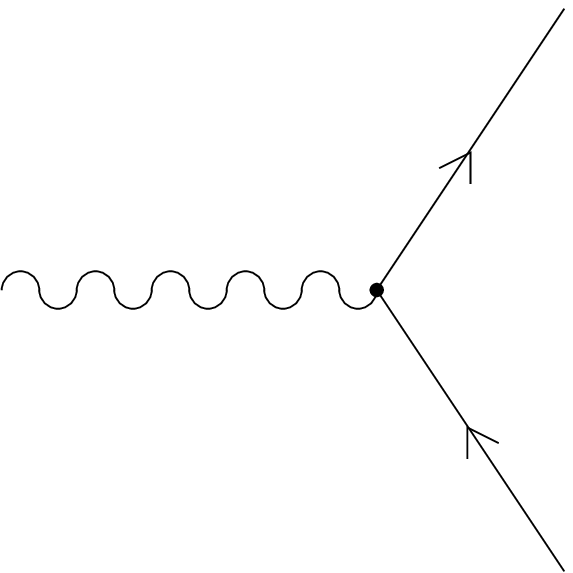}}
 }
\end{picture}
}}}}

\newcommand{\gaugess}[3]{\mbox{$\!\!$\raisebox{-1.65625cm}{
\mbox{
\setlength{\unitlength}{1cm}
\begin{picture}(3.25,3.5)
 \put(0.625,2.125){\mbox{${\displaystyle #1}$}}
 \put(3.0,0.125){\mbox{${\displaystyle #2}$}}
 \put(3.0,3.125){\mbox{${\displaystyle #3}$}}
 \put(-0.25,0.25){
  \mbox{\epsfxsize=3.0cm
   \epsffile{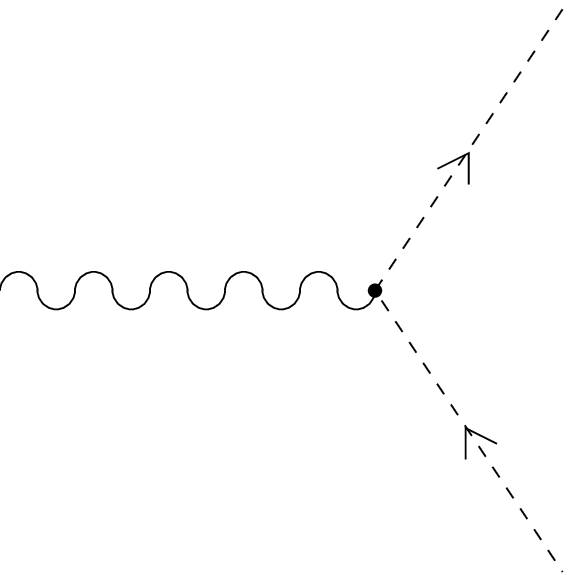}}
 }
\end{picture}
}}}}

\newcommand{\scalarff}[3]{\mbox{$\!\!$\raisebox{-1.65625cm}{
\mbox{
\setlength{\unitlength}{1cm}
\begin{picture}(3.25,3.5)
 \put(0.625,2.125){\mbox{${\displaystyle #1}$}}
 \put(3.0,0.125){\mbox{${\displaystyle #2}$}}
 \put(3.0,3.125){\mbox{${\displaystyle #3}$}}
 \put(-0.25,0.25){
  \mbox{\epsfxsize=3.0cm
   \epsffile{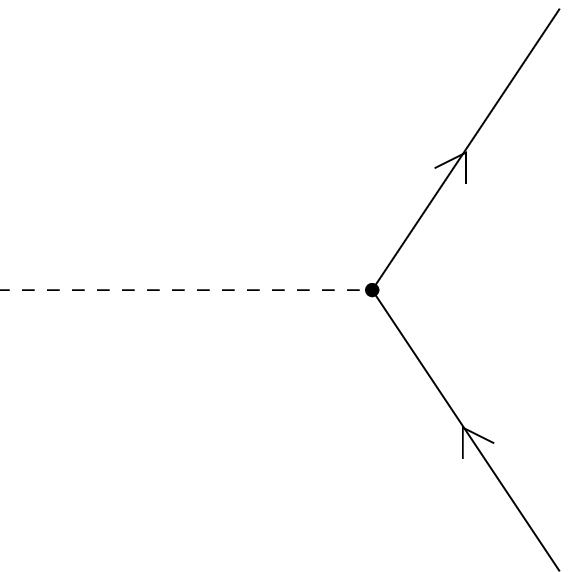}}
 }
\end{picture}
}}}}

\newcommand{\scalarss}[3]{\mbox{$\!\!$\raisebox{-1.65625cm}{
\mbox{
\setlength{\unitlength}{1cm}
\begin{picture}(3.25,3.5)
 \put(0.625,2.125){\mbox{${\displaystyle #1}$}}
 \put(3.0,0.125){\mbox{${\displaystyle #2}$}}
 \put(3.0,3.125){\mbox{${\displaystyle #3}$}}
 \put(-0.25,0.25){
  \mbox{\epsfxsize=3.0cm
   \epsffile{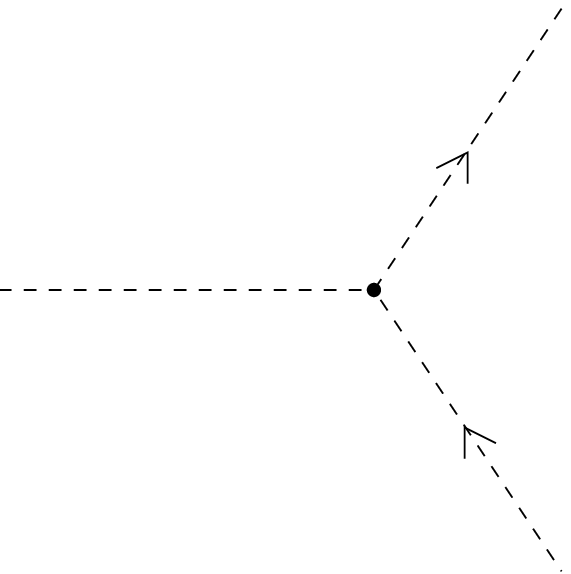}}
 }
\end{picture}
}}}}

\newcommand{\scalarvv}[3]{\mbox{$\!\!$\raisebox{-1.65625cm}{
\mbox{
\setlength{\unitlength}{1cm}
\begin{picture}(3.25,3.5)
 \put(0.625,2.125){\mbox{${\displaystyle #1}$}}
 \put(3.0,0.125){\mbox{${\displaystyle #2}$}}
 \put(3.0,3.125){\mbox{${\displaystyle #3}$}}
 \put(-0.25,0.25){
  \mbox{\epsfxsize=3.0cm
   \epsffile{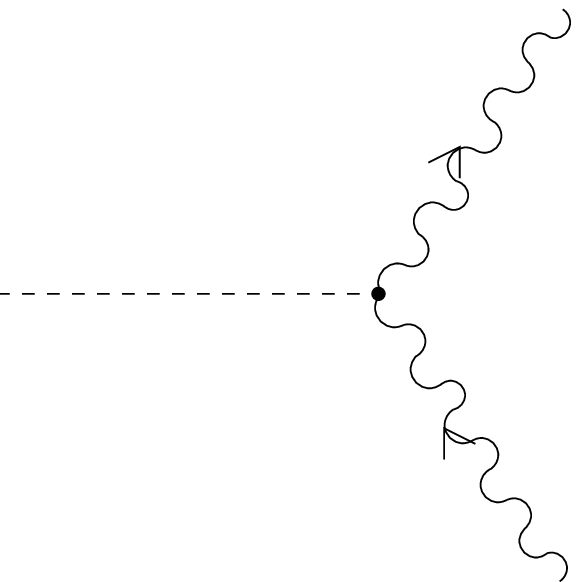}}
 }
\end{picture}
}}}}

\newcommand{\overlaystuff}[4]{\mbox{$\!\!$\raisebox{-1.65625cm}{
\mbox{\setlength{\unitlength}{1cm}
\begin{picture}(4.25,3.5)
 \put(0.0,1.65625){\mbox{{#1}}}
 \put(0.25,1.375){\vector(1,0){1.0}}
 \put(0.625,1.0){\mbox{${\displaystyle #2}$}}
 \put(2.25,1.0){\vector(2,-3){0.5}}
 \put(2.0,0.5){\mbox{${\displaystyle #3}$}}
 \put(2.25,2.5){\vector(2,3){0.5}}
 \put(2.0,2.75){\mbox{${\displaystyle #4}$}}
\end{picture}}
}}}


\newcommand{\eqnpict}[2]{
   \begin{equation}
   \mbox{\raisebox{-1.625cm}{
      \setlength{\unitlength}{1cm}
      \begin{picture}(0.0,4.0)
         \put(-7.5,1.625){
            \begin{tabular}{cc}
               {$\displaystyle #1$}&{$\displaystyle #2$}
            \end{tabular}
         }
      \end{picture}
   }}
   \end{equation}
}

\pagestyle{empty}
\begin{center}
\vfill
{\Large \bf STRING INSPIRED QCD AND $E_6$ MODELS}\\
\vfill
\mbox{\epsfxsize=10cm
      \epsffile{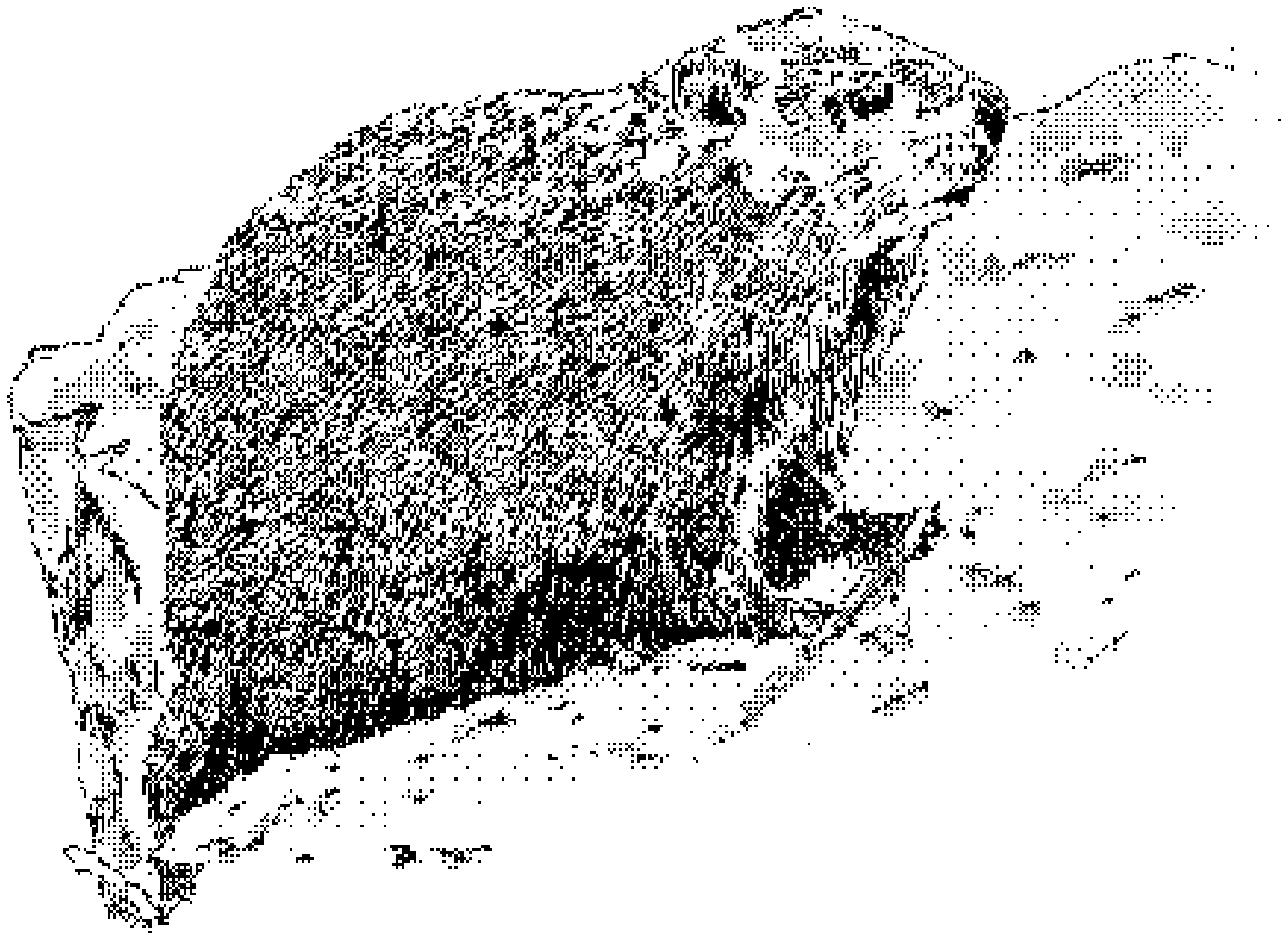}}\\
\vfill
{\Large \bf MICHAEL M. BOYCE}\\
\vfill
{\Large \bf PH.D. THESIS}\\
\vfill
{\Large \bf 1996}\\
\vfill
\end{center}
\clearpage

\pagestyle{empty}

\vspace*{0.5in}
\begin{center}
{\Large \bf String Inspired QCD and $E_6$ Models} \\[0.4in]
%
%
                     by \\[0.3in] 
         {\bf Michael~M.~Boyce, B.Sc., M.Sc.}            \\ [0.3in] 
               A thesis submitted to                     \\ 
     the Faculty of Graduate Studies and Research        \\ 
             in partial fulfillment of                   \\
        the requirements for the degree of               \\ 
              Doctor of Philosophy                       \\[0.3in]
       Ottawa-Carleton Institute for Physics             \\ 
              Department of Physics                      \\[0.3in]
            {\bf Carleton University}                    \\
             Ottawa, Ontario, Canada                     \\
               June 10, 1996                             \\[0.3in]
               \copyright copyright                      \\
              1996, Michael M. Boyce
\end{center}


\clearpage

\pagestyle{plain}
\pagenumbering{roman}
\setcounter{page}{2}

\vspace*{-0.5in}
\begin{center}
The undersigned recommend to \\
the Faculty of Graduate Studies and Research\\
acceptance of the thesis\\[.5in]
{\large \bf  String Inspired QCD and E$_{\bf 6}$ Models}\\[.5in]
submitted by {\bf \bf Michael M. Boyce, B.Sc., M.Sc.} \\
in partial fulfilment of the requirements for \\
the degree of Doctor of Philosophy \\[.5in]
\mbox{\epsfxsize=7.5cm
      \epsffile{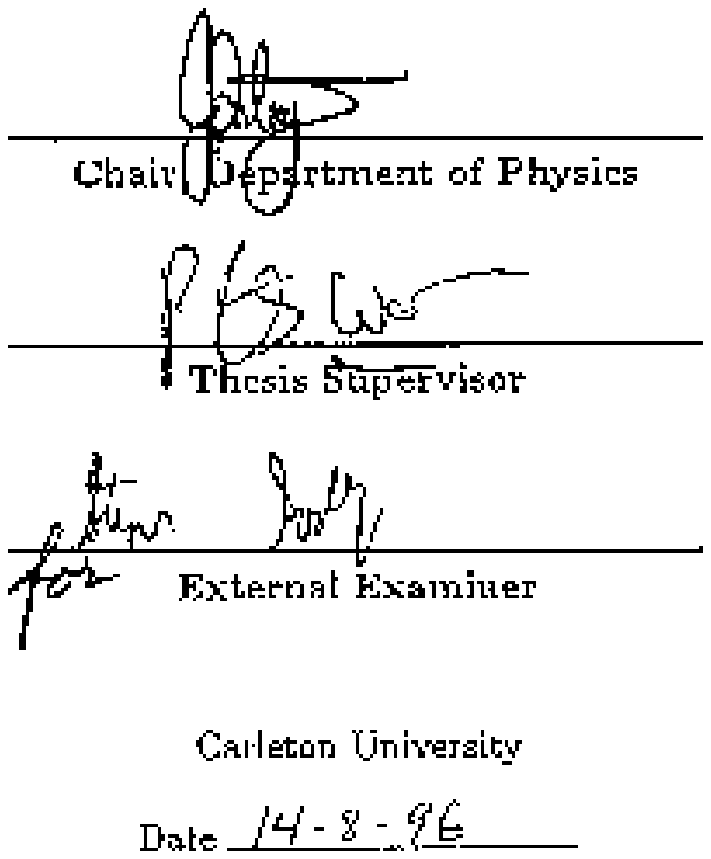}}
\end{center}
\clearpage


\vspace*{\fill}
\begin{center}
{\LARGE\bf Abstract}
\end{center}
The  work in this thesis  consists of two  distinct parts:

 A class of models called, ``String-flip potential models,'' (SFP's) are
studied as a possible candidate for modeling  nuclear matter in terms of
constituent    quarks.   These  models  are     inspired  from   lattice
quantum-chromodynamics (QCD) and  are nonperturbative in  nature.  It is
shown that they are viable candidates  for modeling nuclear matter since
they reproduce most  of the bulk  properties except for nuclear binding.
Their properties are studied in nuclear and mesonic matter.  A new class
of models  is developed, called  ``flux-bubble potential models,'' which
allows   for  the  SFP's to   be  extended  to  include perturbative QCD
interactions.  Attempts to obtain nuclear binding is not successful, but
valuable insight was  gained   towards  possible future directions    to
pursue.

  The possibility of  studying Superstring inspired \Esix  phenomenology
at  high energy hadron colliders   is  investigated.  The production  of
heavy  lepton   pairs {\it   via}   a gluon-gluon  fusion  mechanism  is
discussed.  An enhancement in the parton level cross-section is expected
due to the heavy (s)fermion loops which couple to the gluons.
\vspace*{\fill}

\addcontentsline{toc}{section}{Abstract}
\clearpage

\mbox{}
\vfill
\begin{center}
{\bf TO MY PARENTS}
\end{center}
\addcontentsline{toc}{section}{Dedication}
\vfill
\clearpage

\vspace*{2in}
\begin{center}
{\LARGE\bf Acknowledgements}
\end{center}
\addcontentsline{toc}{section}{Acknowledgements}

  The size of this thesis  represents only a  small part of the mountain
of work, with  its  multitude of crevices,   that went into its  making.
Herein are the remnants of literally thousands of lines of computer code
and many thick binders of hand computations.

  I would  like to thank my supervisor   Dr.~P.J.S.~Watson for giving me
the opportunity to see what real research is like.  Real in the sense of
being innovative  and trying to come up  with original ideas, as opposed
to just grinding out some calculations.

  I would like to thank Drs.  M.A.~Doncheski and H.~K\"{o}nig for giving
me   the opportunity to  collaborate   with  them  on  some $E_6$  work,
contained  in chapter 4 of  this thesis --- OK   this is grinding!  {\tt
:)~RL}

  I would like to thank my Ph.D.  examination committee Dean R.~Blockley
(chair),  Dr.~F.~Dehne,  Dr.~S.~Godfrey,  Dr.~G.~Karl,    Dr.~G.~Oakham,
Dr.~A.~Song,  and  Dr.~P.J.S.~Watson  for giving  me   an enjoyable  but
challenging  thesis   defense.  I   would   especially like    to  thank
Dr.~S.~Godfrey  who  served double duty  by filling  in on Dr.~G.~Karl's
behalf,  who was  ill at time  of the   defense and therefore  unable to
attend (I sincerely  hope all went  well).  I would  also like to  thank
Dr.~S.~Godfrey for filling in after the defense as acting supervisor, as
mine was away at this time, making sure that all of the corrections were
made to my thesis, and for carefully re-reading it for any minor errors.

  I would like to  thank my Ph.D.   committee members, past and present,
Dr.~S.~Godfrey, Dr.~W.~Romo,  Dr.~G.~Slater, Dr.~P.J.S.~Watson for a job
well done.

  I would like to thank  S.~Nicholson for taking  time  out of her  very
busy  schedule  to proofread  my  thesis.  Also, I   would like to thank
H.~Blundell,  A.~Dekok, Dr.~M.A.~Doncheski, and Dr.~H.~K\"{o}nig,    for
proofreading various  documents of mine, such  as parts of  this thesis,
papers, conference proceedings, {\it etc.}

  I  would   like   to  thank   Dr.~M.A.~Doncheski,    Dr.~H.~K\"{o}nig,
H.~Blundell,  M.~Jones, I.~Melo, Dr.~S.~Sanghera,   Dr.~M.K.~Sundaresan,
and Dr.~G.~Oakham for their very useful physics consultations.

  I  would  like to thank  OPAL, CRPP,  and  \mbox{T{\scriptsize HEORY}}
groups,  as well   as  the Department   of  Physics for  usage of  their
computing  facilities.  Also,    I   would like  to   thank   A.~Barney,
J.~Carleton, Dr.~F.~Dehne,  A.~Dekok, W.~Hong, B.~Jack, M.~Jones, and M.
Sperling,  for  their   general  computer  and   computational   related
consultations.

  I would   like to the   lab  and tech.   guys D.~Paterson,  G.~Curley,
G.~Findlay, and J.~Sliwka, well\ldots for just being lab and tech. guys!

  I  would like  to thank R.~Tighe  for  her warm hearted nature towards
graduate students,  especially to those in need.  Also, I  would like to
thank  the  graduate advisors, past   and present,  Drs. P.~Kalyniak and
W.~Romo, who also kept a caring eye on their flocks.

  I would like to  thank the  secretaries,  past and present,  R.~Tighe,
T.~Buckley,  and  E.~Lacelle   for generally being   helpful  and always
bringing a shaft  of light   into the  department for those   especially
gloomy days.

  Finally, I  would like to thank all  of my  friends, past and present,
the bar go-ers, the sports players, the movie  go-ers, the pool players,
{\it  etc.}, Dr.~G.~Bhattacharya, H.~Bundell, G.~Cron, F.~Dalnok-Veress,
A.~Dekok,  Dr.~M.A.~Doncheski,  V.~Dragon,   Dr.~D.J.~Dumas,   L.~Gates,
M.~Gintner, Dr.~I.~Ivanovi\'{c}, M.~Jones,  D.~Kaytar, Dr.~H.~K\"{o}nig,
G.~Laberge,  E.P.~Lawrence,    Dr.~I.~Melo,    S.~Nicholson, J.K.~Older,
Dr.~K.A.~Peterson,    B.F.~Phelps,  Dr.~A.~Pouladdej,     Dr.~P.~Rapley,
M.~Richardson, S.~Sail, Dr.~S.~Sanghera, D.~Sheikh-Baheri,  V.~Silalahi,
Dr.~R.~Sinha,  A.~Turcotte,   S.~Towers,   Dr.~P.M.~Wort,  Y.~Xue,   and
G.~Zhang,   for making my stay   in  Ottawa and  the Carleton University
Department of Physics a pleasant one.

A thousands pardons to those whom I may have missed.
\mbox{}\\
\vfill
\begin{center}
\mbox{\epsfysize=10cm
   \epsffile{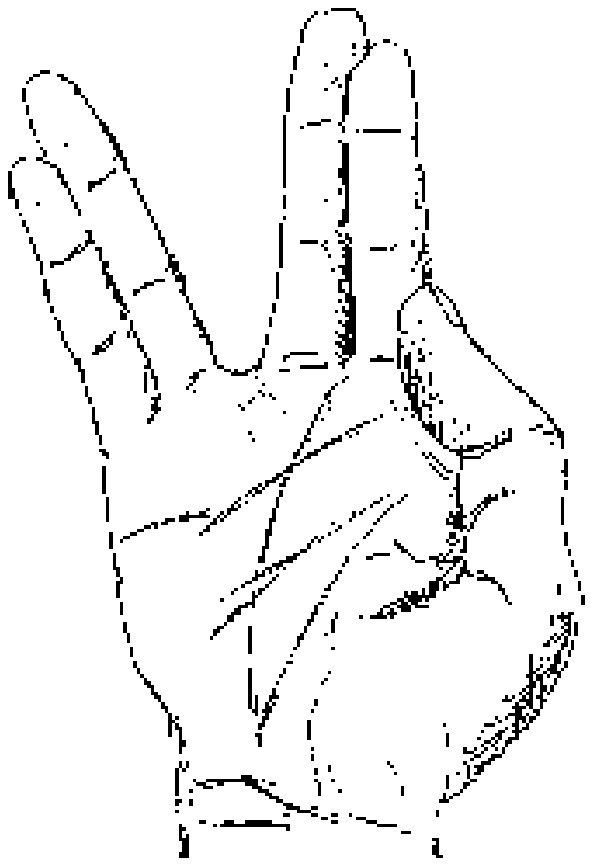}}\\
\mbox{}\\
\vfill
MAY YOU ALL LIVE LONG AND PROSPER!
\end{center}
\vfill


\pagebreak

%
%
\tableofcontents
\pagebreak

%
%
\listoftables
\pagebreak

%
%
\listoffigures
\pagebreak

%
%
\pagenumbering{arabic}
\setcounter{page}{1}
\pagestyle{headings}

%
%
\chapter{Introduction}

  Throughout  the history of the  universe  many 
forces  have   played  a role    in   shaping  it   into  what   it   is
today~\cite{kn:Siemens,kn:Weinberg}.  In this thesis the role of Quantum
Chromodynamics (QCD) in the creation of baryonic matter and the possible
low energy consequences of  the  exotic theory  of superstrings will  be
investigated.
  
   In particular, the viability of a class of  models for nuclear matter
called   string-flip potential models  will  be considered.  String-flip
potential   models attempt  to explain   nuclear  matter  from the  more
fundamental constituent-quark-level picture.  Explaining nuclear  matter
in this way is by no means an easy feat: attempts to do so have met with
varied results.
Unlike nuclear physics, which  is quite successful at explaining nuclear
phenomena from a  nucleon    perspective, constituent quark   models  of
nuclear matter are  far from complete.  The main  stumbling block is the
non-perturbative and many-body nature  of the strong  interaction.  Very
recently some inroads have been made in the area of lattice QCD that may
soon prove to be revolutionary to  this field~\cite{kn:Lapage}.  In fact
the models that will be examined here were inspired by lattice QCD.

  The phenomenology of superstring-inspired   \Esix models will also  be
investigated.  In particular, the possibility of heavy lepton production
at high energy hadron colliders will be studied.  Such a find would help
solve the generational hierarchy problem in  the standard model and lend
support to a theory that unifies all  the known forces of nature, namely
superstrings.

\section{The String-Flip Potential Model}

For the past 30 years several attempts  have been made, with very little
success, to describe nuclear matter  in terms of its constituent quarks.
The main difficulty is due to  the non-perturbative nature  of QCD.  The
most rigorous method for handling  multiquark systems to date is lattice
QCD.  However,  lattice QCD is very  computationally intensive and given
the  magnitude of the problem  it appears unlikely to   be useful in the
near future.\footnote{Some  very recent advancements  have  been made in
the  area of lattice  QCD that have reduced  computation time by several
orders of magnitude.  ``Now   what took hundreds of   Cray Supercomputer
hours can    be done  in  only  a  few  hours  on  a laptop  computer.''
\cite{kn:Lapage}.} As  a  result, more  phenomenological  means must  be
considered.

  The   idea of string-flip potential  models  is  borrowed from certain
results in lattice QCD and  experimental particle physics.  A  potential
derived from computations in lattice QCD is confirmed by fitting mesonic
spectra in  particle physics experiments.   It has  been  found that the
most   consistent   inter-quark potential   model  between    quarks and
antiquarks has the form
\begin{equation}
V(r)\sim\sigma\,r-\frac{4}{3}\,\frac{\alpha_s}{r}\;.
\label{eq:phenpot}
\end{equation}
This formula  is basically   an  interpolation between  the long   range
non-perturbative  ($\sigma    r$)    and   short    range   perturbative
($-\frac{4}{3}\, \frac{\alpha_s}{r}$) parts of  the force between  pairs
of quarks (figure~\ref{fig:qqpair}).   The  string-flip  potential model
ignores the short range part of the potential  and considers an ensemble
of quark-antiquark pairs, $q\bar{q}\,$,  such  that the total amount  of
string, $\sum r_{q\bar{q}}\,$, shared between them is minimized.
\begin{encapfig}{htbp}
   \centering
   \mbox{\epsfxsize=10cm
      \epsffile{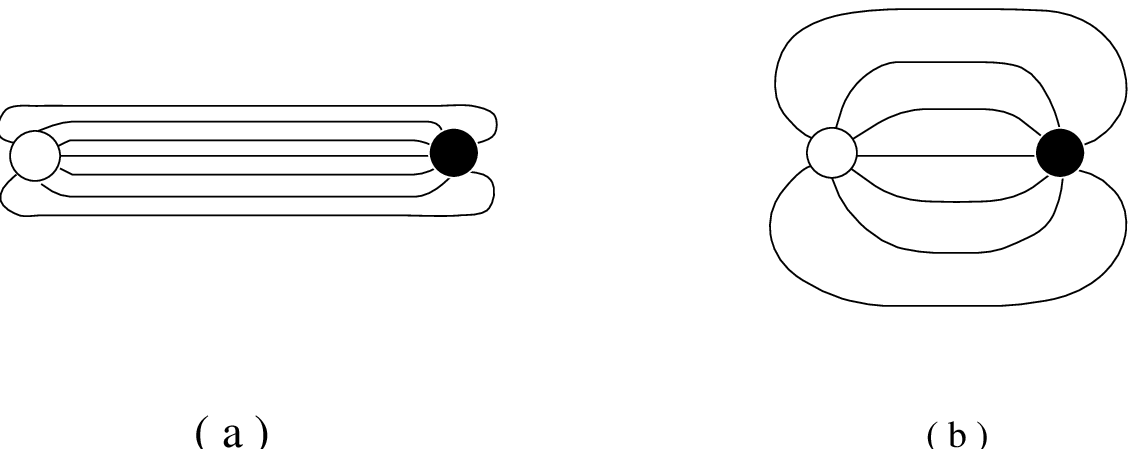}}
\caption[$q\bar q$ flux-tube  diagram]{\footnotesize The colour field
     lines  between  quarks collapse  upon themselves,  due to  the self
     interacting   nature  of the   gluons,  to   form  a flux-tube-like
     structure.  At  long  distances (a) the  fields  lines  collapse to
     become  almost  string-like and at   short distances (b) the fields
     lines expand to become almost QED-like.}  \label{fig:qqpair}
\end{encapfig}

  This  particular model has been  used  in an attempt  to model nuclear
matter.  Although it  has  an obvious shortcoming,  in  that it is  more
applicable  to a pion gas, it  does surprisingly well at predicting some
of     the        overall      bulk     properties      of       nuclear
matter~\cite{kn:HorowitzI,kn:Watson}.

  It is fairly straightforward to generalize this simple model to a more
realistic  one which involves  triplets of  quarks.  Here the flux-tubes
leaving each quark  meets at a central vertex  such that overall length,
$r$, of       ``flux-tubing''    is     minimized         ({\it     cf}.
figure~\ref{fg:figa}.a).   The  potential energy is simply  $\sigma r\,$
\cite{kn:CarlsonB}.

  In  a more general  setting one could consider  a full many-body quark
potential in which large clusters of quarks may be connected by a single
network of flux-tubes  \cite{kn:HorowitzI}; in general,  this ``gas'' is
assumed to consist of colourless objects.  Again the potential energy is
simply $\sigma r\,$, where $r$ is now the  minimal amount of flux-tubing
used for a given cluster of quarks.  In such  a model, there could exist
very complex   topological  configurations of  flux-tubes, such  as long
strands or web-like structures (figure~\ref{fig:gqpot}).
\begin{encapfig}{htbp}
   \centering
   \mbox{\epsfxsize=14cm
      \epsffile{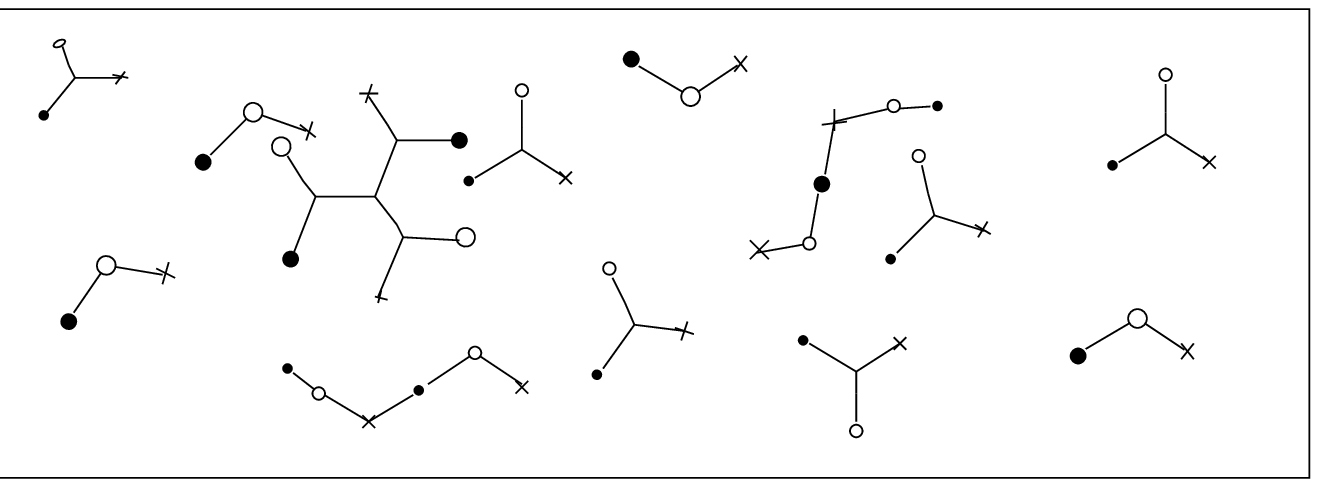}}
   \caption[Generalized quark gas]{\footnotesize Possible flux-tube
      configurations for an $SU_{\rm c}(3)$ quark gas in a box.}
   \label{fig:gqpot}
\end{encapfig}

  All of  these models are  completely motivated by results from lattice
QCD,  where     variations  are  taken about   minimal     lattice field
configurations between quarks.

\subsection{Possible Phases of Nuclear Matter}

  There is a major advantage to understanding nuclear matter in terms of
its constituent quarks.   Not only is  a deeper understanding of nuclear
physics likely to be achieved, but  also a more general understanding of
the nature  of quark matter.  This understanding  could possibly lead to
the prediction of more exotic forms of matter.  To illustrate this point
let us now ``hypothesize'' some of the  possible phases of nuclear/quark
matter (figure~\ref{fig:phases}) in the context of a general string-flip
potential picture.

\begin{encapfig}{htbp}
   \mbox{\epsfxsize=144mm
      \epsffile{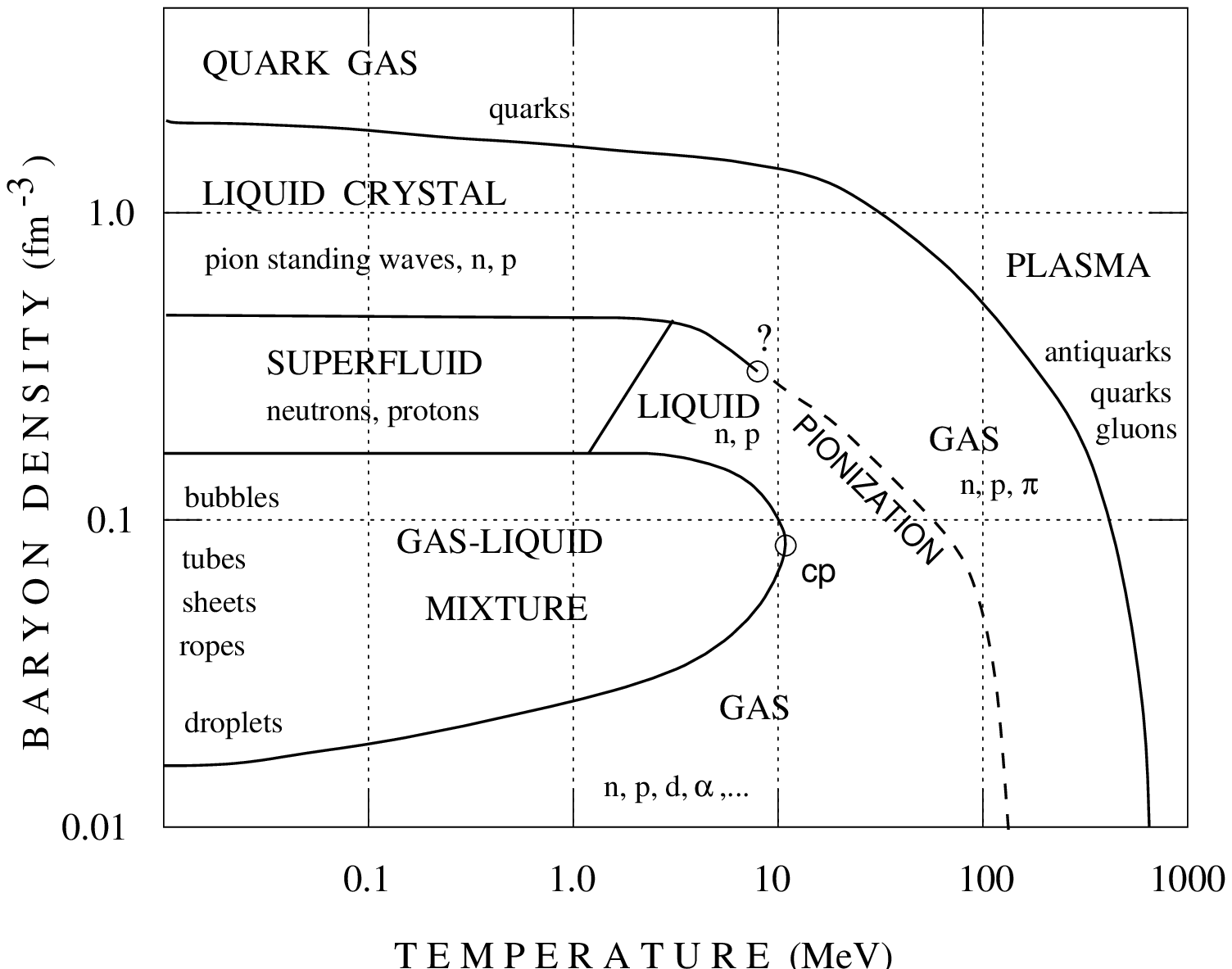}}
   \caption[Phase  diagram of nuclear matter]{\footnotesize   Postulated
      nuclear matter phase  diagram  ({\it   adapted from Siemens  \&
      Jensen \cite{kn:Siemens}.})}
   \label{fig:phases}
\end{encapfig}

   First consider   nuclear matter in   a box at   low temperature ({\it
i.e.}, in  its  ground state)  in  a standard   nuclear model,  with  no
references to quarks.  At  very  low density the   system  is a gas   of
isolated nucleons with a  Fermi degeneracy pressure  on the walls of the
box.   As the box  is gradually squeezed  some of the  nucleons may come
close enough  to start clumping  together.   At this stage the  pressure
becomes negative  due to the clumping forces  which are trying to reduce
the volume.  As the box is squeezed  even further, short-range repulsion
and other  saturation mechanisms \cite{kn:Siemens,kn:Preston}  cause the
pressure to become positive.   Effectively the system behaves  very much
like water vapor; little droplets of  nucleons floating around in a box.
Further squeezing of the box causes liquid (probably a superfluid) to be
formed.

   Now   consider  a very simple   flux-tube   model where the many-body
potential  contains only  three-body  forces  between  red ($r$),  green
($g$), and  blue ($b$)  quarks: each  nucleon  is represented  by  three
valence quarks, $rgb$, connected by a triangular web of flux-tubes ({\it
cf}.  figure~\ref{fg:figa}.a).  In a  flux-tube picture it is  suspected
that the role of meson exchange between  nucleons in the nuclear physics
picture is mimicked by the swapping of flux-tubes as nucleons move close
to  each  other~\cite{kn:WatsonA}.   At sufficiently   low pressure  the
system behaves as  a gas of nucleons because  the clusters of quarks are
essentially isolated from one another: no flux-tubes  are exchanged.  In
addition, as the quarks are  fermions a Fermi pressure is  set up on the
walls of the box but not  within the $rgb$  clusters themselves.  As the
box is squeezed  some of the clusters come  into contact with each other
causing a  clumping effect.  The  saturation effect is perhaps  not that
obvious, however it  has been shown  that this  simple flux-tube picture
does      indeed      lead      to     saturation         of     nuclear
forces~\cite{kn:HorowitzI,kn:Watson,kn:Boyce}.   It  even  produces  the
subtle effect of   nucleon swelling in nuclear matter.\footnote{The  EMC
effect} However it does not appear to produce a strong enough attractive
force to produce binding; speculation  as to ``why?''  will be discussed
later on in this thesis.  ``Further squeezing of the box causes a liquid
(probably a superfluid) to be formed.''  Presumably if spin correlations
were set up between pairs of clusters  of quarks, collective states that
indicate superfluidity may be detected.  This  idea has not been tested.
The  possibility of new  physics is introduced  when the box is squeezed
further.   At  such densities  the  nucleon could perhaps form  a liquid
crystal~\cite{kn:Migdal}, as it  might be more energetically  favourable
for the planes of the $rgb$ clusters to align themselves.  If the box is
squeezed  even further the  flux-tubes   essentially dissolve leaving  a
Fermi gas of quarks.

  If the temperature of the  box is now  increased, even more phases  of
nuclear  matter   become evident.  For   very  low densities  the system
remains a Fermi gas of nucleons, but now  it may become an excited Fermi
gas.  As the box is squeezed the nucleons may clump to form a gas-liquid
mixture provided the temperature is not too high.  If the temperature is
too high it  may   simply stay in  its  gaseous  phase.  As the  box  is
squeezed further each of the aforementioned possible states will go over
into a  liquid.  If  the temperature  and  pressure is just right  it is
possible for all  three phases of nuclear matter  to  coexist: {\it cf}.
figure~\ref{fig:phases}.

   If  the temperature at low density  is pushed higher the gas pionizes
(figure~\ref{fig:phases}): the mesons go on shell.  In order for this to
work in a flux-tube model a meson production mechanism  would have to be
incorporated.  Perhaps the  simplest incorporation would be to introduce
a  string  breaking mechanism,    figure~\ref{fig:break}, then  if   the
flux-tubes became  too long they would break,  producing mesons.  If the
temperature is  further  increased a plasma  of  quarks, antiquarks, and
gluons would be produced.

\begin{encapfig}{htbp}
   \centering                                     \mbox{\epsfxsize=144mm
   \epsffile{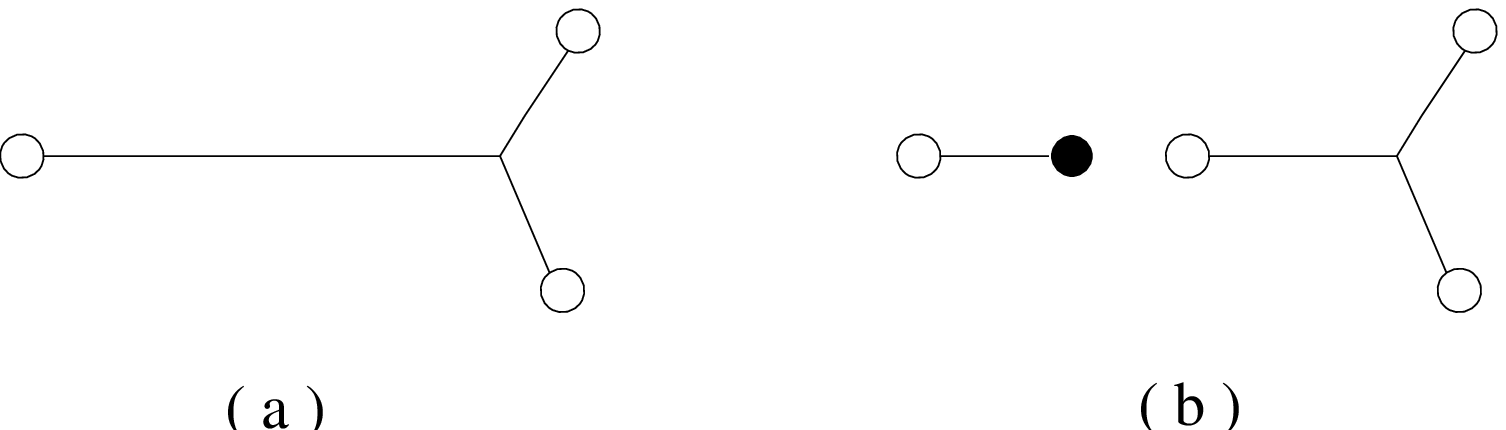}} \caption[A
   na\mbox{\"{\i}ve}   meson      production    scheme]{\footnotesize  A
   na\mbox{\"{\i}ve}   meson  production  scheme.    If  a flux-tube  is
   stretched  long enough such  that  there is  enough mass  energy in a
   segment to produce  a $q\bar{q}$ pair of mass  $2m_q$ then the string
   (flux-tube)  will   break creating  a    pair  of quarks,   with  the
   corresponding segment   vanishing,   with probability  $\sim  e^{-\pi
   m_q^2/\sigma}$ \cite{kn:Barger}.  Note   that  this model   does  not
   include    a  mechanism  for      getting   rid   of  the    mesons.}
   \label{fig:break}
\end{encapfig}

   Finally,  by   varying   the temperature     and  pressure   of   the
aforementioned phases other phases can be reached: if the temperature is
increased in the  liquid crystal phase (see figure~\ref{fig:phases}) the
liquid crystal  will  eventually  become disrupted forming,   perhaps, a
pionized gas  or a plasma. If the  pressure or temperature of the liquid
phase is raised a liquid crystal, pionized  gas, or plasma may be formed
(this suggests perhaps another triple point on the phase diagram).

  A  summary of the  various phases  of  nuclear  matter that  have been
discussed  is illustrated   in  figure~\ref{fig:phases}.  It  should  be
emphasized  that all of  this is  speculative  and model dependent, {\it
i.e.},     the only  experiments  that    exist    are  in the    region
$T\approxle\order{1}MeV$                                             and
$\rho\approxle\order{0.2}fm^{-3}$~\cite{kn:WatsonA}.

\subsection{A Crude Model of Nuclear Matter}

  It should  now be apparent   that attempting to  construct a  model of
nuclear matter in terms of quarks  is not a  very simple task. To reduce
this burden some simplifying assumptions or restrictions will have to be
made.  One restriction  is  to only consider low  temperature phenomena.
Some  simplifying   assumptions  would be  to   require  that the  model
correctly predict  some  of the  very basic bulk  properties  of nuclear
matter:
\begin{itemize}
\item nucleon gas at low densities with no van der Waals forces
\item nucleon binding at higher densities
\item nucleon swelling and saturation of nuclear forces with increasing 
density
\item quark gas at extremely high densities
\end{itemize}
There are many models that attempt  to reproduce all of these properties
but none that reproduces them completely.

   In   this thesis  only one   particular    class of models,    called
string-flip potential    models  \cite{kn:Boyce,kn:HorowitzI,kn:Watson},
will be considered,  in  chapters 2  and 3.  These  models appear  to be
promising because they reproduce most  of the aforementioned  properties
with the exception of nucleon binding.

\section{Superstring Inspired ${\rm E}_6$ Models}

  The \STM standard model (SM)  is a very successful  model. It has thus
far  withstood  rigorous  experimental  testing.  However,  despite  its
success the SM has many problems:
\begin{itemize}
\item no unification of the forces
\item gauge hierarchical and fine tuning problems
\item three generations of quarks and leptons for no particular reason
\item too many parameters to be extracted from experiment
\end{itemize}

  Some of the earlier attempts at  unification tried to unify the strong
and electroweak  forces   by embedding the  \STM  structure  into higher
groups, such as SU(5) and SO(10).   These ``grand unified theories'', or
GUT's, were only  partially successful.  The  simplest of the  GUT's was
SU(5) which seemed promising at the  time because it predicted the ratio
of the \SU{2}{w} and \U{1}{em} couplings
and the proton lifetime \cite{kn:Ross}.  However, the ordinary SU(5) GUT
is no  longer    a  possibility  because more    refined    experimental
measurements  are now in   disagreement  with  its predictions  for  the
couplings
and the proton lifetime    \cite{kn:Hewett}.  In addition, this   simple
model had too many parameters and no explanation for family replication.
The next likely candidate group was SO(10), although the three (or more)
copies of the generational structure still had to be inserted by hand.

   Difficulties with  the SM and GUT  models  concerning gauge hierarchy
and  fine  tuning  problems  led   to   theoretical  remedies  such   as
technicolour and supersymmetry   (SUSY).   The most appealing  of  these
theories    is SUSY,  which  has  generators   that relate  particles of
different  spin   in the same  supermultiplet.   The  locality  of these
generators   leads to  supergravity  models.    SUSY  (and  its extended
versions)  however,   did not  have   enough  room for  all  of   the SM
particles~\cite{kn:Ross}.     To  solve   this   problem direct  product
structures   were made with  SUSY  and  Yang-Mills  gauge groups.  These
structures are now commonly referred  to as ``SUSY'' models.  Of  course
the  price paid for this was  a large particle  spectrum (at least twice
that of the SM) and the problem of family replication still remained.

   In the early 1970's some interest was sparked in  \Esix as a GUT when
it was discovered that all the then  known generations of fermions could
be placed  in  a single   {\bf  27}  dimensional representation.    This
(``topless'') model  was  quite  popular because   the newly  discovered
$\tau$ lepton and $b$  quark could also  be fitted neatly into  the {\bf
27}; there  was no need for a  third generation.  However this model was
quickly disallowed  as it was  experimentally shown that the  $\tau$ and
$b$ belonged to a third generation, and the idea of \Esix as a GUT died.

\begin{encapfig}{htbp}
$${\bf 27}=\left\{\fbox{$\UL  \EL\UR\DR\ER$}\;
  \NER\DpL\DpR\EpL\EpR\NppE\right\}$$
$${\bf 27}=\left\{\fbox{$\CL\MUL\CR\SR\MUR$}\;
  \NMUR\SpL\SpR\MUpL\MUpR\NppMU\right\}$$
$${\bf 27}=\left\{\fbox{$\TL\TAUL\TR\BR\TAUR$}\;
  \NTAUR\BpL\BpR\TAUpL\TAUpR\NppTAU\right\}$$
\caption[\Esix     particle  content]{\footnotesize    \Esix    particle
   content.$^a$ The SM particles are shown in the boxes  on the left and
   their ``exotic'' counterparts   outside   the boxes on  the    right.
   Although the exotics are labeled in a way that suggests they have the
   same quantum numbers as  the non-exotics, in  general they  need not.
   The labeling  for  these particles in  the   literature has not  been
   settled upon and   varies quite  significantly  from  paper to  paper
   \cite{kn:Hewett}.  Here the labeling  scheme was chosen to  reflect a
   specific  \Esix model  that will be  constructed in  this thesis.  In
   particular,   all the  exotics  will  carry  the ``expected'' quantum
   numbers as their  non-exotic   counter parts do, with  the  exception
   being  L=0 ({\it i.e.}, {\bf L}epton  number zero) for the primed and
   double primed ones.}
\vspace{3mm}
   {\footnoterule\footnotesize $\mbox{}^a$Note: Embedded in the  {\bf
   27}'s is the symmetry  group \SU{2}{I} due  to an ambiguity in the
   particle assignments
   {\tiny
   $
    \left\{
      \left(\begin{array}{@{}c@{}}\nu_l\\ l\end{array}\right)_\L
      \DR
    \right\}
    \;\Longleftrightarrow\;
    \left\{
      \left(\begin{array}{@{}c@{}}\nu_l^\prime\\ 
            l^\prime\end{array}\right)_\L
      \DpR
    \right\}
   $}
   and
   $
    \{\nu_{l_{\rm L}}^c\}
    \;\mbox{\tiny $\Longleftrightarrow$}\;
    \{\nu_{l_{\rm L}}^{\prime\prime c}\}\,
   $
   ({\it  cf}. figure~\ref{fig:Wilson}.d). This  ambiguity can easily be
   seen       {\it       via}    the          decomposition         {\bf
   27}=$\sum_\oplus$(SO(10),SU(5)).}
\label{fig:esix}
\end{encapfig}
  In  late 1984  Green  and Schwarz showed   that  10 dimensional string
theory is anomaly free if its gauge group  is either \ExE or SO(32). The
group that had received the most attention was \ExE  as it led to chiral
fermions, similar    to those  in the   SM,    whereas SO(32)   did not.
Furthermore, it  was shown  that compactification  down  to 4 dimensions
(assuming N=1 SUSY) can  lead to \Esix  as  an ``effective''  GUT group.
Each   family   of SM   particles   now  sits   in  its  own  {\bf  27},
figure~\ref{fig:esix}.
The generational problem  may be solved because it  is expected that any
reasonable compactification scheme   should   generate  the  appropriate
number  of  copies  of the {\bf  27}.    For instance  in   a Calabi-Yau
compactification scheme  \cite{kn:Kaku},   $$\ExE\;\longrightarrow\;{\rm
SU(3)}\otimes{\rm  E}_6\otimes{\rm  E}_8^\prime\;,$$  the    number   of
generations is related  to  the topology of  the  compactified space.  A
further  assertion   that the  matter  fields  remain supersymmetrically
degenerate ensures proper management of any  gauge hierarchical and fine
tuning problems.     It  is assumed    that  the  hidden  sector,  ${\rm
E}_8^\prime$,  which  couples    to the  matter    fields of   \Esix  by
gravitational   interactions will provide  a mechanism   for lifting the
degeneracy.

  So  the  inspiration for using  \Esix  is that if   it proves to  be a
possible GUT  then it opens  up the possibility of  finding a  TOE ({\bf
T}heory {\bf O}f  {\bf E}verything). However,  it should  be pointed out
that \Esix is not the only possible stop $en$  $route$ to the SM, but it
is the most  studied~\cite{kn:Hewett}.  It is  for this reason  that the
low energy phenomenology  resulting from \Esix will  be  studied in this
thesis.

\subsection{\Esix Phenomenology}
\subsubsection{An extra \ZE}
  In order to  produce SM phenomenology \Esix must  be broken.   Also to
handle any hierarchical and fine tuning problems, SUSY must be preserved
\cite{kn:Kaku}.  This restriction makes  the  task more difficult, using
most na\mbox{\"{\i}ve} breaking schemes. The solution to the problem was
found   by  using  a  Wilson-loop   mechanism  \cite{kn:Kaku}   over the
non-simply-connected-compactified-string-manifold   to  factor   out the
various subgroups of \Esix.  Figure~\ref{fig:Wilson}  shows some of  the
possible, popular, rank 5 and rank 6 groups that can be produced by this
scheme.
\begin{encapfig}{htbp}
 \begin{tabular}{lc@{$\,$}c@{$\,$}l}
    (a) & \Esix & $\longrightarrow$ & \STM$\otimes$\U{1}{Y_E}   \\
    (b) & \Esix & $\longrightarrow$ & 
            \raisebox{-4.5mm}{$\left.\mbox{\begin{tabular}{@{}l}
    	       	   SO(10)$\otimes$U(1)$_\psi$ \\
    	       	   \downrighthookarrow SU(5)$\otimes$U(1)$_\chi$
    	      	 \end{tabular}}\right\}$
    	       	$\stackrel{\mbox{\scriptsize ER5M}}{\longrightarrow}$
            \begin{tabular}[t]{@{}c}
    	       	 \STM$\otimes$U(1)$_\theta$\\
              {\tiny ($i.e.$, 
              U(1)$_\psi\otimes$U(1)$_\chi\rightarrow$U(1)$_\theta$
              in the large VEV limit.)}
            \end{tabular}
          }                                                      \\
    (c) & \Esix & $\longrightarrow$ &   
            $\SU{3}{c}\otimes\SU{2}{L}\otimes\SU{2}{R}\otimes
            \mbox{\hspace{-14.5mm}}
            \underbrace{\U{1}{L}\otimes\U{1}{R}}_{
            \mbox{\hspace{16mm}}
            \raisebox{3mm}{\rm\scriptsize ER5M}
            \downrighthookarrow\U{1}{V=L+R}
          }$                                                     \\
    (d) & \Esix & $\longrightarrow$ & 
            $\SU{3}{c}\otimes\SU{2}{L}\otimes\U{1}{Y}\otimes
            \mbox{\hspace{-5.5mm}}
            \underbrace{\SU{2}{I}\otimes{\rm U}(1)^\prime}_{
            \mbox{\hspace{8mm}}
            \raisebox{3mm}{\rm\scriptsize ER5M}
            \downrighthookarrow \SU{2}{I}
          }$
 \end{tabular}
 \caption[\Esix  Wilson-loop-breaking   schemes]{\footnotesize     \Esix
   Wilson-loop-breaking  schemes \cite{kn:Hewett}.  (a)  shows a  rank-5
   model and (b) through (d) show rank-6 models. Scheme (a) gives the SM
   plus  an  extra  \U{1}{Y_E}.   Schemes (b)  through   (d) can produce
   effective rank-5 models, ER5M, by taking a large VEV limit.}
 \label{fig:Wilson}
\end{encapfig}
As it  can  be seen, the various   breaking schemes always give  rise to
extra vector bosons beyond  the SM: in fact  it is unavoidable.  In this
thesis only  the  simplest of  these  models (figure~\ref{fig:Wilson}.a)
which generates an extra vector boson, the \ZE, will be considered.

\subsubsection{The Supermatter Fields}

  The  most general superpotential  that  is  invariant under \STM   and
renormalizable  for the fields given  in figure~\ref{fig:esix} is of the
form (neglecting various  isospin contractions and generational indices)
\cite{kn:Hewett},
\begin{equation}
W=W_0 + W_1 + W_2 + W_3
\label{eq:supot}
\end{equation}
$$
\begin{array}{@{}l@{}l@{}l@{}}
W_0 &=& \lambda_1\Super{\Rp}\Super{Q}\Super{\UR} + 
        \lambda_2\Super{\Lp}\Super{Q}\Super{\DR} + 
        \lambda_3\Super{\Lp}\Super{L}\Super{\ER} + 
        \lambda_4\Super{\Rp}\Super{\Lp}\Super{\NppE} +
        \lambda_5\Super{\DpL}\Super{\DpR}\Super{\NppE}      \\
W_1 &=& \lambda_6\Super{\DpL}\Super{\UR}\Super{\ER} + 
        \lambda_7\Super{L}\Super{\DpR}\Super{Q} + 
        \lambda_8\Super{\NER}\Super{\DpL}\Super{\DR}     \\
W_2 &=& \lambda_9 \Super{\DpL}\Super{Q}\Super{Q} + 
        \lambda_{10}\Super{\DpR}\Super{\UR}\Super{\DR}   \\
W_3 &=& \lambda_{11}\Super{\Rp}\Super{L}\Super{\NppE}\,.   
\end{array}
$$
$\Super{A}    = \Phi(A,\tilde{A})$     is  the   superfield,   such that
$A=\Rp,Q,\UR,$\ldots$\,$, and
$$
 \begin{array}{cccc}
   \Super{Q}   = \Doublet{\Super{ u  }}{\Super{ d  }}_\L\,,  &
   \Super{L}   = \Doublet{\Super{\NE }}{\Super{ e  }}_\L\,,  &
   \Super{\Lp} = \Doublet{\Super{\NpE}}{\Super{\Ep }}_\L\,,  &
   \Super{\Rp} = \Doublet{\Super{\Ep }}{\Super{\NpE}}_\L^c\,, 
 \end{array}
$$
for the first generation of the {\bf 27}'s,  and similarly for the other
generations.   The   Yukawa     couplings,  $\lambda_i$'s, also    carry
generational   indices  which have  been  suppressed;  the couplings are
inter-generational as well  as  intra-generational.  The superpotential,
W, summarizes the  entire possible spectrum of  low energy physics which
can occur within the context of an \Esix framework.

  Notice that $W$  was only required to be  invariant under the SM gauge
group.  Further constraints from \Esix  model building may cause some of
the $\lambda_i$ terms to disappear.  Furthermore, not all of these terms
can simultaneously exist   without giving rise to   $\Delta$L$\neq0$ and
$\Delta$B$\neq0$  interactions;  \Esix  models  say  nothing  about  the
assignments of baryon (B) and lepton (L) number until they are connected
to SM representations.  As a result various scenarios may occur,

\vspace{1ex}
\begin{tabular}{@{ $\bullet$ }l@{: }l@{}r@{ $\;$\implies$\;$ }l}
Leptoquarks & B($q^\prime_\L$)=&$\onethree\,$, L($q^\prime_\L$)=1 &
              \Lam{9}=\Lam{10}=0 \\
Diquarks    & B($q^\prime_\L$)=&$-\,\twothree\,$, L($q^\prime_\L$)=0 &
              \Lam{6}=\Lam{7}=\Lam{8}=0 \\
Quarks      & B($q^\prime_\L$)=&$\onethree\,$, L($q^\prime_\L$)=0 &
              \Lam{6}=\Lam{7}=\Lam{8}=\Lam{9}=\Lam{10}=0
\end{tabular}
\vspace{1ex}

\noindent
where   it  has  been   assumed that   L($\nu_{l_\L}^c$)= $-1\,$  (these
scenarios  assume that there  exist only three  copies of  the {\bf 27};
more complicated ones  can be constructed  by adding extra copies).   In
this thesis   the least  exotic  of   these   models, {\it   i.e.},  the
``Quarks,'' will be investigated.  Furthermore, to avoid any fine tuning
problems  with       the       neutrino masses,      $$m_{\NER}\;<<\;m_e
\;\Longleftrightarrow\;
\Lam{11}<<\Lam{3}\,,$$ it will be assumed \Lam{11}=0.

  In this model the masses of the particles are generated by letting the
role of the Higgs fields be played by
$$
\begin{array}{ccc}
\tLp=\Doublet{\tilde\nu^{\prime}_{e_\L}}{\tilde \Ep_\L}\,, &
\tRp=\Doublet{\tilde e^{\prime c}_\L}{\tilde\nu^{\prime c}_{e_\L}}
\,,&
\tilde\nu^{\prime\prime c}_{e_\L}\,,
\end{array}
$$
for each generation.  It  is possible to work  in a basis where only the
third  generation of  Higgses  acquire  a  VEV;  the   remainder  become
``unHiggses''   \cite{kn:Hewett,kn:EllisI}.   In  this basis  the Yukawa
couplings,      $$\lambda_4^{ijk}        \Super{\Rp_i}     \Super{\Lp_j}
\Super{\nu^{\prime\prime}_{l_k}}  \,,$$    where      $i,j,k=1,2,3$  are
generational indices, takes on a much simpler form,
$$\mbox{\footnotesize       {$\lambda_4   \;\in\;     \{\lambda_4^{ijk}|
\lambda_4^{i33}= \lambda_4^{3i3}= \lambda_4^{33i}=0\,,\;\lambda_4^{333}=
\lambda_4^{3jk}= \lambda_4^{j3k}=   \lambda_4^{jk3} \neq0  \;{\rm  s.t.}
\;i=1,2\; \&  \;j,k=1,2,3\}  \,.$}}$$  This  basis  also  eliminates the
potential   problem of flavour  changing   neutral currents at  the tree
level.  It is also assumed that the  $\lambda_i$'s are real and that the
couplings to the unHiggses are  very small.  The former assumption helps
to further  simplify the model  and reduce any effects  it might have in
the CP violating sector
\cite{kn:Hewett}.

\subsection{Heavy Lepton Production} 

   \Esix models are  very rich in their spectrum  of possible low energy
phenomenological  predictions.  If any new particles  are found that fit
within this  framework then  perhaps it  will  lead the  way  to a  more
unified theory of the fundamental forces of nature.   However this is no
small task, for a full theory would have to  be able to actually predict
the mass spectrum of the particles and the relationships between various
couplings, and yet   require very few parameters.   Superstring inspired
\Esix models  are far from being able  to  complete this task.  However,
proof that  \Esix is an effective GUT  would be a  good first step.  But
even this would not necessarily qualify superstrings to be the next step
for  it is not totally inconceivable  that some  other theory might give
rise to \Esix as an effective GUT --- $caveat$ $emptor$.

   A natural question  to   ask would be,  ``Where   to look for   \Esix
phenomenology?''  High  energy hadron colliders,  such as the $\teva$ at
Fermilab (1.8 TeV c.o.m.,  ${\cal L}\sim 10^2 pb^{-1}/yr\,$, $p\bar{p}$)
or   the LHC (14 TeV  c.o.m.,   ${\cal L}\sim 10^5 pb^{-1}/yr\,$, $pp$),
offer possibilities of observing phenomena beyond  the SM by looking for
the production of heavy leptons through a mechanism known as gluon-gluon
fusion, see figure~\ref{fig:ggfuse}.
\begin{encapfig}{htbp}
  \begin{center}
    \mbox{
      \epsfysize=5.0cm
      \epsffile{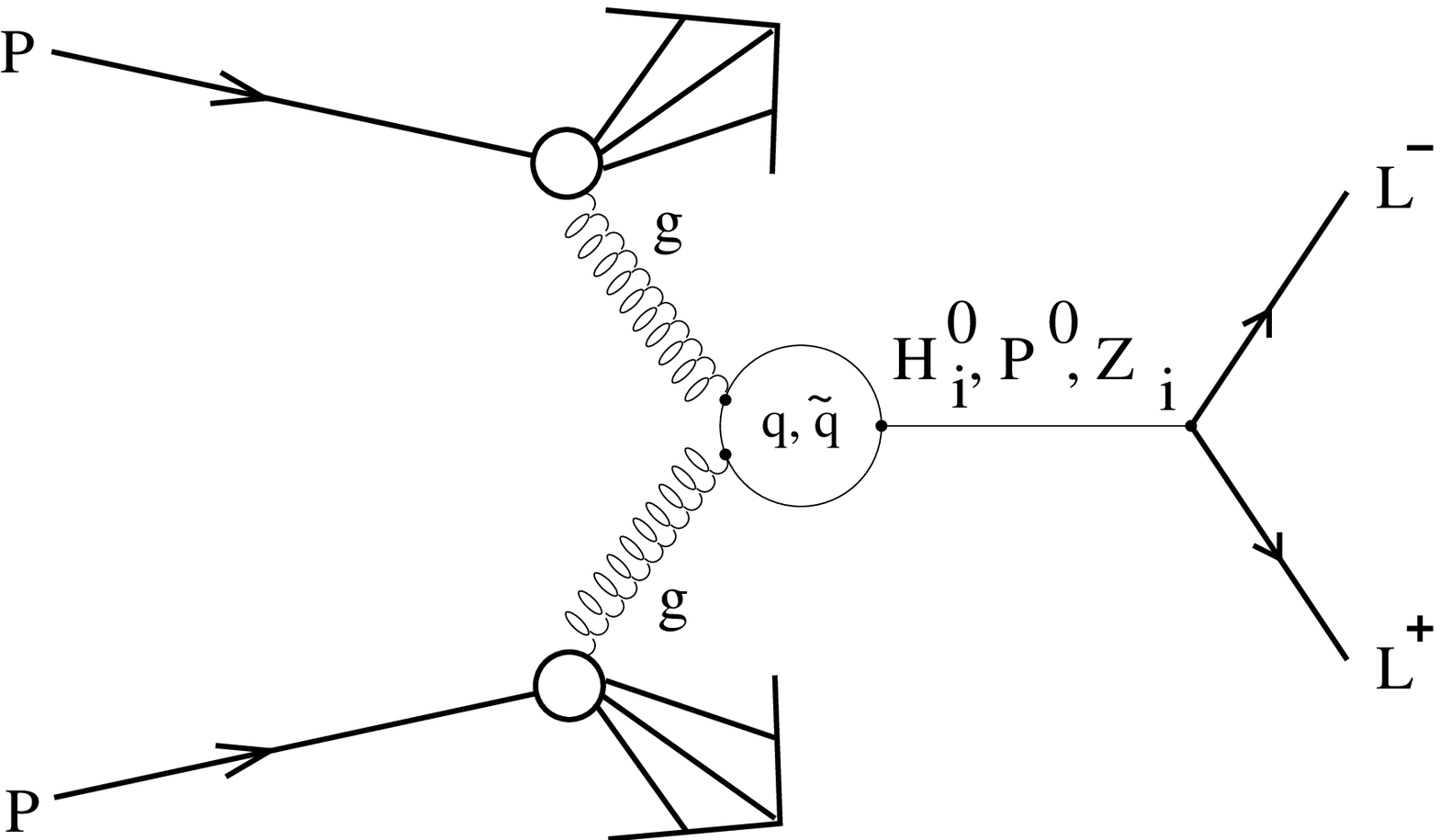}
    }
\caption[Gluon-gluon fusion]{\footnotesize Gluon-gluon fusion to two
    heavy leptons, $gg$ $\longrightarrow$  L$^+$L$^-$. The loop contains
    quarks, $q\,$,   which    couple to    vector   bosons, $Z_{i=1,2}$,
    scalar-Higgses, $H^0_{i=1,2,3}$, and a pseudo-scalar-Higgs, $P^0\,$,
    and       squarks,     $\tilde       q\,$,       which     couple to
    scalar-Higgses. \label{fig:ggfuse}}
\end{center}
\end{encapfig}
This is  an interesting process because  there  are enhancements in  the
cross-sections related  to the heavy (s)fermions  running  around in the
loop.  The computation  was done in  the minimal supersymmetric standard
model (MSSM) by Cieza Montalvo,  $et$ $al.$, \cite{kn:Montalvo} in which
they  predict ${\cal  O}(10^5)\,events/yr$.  Therefore for   \Esix it is
expected that  the production rate should in  principle be  higher since
there are more particles running around in  the loop.  This process will
be investigated in this thesis.

\section{Summary}

   String-flip potential    models and superstring-inspired-\Esix models
have been discussed in a general setting. In this thesis various aspects
of    these models  will be   discussed.    In chapter   2 the \SU{3}{c}
string-flip-potential model   will     be investigated and    put   into
perspective with more generalized models.  In chapter 3 modifications to
the general string-flip-potential will   be investigated.  In chapter  4
heavy  lepton production   {\it   via}   gluon-gluon  fusion  will    be
investigated in an \Esix framework.   Chapter 5 will contain an  overall
summary of the work done in this thesis.

\newcommand{\figa}
    {
     \begin{encapfig}{ht}
     \begin{center}
     \mbox{\epsfxsize=10cm
        \epsffile{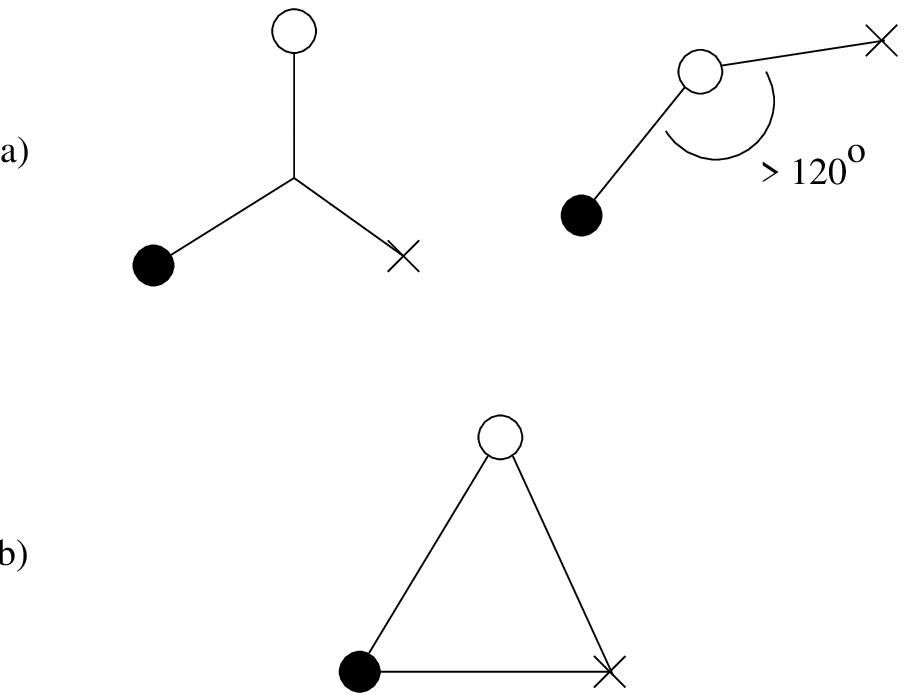}}\\
     \end{center}
     \caption[Flux-tube  arrangements for   the 3q cluster  potentials]{
        \footnotesize   Flux-tube    arrangements  for  the  3q  cluster
        potentials a) ${\mathrm{v}}_\ell$ and b) ${\mathrm{v}}_h$.}
     \label{fg:figa}
     \end{encapfig}
    }
\newcommand{\figb}
    {
     \begin{encapfig}{p}
     \begin{center}
     \mbox{\epsfxsize=10cm
        \epsffile{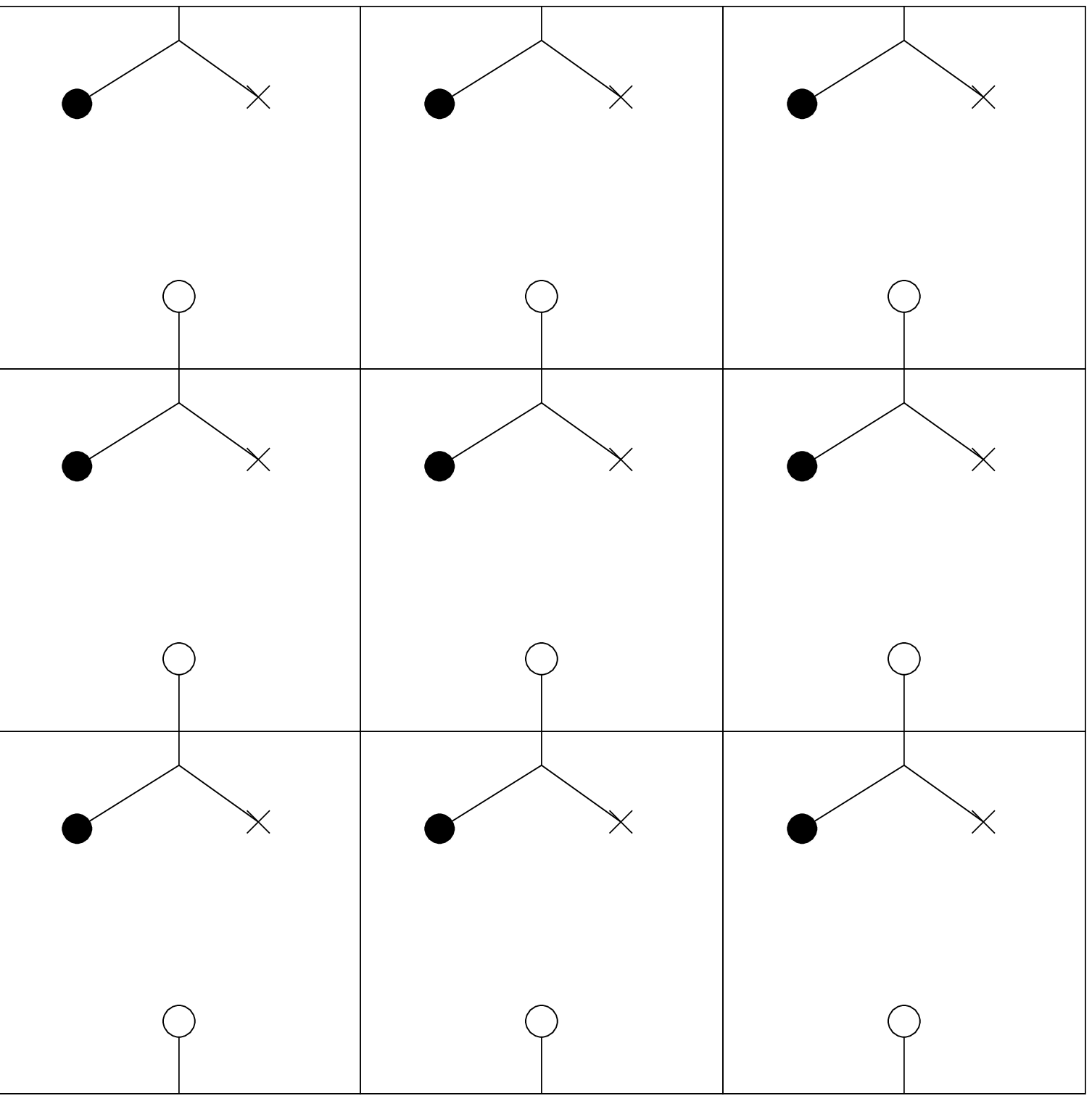}}\\
     \end{center}
     \caption[3q periodic boundary conditions]{\footnotesize A 2-D slice
        showing a typical flux-tube arrangement for three quarks, placed
        inside   a  central cube,      subjected to  periodic   boundary
        conditions.}
     \label{fg:figb}
     \end{encapfig}
    }
\newcommand{\figc}
    {
     \begin{encapfig}{p}
     \begin{center}
     \mbox{\epsfxsize=12.5cm
        \epsffile{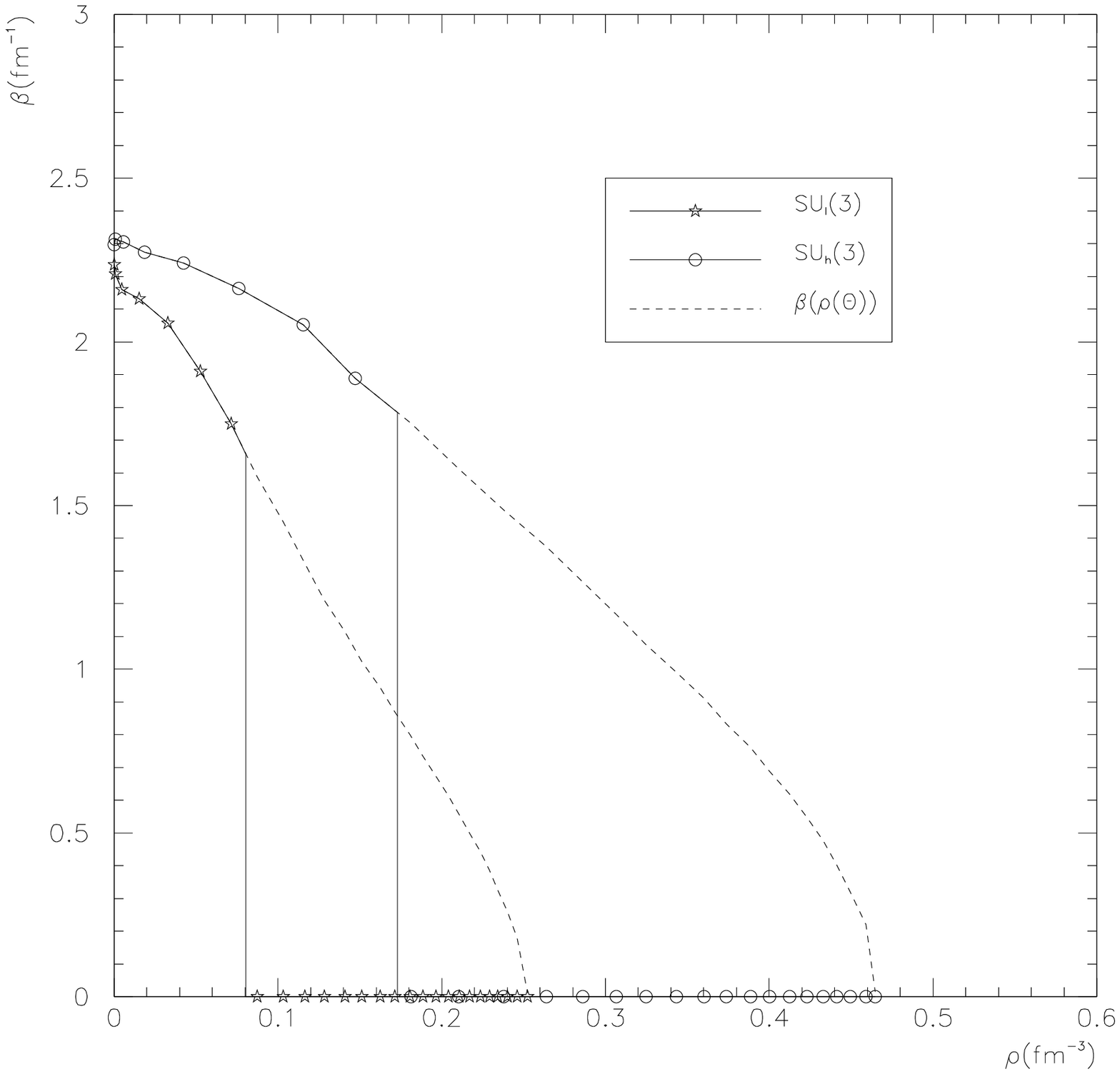}}\\
     \end{center}
     \caption[Graph of   $\beta(\rho)$   for $\mathrm{SU}_\ell(3)$   and
        $\mathrm{SU}_h(3)$]{\footnotesize   Graph  of  $\beta(\rho)$ for
        $\mathrm{SU}_\ell(3)$ and $\mathrm{SU}_h(3)\,$.}
     \label{fg:figc}
     \end{encapfig}
    }
\newcommand{\figd}
    {
     \begin{encapfig}{p}
     \begin{center}
     \mbox{\epsfxsize=12.5cm
        \epsffile{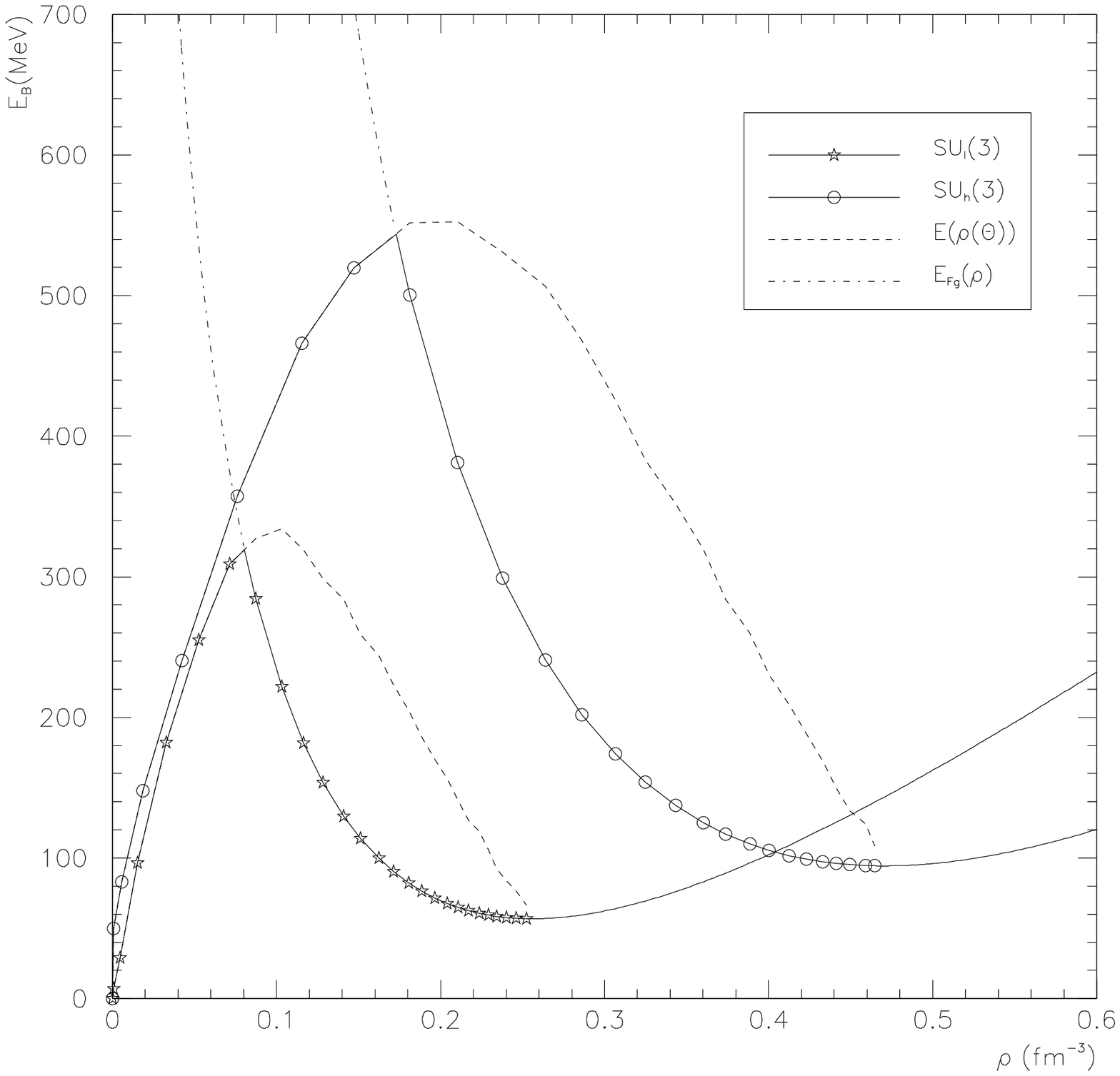}}\\
     \end{center}
     \caption[Graph      of $E_B(\rho)$   for $\mathrm{SU}_\ell(3)$  and
        $\mathrm{SU}_h(3)$]{\footnotesize Graph   of    $E_B(\rho)$  for
        $\mathrm{SU}_\ell(3)$ and $\mathrm{SU}_h(3)\,$.}
     \label{fg:figd}
     \end{encapfig}
    }
\newcommand{\fige}
    {
     \begin{encapfig}{p}
     \begin{center}
     \mbox{\epsfxsize=12.5cm
        \epsffile{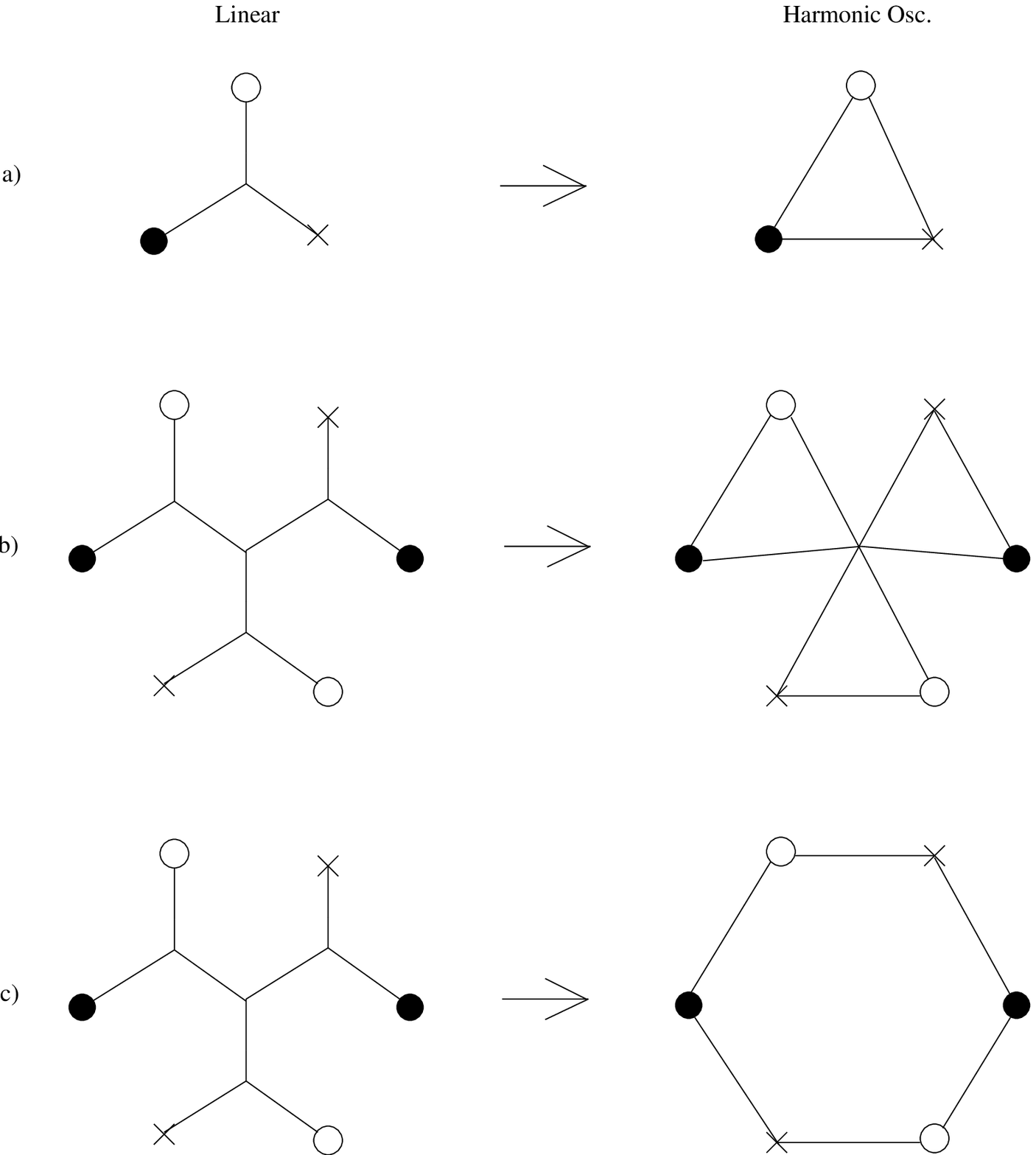}}\\
     \end{center}
     \caption[Different    $\mathrm{SU}_h(3)$    flux-tube  construction
        schemes]{\footnotesize   Different  $\mathrm{SU}_h(3)$ flux-tube
        construction schemes  (RHS)   motivated  by  their corresponding
        linear cousins  (LHS).  Figures  a)  and c)  represented the  HP
        construction. Figures a) and  b) show a construction scheme with
        a more consistent weighting for $s$-states.}
     \label{fg:fige}
     \end{encapfig}
    }
%

\chapter{The SU(3) String-Flip Potential Model}
\label{sec-strflp}
\section{Introduction}

In this  chapter a string-flip potential  model for 3-quark systems will
be    constructed.   Some   simplifying   assumptions  about   flux-tube
minimization will be made in order to reduce the Monte Carlo computation
time.  The results  for a linear potential model, $\mathrm{SU}_\ell(3)$,
and a harmonic oscillator  potential model, $\mathrm{SU}_h(3)$, in which
the  colour has been fixed  to  a given quark,   will be presented.  The
results  will be compared with  an  $\mathrm{SU}_h(3)$ model proposed by
Horowitz  and  Piekarewicz~\cite{kn:HorowitzI}  in which       different
simplifying   assumptions,  about the  minimal flux-tube  topology, were
made.  Also a comparison of their  results \cite{kn:HorowitzI} with some
earlier work done by Watson~\cite{kn:Watson} on $\mathrm{SU}(2)$ will be
made.  The   chapter concludes with   a  discussion on  possible  future
directions to pursue in order to obtain bound state nuclear matter.

\section{The General String-Flip Potential Model}

As   mentioned in the  previous chapter  a  crude quark model of nuclear
matter  would be expected to have  to at least the following properties:
at   low  densities the  quarks  should  condense out   to form isolated
baryons; at a higher density, the interaction between quarks should lead
to positive binding energy between nucleons  and a swelling of nucleons;
and  at still higher  densities, it is expected  that the hadrons should
dissolve into a quark-gluon plasma.  This last assumption is in contrast
to  the traditional  nucleon  models,  which require  the  forces  to be
carefully   adjusted so    that they   saturate  at   infinite  density,
effectively implying a hard core.  Some simple models which appear to be
likely candidates are string-flip potential models
\cite{kn:HorowitzI,kn:Watson}, and  to   some   extent   linked  cluster
expansion models \cite{kn:Nzar}.

The cluster models are based on one-gluon exchange potentials and use an
N-body harmonic oscillator potential, {\it i.e.},
\begin{equation} 
V_{\mathrm{conf}}=\frac{1}{2}k\sum_{i<j}(\vri-\vrj)^2\,,
\end{equation} 
to mimic quark confinement. These  models are mainly used for describing
short range nuclear  effects, as they suffer from  van der Waals  forces
due to the nature of  the confining potential. Despite this shortcoming,
they  appear to be  quite useful  in  explaining local  effects such  as
nucleon  swelling (fat nucleons) \cite{kn:Arifuzzaman}, quark clustering
preferences, and  relative strengths of   the various one-gluon exchange
potentials \cite{kn:Nzar}.

The string-flip potential  models are, on the  other hand,  motivated by
lattice QCD. These models  postulate how flux-tubes should form  amongst
the quarks at zero temperature based on some input from lattice QCD.  An
adiabatic assumption is made, in which the quarks move slowly enough for
their fields to reconfigure  themselves, such that the overall potential
energy is minimized: {\it i.e.},
\begin{equation}
V=\min
   \{
     \sum_{\{q_m\ldots q_n\}}
     \mathrm{v}({\vec{\mathrm{r}}}_m\ldots {\vec{\mathrm{r}}}_n)\,|\,
     \bigcup_{\{m\ldots n\}}^\sim\{q_m\ldots q_n\}=\{q_1\ldots q_{N_q}\}
   \}\;,
\label{eq:cpot}
\end{equation}
where the $N_q$ quarks are placed in a cube of side $L$ and subjected to
periodic boundary conditions to  simulate continuous quark matter.   The
sum is over all gauge invariant sets $\{q_m\ldots q_n\}$ of quarks, such
that at least one element from each set lies inside  a common box, whose
disjoint  union, $\stackrel{\mbox{\footnotesize$\sim$}}{\cup}\,$,  makes
up  the  complete colour singlet  set  $\{q_1\ldots q_{N_q}\}$  of $N_q$
quarks.  It is  easy  to see that  this  potential  allows for  complete
minimal quark clustering separability at low densities without suffering
from   van       der  Waals    forces.      At      present,       these
models~\cite{kn:HorowitzI,kn:Watson} are quite crude in that they do not
include short  range one gluon exchange  phenomena and  spin effects and
are  flavour degenerate.  Despite  their shortcomings, in general, these
models  seem quite  capable of   correctly describing  most of  the bulk
nuclear properties, with the exception of nuclear binding.

It  is known that  the $\mathrm{SU}(2)$  string-flip potential models do
show these  properties,  except for the   positive binding  energy which
probably   arises from short  range forces.   However, the only existing
extension to an $\mathrm{SU}(3)$ model  \cite{kn:HorowitzI} leads to the
rather surprising  result that the nucleon appears  to shrink in nuclear
matter.   It  is therefore of some  interest   to repeat the calculation
of~\cite{kn:HorowitzI},   in  an   attempt  to determine     whether the
approximations made there alter qualitatively the solution.

\section{$\mathrm{SU}(3)$ String-Flip Model}

The string-flip model involves solving  a Hamiltonian system of fermions
governed by the  potential  given in equation~(\ref{eq:cpot}). To  solve
this   system requires the   use  of variational Monte Carlo  techniques
\cite{kn:Ceperley}.  In order to  compute   any observable  in a  finite
amount of time further assumptions about the  form of the potential must
be made.

\clearpage
In this model the potential is restricted 
\begin{itemize}
\item[$a$)] to summing over sets of colour
singlet clusters of three quarks,
\begin{equation}
V=\min\{\sum_{\{q_rq_gq_b\}}\mathrm{v}(\vrr,\vrg,\vrb)|\,
        \bigcup_{\{rgb\}}^\sim\{q_rq_gq_b\}=
        \{q_1\ldots q_{N_q}\}
      \}\;,\label{eq:pot}
\end{equation}
\item[$b$)] such that the colour of  a given quark is   fixed. 
\end{itemize}
Assumption $a$ does have some validity as it has been shown, {\it via} a
linked quark cluster model, that it is energetically more favourable for
$6q$  systems to dissociate into two  nucleons as  a result of hyperfine
interactions \cite{kn:Nzar,kn:MaltmanA,kn:MaltmanB}.  However,  this  is
not necessarily the  case at  lower   densities, as the  linked  cluster
models are unreliable  here.   Assumption  $b$,  that of fixed   colour,
greatly   restricts the number of    possible  field configurations  and
therefore reduces  the chance  of finding  an absolute  minimum.  At low
densities this  should  not have any effect   on the potential,   as the
system  consists of isolated  nucleons.  Similarly at  high densities no
effect  is expected, as the system  consists of uncorrelated quarks.  At
intermediate densities  some  effects might   be expected,  particularly
around any regions in which a phase transition might occur.

For the $\mathrm{SU}_\ell(3)$ model, the potential, V, has the
components \cite{kn:CarlsonB}
\begin{equation}
{\mathrm{v}}_\ell(\vrr,\vrg,\vrb)=\sigma
\left\{
 \begin{array}{lr}
    \rbr+\rrg & \mathrm{if}\;\angle\,brg\,\ge\,120^\circ\\
    \rrg+\rgb & \mathrm{if}\;\angle\,rgb\,\ge\,120^\circ\\
    \rgb+\rbr & \mathrm{if}\;\angle\,gbr\,\ge\,120^\circ\\ \\
    \frac{1}{\sqrt{2}}\sqrt{3\xi^2+\sqrt{3}A} & \mathrm{otherwise}
 \end{array}
\right.\;,
\end{equation}
where $\vrij=\vri-\vrj\,$, $\xi$ or
\begin{equation}
\xi_{\mathrm{rgb}}=\frac{1}{\sqrt{3}}\sqrt{\rrg^2+\rgb^2+\rbr^2}
\label{eq:xi}\,,
\end{equation}
and
\begin{equation}
A=\frac{1}{4}\sqrt{(\rrg+\rgb+\rbr)
                   (-\rrg+\rgb+\rbr)(\rrg-\rgb+\rbr)(\rrg+\rgb-\rbr)}
\end{equation}
is the area inclosed by the triangle $\triangle rgb\,$ (see figure
\ref{fg:figa}.a).  For $\mathrm{SU}_h(3)$ the components are
\begin{equation}
{\mathrm{v}}_h(\vrr,\vrg,\vrb)=\frac{1}{2}k\xi_{\mathrm{rgb}}^2\;
\end{equation}
(see figure \ref{fg:figa}.b).  This potential was  obtained by replacing
the linear segments  of ${\mathrm{v}}_\ell$ by  springs when the quarks,
which were assumed to be of equal  mass, formed a triangle with interior
angles less than $120^\circ\,$ ({\it cf}.  figure \ref{fg:fige}.a). This
analogue model   is    expected    to   have   similar    features    to
$\mathrm{SU}_\ell(3)$ for $s$-wave $(qqq)$ states.
\figa

The  sum in equation (\ref{eq:pot}) can  be re-ordered by restricting it
to run over all  sets, $\{\mathrm{rgb}\}_L$, of quark triplets contained
in a central  box, such that  the potential $\mathrm{v}(\vrr,\vrg,\vrb)$
is minimized with respect to  all possible periodic permutations of  the
vectors  $\{\vrr,\vrg,\vrb\}$, with the constraint that  at least one of
the vectors lies inside the central box: {\it i.e.},
\begin{equation}
V=\min\{\sum_{\{\mathrm{rgb}\}_L}v(\vrr,\vrg,\vrb)\}\,,\label{eq:mpot}
\end{equation}
where
\begin{eqnarray}
v(\vrr,\vrg,\vrb)&=&
\min\{
      \mathrm{v}(\vrr+L\vec{k}_r,\vrg+L\vec{k}_g,\vrb+L\vec{k}_b)|
      \;(\vec{k}_q)_a=-1,0,1\nonumber\\
      & &\;\;\;\;\;\;\mbox{ \& at least one }\vec{k}_q=\vec{0}
    \}\,.\label{eq:metric}
\end{eqnarray}
This means  that for quark triplets a  search of one  box  deep from the
central box is required, giving a total of $27^2$ possible permutations,
in order  to minimize a  given $\mathrm{v}(\vrr,\vrg,\vrb)$  (see figure
\ref{fg:figb}). These permutations can be  reduced to $3$ by requiring
that at least  two sides of the  triangle $\triangle rgb\,$, formed by a
given permutation of quarks, be a minimum: {\it i.e.},
\begin{equation}
v(\vrr,\vrg,\vrb)=\min\{\mathrm{v}(\vdrg,\vdgb,\vrbr),
                             \mathrm{v}(\vrrg,\vdgb,\vdbr),
                             \mathrm{v}(\vdrg,\vrgb,\vdbr)\}\,,
\end{equation}
where  $\vdij$ is the minimum distance  vector between the points $\vri$
and  $\vrj$, in a  box of  side $L$ with  periodic  boundary conditions,
which is given by
\begin{equation}
(\vdij)_a=
      \left\{
       \begin{array}{ll}
       (\vri-\vrj)_a+L&\mathrm{if}\;\;(\vri-\vrj)_a\;\,<\,-L/2\\
       (\vri-\vrj)_a  &\mathrm{if}\;|(\vri-\vrj)_a|\,<\,\;\;L/2\\
       (\vri-\vrj)_a-L&\mathrm{if}\;\;(\vri-\vrj)_a\;\,>\,\;\;L/2
       \end{array}
      \right.\,,
\end{equation}
where $a=x,y,z\,$.    This   is   exact  for  $\mathrm{SU}_h(3)$;    for
$\mathrm{SU}_\ell(3)$, classical Monte Carlo shows  that about $19\%$ of
the events deviate from the exact answer by $\sim\,0.3\%$, on average.
\figb

The    number    of      different     elements    in     the       set,
$\{\mbox{\footnotesize$\sum$}v\}$,  from   which  the  minimum   must be
extracted in order to obtain  V, defined by equation (\ref{eq:mpot}), is
$(N_n!)^2$  (where $N_n=N_q/3$ is the number  of nucleons in the central
box).  For example, if $N_n=7$ this would yield $25\,401\,600$ elements!
These  elements  can  be  reduced by  fragmenting  the set  $\{q_1\ldots
q_{N_q}\}$ into smaller  pieces, or subclusters,  such that each element
can  find  $N_{br}$  complementary coloured  pairs   of quarks that  are
``closest''  to it.  These subclusters  can  be  further fragmented,  by
``softening'' the requirement that at least $N_{br}$ complementary pairs
exist: {\it   i.e.},   by searching   for   disjoint subclusters.  These
subclusters are referred to as  softened subclusters.  The ``closeness''
of quark $q_r$ to  the complementary pair $(q_gq_b)$  is defined by  the
function  $\delta_{r,(gb)}=v(\vrr,\vrg,\vrb)$,     given   in   equation
(\ref{eq:metric}). The fragmented sets are thus constructed by computing
an $N_n\times N_n^2$  matrix ($\Delta$) with elements $\delta_{r,(gb)}$,
and then converting it into block  diagonal form ($\Delta^d$) increasing
in size from top to bottom, by swapping rows  and columns such that each
block diagonal element contains the elements of a fragmented set and all
the   off block diagonal  elements  are set  to  zero.   The elements of
$\{\mbox{\footnotesize$\sum$}v\}$  are now   constructed  by  extracting
permutations   of  elements $\delta_{r,(gb)}$, from   unique columns and
rows,  of    the block  diagonal  elements   of   $\Delta^d\,$.  Further
computational speed  is gained by  discarding sums that start  to exceed
the current minimum.   In general, the  fragmented sets constructed from
$\{q_1\ldots  q_{N_0}\}$, are not   all disjoint from  one another,  and
therefore the  block diagonal elements of $\Delta^d$   may overlap.  The
degree  of overlap increases  with increasing density, causing the Monte
Carlo to slow down.

This  fragmentation procedure, or   nearest  neighbour search  of  depth
$N_{br}$ ({\it  cf}.  \cite{kn:Watson}), reduces computation  time quite
significantly. The cost is that rare configurations with flux-tubes that
stretch across  the box or across   unsoftened subclusters, that  give a
global minimum,  might be missed.   Preliminary Monte Carlo results show
that the inclusion of softened  subclusters yields no noticeable change.
However, for  a full $(N_n!)^2$ brute  force search, performing  a Monte
Carlo becomes  virtually impossible.  A few  brute force computations of
the potential were made, for particles randomly thrown into a box, which
seem to   suggest  that the fragmentation  procedure   is good to  about
$1\%\,$, with $N_{br}\approx 4$.  $\mathrm{SU}_\ell(2)$ models also give
similar results \cite{kn:Watson}.

The   validity of the fragmentation  procedure   can also  be argued  on
physical  grounds, for  it is  reasonable to assume  that long flux-tube
configurations would tend to dissociate into $q\bar q$ pairs. Therefore,
the fragmentation procedure can be considered as a valid low temperature
approximation.

The  choice of variational wave function  should attempt  to reflect the
overall bulk properties of the system. Here the wave function was chosen
to be of the form,
\begin{equation}
\Psi(\alpha,\beta,\rho)=
\mathrm{e}^{-
            {\displaystyle \sum_{\{\mathrm{rgb}\}}
            (\beta\xi_{\mathrm{rgb}})^\alpha}
           }
\prod_{c\,\varepsilon\,\{\mathrm{rgb}\}}|\Phi_{\mathrm{S}_c}(\rho)|\;,
\label{eq:waves}
\end{equation}
where $\alpha$,  $\beta$, and $\rho$  ($=N_n/L^3$ s.t.  $N_n=N_q/3$) are
variational parameters, $\sum_{\{\mathrm{rgb}\}}$   is over the set   of
quarks   $\{\mathrm{rgb}\}$ which  gives the   minimal  potential V, and
$|\Phi_{\mathrm{S}_c}(\rho)|$  is  the   Slater determinant  which  is a
function   of   density,  $\rho$.\footnote{\S~\ref{sec-gslb}   gives  an
algorithm for generating  an arbitrary dimensional Slater  determinant.}
The    $\Phi_{\mathrm{S}_c}(\rho)$        contain       the     elements
$\phi_{ij}=\phi_i(\vrj)$, which are composed of the plane wave states
\begin{equation}
\phi_i(\vrj)=
\sin(
     \frac{2\pi}{\mathrm{L}}\,\vn_i\cdot{\vec{\mathrm{r}}}_j+\delta_i
    )\,,\label{eq:slaterwave}
\end{equation}
where  $\delta_i=0$ or $\pi/2\,$,  and  $(\vn_i)_a=0,\pm 1,\pm 2,\ldots$
are  the components of the  Fermi  energy level packing vector, $\vn_i$,
for particles in a cube of side $L$ (with  ordinates ranging from $-L/2$
to $L/2$)  subjected to periodic boundary  conditions. The exponent part
to the left Slater  part of equation~\ref{eq:waves}  yields a product of
three-body   harmonic   oscillator   wave  functions  when   $\alpha=2$.
Therefore,   the parameter  $\beta$ is  related   to  the inverse r.m.s.
radius of the nucleons in the  system ({\it cf}. \cite{kn:Watson}).  Our
particular    choice   of  variational   wave    function,   {\it i.e.},
equation~\ref{eq:waves}, mimics    the overall gross  features  of quark
matter,   yielding  highly correlated behaviour  at    low densities and
uncorrelated behaviour at high densities.

The total energy for this many body system is,
\begin{equation}
E(\alpha,\beta,\rho)=T_{\mathrm{-s}}+V\,,
\end{equation}
where 
\begin{equation}
T_{\mathrm{-s}}=\frac{-\hbar^2}{4m_q}(\nabla^2\ln\Psi)
\end{equation} 
is the  kinetic energy, obtained by  eliminating  the surface terms from
the integral   $\int\Psi^*\nabla^2\Psi\,$.      Thus,  the    many  body
Hamiltonian system can  be solved  by  varying the parameters  $\alpha$,
$\beta$, and  $\rho$,  and evaluating  the expectation   values by Monte
Carlo integration at each step, until a minimum $E$ is found.
\section{Monte Carlo Calculation}
\label{sec-metrop}
The   Monte    Carlo    procedure   uses     the  Metropolis   algorithm
\cite{kn:Ceperley,kn:Metropolis,kn:Feller,kn:Kalos}   to    generate   a
distribution   in $|\Psi|^2$.   The  Monte   Carlo   procedure  involves
computing the average of an observable ${\cal O}$ such that
\begin{equation}
\bar{\cal O}=\frac{1}{\langle\Psi|\Psi\rangle}
             \int{\cal O}(\vx)|\Psi(\vx)|^2\mathrm{d}\vx
          \approx\frac{1}{N}\sum_{n=N_0}^{N+N_0}{\cal O}(\vx_n)\;,
          \label{eq:monte}
\end{equation}
where the  summation   is  taken over   $N$ sequential   samples  of the
distribution    $|\Psi(\alpha,\linebreak[0]  \beta,\linebreak[0]   \rho;
\linebreak[0]\vx)|^2$ (s.t.  $\alpha$,  $\beta$, and $\rho$  are fixed),
after $N_0$  iterations have been made.   The distribution in $|\Psi|^2$
is generated  by the following  algorithm: from a given configuration of
particles $\vx$, change all their  positions randomly to a new  position
$\vx+\mathrm{d}\vx\,$. Next compute the transition probability function
\begin{equation}
\tau=\min\{(|\Psi(\vx+\mathrm{d}\vx)|/|\Psi(\vx)|)^2,1\}\;,
\end{equation}
and   compare it with    a random number  $r\,\varepsilon\,[0,1)\,$.  If
$\tau>r$  then     accept    the   move by     replacing     $\vx$  with
$\vx+\mathrm{d}\vx$,  otherwise   reject  the move  by   keeping the old
$\vx\,$.   Finally, repeat  the procedure  until  a desired error level,
$\delta\bar{\cal O}\,$, has been reached.   The initial configuration of
particles, $\vx=\vx_0$,  is generated by throwing   them randomly into a
box with side  of length $L$.   All subsequent moves are constrained  to
the box such that, if a particle randomly  moves outside of the box, its
periodic   image enters from the   opposite side.  This algorithm, which
satisfies detailed balance,  is   called the Metropolis   algorithm  and
converges  to the distribution $|\Psi|^2$  after  $N_0$ moves have  been
made.   The  value of $N_0$   is determined by  the   point at which the
statistical fluctuations in $\sum_n{\cal O}_n$ have become substantially
reduced.  A general  rule of thumb is that  convergence is  more rapidly
achieved         if        the   step          size,       $\delta_s  x$
({\footnotesize$=|\mathrm{d}\vx|/\sqrt{3N_q}$}), is chosen such that, on
average, $\tau\approx 1/2\,$.    A natural  length  scale  to  use, when
considering an appropriate step size, is ({\it cf}.~\cite{kn:Watson})
\begin{equation}
\delta_s x\sim\frac{rf}{\beta+\rho^{1/3}}\,,
\end{equation}
where the constant $f\approx1/4$. This  is determined by taking  several
small  samples  from  the probability   distribution  $|\Psi|^2$  and by
restarting the Monte  Carlo for different $f$  values, until the desired
value of $\bar\tau$ is reached. To ensure convergence in a finite amount
of cpu time, particularly at low densities, $rgb$ clusters of quarks (of
radius order $\delta_s x/2$) are thrown into the box randomly.

The   Monte   Carlo        evaluation   of     the    total      energy,
$\bar{E}=\overline{T_{\mathrm{-s}}+V}$ in its  current form, can produce
a significant amount of error \cite{kn:Ceperley}. This can be reduced by
introducing a mean square ``pseudoforce'',
\begin{equation}
F^2=\frac{\hbar^2}{4m_qN_n}(\grad\ln\Psi)^2\,,
\end{equation}
and re-expressing the kinetic energy as
\begin{equation}
T=2T_{\mathrm{-s}}-F^2\,.
\end{equation}
In this form, the variance of the total energy,
\begin{equation}
\bar{E}=\overline{2T_{\mathrm{-s}}-F^2+V}\,,
\end{equation}
goes to zero as the wave function approaches an eigenstate of the
Hamiltonian.

The variational wave function  $\Psi$  is made  up   of a product  of  a
correlation piece,  $\chi$, and a Slater   piece, $\Phi$. Therefore, the
kinetic  energy expression can be  split  up  into three separate  terms
involving pure and mixed, correlation and Fermi energies: {\it i.e.},
\begin{equation}
\bar{T}=\bar{T}_{\mathrm{C}}+\bar{T}_{\mathrm{F}}+\bar{T}_{\mathrm{CF}}
\end{equation}
The explicit forms for these terms are: the correlation energy
\begin{equation}
\bar{T}_{\mathrm{C}}=\frac{\alpha\beta^\alpha}{2m_qN_n}\langle
\sum_{\{\mathrm{rgb}\}}\xi_{\mathrm{rgb}}^{\alpha-2}
[\alpha(1-(\beta\xi_{\mathrm{rgb}})^\alpha)+4]\rangle\;,
\label{eq:tcorr}
\end{equation}
the Fermi energy
\begin{equation}
\bar{T}_{\mathrm{F}}=\frac{2\pi^2}{m_qN_nL^2}\sum_{q=1}^{N_q}\vn^2_q\,,
\end{equation}
and the mixed correlation-Fermi energy
\begin{eqnarray}
\bar{T}_{\mathrm{CF}}&=&\frac{\pi\alpha\beta^\alpha}{m_qN_nL}\langle
\sum_{i=1}^{N_n}      
   \sum_{\{\mathrm{rgb}\}}
     \xi_{\mathrm{rgb}}^{\alpha-2}
      [
        \bar{\phi}_{i\mathrm{r}}(\vrrg-\vrbr)\phi_{i\mathrm{r}}^\prime
       +\bar{\phi}_{i\mathrm{g}}(\vrgb-\vrrg)\phi_{i\mathrm{g}}^\prime
        \nonumber\\ & &
       +\bar{\phi}_{i\mathrm{b}}(\vrbr-\vrgb)\phi_{i\mathrm{b}}^\prime
    ]\cdot\vn_i\rangle\;,\label{eq:mixed}
\end{eqnarray}
where              $\bar{\phi}_{ij}=(\phi^T)^{-1}_{ij}\,$,           and
$\phi_{ij}^\prime=\phi_i(\vrj+\frac{L}{4\vn_i^2}\vn_i)\,$.   A  detailed
derivation of  the     above    expressions    can  be  found         in
appendix~\ref{sec-appa}.

  The value of $\alpha$ is fixed for  free nucleons at $\rho=0$. For the
$\mathrm{SU}_h(3)$ model $\alpha=2$,   as the wave  function $\Psi$ must
become that of a free 3-body harmonic  oscillator.  Therefore, the total
energy for this system is simply
\begin{equation}
E_{\mathrm{free}}^{(h)}(\beta)=
\frac{3\hbar^2}{m_q}\beta^2+\frac{3k}{4\beta^2}\,.
\end{equation}
Minimizing this gives,
\begin{equation}
E_0^{(h)}=3\hbar\sqrt{\frac{k}{m_q}}\,,\label{eq:spc}
\end{equation}
where $E_0^{(h)}=E_{\mathrm{free}}^{(h)}(\beta_0^{(h)})$, and
\begin{equation}
\beta_0^{(h)}=\left(\frac{m_qk}{4\hbar^2}\right)^{1/4}\,.
\end{equation}
$E_0^{(h)}$  and $\beta_0^{(h)}$ can be  used to check  the Monte Carlo.
However,   for   the $\mathrm{SU}_\ell(3)$ model   such   a check is not
possible, as it  is impossible to find  $V$ analytically  at $\rho=0\,$.
However, by fitting the results to the expression
\begin{equation}
E_{\mathrm{free}}^{(\ell)}(\beta)
=g_{\mbox{\tiny $T$}}\,\frac{\hbar^2}{m_q}\beta^2
+g_{\mbox{\tiny $V$}}\,\frac{\sigma}{\beta}\,,\label{eq:anyla}
\end{equation}
we      find $g_{\mbox{\tiny    $T$}}\approx1.07$    and $g_{\mbox{\tiny
$V$}}\approx3.09\,$.      Also   the     virial   relation      $\langle
T_\ell\rangle=\langle  V_\ell\rangle/2\,$ can be  verified, which should
hold  at  all  densities.   A similar    check can  also  be   done  for
$\mathrm{SU}_h(3)$, with the virial relation $\langle T_h\rangle=\langle
V_h\rangle\,$.   For $\mathrm{SU}_\ell(3)$ the  parameters  $\alpha$ and
$\beta_0^{(\ell)}$     can  be       obtained         by       computing
$E_{\mathrm{free}}^{(\ell)}$ for  different  values of  $(\alpha,\beta)$
until a minimum is found.

A  further reduction   of  the variational   parameters  is  obtained by
introducing the scaling transformation \cite{kn:Watson},
\begin{equation}
(\beta,\rho^{1/3})\;\longrightarrow\;\zeta(\cos\theta,\sin\theta)\,,
\label{eq:scaletran}
\end{equation}
where  $\zeta\,>\,0$,   and  $\theta$ is   restricted    to the interval
$(0,\pi/2)$. This allows    the  total energy   to be    expressed as  a
polynomial in $\zeta$, which  can subsequently be minimized to eliminate
$\zeta$: {\it i.e.}, $\bar{E}(\rho,\beta)$ becomes
\begin{equation}
\bar{E}(\zeta,\theta)=
\tilde{\bar{T}\,}(\theta)\zeta^2+\tilde{\bar{V}}(\theta)/\zeta^\kappa\,,
\end{equation}
such that $\bar{T}(\zeta,\theta)=\tilde{\bar{T}\,}(\theta)\zeta^2\,$,
$\bar{V}(\zeta,\theta)=\tilde{\bar{V}}(\theta)/\zeta^\kappa$ and
\begin{equation}
\kappa=\left\{
         \begin{array}{ll}
          1  & \mathrm{if}\;\mathrm{SU}_\ell(3)\\
          2  & \mathrm{if}\;\mathrm{SU}_h(3)
         \end{array}
       \right.\,,
\end{equation}
which can be minimized with respect to $\zeta$ to give
\begin{equation}
\bar{E}(\theta)=(\kappa+2)
\left(
  \frac{\tilde{\bar{V}}^{\,2}(\theta)
        \tilde{\bar{T}\,}^\kappa(\theta)}{4\kappa^\kappa}
\right)^\frac{1}{\kappa+2}\,,
\end{equation}
with
\begin{equation}
\zeta(\theta)=
\left(
      \frac{\kappa\tilde{\bar{V}}(\theta)}{2\tilde{\bar{T}\,}(\theta)}
\right)^\frac{1}{\kappa+2}\,.
\end{equation}
Notice that  the elimination of the  parameter $\zeta$  is equivalent to
imposing the virial theorem, which implies
\begin{equation}
\langle T\rangle=\frac{\kappa}{2}\langle V\rangle\,.\label{eq:virial}
\end{equation}

Therefore,  the Monte  Carlo  only has to  be run  for different $\theta$
values extracted from the  ``open'' interval $(0,\pi/2)$. The end points
are obtained  by taking a limit. The  $\theta=0$ limit is  equivalent to
taking $\rho=0\,$, which has already  been discussed. The $\theta=\pi/2$
limit is   equivalent  to taking $\beta=0\,$,   which  corresponds to an
uncorrelated Fermi gas, with energy
\begin{equation}
E_{\mbox{\tiny Fg}}(\rho)=\left(\frac{3^5\pi^4}{2}\right)^{1/3}
\frac{3\hbar^2}{5m_q}\rho^{2/3}+V_{\mbox{\tiny Fg}}(\rho)\,,
\end{equation}
where
\begin{equation}
V_{\mbox{\tiny Fg}}(\rho)=
\left\{
 \begin{array}{ll}
  {\displaystyle c_\kappa\,\frac{\sigma}{\rho^{1/3}}}
   &\mathrm{for}\;\mathrm{SU}_\ell(3)\\ \\
  {\displaystyle c_\kappa\,\frac{k}{2\rho^{2/3}}}    
   &\mathrm{for}\;\mathrm{SU}_h(3)
 \end{array}
\right.\,,
\end{equation} 
and   $c_\kappa$ is obtained   by  a fit    to the  Monte  Carlo  in the
$\theta=\pi/2$ limit. Thus the $\beta=0$ limit is described by the curve
$E_{\mbox{\tiny Fg}}(\rho)\,$. This   curve is compared  with  the Monte
Carlo results  for $\bar{E}(\rho(\theta))$ from  which  a minimum energy
curve $\bar{E}(\rho)$ is obtained.

\figc
\figd
Figures \ref{fg:figc} and \ref{fg:figd} show the variational Monte Carlo
results    for    $\beta(\rho)$     and    the       binding     energy,
$E_B(\rho)\equiv\bar{E}(\rho)-E_0\,$, respectively.  The dashed lines on
these   graphs show   the   remnants  of  the   minimal   $\rho(\theta)$
trajectories,  for  $\beta$ and $E_B\,$,  after  a  phase transition, at
$\rho=\rho_c$, from   a correlated system  of quarks  to an uncorrelated
Fermi gas.  The slight roughness of these lines  is because the data was
not fitted.  In plotting  these graphs it  was assumed  that: $N_n=7\,$,
$m_q=330MeV\,$, $\sigma=910MeV/fm\,$, and $k\approx3244MeV/fm^2\,$.  The
value of   $k$  was determined   by  setting  $E^{(h)}_0\,$  in equation
(\ref{eq:spc}) to equal  $E^{(\ell)}_0\,$,  in the limit  $N_n=1\,$  and
$\theta=0.0001\,$. The  other parameters  that  were determined   by the
Monte Carlo are given in table (\ref{tb:taba}).
\begin{table}
\begin{center}
\caption[Parameters determined by Monte Carlo, with $N_n=7\,$]{
        \footnotesize Parameters determined    by   Monte  Carlo,   with
        $N_n=7\,$.}
\label{tb:taba}
\begin{tabular}{|lr|c|c|} 
 \hline Parameters&&$\mathrm{SU}_\ell(3)$&$\mathrm{SU}_h(3)$ \\ \hline
$\alpha$                     &             & $1.75$    &$2.00$\\
$c_\kappa(\rho_{\mbox{\tiny Fg}},\beta_{\mbox{\tiny Fg}})$
                             &             &$0.908$    &$0.347$ \\
$\rho_{\mbox{\tiny Fg}}$     & $(fm^{-3})$ & $0.2524$  & $0.465$ \\
$\beta_{\mbox{\tiny Fg}}$    & $(MeV)$     & $0.01264$ & $0.01549$\\
$E_0(\rho_0,\beta_0)$&$(MeV)$& $1895$      & $1856$                \\
$\rho_0$                     & $(fm^{-3})$ & $1.119\times10^{-11}$ 
  & $1.211\times10^{-11}$\\
$\beta_0$                    & $(fm^{-1})$ & $2.237$   & $2.297$     \\
\hline
\end{tabular}
\end{center}
\end{table}
\section{Discussion}
The parameter $\beta$  is related to the  confinement scale for triplets
of   quarks \cite{kn:HorowitzI,kn:Watson}.  Figure \ref{fg:figc}   shows
that the quarks become less confined as $\rho$ increases, and completely
deconfined beyond the phase transition point, $\rho_c$.  Thus, as $\rho$
increase from $0$ to $\rho_c$  the nucleon swells producing an  EMC-like
effect.  Figure 11.b, of  reference \cite{kn:HorowitzI}, shows a plot of
$\lambda$   ($\sim\beta^2$)   $vs.$ $\rho$,  obtained   by Horowitz  and
Piekarewicz, which in general   indicates  that the quarks   become more
confined as  $\rho$  approaches   $\rho_c$, and  completely   deconfined
beyond.  Therefore,   their   model    does   not  explain   the     EMC
effect~\cite{kn:Close,kn:NachtmannA,kn:NachtmannB}  and  is inconsistent
with what we have found here.

The Horowitz  and Piekarewicz (HP) model  approximates the  higher order
flux-tube  topologies  of  equation (\ref{eq:cpot}) with  long  harmonic
oscillator chains that close upon themselves: {\it i.e.},
\begin{equation}
V= \min\{\sum_{\{\mathrm{rg}\}}v_{\mathrm{rg}}\}
  +\min\{\sum_{\{\mathrm{gb}\}}v_{\mathrm{gb}}\}
  +\min\{\sum_{\{\mathrm{br}\}}v_{\mathrm{br}}\}\,,
\end{equation}                              
where    $v_{ij}=\frac{1}{2}kr_{ij}^2\,$,     with the    wave  function
$\Psi=\mathrm{e}^{-\lambda  V}\Phi$ ({\it cf}. equation (\ref{eq:waves})
with $\alpha=2$).   They   have shown  for $\rho<\rho_c$   that  3-quark
clusters (chains)  make up more than $90\%$   of nuclear matter,  with a
large remainder of these being 6-quark clusters.  A closer look at their
$\lambda$ $vs.$   $\rho$  plot shows  a small   dip in  $\lambda$ around
$\rho=0.2\,$.  This indicates a slight swelling  of the nucleon for very
small  $\rho\,$.  In  fact,   in this  density  regime  3-quark clusters
completely  dominate   ($>99\%$).  Therefore,  the evidence   from these
graphs seems  to suggest that  too much weight is  being given to higher
order flux-tube topologies at intermediate densities.

Lattice  QCD shows  that quarks  like to   cluster together {\it  via} a
linear potential. However,  most  phenomenological models that  describe
isolated hadronic matter using a harmonic oscillator potential work just
as well.  As can been  seen from figures \ref{fg:figc} and \ref{fg:figd}
the harmonic oscillator model gives the same overall shape as the linear
one. This model was motivated by replacing each linear segment of string
in a 3-quark state by a spring, with spring  constant $k$. For quarks of
equal mass this reduces to a triangle of springs (see figure
\ref{fg:fige}.a). Similarly a 6-quark   state would give an object  that
simplifies to three triangles with one of  the tips from each meeting at
a common vertex (see figure \ref{fg:fige}.b).  The corresponding 6-quark
state for the HP model  forms a closed ring  which, in general, requires
less energy to form  (see  figure \ref{fg:fige}.c).  Thus QCD  motivated
models would also seem to support the  aforementioned claim, that HP are
giving too much weight to higher order flux-tube topologies.
\fige

The  $\mathrm{SU}_h(2)$  model, by Watson  \cite{kn:Watson}, agrees with
the HP model \cite{kn:HorowitzI}.  These  graphs appear to be similar to
those shown in figures  \ref{fg:figc}  and \ref{fg:figd}. Of  course the
fact that these models  agree is only a  check  of consistency, as  they
both  have the same potential, which  only includes interactions between
$q\bar        q$       pairs.     Also          the   $\mathrm{SU}_h(2)$
models~\cite{kn:HorowitzI,kn:Watson}     when           compared    with
$\mathrm{SU}_\ell(2)$~\cite{kn:Watson} gives similar contrasting figures
to the ones presented here.

Figure \ref{fg:figd}, along   with similar figures  given  in references
\cite{kn:HorowitzI,kn:Watson}, show  a saturation of  nuclear forces  as
$\rho\rightarrow\rho_c$, followed by  a phase transition to quark matter
at $\rho_c$.  All  of these models,  however, fail  to give any  nuclear
binding below $\rho_c$,  which would seem to  suggest that the flux-tube
models     are incapable of  obtaining      nuclear binding.  Even   the
$\mathrm{SU}_h(3)$  HP  model with   its   long chains, which  tends  to
underestimate the potential, indicates that  this would appear to be the
case  \cite{kn:HorowitzI}.  HP have a  2q model \cite{kn:HorowitzI} that
would  seem to suggest that  even if  colour  were not fixed  to a given
quark, no nuclear binding would occur: albeit this model is for $p$-wave
(qq) states.  Thus, it would appear  that string-flip models, even those
that include  higher order flux-tube topologies, or  allow the colour to
move  from  quark   to   quark,  are   insufficient to  obtain   nuclear
binding. Therefore, another mechanism for obtaining nuclear binding must
be included in these models.

  One  possibility  is to  include  one-gluon exchange  interactions. As
suggested by  Nzar and Hoodbhoy  \cite{kn:Nzar}, the most significant of
these are the  hyperfine interactions.   In  a relativistic  setting one
could also  consider the possible effect  of chiral symmetry breaking in
which   the  constituent      quark   mass   changes    with    momentum
scale~\cite{kn:Politzer,kn:Drukarev}.\footnote{This          interesting
possibility   was     pointed   out  to     us    by   the  referee   of
reference~\cite{kn:Boyce}.}
Finally, other effects  such as quark mass  differences and isospin  are
expected to be negligible.
 
\section{Conclusion}
Various string-flip potential models have been discussed in general, and
have  been shown to     adequately  describe the  bulk  properties    of
nuclear/quark  matter  with the exception of  nuclear   binding.  At low
densities they yield free  nucleon matter and at  high densities a phase
transition to free  quark matter.  They show   an overall saturation  of
nuclear  forces as nucleon densities    are increased.  At  intermediate
densities these models, with  the exception of the HP $\mathrm{SU}_h(3)$
linked  chain  model~\cite{kn:HorowitzI},   give   an overall   EMC-like
swelling of the  nucleon.  It  is believed  that  these models with  the
addition of  one-gluon exchange effects should  be capable of predicting
nuclear binding.  This is the topic of the next chapter.

%
%
\chapter{Flux-Bubble Models and Mesonic Molecules}
\label{chap-fbbl}

\section{Introduction}

  In   the    previous  chapter   string-flip   potential  models   were
investigated.  It was hypothesized  that these models by themselves were
incapable of  producing nuclear binding and would  therefore  have to be
extended.    The suggested extension was  to  include one-gluon exchange
interactions.  In this chapter a new class of models, called flux-bubble
models, is proposed  which allows  for the  extension of  the  flux-tube
model to include these interactions.
 
\section{Flux-Bubble Models}

   The primary objective  is to  construct  a model  which combines both
nonperturbative   (flux-tubes)  and   perturbative (one-gluon  exchange)
aspects of QCD in a consistent fashion.   In order to simplify this task
only the colour Coulomb  extensions to  an $SU_\ell(2)$ potential  model
will be considered.

   The extension of the linear potential model for $q\bar q$ pairs is
\vspace{1ex}
$$
  \setlength{\unitlength}{1in}
  \begin{picture}(5.875,0.375)
     \put(0.5625,0.125)
        {
         $
          {\rm v}_{ij}\sim\left\{
          \begin{array}{ll}
             {\displaystyle\sigma({\rm r}_{ij}-{\rm r}_0)} & 
                {\displaystyle{\rm if}\;{\rm r}_{ij}>{\rm r}_0} \\
             {\displaystyle\alpha_s\lambda_{ij}
                \left(\frac{1}{\rm r_{ij}}-\frac{1}{{\rm r}_0}\right)} &
                {\displaystyle\mbox {\rm if}\;{\rm r}_{ij}<{\rm r}_0}
          \end{array}
          \right.
         $
        }
     \put(3.8125,-0.125)
        {
         \mbox{\epsfxsize=2.5cm
	\epsffile{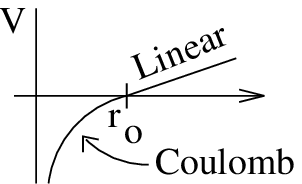}}
        }
     \put(5.59375,0.125){\addtocounter{equation}{1}(\theequation)}
  \end{picture}
$$
where $\lambda_{ij}=-3/4,1/4$  for unlike and like colours respectively,
and  $\alpha_s\approx0.1\,$. This   is      simply a variant    of   the
phenomenological potential,

\vspace{1ex}
\hfill ${\displaystyle
V(r)\sim\sigma\,r-\frac{4}{3}\,\frac{\alpha_s}{r}\;}$,
\hfill(\ref{eq:phenpot})
\vspace{1ex}

\noindent
mentioned in  chapter  1;    the   major  difference  being   that   the
nonperturbative and perturbative parts   are completely isolated  in the
former as opposed  to the latter.  When  the quarks  are separated at  a
distance greater than r$_0$ the potential is purely linear and when they
are inside this  radius it is purely Coulomb.   In effect, for distances
less than r$_0$, a ``bubble'' is formed in  which the quarks are free to
move around,  in an  asymptotically  free  fashion.  In  both   distance
regimes the net colour of the system is neutral.

\begin{figure}[ht]
\begin{center}
\framebox{\setlength{\unitlength}{1in}
\begin{picture}(5.75,2.875)\put(-0.125,-0.0625){
\setlength{\unitlength}{1in}
\begin{picture}(5.75,2.625)
\put(0,2.25){\begin{minipage}{3.875in}\caption[The flux-bubble model]{ 
   \footnotesize   Consider configuration (a)  of  quarks, with r $>{\rm
   r}_0$,  about to move to (b),  s.t.  two of  them are within r $<{\rm
   r}_0$.   Then the procedure  is to draw a bubble  of ${\rm r}_0$ away
   from the  two, (b), and  to cut  the flux-tubes at  the boundary  and
   insert virtual $q\bar q$ pairs,  (c).  Once the potential is computed
   the configuration   is  restored  to  (b)  before   the next move  is
   made.}\end{minipage}}
\put(0,0.125){\mbox{\epsfxsize=146mm
	\epsffile{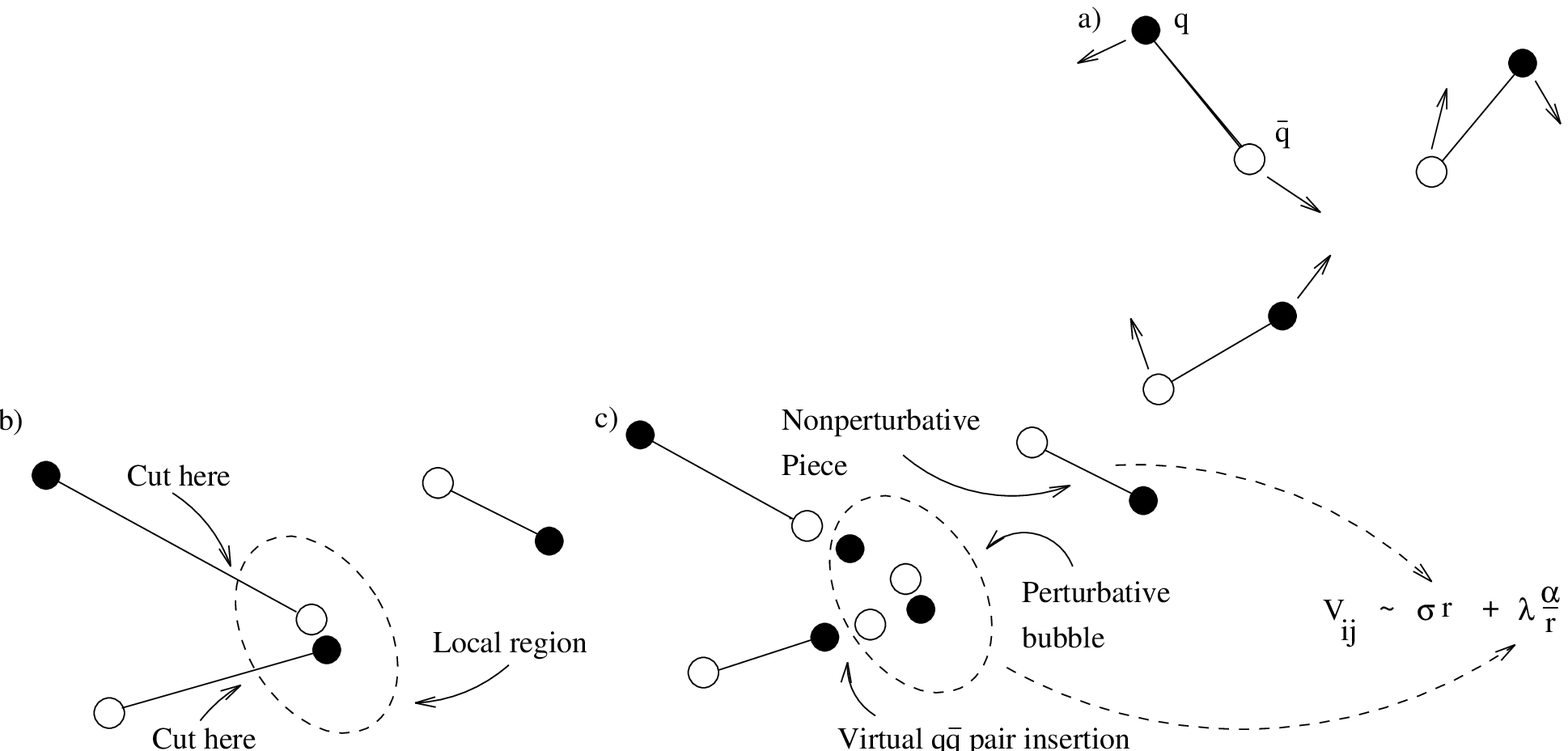}}}
\end{picture}}
\end{picture}}
\end{center}
\label{fig:fluxbubble}
\end{figure}

  The extension of the  linear potential although simple  for a pair of
quarks becomes  more complex when  considering extensions for many pairs
of quarks.
In  particular, how can  a potential model be  constructed in which some
quarks are close enough to  be inside perturbative  bubbles while at the
same  time  a  subset of  them  are still  connected  to nonperturbative
flux-tubes that  extend outside of these  bubbles.  The solution to this
problem is  easily remedied by inserting  virtual $q\bar q$ pairs across
any of the intersection  boundaries formed  by  the flux-tubes  with the
bubbles.   Now the segments of flux-tubes   that lie outside the bubbles
remain intact while the    segments inside simply dissolve;   giving the
desired result.  Figure~\ref{fig:fluxbubble} illustrates the dynamics of
this model.
Notice  that  this model allows   the construction of colourless objects
because of the insertion of the virtual $q\bar q$ pairs.
These virtual quarks are used as a tool  to calculate the overall length
of the flux-tube correctly.  They are  not used in computing the Coulomb
term however, as the  field energy is  already taken into account by the
``real'' quarks inside the  bubbles.  In general,  once the bubbles have
been determined,    the  flux-tubes must be    reconfigured  in order to
minimize the linear part of the potential.

  Although the model is currently for  $SU_\ell(2)$ it should be easy to
extend it to a full  $SU_\ell(3)$ model with  all the one-gluon exchange
phenomena.

\section{In Search of a Wave Function}

  Figure~\ref{fig:preliminary} shows some preliminary results using the
$SU_\ell(2)$ flux-bubble model described in the previous section.
\begin{encapfig}{ht}
\mbox{\epsfxsize=144mm
	\epsffile{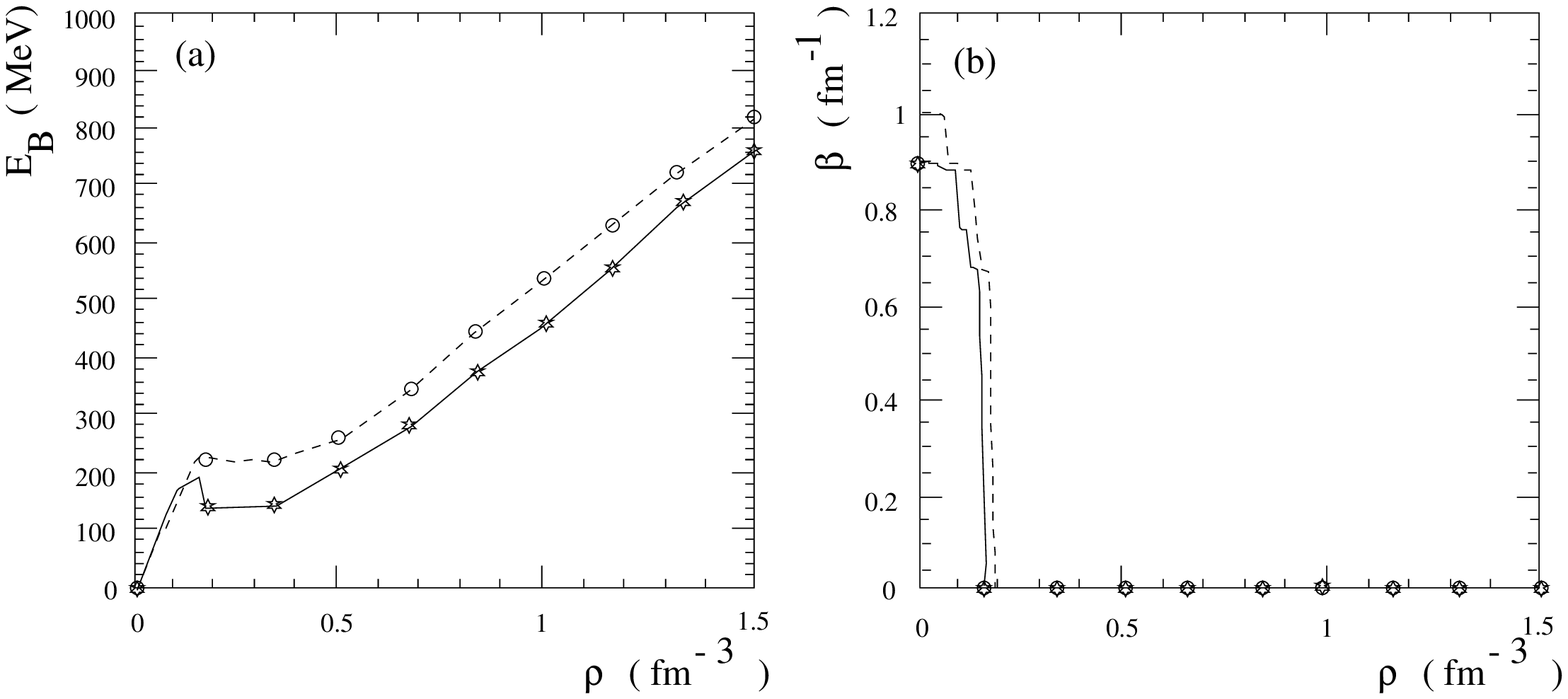}}
\caption[$E_B(\rho)$ and $\beta(\rho)$ for ${\rm r}_0=0\,$fm  and  ${\rm
     r}_0=0.1\,$fm]{\footnotesize   $E_B(\rho)$, (a), and $\beta(\rho)$,
     (b),  for   ${\rm     r}_0=0\,$fm    ({\tiny$-\,-$})   and    ${\rm
     r}_0=0.1\,$fm(---), with $m_q=330MeV\,$,  $\sigma=910MeV/fm\,$, and
     $\alpha_s=0.1\,$.   These graphs  were created   by using a  coarse
     $10\times10$ mesh of points  in  $\rho$ and $\beta$.  The   minimal
     curves were extracted  by  using linear interpolation  between  the
     minimum data points on the $E_B(\rho,\beta)$-mesh surface.}
\label{fig:preliminary}
\end{encapfig}
For ${\rm r}_0=0\,fm$ the Monte Carlo recovers the  same results for the
$SU_\ell(2)$ string-flip  potential model \cite{kn:Watson}, as expected,
and for ${\rm r}_0=0.1\,fm$ the result differs only slightly.

  These results are questionable, as the wave function  that was used is
not ideal.  It consisted of a slight modification to an old $SU_\ell(2)$
wave function \cite{kn:Watson},
\begin{equation}
\Psi_{\alpha\alpha^\prime\beta\rho}=
\underbrace{\mathrm{e}^{
            -{\displaystyle \sum_{{\rm min}\{\mathrm{q\bar{q}}\}}
            (\beta r_{\rm q\bar{q}})^\alpha}
           }}_{\chi^{\rm (Linear)}_{\alpha\beta}}\,
\Phi^{\rm (Fermi)}_\rho\label{eq:watson}
\end{equation}
({\it cf}.  equation~\ref{eq:waves}),  in which a new correlation piece,
$\chi^{\rm (Coulomb)}_{\alpha^\prime\beta}\,$,  was added to account for
the local attractive Coulomb interactions as they occurred: {\it i.e.},
\begin{equation}
\Psi_{\alpha\alpha^\prime\beta\rho}\;\sim
\;\chi^{\rm (Linear)}_{\alpha\beta}\;\times\;
\chi^{\rm (Coulomb)}_{\alpha^\prime\beta}\;\times\;
\Phi^{\rm (Fermi)}_\rho\;,
\end{equation}
where
\begin{equation}
\chi^{\rm (Coulomb)}_{\alpha^\prime\beta}=
\mathrm{e}^{
 -{\displaystyle\sum_{\overline{{\rm min}\{{\rm q\bar{q}}\}}}
 \{(\beta r_{\rm q\bar{q}})^{\alpha^\prime}
 \theta(r_0-r_{\rm q\bar{q}})\}}
} 
\label{eq:sick}
\end{equation}
with  $\alpha=1.75$  and $\alpha^\prime=1\,$ (the repulsive interactions
were assumed to  be taken care  of by the  presence  of the Slater  wave
function, $\Phi^{\rm (Fermi)}_\rho\,$).  However, because the flux-tubes
and  bubbles can now  be created or destroyed   the wave function is, in
general,  no  longer continuous   to  order $\vec{\nabla}^2$  ({\it cf}.
equation~\ref{eq:sick}), and  so a variational lower  bound is no longer
guaranteed. Therefore, a new wave  function is needed. Unfortunately, it
is a rather difficult task  to come up with  a wave function that  takes
into  account  the locality  of these flux-bubble  interactions which is
smooth and continuous, and involves very few parameters.

  The aforementioned  wave function  has two ``independent'' parameters,
$\rho$ and $\beta$; the $\alpha$'s were assumed to be fixed.  This is in
contrast to   the   previous  case   for the   $SU_\ell(3)$   model,  in
chapter~\ref{sec-strflp},  in   which     $\rho$ and  $\beta$     varied
parametrically with a  single parameter,  $\theta\,$.   The reason  this
does not   apply here is  because  the flux-bubble  potential breaks the
scaling transformation

\vspace{1ex}
\hfill 
${\displaystyle
(\beta,\rho^{1/3})\;\longrightarrow\;\zeta(\theta)\,
(\cos\theta,\sin\theta)\,}$. 
\hfill (\ref{eq:scaletran})
\vspace{1ex}

\noindent
This extra degree of freedom greatly increases the computation time.

  The results in figure~\ref{fig:preliminary} were generated on a coarse
$10\times10$ mesh of  points, in $\rho$  and $\beta$, and required 18hrs
of CPU   time on  an  8  node farm.   Clearly  this  procedure  would be
ridiculously slow if more parameters were to be added: 
\begin{equation}
\tau_{\mbox{\tiny CPU}}\;\sim\;{\cal O}(R^{\vspace{1ex}p})
\end{equation}
where $R$ is the mesh resolution in each ordinate, and $p$ is the number
of parameters.

   Therefore, a  way   of   checking  different wave     functions   and
minimization schemes which does not consume large amounts of CPU time is
desirable.  In particular, a  mini-laboratory is needed in which various
aspects of the string-flip  and flux-bubble potential models,  from wave
functions to minimization schemes, can be investigated.

\section{Mesonic Molecules}
  
  Some  work  was  done with  mesonic-molecules  \cite{kn:Weinstein}  by
Treurniet and Watson~\cite{kn:Treurniet},  using $SU_\ell(2)$, which has
shown that  these molecules make  useful  mini-laboratories for studying
string-flip potential  models.   They used  a  mesonic-molecule,  $Q_2$,
consisting of two heavy quarks and  two relatively light antiquarks: see
figure~\ref{fig:mesmol}.
\begin{encapfig}{ht}
\begin{center}
\mbox{\epsfxsize=10cm
   \epsffile{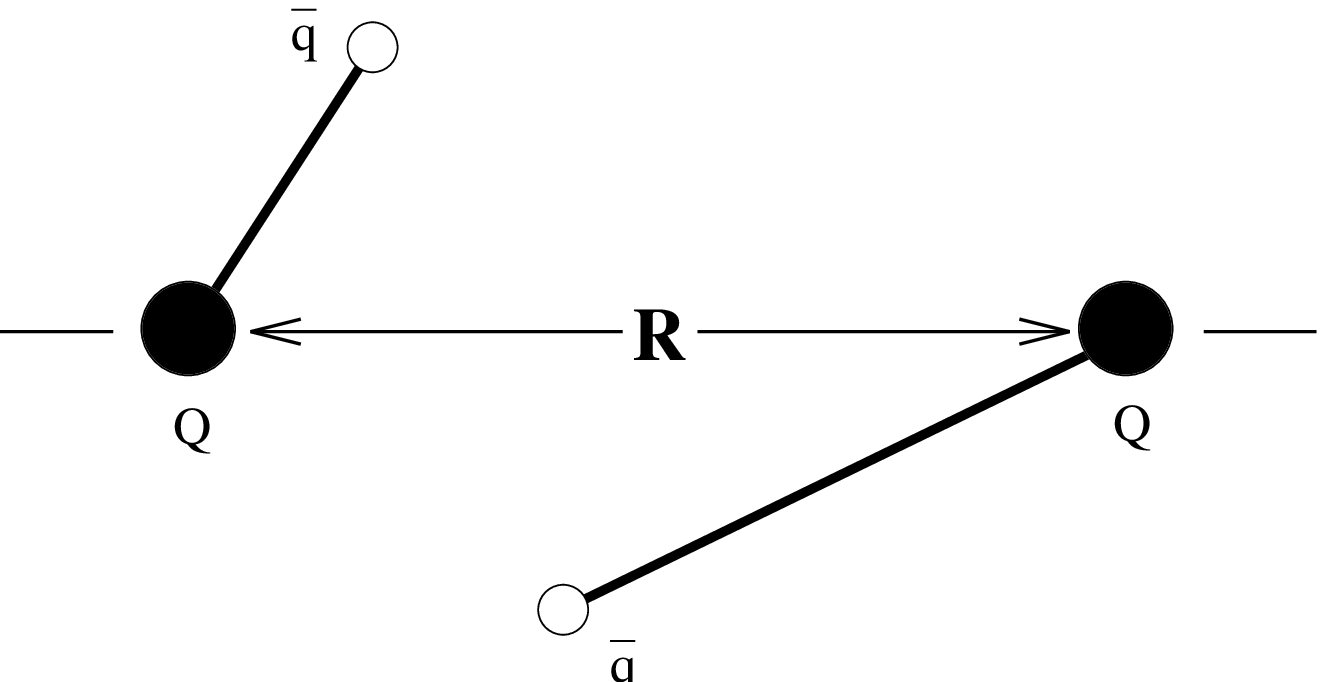}}
\end{center}
\caption[A mesonic-molecule]{
         A  mesonic-molecule,  $Q_2\,$, with  two  heavy quarks  and two
         light antiquarks.}
\label{fig:mesmol}
\end{encapfig}
\noindent The quarks are assumed to be heavy so that the light 
antiquarks  can move around  freely  without disturbing their positions.
By varying the distance, R, between the heavy quarks a mesonic-molecular
potential,          $U(R)\,$,       can      be          computed  ({\it
cf}. equation~\ref{eq:Schiff}).  The $Q_2$ system provides a good way of
checking   potential models for   the  possibility of nuclear  (mesonic)
binding.  Moreover, because  of  its simplicity, it allows  for checking
various wave functions and  minimization schemes without being concerned
about CPU overhead.

\subsection{The Distributed Minimization Algorithm}
\label{sec-mindist}

  There are many ways of determining the minimum  of a function, $f(x)$,
of    several   variables, $x=(x_1$,\ldots,$x_n)\,$   \cite{kn:NumRecF}.
However, for the case where the  function is approximated by Monte Carlo
most of these methods,  in general, will not work.   The reason for this
is  because  Monte   Carlo  calculations   produce  results  that   have
statistical uncertainties;  most of the  methods  are for functions that
give ``exact'' answers. For example, if gradient methods were used, then
the error would  propagate into  the  gradient calculations which  would
effectively add noise to the search.
\begin{encapfig}{htbp}
\begin{center}
\mbox{\epsfxsize=144mm
   \epsffile{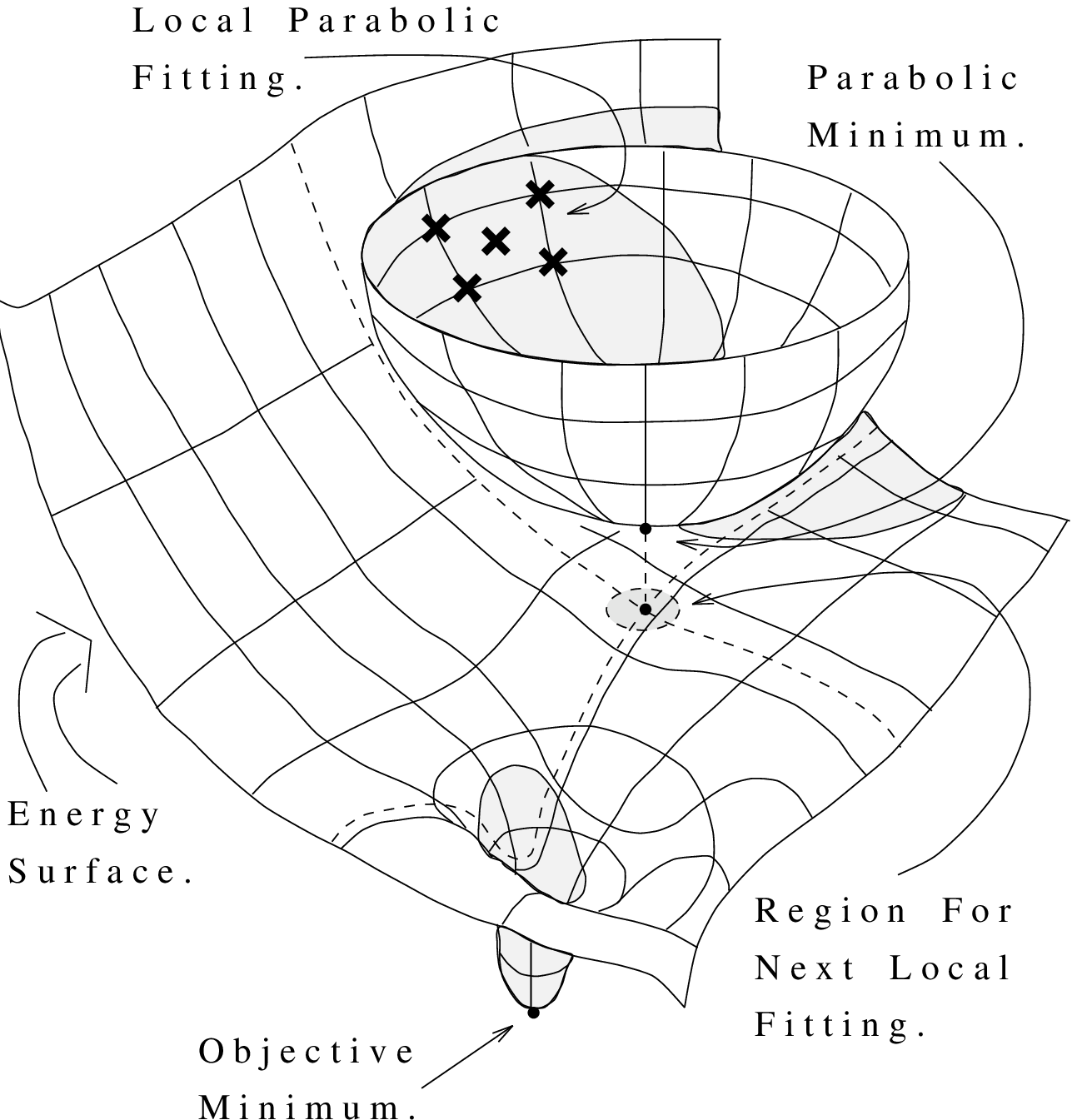}}
\end{center}
\caption[The Parabolic  Minimization Algorithm]{\footnotesize
	The parabolic   minimization   algorithm    for   an   arbitrary
        $d$-dimensional,  $d-D\,$,    energy         surface,      where
        $p=\frac{1}{2}d(d-1)+2d+1$  is the  minimum number of parameters
        required for fitting locally to a $d-D$ parabola.}
\label{fig:distmin}
\end{encapfig}

  Of the  viable methods the    most  promising one  is the    parabolic
minimization algorithm  \cite{kn:Minuit}. The  algorithm is  as follows
(see figure~\ref{fig:distmin}):
\begin{itemize}
\item Pick an initial starting point, $x_i\,$.
\item Pepper its neighborhood  with $p$ points, $x_p\,$.
\item Evaluate the $d-D$ function, $f(x)$, at these points.
\item Fit a $d-D$ parabola to these evaluated points, 
      $(f(x_{p,i}),x_{p,i})\,$.
\item Find the critical point, $x_c\,$, of the parabola.
\item Use this point to repeat the procedure.
\end{itemize}
This algorithm  will, in  general, converge  to a critical  point on the
surface that is being  searched.  However, this  does not guarantee that
the point will be a minimum, it could very well be a maximum or a saddle
point.      Therefore,  it   is  important    to   check  the  curvature
\cite{kn:Minuit} and then take  appropriate measures to drive the search
away  from this point if  is indeed not a  minimum.  Additionally, it is
probably a good  idea to bound the search  to a box  so that it does not
drift out to infinity.  It is important to note that this algorithm does
not guarantee convergence to a global minimum but  then neither does any
other algorithm.

  This algorithm must now be adapted  to take advantage of a distributed
computing environment.

  The first step is simply to submit  $p$ points to $m$ computers; where
if  $p<m$ use $p$  computers and if $p\ge  m$ distribute  $p/m$ jobs per
computer (a slight improvement on this last  step would be to distribute
$m$ points and then  dole  out the remaining  $m+i$ ($i$=1,\ldots,$p-m$)
points to the machines that finish first).  However, this method is very
inefficient because each iteration runs in time
\begin{equation}
\tau\sim{\cal O}(\left\lceil\frac{p}{m}\right\rceil)\,.
\end{equation}
A much  more efficient  method would   be to  develop  a procedure  that
analyzes  the data in a continuous  fashion as it  streams  in, point by
point.

  Such an algorithm  was developed over the course  of time that it took
to produce   the results in  the   later sections  of this chapter.  The
details of  the algorithm can be found  in \S~\ref{sec-distmin}. A basic
outline of the procedure is as follows:
\begin{enumerate}
\item Create an initial data base of $\kappa-m$ sample points, 
      $(f(x_{\kappa-m}),x_{\kappa-m})\,$, with $m$ points pending.
\item Fit the neighborhood of each sample point to a parabola.
\item Submit the critical points, $x_c\,$, that correspond to 
      parabolic minima.
\item Keep the $m$ computers occupied by submitting extra sampling 
      points if necessary.
\item Wait for a point, $(f(x),x)\,$, to arrive.
\item Update the data base.
\item Fit the point about its neighborhood to a parabola.
\item Submit the newly predicted point, $x_c\,$, if it corresponds to a 
      parabolic minimum.
\item Go back to step 4.
\end{enumerate}
This algorithm is effectively performing several parabolic minimizations
all on different regions of the  $d-D$ surface $f(x)$ by interlacing its
searches.  For the work done here the convergence turned out to be quite
rapid (due to cross-talking) with a fairly small start up cost,
\begin{equation}
\tau_s\sim{\cal O}(\left\lceil\frac{\kappa}{m}\right\rceil)\,,
\end{equation}
where typically
\begin{equation}
\kappa\sim{\cal O}(2p)+m\,:
\end{equation}
{\it i.e.}, there were $m$ points pending at the time of creation of the
data  base in  step  1, above.  Also   the time between iterations ({\it
i.e.}, ``events'') was observed to be
\begin{equation}
\tau_i\sim{\cal O}(\frac{1}{m})\,,\label{eq:lintime}
\end{equation}
where $m\le 8\,$.  Ultimately this algorithm will saturate when $\tau_i$
approaches   the    time   required     to process   each   calculation,
$\tau_p\,$. $\tau_p$ is not a constant but a function of the size of the
ever  building data base and therefore  will grow with time.  Regardless
though    $\tau_p<<\tau_i$  and therefore    for all  practical purposes
equation \ref{eq:lintime}  holds.  At some point   $m$ will become large
enough that this algorithm would begin to have  the appearance of simple
grid search.  In fact, it was designed with this in mind.

   For the Monte Carlo results contained herein the number of iterations
required to  obtain an accuracy well below  the 1\% level  was about 30,
60, and 150  for the  1, 2, and  3  dimensional searches,  respectively.
Upon further investigation it  was found that the convergence  criterion
that was used was too  weak.  The minimum  was found after about 65\% of
the  iterations   needed  to meet  the  convergence    criterion.  It is
important to point out  that this algorithm  is still in its infancy and
requires  further  development.   Indeed,   later work   with the  $1-D$
searches have reduced the iterations by about 50\% and is expected to do
the same for higher dimensional  searches.  The improvements were mainly
due to establishing good convergence  criterion and weighting techniques
(see \S~\ref{sec-distmin}).

\subsection{A General Survey of Extensions to $SU_\ell(2)$}
\label{sec-general}
   In  this    section the effects  of   extending  the old $SU_\ell(2)$
\cite{kn:Watson} model to  include flux-bubbles, with and without  fixed
colour, will be  investigated  in the  context of the  mesonic-molecular
system, $Q_2\,$.  To simplify   the situation these extensions  will  be
considered in the frame work of $SU(2)$ colour, $SU_c(2)$.

   A good place  to  start is by studying  a  $Q_2$ effective potential,
$U(R)\,$, that  was generated by interactions  with the light antiquarks
through a linear potential,
\begin{equation}
V=\sigma\,\sum_{\min\{Q\bar q\}}r_{Q\bar q}\,,\label{eq:stringy}
\end{equation}
where the sum, ${\small\displaystyle \sum_{\min\{Q\bar q\}}}\,$, is over
the  set of quark-antiquark pairs, $\{Q\bar  q\}\,$, which minimizes the
potential, $r_{Q\bar{q}}$ represents the distance  between a given light
antiquark, $\bar q\,$, and a  given -- fixed  -- heavy quark, $Q\,$, and
$\sigma\approx910MeV/fm\,$.   The Schr\"odinger  equation that describes
the effective potential  between the two heavy  quarks, in the adiabatic
approximation, is as follows
\cite{kn:Schiff}
\begin{equation}
\left(\frac{1}{2m_q}
\sum_{\bar q}\vec\nabla^2_{\bar q}+V\right)\Psi=U(R)\Psi\,,
\label{eq:Schiff}
\end{equation}
with   $m_q\approx330MeV\,$.    For  the   old   $SU_\ell(2)$ model  the
variational  wave   function   was assumed      to    be of   the   form
\cite{kn:Watson},
\begin{equation}
\Psi_{\alpha\beta}=
\mathrm{e}^{
            -{\displaystyle \sum_{\min\{Q\bar q\}}
            (\beta r_{Q\bar{q}})^\alpha}
           }\,.\label{eq:watwav}
\end{equation}
Therefore,  following    a    similar     procedure    to     that    of
section~\ref{sec-metrop} the effective potential is found by evaluating
\begin{equation}
\bar U=\bar T_C + \bar V\label{eq:effective}
\end{equation}
where 
\begin{equation}
\bar T_C = \alpha\beta^\alpha\,\frac{1}{2m_q}\,
  \langle 
     \sum_{\min\{Q\bar q\}}[\alpha(1-(\beta r_{Q\bar q})^\alpha)+1]\,
     r_{Q\bar q}^{\alpha-2}
  \rangle\,,\label{eq:effectr}
\end{equation}
({\it cf}. equation~\ref{eq:tcorr}), at different values of $R\,$.

   Figure~\ref{fig:modelaa} shows a plot of  the Monte Carlo results for
$\bar U(R)$ where $\alpha$ and $\beta$ have both been allowed to vary.
\begin{encapfig}{htbp}
\begin{center}
\mbox{}\vspace{-1cm}\\
\mbox{\epsfxsize=144mm
   \epsffile{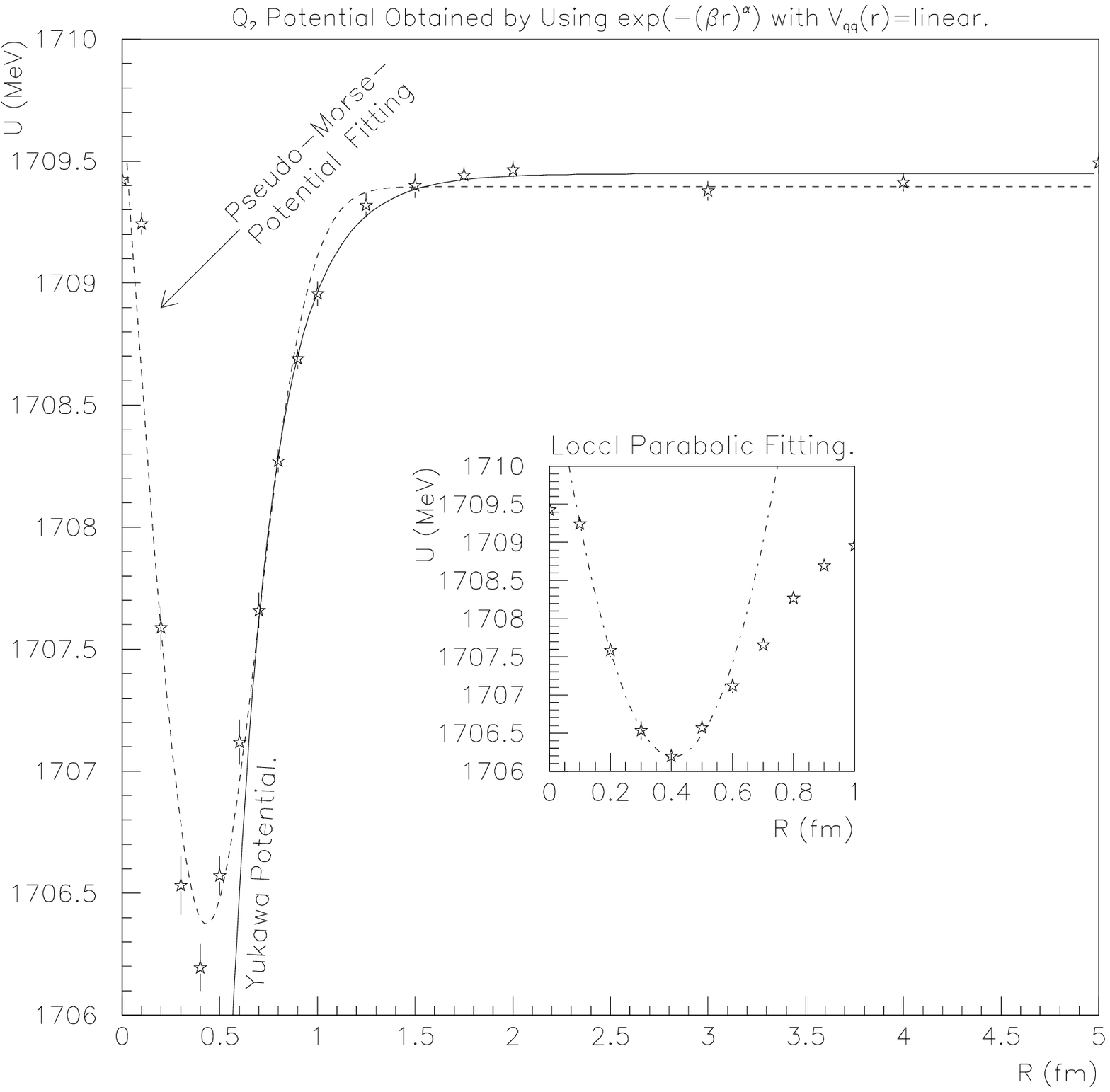}}
\end{center}
\mbox{}\vspace{-2cm}
\caption[$Q_2$ Potential for $V\sim\sigma r$ and 
         $\psi\sim   e^{-(\beta  r)^\alpha}$]{\footnotesize  The   $Q_2$
         potential   obtained   using    a linear   interaction   with a
         pseudo-hydrogen wave function, equation~\ref{eq:watwav}.}
\label{fig:modelaa}
\end{encapfig}
This potential has been parameterized by a pseudo-Morse potential of the
form
\begin{equation}
\tilde U(r)=\left\{e^{-2\eta[\beta(r-r_0)]^\alpha}-
2\eta e^{-[\beta(r-r_0)]^\alpha}\right\}\,\tilde\sigma\,(r-r_0^\prime)
+\tilde U_\infty\,.\label{eq:pseudo}
\end{equation}
The term in the braces, ``\{\}'', on  the left along with $\tilde\sigma$
is   the  Morse  potential  \cite{kn:Schiff,kn:Flugge}  with   two extra
parameters  included; $\eta$ and $\alpha$.  Outside  the brackets to the
right is a  linear term  with an   offset  parameter $r_0^\prime$  (this
mainly takes effect in  cases where the  $Q_2$ potential blows up at the
origin).  Finally  $\tilde  U_\infty$ is  the total  energy  at infinite
separation.  For these figures and the  figures which follow, it will be
assumed that $\eta=1$ and $r_0^\prime=0$ unless  specified otherwise.  A
summary of    the   values of  these   parameters     can be  found   in
table~\ref{tb:tfmlaa}.  This parameterization was used mainly because it
gave a fairly universal description  of the (fully minimized) potentials
in  this  chapter, however  its  physical  significance  should be taken
lightly.

\begin{encaptab}{htbp}
\caption[Summary of Parameters for the Fits in 
        Figure~3.5]{\footnotesize Summary of parameters for the fits
        in figure~\ref{fig:modelaa}.}
\label{tb:tfmlaa}
{
\def\BETA{\beta\;(fm^{-1})}
\def\RO{r_0\;(fm)}
\def\ALPH{\alpha}
\def\ROP{r_0^\prime\;(fm)}
\def\UOI{\tilde U_\infty\;(MeV)}
\def\SIGA{\tilde\sigma\;(MeV/fm)}
\def\VOY{V_0\;(MeV)}
\def\VOI{V_\infty\;(MeV)}
\def\EH{a\;(fm)}
\def\SEE{{\cal C}\;(MeV/fm^2)}
\def\YO{y_0\;(MeV)}
\begin{tabular}{|l|c|c|c|}\hline
Potential & \multicolumn{3}{c|}{Parameters}\\\hline\hline
Pseudo-Morse
& $\BETA$         &   $\RO$         & $\ALPH$        \\\cline{2-4}
& $1.162\pm0.083$ & $-0.41\pm0.18$  &  $3.0\pm1.2$   \\\cline{2-4}
& $\SIGA$         & $\UOI$          & $\ROP$         \\\cline{2-4}
&  $11.9\pm7.1$   & $1709.4\pm0.4$ & $0.033\pm0.060$\\\hline\hline
Yukawa
& $\VOY$          &   $\EH$           & $\VOI$       \\\cline{2-4}
&  $39.8\pm3.2$ & $0.3092\pm0.0075$&$1709.40\pm0.02$\\\hline\hline
Parabolic
& $\SEE$  & $\RO$     & $\YO$   \\\cline{2-4}
& $32.6\pm2.8$  & $0.4060\pm0.015$   & $1706.2\pm0.1$\\\hline
\end{tabular}
}
\end{encaptab}

   The values of  Monte Carlo results for $\bar  U(R)$ at the end points
of the curve, from $R=0fm$ and out to $R=5fm$,  were checked against the
analytic        solution      ({\it       cf}.  \cite{kn:Watson}     and
equation~\ref{eq:anyla})
\begin{equation}
E_{free}=2\left(\frac{g_T^{\mbox{}}(\alpha)}{2\mu}\,\beta^2+
g_L^{\mbox{}}(\alpha)\,\frac{\sigma}{\beta}\right)\,,\label{eq:anaa}
\end{equation}
where
\begin{eqnarray}
g_T^{\mbox{}}(\alpha)&=&
\frac{\alpha^2\,\mbox{2\raisebox{1.5ex}{$\frac{2}{\alpha}-2$}}
\,\Gamma(2+1/\alpha)}{\Gamma(3/\alpha)}\label{eq:anaaa}\,,\\
g_L^{\mbox{}}(\alpha)&=&
\frac{\Gamma(4/\alpha)}{
\mbox{2\raisebox{1.5ex}{$\frac{1}{\alpha}$}}\,\Gamma(3/\alpha)}
\label{eq:anaab}\,,
\end{eqnarray}
and $\mu$ ($\approx m_q$)  is the reduced mass.   This solution was easy
to obtain because at the two extremes, $R=0$ and $R=\infty\,$, V becomes
separable;  the   $Q_2$  system  becomes  equivalent    to two  isolated
mesons. Equation~\ref{eq:anaa} can be minimized  in $\alpha$ and $\beta$
easily by using Mathematica \cite{kn:Wolfram}.  The results can be found
in table~\ref{tb:chpa} of \S~\ref{sec-mindist}.

   It is interesting  to  investigate  whether or not  the   string-flip
potential,   equation~\ref{eq:stringy},   actually mimics pion exchange.
Therefore, the asymptotic part of $U(R)\,$, in figure~\ref{fig:modelaa},
has been fitted to a Yukawa potential:
\begin{equation}
V(r)=-V_0\,\frac{e^{-r/a}}{r/a}\,+V_\infty\,,
\end{equation}
and so the mass of the exchange particle is
\begin{equation}
\fbox{$\displaystyle m_{\mbox{\tiny ex}}= \frac{\hbar c}{a}$}\,
\label{eq:exchange}.
\end{equation}
A summary of the values of the parameters $V_0\,$, $a\,$, and $V_\infty$
can be   found   in  table~\ref{tb:tfmlaa}.  Therefore,   
\begin{equation}
m_{\mbox{\tiny ex}}\approx(638\pm15)MeV\,,
\end{equation} 
which is about $4.6$ times too big.

  Finally a test of whether or not the  system is capable of binding can
be  done by  expanding  about the  minimum  of $\bar  U(R)$ by using the
parabolic approximation \cite{kn:Flugge}
\begin{equation}
y(r)={\cal C}(r-r_0)^2+y_0\,,\label{eq:parabolic}
\end{equation}
which can be transformed into the harmonic oscillator potential
\begin{equation}
V_h(r)= -D+\txtfrac{1}{2}\mu_Q\omega^2(r-r_0)^2\,,
\end{equation}
with
\begin{equation}  
\fbox{$\displaystyle D=U_\infty -  y_0$}\,,\label{eq:depth}
\end{equation}
$\omega=c\,\sqrt{2{\cal    C}/\mu_Q}\,$,     and  $\mu_Q=m_Q/2\,$, where
$U_\infty$ shall   be taken as the  analytic  result for $U(R)$   out at
$\infty$.  Therefore, the binding energy is
\begin{equation}
E_h^\nu= -D+\hbar\omega(\nu+\txtfrac{1}{2})\,,
\end{equation}
which implies
\begin{equation}
\fbox{$\displaystyle \mu_Q\ge
   \frac{{\cal C}(\hbar c)^2}{2(U_\infty - y_0)^2}$}
\label{eq:parbnd}
\end{equation}
in order to  obtain binding.  A summary of  the parameters ${\cal C}\,$,
$r_0\,$, and $y_0\,$, and the analytic result $U_\infty$ can be found in
tables  \ref{tb:tfmlaa}  and \ref{tb:chpa}, respectively.  So  to obtain
binding
\begin{equation}
\mu_Q\approxge(55\pm6)GeV\,;
\end{equation}
which is not surprising as the well is quite shallow,
\begin{equation}
D\approx(3.4\pm0.1)MeV\,.
\end{equation}

  The next step is  to add  an  $SU_c(2)$ term to the  linear potential,
equation~\ref{eq:stringy},
\begin{equation}
V=\sum_{\min\{Q\bar q\}}
\left[
      \sigma\,(r_{Q\bar q}-r_0)\,\theta(r_{Q\bar q}-r_0)-
      \frac{3}{4}\,\alpha_s\,
      \left(\frac{1}{r_{Q\bar q}}-\frac{1}{r_0}\right)
      \theta(r_0-r_{Q\bar q})
\right]
\,,\label{eq:stringya}
\end{equation}
where $\alpha_s\approx0.1\,$, and
$r_0\approx0.1fm\,$. Figure~\ref{fig:modelba} shows the results of the
Monte Carlo for this potential.
\begin{encapfig}{htbp}
\begin{center}
\mbox{}\vspace{-1cm}\\
\mbox{\epsfxsize=144mm
   \epsffile{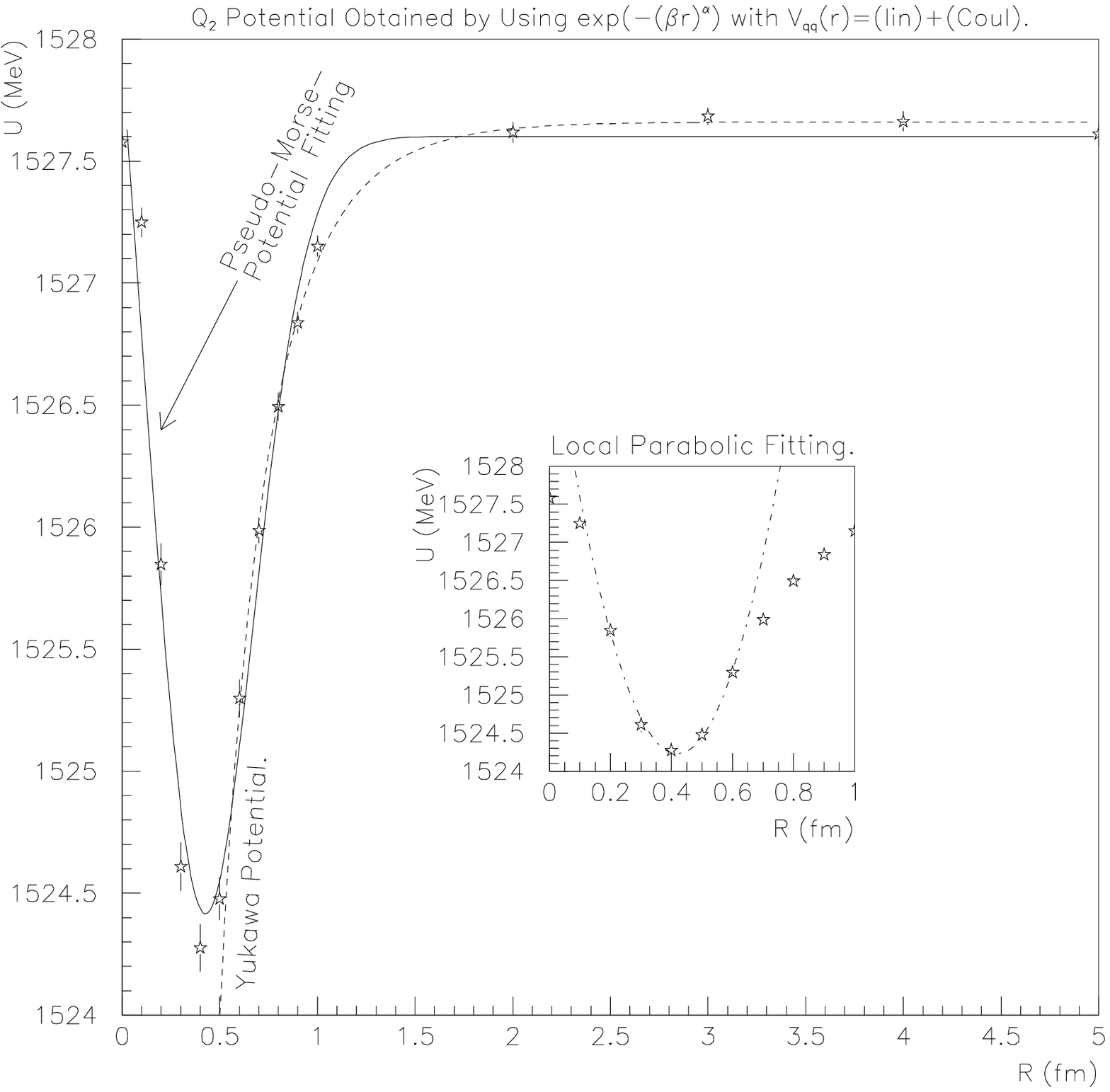}}
\mbox{}\vspace{-2cm}
\end{center}
\caption[$Q_2$ Potential for $V\sim\sigma r-\alpha_s/r$ and 
         $\psi\sim   e^{-(\beta r)^\alpha}$]{\footnotesize  The    $Q_2$
         potential obtained using a linear-plus-coulomb interaction with
         a pseudo-hydrogen wave function.}
\label{fig:modelba}
\end{encapfig}

  The  Monte Carlo results for  $\bar U(R)$ were  checked at $R=0fm$ and
$R=5fm$ against the analytic solution
\begin{eqnarray}
E_{free}&=&2
\left[
   \frac{g_T^{\mbox{}}(\alpha)}{2\mu}\,\beta^2+
   \sigma
    \left(
      \frac{g_L^{\mbox{}}(\alpha,\beta)}{\beta}-
      g_0^{\mbox{}}(\alpha,\beta)\,r_0
   \right)\right.\nonumber\\\nonumber\\&&\left.-
   \frac{3}{4}\,\alpha_s
    \left(
      g_C^{\mbox{}}(\alpha,\beta)\,\beta-
      \frac{(1-g_0^{\mbox{}}(\alpha,\beta))}{r_0}
   \right)
\right]\,,\label{eq:anab}
\end{eqnarray}
where $g_T^{\mbox{}}(\alpha)$ is given by equation~\ref{eq:anaaa},
\begin{eqnarray}
g_L^{\mbox{}}(\alpha,\beta)&=&
\frac{\Gamma(4/\alpha,2(\beta r_0)^\alpha)}{
\mbox{2\raisebox{1.5ex}{$\frac{1}{\alpha}$}}\,\Gamma(3/\alpha)}
\;\;\;\;\;\;\;\;\;\;\;\;\;\;\;\;=\;
(1-{\rm P}(4/\alpha,2(\beta r_0)^\alpha)\,)\,g_L^{\mbox{}}(\alpha)\,,
\;\;\;\mbox{}\\
g_C^{\mbox{}}(\alpha,\beta)&=&
\frac{\Gamma(2/\alpha)-\Gamma(2/\alpha,2(\beta r_0)^\alpha)}{
\mbox{2\raisebox{1.5ex}{$-\frac{1}{\alpha}$}}\,\Gamma(3/\alpha)}
\;=\;{\rm P}(2/\alpha,2(\beta r_0)^\alpha)\,g_C^{\mbox{}}(\alpha)\,,\\
g_0^{\mbox{}}(\alpha,\beta)&=&
\frac{\Gamma(3/\alpha,2(\beta r_0)^\alpha)}{\Gamma(3/\alpha)}
\;\;\;\;\;\;\;\;\;\;\;\;\;\;\;\;=\;
1-{\rm P}(3/\alpha,2(\beta r_0)^\alpha)\,,
\end{eqnarray}
where $g_L^{\mbox{}}(\alpha)$ is given by equation~\ref{eq:anaab}, 
\begin{equation}
g_C^{\mbox{}}(\alpha)=
\frac{\Gamma(2/\alpha)}{
\mbox{2\raisebox{1.5ex}{$-\frac{1}{\alpha}$}}\,\Gamma(3/\alpha)}\,,
\end{equation}
${\rm  P}(a,z)=1-\Gamma(a,z)/\Gamma(a)\,$,   and  $\Gamma(a,z)$ is   the
incomplete gamma  function \cite{kn:Wolfram,kn:Abramowitz}.  At  $R=0fm$
and  $R=5(\approx\infty)$   the  system  effectively  decouples into two
isolated  mesons, for which equation~\ref{eq:anab}  was derived.  Notice
that in   the  limits   $r_0\rightarrow0$  and    $r_0\rightarrow\infty$
equation~\ref{eq:anab} simplifies    to  the  earlier   linear  solution,
equation~\ref{eq:anaa},
and to the Coulomb solution,
\begin{equation}
E_{free}\;\;
\mbox{\raisebox{-0.6ex}{$
   \stackrel{\longrightarrow}{
   \mbox{\tiny$r_0\rightarrow\infty$}}$}
}
\;\;2\left(\frac{g_T^{\mbox{}}(\alpha)}{2\mu}\,\beta^2-
\frac{3}{4}\,\alpha_s g_C^{\mbox{}}(\alpha)\,\beta\right)\,,
\end{equation}
respectively.    The    analytic    ({\it   i.e.},  the      minimum  of
equation~\ref{eq:anab},  {\it via} Mathematica \cite{kn:Wolfram}) verses
the Monte Carlo results are summarized in table~\ref{tb:chpaaa}.
\begin{encaptab}{H}
\caption[MC vs. analytic results for a linear-plus-coulomb 
        potential]{\footnotesize Monte Carlo  (MC)  vs. analytic results
        for a linear-plus-coulomb potential.}
\label{tb:chpaaa}
\begin{tabular}{|ll|c|c|c|}\hline
Parameters         &           
&  Analytic & MC @ $R=0fm$   & MC @ $R=5fm$
\\\hline\hline			
$E_{min}(\alpha,\beta)$&$(MeV)$
& 1527.07   & $1527.58\pm0.04$  & $1527.61\pm0.05$
   \\\hline			
$\alpha$           &           & 1.74      & 1.74    & 1.72
   \\\hline			
$\beta$            &$(fm^{-1})$& 1.37      & 1.37    & 1.38
   \\\hline
\end{tabular}
\end{encaptab}

   The results for the pseudo-Morse, Yukawa, and parabolic fits are
summarized in table~\ref{tb:tfmlab}.
\begin{encaptab}{H}
\caption[Summary of Parameters for the Fits in 
        Figure~3.6]{\footnotesize Summary  of   parameters
        for the fits in figure~\ref{fig:modelba}.}
\label{tb:tfmlab}
{
\def\BETA{\beta\;(fm^{-1})}
\def\RO{r_0\;(fm)}
\def\ALPH{\alpha}
\def\ROP{r_0^\prime\;(fm)}
\def\UOI{\tilde U_\infty\;(MeV)}
\def\SIGA{\tilde\sigma\;(MeV/fm)}
\def\VOY{V_0\;(MeV)}
\def\VOI{V_\infty\;(MeV)}
\def\EH{a\;(fm)}
\def\SEE{{\cal C}\;(MeV/fm^2)}
\def\YO{y_0\;(MeV)}
\begin{tabular}{|l|c|c|c|}\hline
Potential & \multicolumn{3}{c|}{Parameters}\\\hline\hline
Pseudo-Morse
& $\BETA$       &   $\RO$         & $\ALPH$        \\\cline{2-4}
& $2.01\pm1.3$  & $-0.01\pm0.41$  &  $2.0\pm1.3$   \\\cline{2-4}
& $\SIGA$       & $\UOI$          & $\ROP$         \\\cline{2-4}
&  $11.2\pm6.0$ & $1527.6\pm0.5$& $0.027\pm0.072$\\\hline\hline
Yukawa
& $\VOY$        &   $\EH$         & $\VOI$         \\\cline{2-4}
& $13.61\pm0.55$&$0.4299\pm0.0082$& $1527.70\pm0.02$\\\hline\hline
Parabolic
& $\SEE$        & $\RO$           & $\YO$          \\\cline{2-4}
& $33.1\pm1.6$  & $0.419\pm0.053$ & $1524.2\pm0.1$\\\hline
\end{tabular}
}
\end{encaptab}
The Yukawa potential gives an  exchange particle mass of  
\begin{equation}
m_{\mbox{\tiny ex}}\approx(459\pm4)MeV\,.
\end{equation} 
which is about $(3.3)$  times too big.  The parabolic  fit yields a well
depth of
\begin{equation}
D\approx(2.9\pm0.1)MeV
\end{equation}
with the mass constraint
\begin{equation}
\mu_Q\approxge(77\pm6)GeV\,,
\end{equation}
in order to obtain binding.

  Next the  potential   in equation~\ref{eq:stringya}   is  extended  to
include all  of the flux-bubble interactions,  where the colour is fixed
to each of the quarks:
\begin{equation}
V=\sigma\,\sum_{\min\{Q\bar q\}}
      (r_{Q\bar q}-r_0)\,\theta(r_{Q\bar q}-r_0)+
      \alpha_s
      \sum_{i<j}\lambda_{p_ip_j}
      \left(\frac{1}{r_{p_ip_j}}-\frac{1}{r_0}\right)
      \theta(r_0-r_{p_ip_j})
\,,\label{eq:stringyb}
\end{equation}
with   particle  index  $p_k\;\varepsilon\;\{Q_i,\bar  q_j|i,j=1,2\}\,$,
such that $k=1,2,3,4\,$, and $SU_c(2)$ colour factor
\begin{equation}
\lambda_{p_ip_j}=\left\{\begin{array}{rl}
-\frac{3}{4}& {\rm if}\;(p_i,p_j)\,
              \subseteq\,\{(\bar q_i,Q_j)\}\\
 \frac{1}{4}& {\rm if}\;(p_i,p_j)\,
              \subseteq\,\{(\bar q_i,\bar q_j),(Q_i,Q_j)\}
\end{array}\right.\,.
\end{equation}
Notice    that  V   can be    rewritten  so  that    its extension    to
equation~\ref{eq:stringya} is more apparent:
\begin{eqnarray}
V&=&\sum_{\min\{Q\bar q\}}
\left[
      \sigma\,(r_{Q\bar q}-r_0)\,\theta(r_{Q\bar q}-r_0)-
      \frac{3}{4}\,\alpha_s\,
      \left(\frac{1}{r_{Q\bar q}}-\frac{1}{r_0}\right)
      \theta(r_0-r_{Q\bar q})
\right]\nonumber\\&&
+\,\alpha_s
\sum_{p_ip_j\,\varepsilon\,\overline{\min\{Q\bar q\}}}\lambda_{p_ip_j}
\left(\frac{1}{r_{p_ip_j}}-\frac{1}{r_0}\right)
\theta(r_0-r_{p_ip_j})\,,
\end{eqnarray}
where  $\overline{\min\{Q\bar  q\}}$    is the complement   of   the set
$\min\{Q\bar q\}\,$, and therefore defining  more  precisely the sum  in
the exponent of the variational wave function, equation~\ref{eq:watwav}.
Figure~\ref{fig:modelca}  shows the results of the  Monte Carlo for this
potential.
\begin{encapfig}{htbp}
\begin{center}
\mbox{}\vspace{-1cm}\\
\mbox{\epsfxsize=144mm
   \epsffile{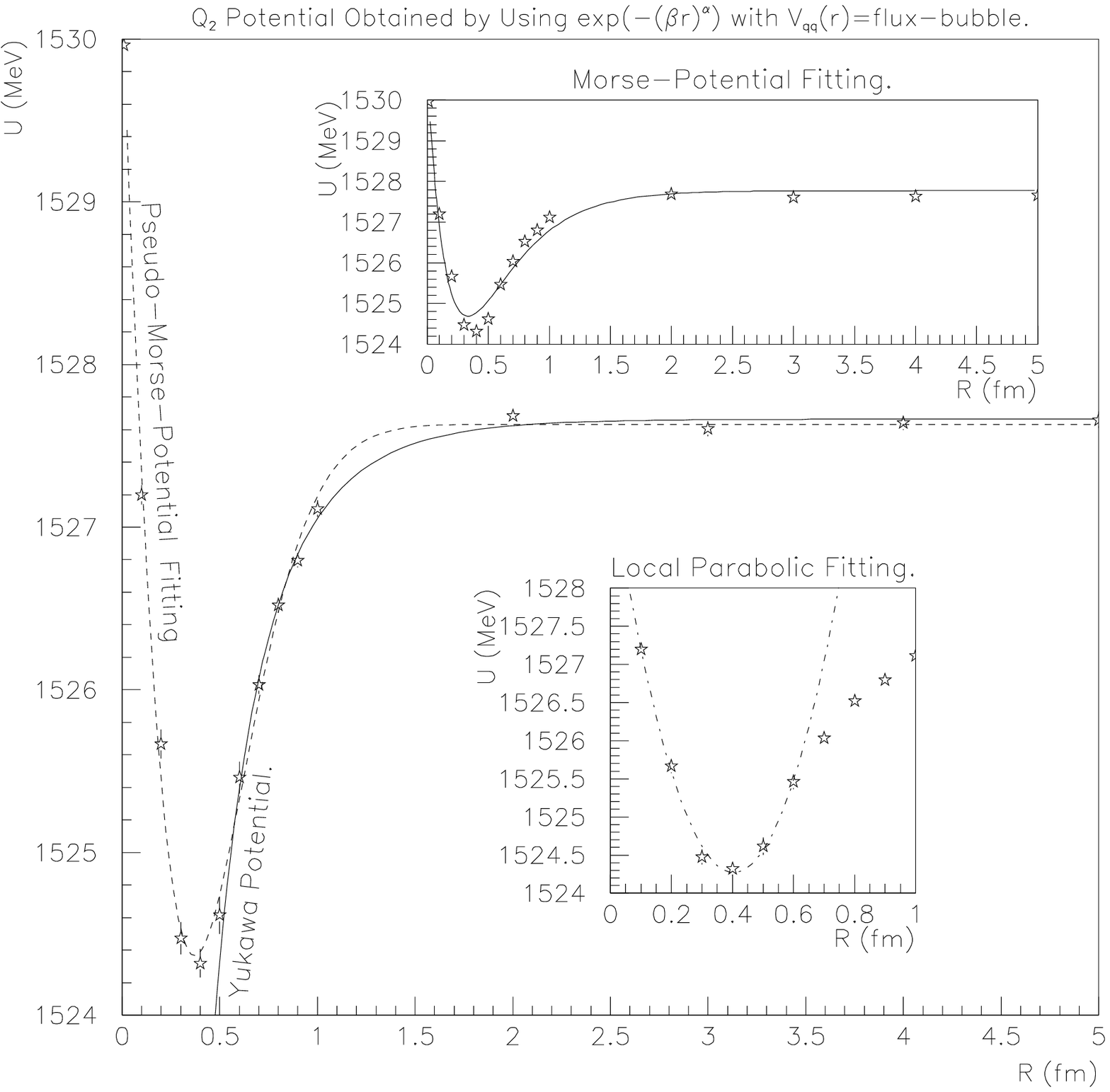}}
\end{center}
\mbox{}\vspace{-2cm}
\caption[$Q_2$ Potential for $V\sim$(flux-bubble) and 
         $\psi\sim e^{-(\beta  r)^\alpha}$]{  \footnotesize The    $Q_2$
         potential obtained using the   flux-bubble interaction  with  a
         pseudo-hydrogen wave function.}
\label{fig:modelca}
\end{encapfig}

  The  Monte Carlo results  were  checked against  the analytic solution
given by equation~\ref{eq:anab}
at $R=5fm$ only,  since  the system no longer  decouples  at the origin.
The results are summarized in table~\ref{tb:chpaab}.
\begin{encaptab}{htbp}
\caption[MC vs. analytic results for a flux-bubble
        potential]{\footnotesize Monte Carlo  (MC)  vs. analytic results
        for a flux-bubble potential.}
\label{tb:chpaab}
\begin{tabular}{|ll|c|c|}\hline
Parameters         &           &   Analytic      & MC @ $R=5fm$
\\\hline\hline
$E_{min}(\alpha,\beta)$&$(MeV)$&  1527.07        & $1527.66\pm0.05$
   \\\hline
$\alpha$           &           &   1.74          & 1.71
   \\\hline
$\beta$            &$(fm^{-1})$&   1.37          & 1.38
   \\\hline
\end{tabular}
\end{encaptab}

  An    attempt   was made  to   fit  $U(R)$    to  the  Morse potential
\cite{kn:Schiff,kn:Flugge},
\begin{equation}
U_M(r)=U_0(e^{-2\beta(r-r_0)}-2e^{-\beta(r-r_0)})+U_\infty\,,
\end{equation}
with binding energy
\begin{equation}
E^\nu_M= -D +\hbar\omega\left\{(\nu+\txtfrac{1}{2})-
\frac{\beta\hbar c}{2\sqrt{2\mu_QD}}(\nu+\txtfrac{1}{2})^2\right\}\,,
\end{equation}
and   depth    $D=U_0\,$,    where  $\omega=\beta c\,\sqrt{2D/\mu_Q}\,$.
For this potential
\begin{equation}
\mu_Q\ge\frac{(\beta\hbar c)^2}{8U_0}\,
\end{equation}
is the corresponding binding  constraint.  
\begin{encaptab}{htbp}
\caption[Summary of Parameters for the Fits in 
        Figure~3.7]{\footnotesize Summary  of   parameters
        for the fits in figure~\ref{fig:modelca}.}
\label{tb:tfmlac}
{
\def\BETA{\beta\;(fm^{-1})}
\def\RO{r_0\;(fm)}
\def\ALPH{\alpha}
\def\ROP{r_0^\prime\;(fm)}
\def\UOI{\tilde U_\infty\;(MeV)}
\def\UO{U_0\;(MeV)}
\def\UI{U_\infty\;(MeV)}
\def\SIGA{\tilde\sigma\;(MeV/fm)}
\def\VOY{V_0\;(MeV)}
\def\VOI{V_\infty\;(MeV)}
\def\EH{a\;(fm)}
\def\SEE{{\cal C}\;(MeV/fm^2)}
\def\YO{y_0\;(MeV)}
\begin{tabular}{|l|c|c|c|}\hline
Potential & \multicolumn{3}{c|}{Parameters}\\\hline\hline
Pseudo-Morse
& $\BETA$       &   $\RO$          & $\ALPH$        \\\cline{2-4}
& $1.34\pm0.20$ & $-0.63\pm0.19$   & $2.10\pm0.37$  \\\cline{2-4}
& $\SIGA$       & $\UOI$           & $\ROP$         \\\cline{2-4}
& $40\pm15$     & $1527.6\pm0.5$ & $0.088\pm0.031$\\\hline\hline
Yukawa
& $\VOY$        &   $\EH$          & $\VOI$         \\\cline{2-4}
& $9.16\pm0.42$ &  $0.498\pm0.012$ & $1527.70\pm0.02$\\\hline\hline
Parabolic
& $\SEE$        & $\RO$            & $\YO$          \\\cline{2-4}
& $31.7\pm1.4$  & $0.4042\pm0.0044$& $1524.3\pm0.1$\\\hline\hline
Morse
& $\BETA$       &  $\RO$           & $\UO$          \\\cline{2-4}
& $2.62\pm0.38$ &  $0.333\pm0.039$ & $3.09\pm0.73$  \\\cline{2-4}
& $\UI$   & \multicolumn{2}{c}{\mbox{}}\\\cline{2-2}
& $1527.8\pm0.5$  & \multicolumn{2}{c}{\mbox{}}\\\cline{1-2}
\end{tabular}
}
\end{encaptab}
Table~\ref{tb:tfmlac}  contains a
summary of the  fits for   the  various potentials. The Morse   and
parabolic fits give,
\begin{eqnarray}
(\mu_Q)_{\mbox{\tiny Morse}}\;\;\; &\approxge& (11\pm4)GeV\,, \\
(\mu_Q)_{\mbox{\tiny parabolic}}   &\approxge& (79\pm7)GeV\,
\end{eqnarray}
with   well  depths 
\begin{eqnarray}
(D)_{\mbox{\tiny Morse}}\;\;\; & \approx &  (3.1\pm0.7)MeV\,,\\
(D)_{\mbox{\tiny parabolic}}   & \approx &  (2.8\pm0.1)MeV\,.
\end{eqnarray}  
Although the Morse gives a lower value it is a  rather dubious one since
the fit  did not describe  the shape of the  knee or the  minimum of the
potential (see Morse  fitting  insert in figure~\ref{fig:modelca})  very
well. Regardless, there is no binding.

   Referring to  equation~\ref{eq:exchange} and to table~\ref{tb:tfmlac}
yields an exchange mass of
\begin{equation}
m_{\mbox{\tiny ex}}\approx{\cal O}(396\pm10)MeV\,,
\end{equation}
which is ${\cal O}(2.8)$ times too big.

  Finally the flux-bubble potential is  extended to allow the colour  to
move     around.   The     potential    is   similar     to     that  of
equation~\ref{eq:stringyb} except that  now the particle indices, $p_k$,
carry colour degrees of freedom,
\begin{eqnarray}
V&=&\sum_{\min\{q_i\bar q_j\}}
\left[
      \sigma\,(r_{q_i\bar q_j}-r_0)\,\theta(r_{q_i\bar q_j}-r_0)-
      \frac{3}{4}\,\alpha_s\,
      \left(\frac{1}{r_{q_i\bar q_j}}-\frac{1}{r_0}\right)
      \theta(r_0-r_{q_i\bar q_j})
\right]\nonumber\\&&
+\,\alpha_s
\sum_{p_ip_j\,\varepsilon\,\overline{\min\{q_i\bar q_j\}}}
\lambda_{p_ip_j}\left(\frac{1}{r_{p_ip_j}}-\frac{1}{r_0}\right)
\theta(r_0-r_{p_ip_j})\,,
\end{eqnarray}
where $\lambda_{p_ip_j}$ is defined just below equation~3.1, and
\begin{eqnarray}
     q_k&\varepsilon& \{R_i,\bar B_i,r_i,\bar b_i|i=1,2\}
        \;\subseteq\;SU_c(2)\,,\\
\bar q_k&\varepsilon& \{\bar R_i,B_i,\bar r_i,b_i|i=1,2\}
        \;\subseteq\;\overline{SU_c(2)}\,,
\end{eqnarray}
{\it i.e.},           
\hfill         
$b\;\sim\;\bar      r\,,\,\mbox{\ldots,     etc.}\;\;\Leftrightarrow\;\;
SU_c(2)\;\sim\;\overline{SU_c(2)}\,,\;\;\;\;\;\;$\hfill\mbox{}\\
\noindent 
where the capital case letters are for the  heavy quarks, the lower case
letters are for the light quarks, the letters $r$  and $R$ represent the
red quarks,  and the letters $b$  and $B$ stand for the  blue quarks.  A
further expansion of V leads to the more useful form
\begin{eqnarray}
V&=&\sum_{\min\{q_i\bar q_j\}}
\left[
      \sigma\,(r_{q_i\bar q_j}-r_0)\,\theta(r_{q_i\bar q_j}-r_0)-
      \frac{3}{4}\,\alpha_s\,
      \left(\frac{1}{r_{q_i\bar q_j}}-\frac{1}{r_0}\right)
      \theta(r_0-r_{q_i\bar q_j})
\right]\nonumber\\&&
-\frac{3}{4}\,\alpha_s
\sum_{q_i\bar q_j\,\varepsilon\,\overline{\min\{q_i\bar q_j\}}_+}
\left(\frac{1}{r_{q_i\bar q_j}}-\frac{1}{r_0}\right)
\theta(r_0-r_{q_i\bar q_j})
\nonumber\\&&
+\frac{1}{4}\,\alpha_s
\sum_{q_iq_j\,\varepsilon\,\overline{\min\{q_i\bar q_j\}}_-}
\left(\frac{1}{r_{q_iq_j}}-\frac{1}{r_0}\right)
\theta(r_0-r_{q_iq_j})\,,
\end{eqnarray}
where   $$     \overline{\min\{q_i\bar   q_j\}}= \overline{\min\{q_i\bar
q_j\}}_+  \cup  \overline{\min\{q_i\bar   q_j\}}_-\,,  $$ plus, ``{\tiny
+}'',  means    attractive,   and  minus,   ``-'',   means    repulsive.
Figure~\ref{fig:modelda} shows  the results of  the Monte Carlo for this
potential.
\begin{encapfig}{htbp}
\begin{center}
\mbox{}\vspace{-1cm}\\
\mbox{\epsfxsize=144mm
   \epsffile{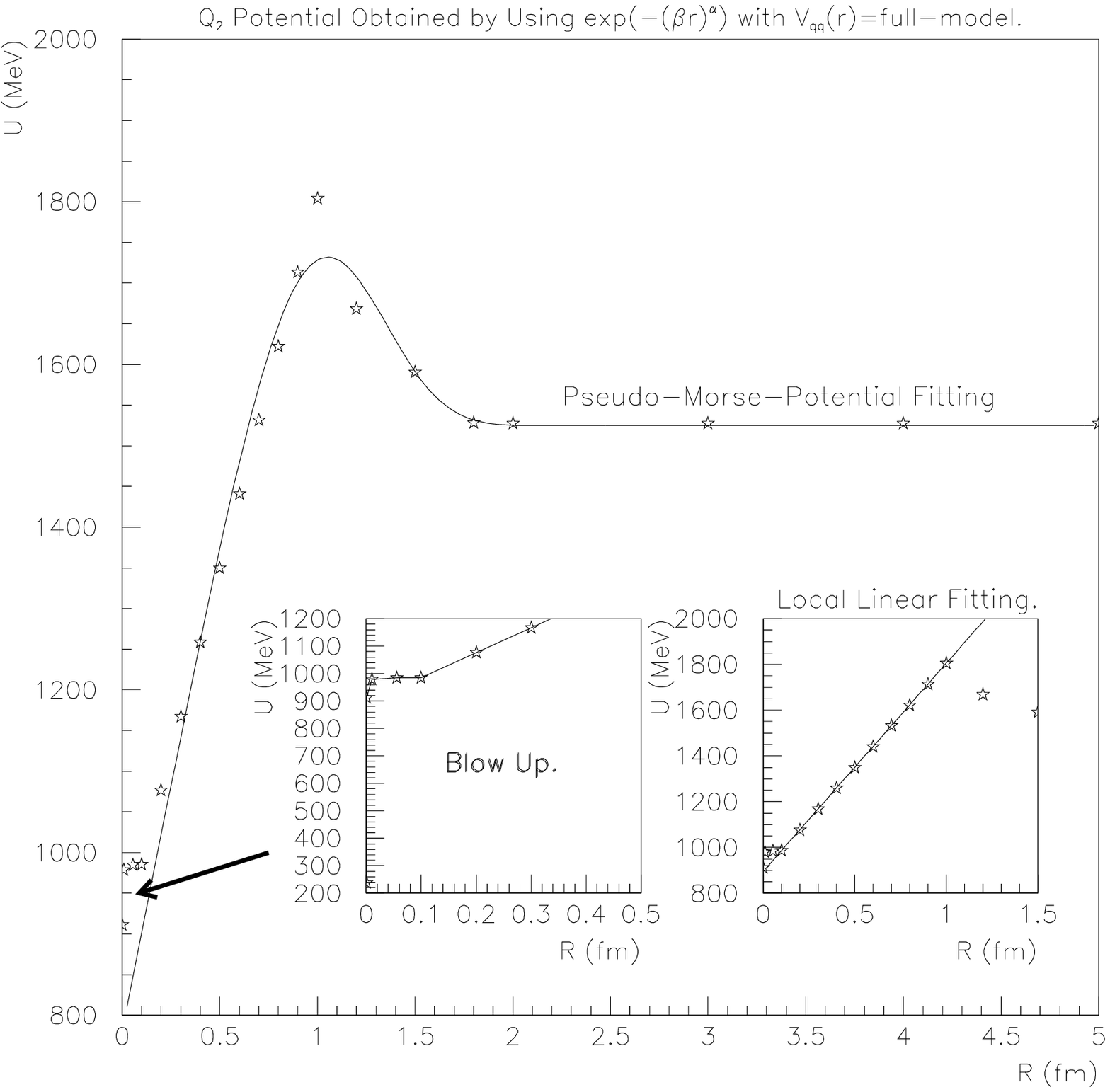}}
\end{center}
\mbox{}\vspace{-2cm}
\caption[$Q_2$ Potential for $V\sim$(flux-bubble)+(moving-colour) and 
         $\psi\sim  e^{-(\beta  r)^\alpha}$   ]{\footnotesize  The $Q_2$
         potential obtained   using  the  full  flux-bubble interaction,
         in which  the  colour is allowed  to  move around, with  a
         pseudo-hydrogen wave function.}
\label{fig:modelda}
\end{encapfig}

  The Monte Carlo results   were checked against the  analytic  solution
given by equation~\ref{eq:stringyb} at  $R=5fm$  ($\approx\infty$).  The
results are summarized in table~\ref{tb:chpaad}.
\begin{encaptab}{htbp}
\caption[MC vs. analytic results for a flux-bubble
        potential]{\footnotesize Monte Carlo  (MC)  vs. analytic results
        for a flux-bubble potential.}
\label{tb:chpaad}
\begin{tabular}{|ll|c|c|}\hline
Parameters         &           &   Analytic      & MC @ $R=5fm$
\\\hline\hline
$E_{min}(\alpha,\beta)$&$(MeV)$&  1527.07        & $1527.62\pm0.04$
   \\\hline
$\alpha$           &           &   1.74          & 1.73
   \\\hline
$\beta$            &$(fm^{-1})$&   1.37          & 1.37
   \\\hline
\end{tabular}
\end{encaptab}

  Table~\ref{tb:tfmlad} gives a summary of the global  and local fits to
the potential.
\begin{encaptab}{htbp}
\caption[Summary of Parameters for the Fits in 
        Figure~3.8]{\footnotesize Summary  of   parameters
        for the fits in figure~\ref{fig:modelda}.}
\label{tb:tfmlad}
{
\def\BETA{\beta\;(fm^{-1})}
\def\RO{r_0\;(fm)}
\def\ALPH{\alpha}
\def\ROP{r_0^\prime\;(fm)}
\def\UOI{\tilde U_\infty\;(MeV)}
\def\SIGA{\tilde\sigma\;(MeV/fm)}
\def\BEE{b\;(MeV)}
\def\SLOP{m\;(MeV/fm)}
\begin{tabular}{|l|c|c|c|}\hline
Potential & \multicolumn{3}{c|}{Parameters}\\\hline\hline
Linear
& $\SLOP$        &  $\RO$              & $\BEE$   \\\cline{2-4}
&  $910.0\pm1.0$ &  $-1.5706\pm0.0047$ & $-534.6\pm4.2$\\\hline\hline
Pseudo-Morse
& $\BETA$            &   $\RO$         &$\ALPH$  \\\cline{2-4}	
& $0.18834\pm0.00033$& $-4.423\pm0.012$&$11.29\pm0.093$\\\cline{2-4}
& $\SIGA$            & $\UOI$          &$\ROP$   \\\cline{2-4}
& $-1310.2\pm5.5$    & $1524.8\pm0.5$  &$0.6470\pm0.0011$\\\cline{2-4}
& $\eta$             &\multicolumn{2}{c}{\mbox{}}\\\cline{2-2}
& $0.9447\pm0.0024$  &\multicolumn{2}{c}{\mbox{}}\\\cline{1-2}
\end{tabular}
}
\end{encaptab}

   One of the  most noticeable peculiarities of figure~\ref{fig:modelda}
is the apparent linearity of  the inside of  the  potential. In fact,  a
linear fit to
\begin{equation}
y(r)=m(r-r_0)+b
\end{equation}
in    this region (see  the   insert in figure~\ref{fig:modelda} and the
fitted  results    in   table~\ref{tb:tfmlad})   yields   a     slope of
$m\approx910.0MeV/fm\,$!   This  corresponds  to  the  linear  potential
$U\sim\sigma  R\,$.  A  more  subtle feature is   at the origin (see the
``blow up''  insert in   figure~\ref{fig:modelda}) where  the  potential
starts to  plummet  to $-\infty$.   This  region is  due to  the coulomb
attraction between the two heavy quarks.  Out near $R=1fm$ there appears
to be  a barrier and  beyond this no more  structure.  Therefore, as two
mesons, each containing   a heavy  quark and  a  light  anti-quark,  are
brought together  from infinity they  feel a repulsive force.  When near
enough, flux-tubes are exchanged, and the two mesons dissociate into one
meson containing two heavy    quarks and another containing   two  light
anti-quarks.  This situation shows that $SU_c(2)\,$, with moving colour,
does not make a good model of nuclear matter.

  Table~\ref{tb:sumrela} gives a summary  of  all the properties of  the
$Q_2$ potential  for the  linear, linear-plus-Coulomb,  and  flux-bubble
(with fix colour) models that were studied in this section.
\begin{encaptab}{htbp}
\caption[Summary of the properties of $U(R)$]{\footnotesize 
        Summary of the properties of $U(R)$.}
\label{tb:sumrela}
{
\begin{tabular}{|lr|c|c|c|}\cline{3-5}
\multicolumn{1}{c}{\mbox}
          & &Fig.~\ref{fig:modelaa}&Fig.~\ref{fig:modelba}
            &Fig.~\ref{fig:modelca}\\
\hline
$D$                     &$(MeV)$& $3.4\pm0.1$ & $2.9\pm0.1$ 
                                & $2.8\pm0.1$\\\hline
$\mu_{\mbox{\tiny min}}$&$(GeV)$& $55\pm6$    & $77\pm6$ 
                                & $79\pm7$\\\hline
$m_{\tiny ex}$          &$(MeV)$& $638\pm15$  & $459\pm4$ 
                                & $396\pm10$\\\hline
\end{tabular}
}
\end{encaptab}
In general the extensions of the basic linear potential model to include
the  colour-Coulomb interactions did     not alter the  $Q_2$  potential
significantly enough  to achieve binding.   In fact, it only exacerbated
the  problem.  However, these extensions  did lead to a slight softening
of the  potential which  perhaps suggests  that there  might be  more to
meson exchange  than just  flux-tube   swapping.  When the  colour   was
allowed  to  move around  a rather   unphysical situation occurred which
suggested that there was perhaps a  problem with using $SU_c(2)$ or with
the variational wave function itself --- perhaps even both.  In the next
section a more detailed study of the properties  of the wave function is
carried out.

\subsection{Back to the $SU_\ell(2)$ String-Flip Potential Model}
\label{sec-bkto}
   When  the $SU_\ell(2)$ string-flip  potential  model was investigated
\cite{kn:Watson} the   parameter  $\alpha$  in   the  variational   wave
function, given in equation~\ref{eq:watson}, was fixed by requiring that
it minimize the total energy at zero  density.  Since this could be done
analytically  it allowed for a  reduction  in the  number of variational
parameters used in the Monte Carlo.  It was  assumed that the constraint
would have  very little affect on the  physics as a function  of density
since  the results varied by about  1\% for $1.5\,<\,\alpha\,<2.1\,$, at
zero density. The   validity  of this  claim  can now  be checked   more
thoroughly by using the $Q_2$ mini-laboratory.

   Figure~\ref{fig:wavetsta}  shows a plot of  the  Monte Carlo results,
obtained     {\it      via} equations   \ref{eq:stringy} \ref{eq:watwav}
and~\ref{eq:effective},  for $\bar U(R)$  where  $\alpha$ is allowed  to
vary, $\alpha=2.00\,$, and $\alpha=1.74\,$.
\begin{encapfig}{hbtp}
\begin{center}
\mbox{}\vspace{-1cm}\\
\mbox{\epsfxsize=144mm
   \epsffile{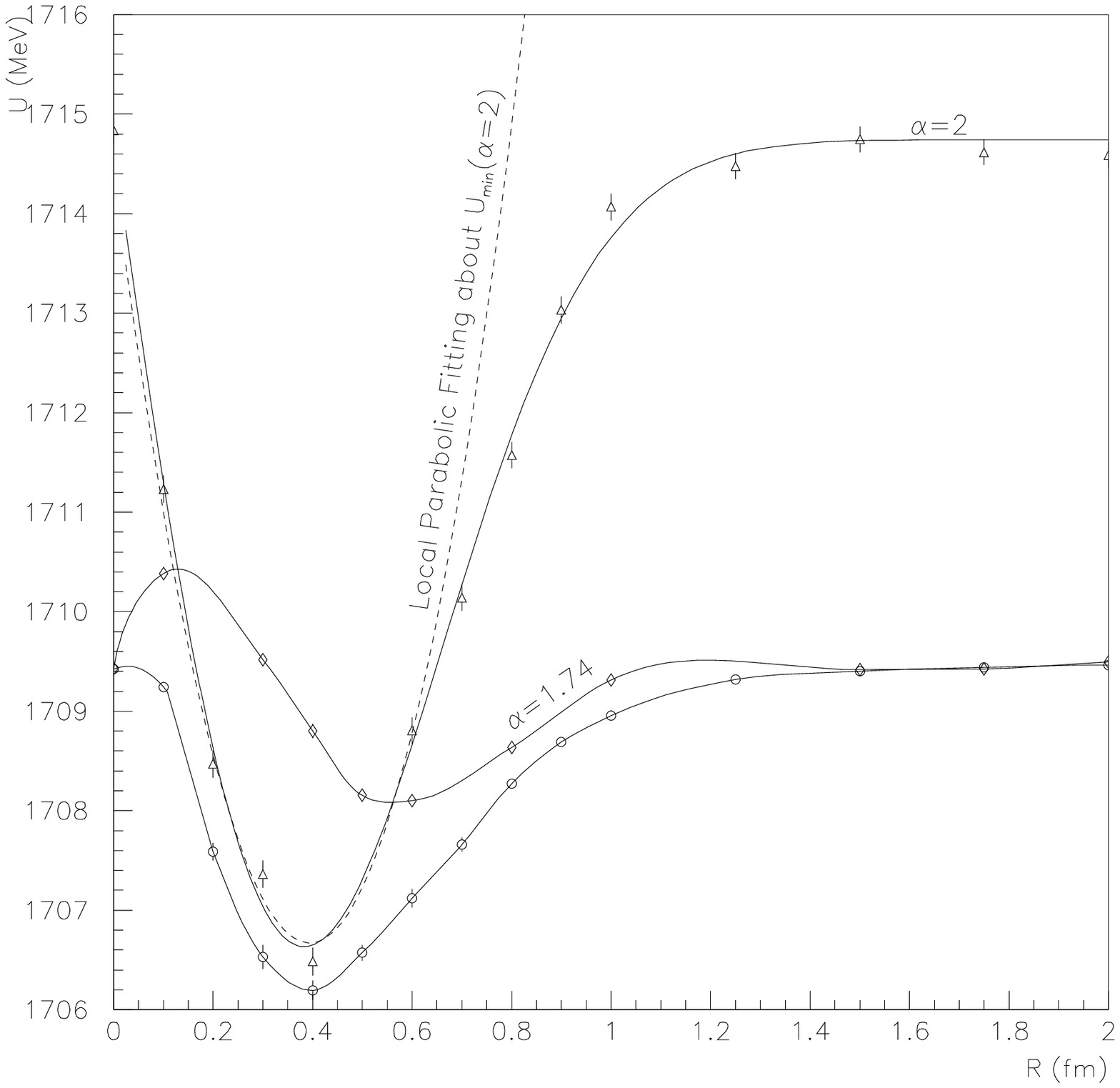}}
\end{center}
\mbox{}\vspace{-2cm}
\caption[Graph of $SU_\ell(2)$ $Q_2$ potentials with various 
         constraints on      $\alpha$]{\footnotesize  $\bar U(R)$  where
         $\alpha$  is, allowed to vary, fixed  at  2, and fixed at 1.74.
         These  potentials      were   found   by     minimizing   $\bra
         U(R)\ket_{\alpha,\beta}\,$, for fixed values  of $R$, using the
         distributed minimization algorithm (\S~\ref{sec-mindist}).  The
         $\bra  U(R)\ket_{\alpha,\beta}\,$,  were  evaluated   using the
         Metropolis algorithm (\S~\ref{sec-metrop}).}
\label{fig:wavetsta}
\end{encapfig}
\noindent
$\alpha=2.00\,$  was  the  value used    in the  old $SU_\ell(2)$  model
\cite{kn:Watson},  and $\alpha\approx1.74$ was  the  value that minimized
$\bar  U(R)$ at zero separation.  The  values of $\bar  U(R)$ at the end
points of  the curves,  from $R=0fm$ and  out to  $R=5fm$,  were checked
against the analytic solution given  by equation~\ref{eq:anyla} and have
been tabulated in Table~\ref{tb:chpa}.
\begin{encaptab}{htbp}
\caption[MC vs. analytic results for the $SU_\ell(2)$ 
         $Q_2$-potential]{\footnotesize  Monte  Carlo (MC)  vs. analytic
         results for the $SU_\ell(2)$ $Q_2$-potential.}
\label{tb:chpa}
\begin{tabular}{|ll|c|c|c|}\hline
Parameters &            & Analytic& MC @ $R=0fm$    & MC @ $R=5fm$
\\\hline\hline
$E_{min}(\alpha,\beta)$&$(MeV)$&
   1709.61 & $1709.42\pm0.04$ & $1709.49\pm0.04$
   \\\hline
$\alpha$   &            & 1.75    &  1.74            &  1.72
   \\\hline
$\beta$    &$(fm^{-1})$ & 1.37    &  1.37            &  1.38
   \\\hline\hline
$E_{min}(2.00,\beta)$&$(MeV)$&
   1714.38 & $1714.83\pm0.13$ & $1714.72\pm0.12$
   \\\hline
$\beta$ @ $\alpha=2.00$&$(fm^{-1})$       & 1.27 & 1.25 & 1.28
   \\\hline\hline
$E_{min}(1.74,\beta)$&$(MeV)$&
   1709.62 & $1709.43\pm0.04$ & $1709.43\pm0.04$\\\hline
$\beta$ @ $\alpha\approx1.74$&$(fm^{-1})$ & 1.37 & 1.38 & 1.38
   \\\hline
\end{tabular}
\end{encaptab}

  Figure~\ref{fig:wavetstb} shows   the  variations in    $\alpha\,$ and
$\beta$ as a function of $R$.
\begin{encapfig}{hbtp}
\begin{center}
\mbox{\epsfxsize=144mm
   \epsffile{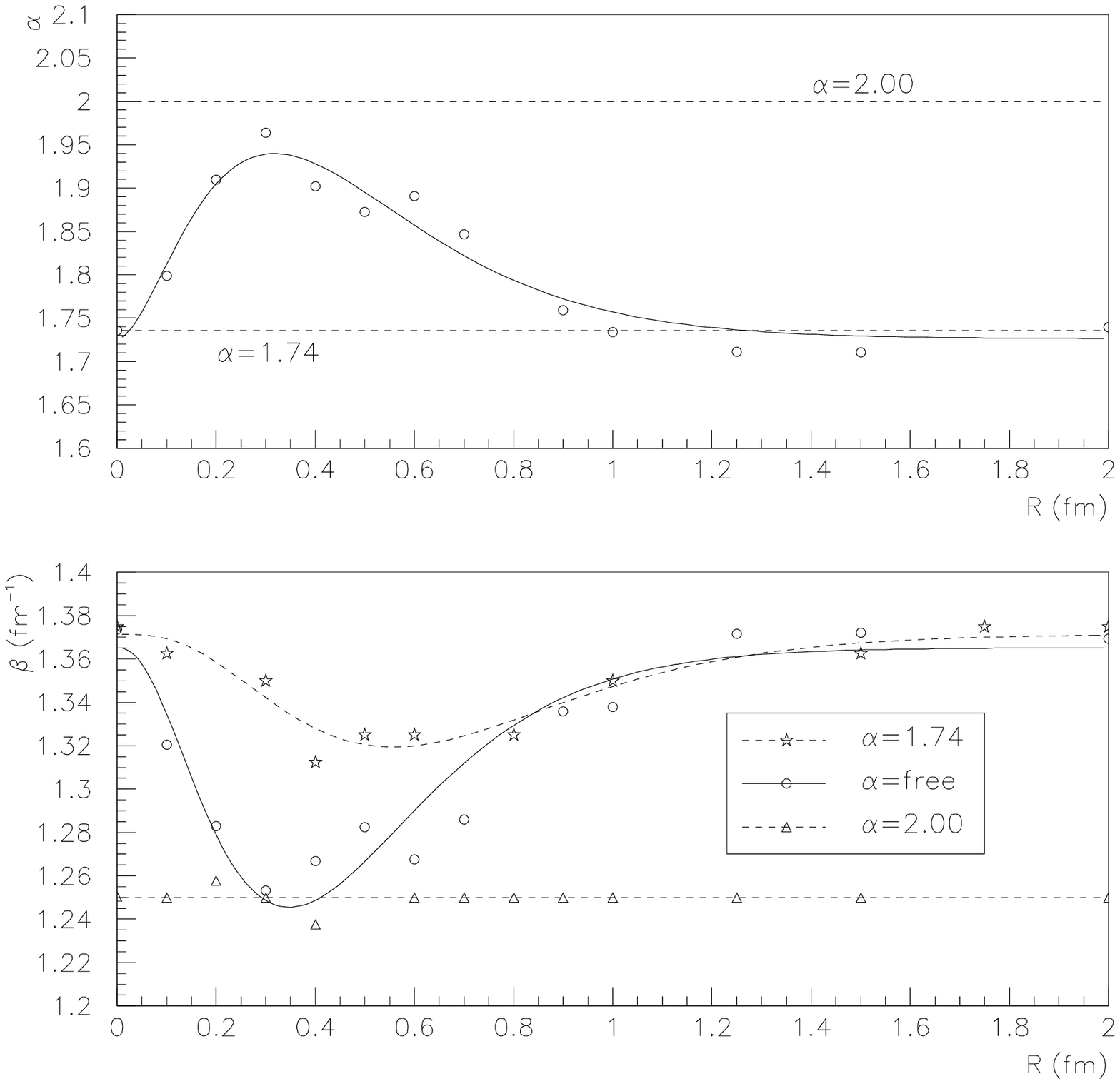}}
\end{center}
\caption[Graphs of $\alpha(R)$ and $\beta(R)$ for the 
         $SU_\ell(2)$ $Q_2$-potential   model]{\footnotesize    Graph of
         $\alpha(R)$ \&  $\beta(R)$ for the $SU_\ell(2)$ $Q_2$-potential
         model.    The    dashed   lines  are   for  $\alpha=2.00$   and
         $\alpha\approx1.74$. The noise is mainly  due to shallow minima
         on  the energy surfaces,  $\bar U(R)$.  The  curve on the upper
         plot and the two  upper  curves on   the lower plot  have  been
         parameterized by  $a\,e^{-b\,R}\,R^c\,$, which  are provided to
         guide the eye; they have no physical significance.}
\label{fig:wavetstb}
\end{encapfig}
\noindent
It would  appear  from this figure that  when  $\alpha$ is left free  to
vary, $\alpha$ and   $\beta$ become  extremely correlated.   Regardless,
looking   at the maximum  fluctuations about  the central  values of the
parameters $\alpha\,$, $\beta\,$,   and $U(R)\,$, from   the information
given   in   tables    \ref{fig:wavetstb}        and    \ref{tb:tfmlaa},
$\alpha\approx(1.88\pm0.13)\,$,  $\beta\approx(1.32\pm0.05)fm\,$,    and
$\bra\bar U(R)\ket_R\approx(1710\pm4)MeV$\footnote{{\it i.e.}, $\bra\bar
U(R)\ket_R\approx(E_{\mbox{\tiny  min}}(\alpha=2)|_{R=\infty}-y_0)/2 \pm
(\bra\bar U(R)\ket_R-y_0)/2$ where $y_0$ is from table~\ref{tb:tfmlaa}.}
which is certainly less then a 1\% effect in the energy
\cite{kn:Watson}.  However for  $\beta$ it would  appear to be an ${\cal
O}(4)$\% effect which may have  a slight affect  on the rate of  nuclear
swelling.
  
  Surprisingly the $\bar U(R)|_{\alpha=2}\,$   gives a much  deeper well
than  expected if  $\alpha$ was left  as a  free parameter:  {\it i.e.},
$D\approx{\cal O}(8)MeV\,$,  {\it via}  table~\ref{tb:tfmlba}.  However,
this  well is not deep enough  to give binding:  {\it i.e.}, a parabolic
approximation about the   minimum requires $\mu_Q\ge{\cal  O}(15)GeV\,$,
{\it via} equation~\ref{eq:parbnd} and table~\ref{tb:tfmlba}.  For equal
mass constituent quarks this is expected to be a much graver situation.

\begin{encaptab}{htbp}
\caption[Summary of Parameters for the Fits in 
        Figure~3.9]{\footnotesize Summary of the  parameters used in the
        fits for the $\alpha=2$ curve in
        figure~\ref{fig:wavetsta}.    See equations \ref{eq:pseudo}  and
        \ref{eq:parabolic} for the definitions of these parameters.\\}
\label{tb:tfmlba}
{
\def\BETA{\beta\;(fm^{-1})}
\def\RO{r_0\;(fm)}
\def\ALPH{\alpha}
\def\ROP{r_0^\prime\;(fm)}
\def\UOI{\tilde U_\infty\;(MeV)}
\def\SIGA{\tilde\sigma\;(MeV/fm)}
\def\SEE{{\cal C}\;(MeV/fm^2)}
\def\YO{y_0\;(MeV)}
\begin{tabular}{|l|c|c|c|}\hline
Potential & \multicolumn{3}{c|}{Parameters}\\\hline\hline
Parabolic
& $\SEE$         & $\RO$            & $\YO$          \\\cline{2-4}
& $49.9\pm3.5$   & $0.3944\pm0.0046$& $1706.7\pm0.1$\\\hline\hline
Pseudo-Morse
& $\BETA$        &   $\RO$          & $\ALPH$        \\\cline{2-4}
& $1.062\pm0.013$& $-0.607\pm0.016$ & $2.768\pm0.044$\\\cline{2-4}
& $\SIGA$        & $\UOI$           & 
\multicolumn{1}{c}{\mbox{}}\\\cline{2-3}
& $39.68\pm0.87$ & $1714.70\pm0.04$ & 
\multicolumn{1}{c}{\mbox{}}\\\cline{1-3}
\end{tabular}
}
\end{encaptab}

  For   certain fixed   values  of  $\alpha\,$,   $\alpha\approx1.74$ in
figure~\ref{fig:wavetsta}  for  example, the  potential  gives  a slight
short  range repulsion.   For    the  range of $\alpha$   values   being
considered here it is well inside the potential.  However, for values of
$\alpha\approx1.0$ the effect becomes quite dramatic.

  In general, it is fairly safe to assume that  the predicted outcome of
the results of  past  papers \cite{kn:Watson,kn:Boyce} will  not  change
significantly if $\alpha$ is allowed to vary.  However, because the wave
function appears  to be correlated  in both $\alpha$ and $\beta\,$, this
seems  to suggest  that another  wave function  should  be considered in
these models.

\subsection{A New Wave Function}

  Wave functions  for the old  $SU_\ell(2)$ string-flip potential models
have been  examined.  For the analysis based  on  the $Q_2$ system there
appeared to be  something pathological about the  wave function that was
used in these models.  Further, despite this shortcoming, it was thought
that this  would not have any  significant effect on  the outcome of the
physics   predicted by past models.  However,   the purpose of using the
$Q_2$ system was not only to study the properties  of this wave function
but to ascertain  the possibility of devising a  wave function with very
few  parameters  that  would  take  into  account the    locality of the
flux-bubble interactions.

  An   interesting place to  look for  such a   wave  function is in the
similarity between  the   $Q_2$   mesonic-molecular  system  and   $H_2$
molecular system.  Recall that the key reason for  the atomic binding is
the screening effect caused by the electrons which are for the most part
``localized'' in between the  protons.  This localization is achieved by
using  a  variational wave  function that  was  the superposition of the
direct product of two ground state hydrogen atoms,
\cite{kn:Schiff,kn:Heitler}
\begin{equation}
\Psi\sim\dspexp{-\beta(r_{e_1P_1}+r_{e_2P_2})}+
\dspexp{-\beta(r_{e_2P_1}+r_{e_1P_2})}\,.
\end{equation}
Although the $Q_2$  system is far removed from  its $H_2$ cousin from  a
dynamical point of view, and the  motivations for achieving localization
are quite different,  this  variational wave   function does solve   the
problem  for the  $H_2$ system.  Therefore,   it would seem plausible to
make the following $ansatz$:
\begin{equation}
\Psi_{\alpha,\beta}=
\dspexp{-\beta^\alpha(r_{\bar q_1Q_1}^\alpha+r_{\bar q_2Q_2}^\alpha)}+
\dspexp{-\beta^\alpha(r_{\bar q_2Q_1}^\alpha+r_{\bar q_1Q_2}^\alpha)}\,.
\label{eq:psdo}
\end{equation}
If  $\bar q_1Q_1$ and  $\bar q_2Q_2$ represent  two separate mesons then
the  first term represents the   internal  meson interactions while  the
second term represents the external meson interactions.  Notice that the
external interactions shut off as the separation, $R\,$, between the two
heavy quarks becomes large,
\begin{equation}
\lim_{R\rightarrow\infty}\bar \Psi(R)=
\dspexp{-\beta^\alpha(r_{\bar q_1Q_1}^\alpha+r_{\bar q_2Q_2}^\alpha)}
\label{eq:new}
\end{equation}
which is the desired property.

    Using equation~\ref{eq:Schiff} and equation~\ref{eq:new} the kinetic
energy    contribution  to  the  effective   potential     $U(R)$     in
equation~\ref{eq:effective} now becomes,
\begin{eqnarray}
\bar T &=& 2\,\bar T_{-s}-\bar F^2\\
&&\nonumber\\
&=&\frac{\alpha\beta^{\alpha}}{2m_q}\left\bra\{
   [(\alpha+1)\,(r_{\bar q_1Q_1}^{\alpha-2}+r_{\bar q_2Q_2}^{\alpha-2})-
   \alpha\beta^\alpha\,(
   r_{\bar q_1Q_1}^{2\alpha-2}+r_{\bar q_2Q_2}^{2\alpha-2})]\,
   \dspexp{-2\beta^\alpha(r_{\bar q_1Q_1}^\alpha+r_{\bar q_2Q_2}^\alpha)}
   \right.\nonumber\\\nonumber\\&&+\,
   [(\alpha+1)\,(r_{\bar q_2Q_1}^{\alpha-2}+r_{\bar q_1Q_2}^{\alpha-2})-
   \alpha\beta^\alpha\,(
   r_{\bar q_2Q_1}^{2\alpha-2}+r_{\bar q_1Q_2}^{2\alpha-2})]\,
   \dspexp{-2\beta^\alpha(r_{\bar q_2Q_1}^\alpha+r_{\bar q_1Q_2}^\alpha)}
   \nonumber\\\nonumber\\&&+\,
   [(\alpha+1)\,(r_{\bar q_1Q_1}^{\alpha-2}+r_{\bar q_2Q_2}^{\alpha-2}
   +r_{\bar q_2Q_1}^{\alpha-2}+r_{\bar q_1Q_2}^{\alpha-2})
   \nonumber\\\nonumber\\&&-
   \alpha\beta^\alpha\,(
   r_{\bar q_1Q_1}^{2\alpha-2}+r_{\bar q_2Q_2}^{2\alpha-2}
   +r_{\bar q_2Q_1}^{2\alpha-2}+r_{\bar q_1Q_2}^{2\alpha-2})]
   \nonumber\\\nonumber\\&&\left.\times
   \dspexp{-\beta^\alpha(r_{\bar q_1Q_1}^\alpha+r_{\bar q_2Q_1}^\alpha
           +r_{\bar q_1Q_2}^\alpha+r_{\bar q_2Q_2}^\alpha)}
\}/\Psi^2\right\ket\,,
\end{eqnarray}
where
\begin{eqnarray}
\bar T_{-s}&=&\frac{-1}{4m_q}\sum_{\bar q}\nabla_{\bar q}^2\ln\Psi\\
&&\nonumber\\
&=&\frac{\alpha\beta^\alpha}{4m_q}\left\bra
\frac{(\alpha+1)}{\Psi}\left[
   (r_{\bar q_1Q_1}^{\alpha-2}+r_{\bar q_2Q_2}^{\alpha-2})\,
   \dspexp{-\beta^\alpha(r_{\bar q_1Q_1}^\alpha+r_{\bar q_2Q_2}^\alpha)}
   \right.\right.\nonumber\\\nonumber\\&&\left.+\,
   (r_{\bar q_2Q_1}^{\alpha-2}+r_{\bar q_1Q_2}^{\alpha-2})\,
   \dspexp{-\beta^\alpha(r_{\bar q_2Q_1}^\alpha+r_{\bar q_1Q_2}^\alpha)}
\right]\nonumber\\\nonumber\\&&+\,
\frac{\alpha\beta^\alpha}{\Psi^2}\,[
   (r_{\bar q_1Q_1}^2+r_{\bar q_1Q_2}^2-R^2)\,
   r_{\bar q_1Q_1}^{\alpha-2}\,r_{\bar q_1Q_2}^{\alpha-2}+
   (r_{\bar q_2Q_1}^2+r_{\bar q_2Q_2}^2-R^2)\,
   r_{\bar q_2Q_1}^{\alpha-2}\,r_{\bar q_2Q_2}^{\alpha-2}
   \nonumber\\\nonumber\\&&\left.-\,
   (r_{\bar q_1Q_1}^{2\alpha-2}+r_{\bar q_2Q_2}^{2\alpha-2}+
    r_{\bar q_2Q_1}^{2\alpha-2}+r_{\bar q_1Q_2}^{2\alpha-2})
]\,
   \dspexp{-\beta^\alpha(r_{\bar q_1Q_1}^\alpha+r_{\bar q_2Q_1}^\alpha
           +r_{\bar q_1Q_2}^\alpha+r_{\bar q_2Q_2}^\alpha)}
\right\ket\,,
\nonumber\\&&\nonumber\\&&
\end{eqnarray}
and
\begin{eqnarray}
\bar F^2&=&\frac{1}{2m_q}\sum_{\bar q}(\nabla_{\bar q}\ln\Psi)^2\\
&&\nonumber\\
&=&\frac{\alpha^2\beta^{2\alpha}}{2m_q}
\left\bra\frac{1}{\Psi^2}\,\left[
   (r_{\bar q_1Q_1}^{2\alpha-2}+r_{\bar q_2Q_2}^{2\alpha-2})\,
   \dspexp{-2\beta^\alpha(r_{\bar q_1Q_1}^\alpha+r_{\bar q_2Q_2}^\alpha)}
   \right.\right.\nonumber\\\nonumber\\&&\left.+\,
   (r_{\bar q_2Q_1}^{2\alpha-2}+r_{\bar q_1Q_2}^{2\alpha-2})\,
   \dspexp{-2\beta^\alpha(r_{\bar q_2Q_1}^\alpha+r_{\bar q_1Q_2}^\alpha)}
\right]\nonumber\\\nonumber\\&&+\,[
   (r_{\bar q_1Q_1}^2+r_{\bar q_1Q_2}^2-R^2)\,
   r_{\bar q_1Q_1}^{\alpha-2}\,r_{\bar q_1Q_2}^{\alpha-2}+
   (r_{\bar q_2Q_1}^2+r_{\bar q_2Q_2}^2-R^2)\,
   r_{\bar q_2Q_1}^{\alpha-2}\,r_{\bar q_2Q_2}^{\alpha-2}]
   \nonumber\\\nonumber\\&&\left.\times
   \dspexp{-\beta^\alpha(r_{\bar q_1Q_1}^\alpha+r_{\bar q_2Q_1}^\alpha
           +r_{\bar q_1Q_2}^\alpha+r_{\bar q_2Q_2}^\alpha)}
\right\ket\,.
\end{eqnarray}
It is interesting to  note, and it  also serves as  a good check of  the
computation, that in the large  $R$ limit the kinetic energy corresponds
exactly to  the  kinetic energy for  the  old wave function in  the same
limit:
\begin{eqnarray}
\lim_{R\rightarrow\infty}\bar T(R)&=&
\frac{\alpha\beta^{\alpha}}{2m_q}\left\bra
(\alpha+1)\,
(r_{\bar q_1Q_1}^{\alpha-2}+r_{\bar q_2Q_2}^{\alpha-2})-
\alpha\beta^\alpha\,
(r_{\bar q_1Q_1}^{2\alpha-2}+r_{\bar q_2Q_2}^{2\alpha-2})
\right\ket\nonumber\\&&\nonumber\\
&=&\frac{\alpha\beta^{\alpha}}{2m_q}\left\bra\sum_{i=1}^2[
\alpha(1-(\beta r_{\bar q_iQ_i})^\alpha)+1]\,r_{\bar q_iQ_i}^{\alpha-2}
]\right\ket\,,
\end{eqnarray}
{\it cf}.  equation~\ref{eq:effectr}. Also direct  evaluation of the RHS
leads to (assuming $\Psi$ is properly normalized)
\begin{equation}
\lim_{R\rightarrow\infty}\bar T(R)=\frac{g_T(\alpha)}{m_q}\,\beta^2\,,
\end{equation}
which is  just   the kinetic term for  the   analytic solutions given  in
equations \ref{eq:anaa} and \ref{eq:anab}.

  The Monte Carlo computations that were done, using the pseudo-hydrogen
wave function, $\PH$ (equation~\ref{eq:watwav}) in \S~\ref{sec-general},
have been repeated here for the pseudo-hydrogen-molecular wave function,
$\PHH$        (equation~\ref{eq:psdo}),      and    are    shown      in
figures~\ref{fig:modelae}             through         \ref{fig:modelde}.
Table~\ref{tb:tfmlab} contains a summary  of the fits for these figures,
table~\ref{tb:tfmlbaa} contains a summary of the checks done against the
analytic results and table~\ref{tb:sumrelaa} contains  a summary of  the
properties of $U(R)\,$.

\begin{encapfig}{P}
\begin{center}
\mbox{}\vspace{-1cm}\\
\mbox{\epsfxsize=144mm
   \epsffile{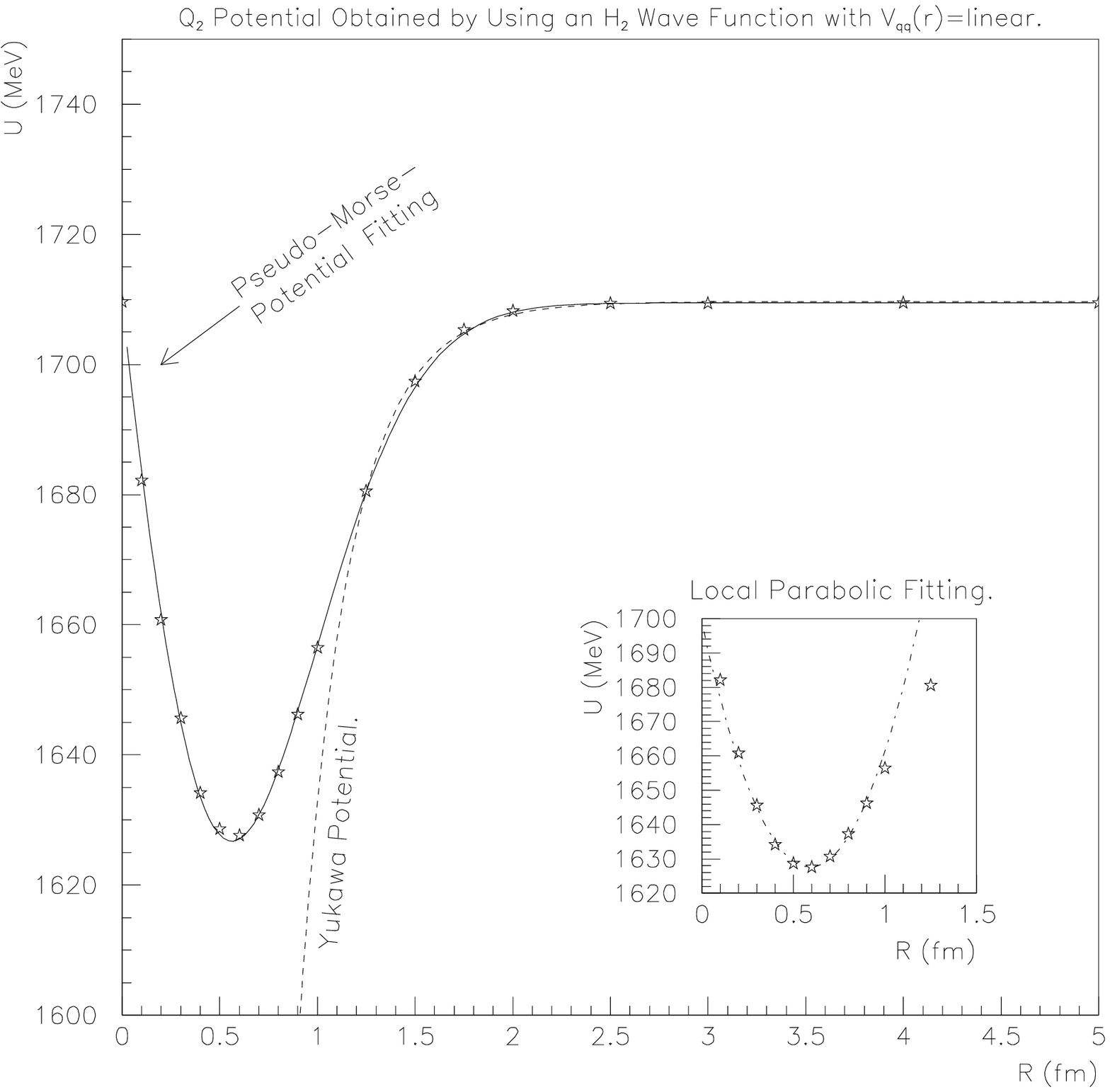}}
\end{center}
\mbox{}\vspace{-2cm}
\caption[$Q_2$ Potential for $V\sim\sigma r$ and 
         $\psi\sim\tilde\psi_{H_2}$]{\footnotesize The  $Q_2$  potential
         obtained using a linear interaction with $\PHH\,$.}
\label{fig:modelae}
\end{encapfig}

\begin{encapfig}{P}
\begin{center}
\mbox{}\vspace{-1cm}\\
\mbox{\epsfxsize=144mm
   \epsffile{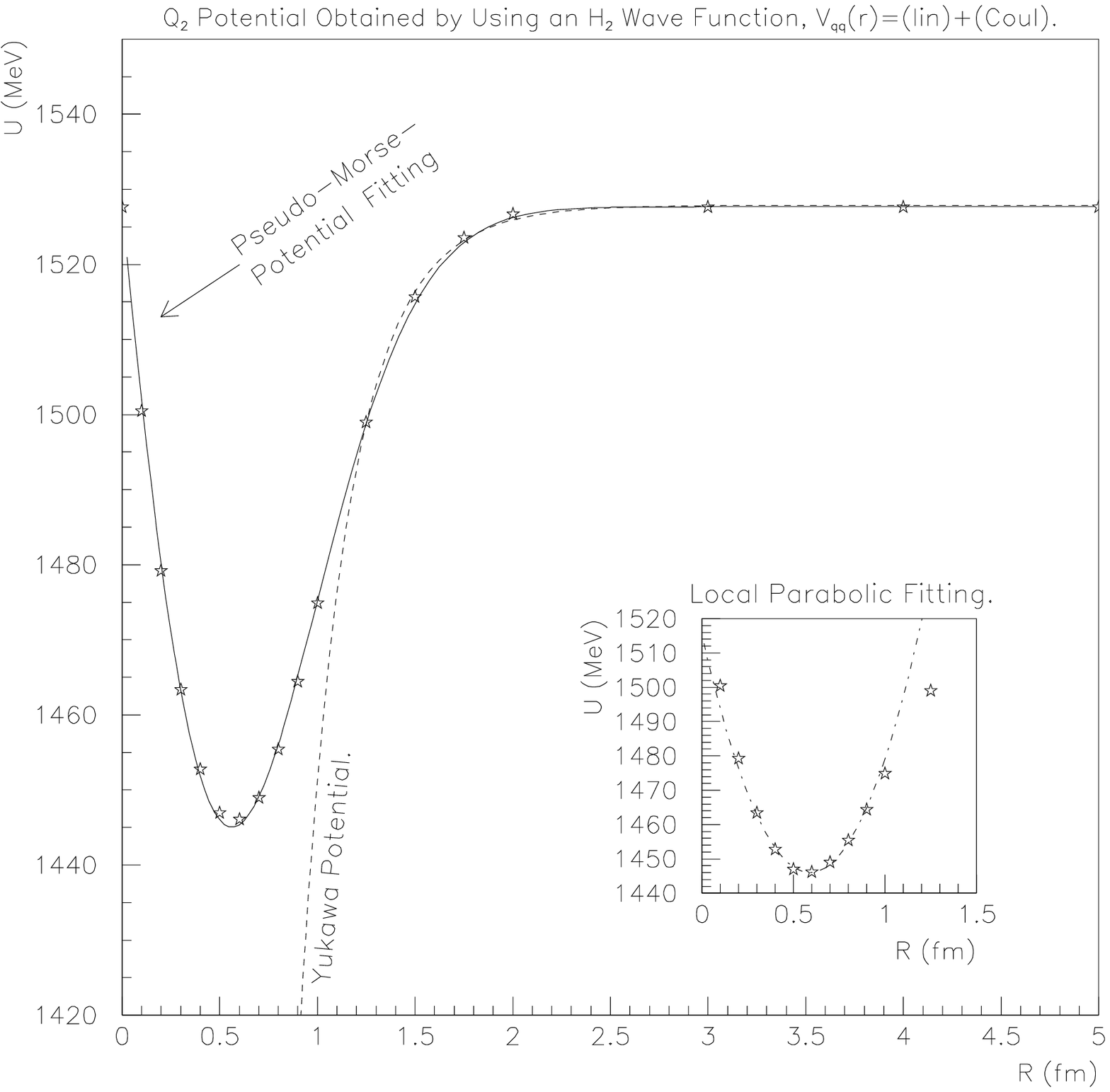}}
\end{center}
\mbox{}\vspace{-2cm}
\caption[$Q_2$ Potential for $V\sim\sigma r-\alpha_s/r$ and 
         $\psi\sim\tilde\psi_{H_2}$]{\footnotesize The $Q_2$   potential
         obtained  using  a     linear-plus-Coulomb   interaction   with
         $\PHH\,$.}
\label{fig:modelbe}
\end{encapfig}

\begin{encapfig}{P}
\begin{center}
\mbox{}\vspace{-1cm}\\
\mbox{\epsfxsize=144mm
   \epsffile{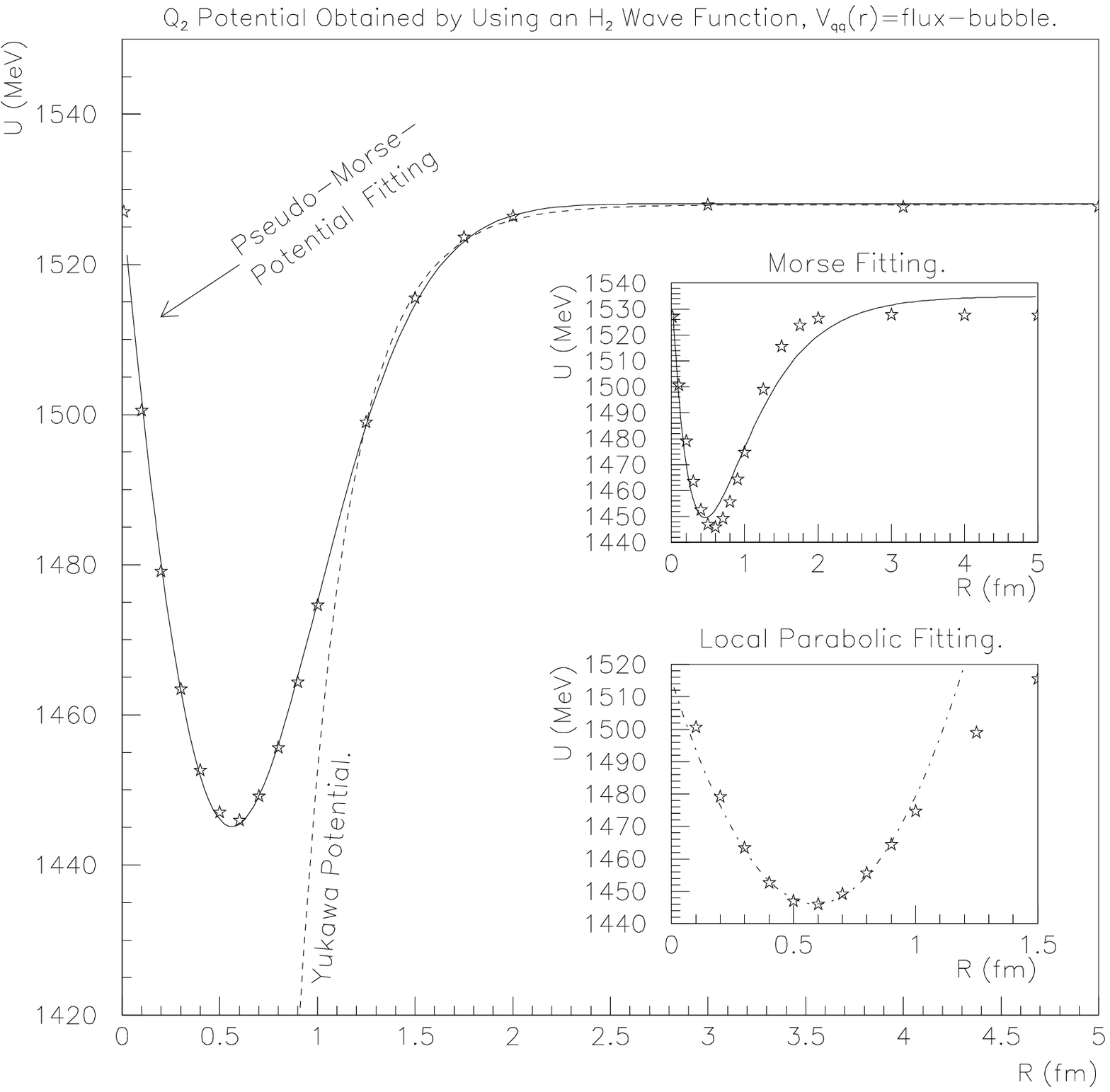}}
\end{center}
\mbox{}\vspace{-2cm}
\caption[$Q_2$ Potential for $V\sim$(flux-bubble) and 
         $\psi\sim\tilde\psi_{H_2}$]{\footnotesize The   $Q_2$ potential
         obtained  using  a     flux-bubble  interaction with    $\PHH\,$.}
\label{fig:modelce}
\end{encapfig}

\begin{encapfig}{P}
\begin{center}
\mbox{}\vspace{-1cm}\\
\mbox{\epsfxsize=144mm
   \epsffile{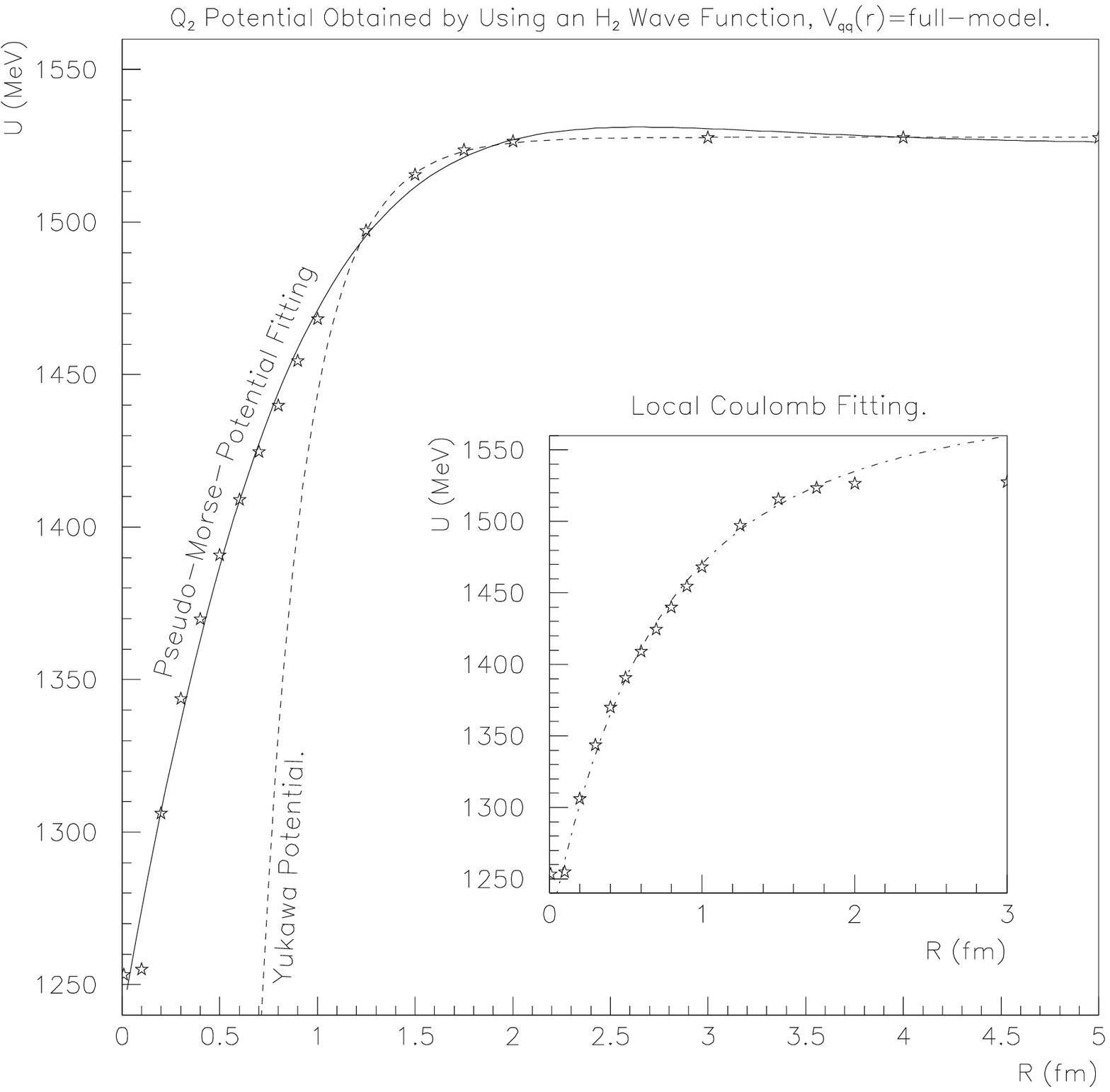}}
\end{center}
\mbox{}\vspace{-2cm}
\caption[$Q_2$ Potential for $V\sim$(flux-bubble)+(moving-colour) and 
         $\psi\sim\tilde\psi_{H_2}$]{\footnotesize The  $Q_2$  potential
         obtained using the full  flux-bubble interaction, in  which the
         colour is allowed to move around, with  $\PHH\,$.  It should be
         noted that there   is a similar  effect in  the neighborhood of
         $R=0$ as shown in the ``blow up'' of figure~\ref{fig:modelda}.}
\label{fig:modelde}
\end{encapfig}

\begin{table}[P]\begin{center}
\caption[All of the MC Fits Revisited]{\footnotesize 
        Fits      for         figures      \ref{fig:modelae} through
        \ref{fig:modelde}.}
\label{tb:tfmlaba}
{
\def\BETA{\beta}
\def\RO{r_0}
\def\ALPH{\alpha}
\def\ROP{r_0^\prime}
\def\UOI{\tilde U_\infty}
\def\UO{U_0}
\def\UI{U_\infty}
\def\SIGA{\tilde\sigma}
\def\VOY{V_0}
\def\VOI{V_\infty}
\def\EH{a}
\def\SEE{{\cal C}}
\def\YO{y_0}
\def\PSM{\mbox{
   \setlength{\unitlength}{1ex}
   \begin{picture}(0.0,0.0)
   \put(-1.5,1.0){\shortstack{P\\s\\e\\u\\d\\o\\$|$\\M\\o\\r\\s\\e}}
   \end{picture}
}}
\def\YUK{\mbox{
   \setlength{\unitlength}{1ex}
   \begin{picture}(0.0,0.0)
   \put(-1.5,0.0){\shortstack{Y\\u\\k\\a\\w\\a}}
   \end{picture}
}}
\def\PAR{\mbox{
   \setlength{\unitlength}{1ex}
   \begin{picture}(0.0,0.0)
   \put(-1.5,1.0){\shortstack{P\\a\\r\\a\\b}}
   \end{picture}
}}
\def\MOR{\mbox{
   \setlength{\unitlength}{1ex}
   \begin{picture}(0.0,0.0)
   \put(-1.5,3.0){\shortstack{M\\o\\r\\s\\e}}
   \end{picture}
}}
\def\CLB{\mbox{
   \setlength{\unitlength}{1ex}
   \begin{picture}(0.0,0.0)
   \put(-1.5,2.0){\shortstack{C\\o\\u\\l}}
   \end{picture}
}}
\def\RSL{\mbox{
   \setlength{\unitlength}{1ex}
   \begin{picture}(0.0,0.0)
   \put(-1.5,-0.5){\bf\scriptsize\shortstack{R\\E\\S\\U\\L\\T\\S}}
   \end{picture}
}}
\begin{tabular}{|c|l|c|c|c|c|}\hline
V  & \multicolumn{5}{c|}{
   Parameters ([Energy, Mass]$\sim MeV\,$, [Length]$\sim fm$)
}\\\cline{3-6}
          & &Fig.~\ref{fig:modelae}&Fig.~\ref{fig:modelbe}
            &Fig.~\ref{fig:modelce}&Fig.~\ref{fig:modelde}\\
\hline\hline
   & $\BETA$ & $0.727\pm0.060$  & $0.758\pm0.027$ 
             & $0.736\pm0.056$  & $0.686\pm0.011$\\\cline{2-6} 
   & $\RO$   &$-0.957\pm0.096$  &$-0.893\pm0.057$ 
             & $-0.94\pm0.12$   & $-1.527\pm0.022$\\\cline{2-6} 
   & $\ALPH$ & $2.50\pm0.11$    & $2.429\pm0.062$ 
             & $2.45\pm0.13$    & $1.367\pm0.034$\\\cline{2-6} 
   & $\SIGA$ & $308.3\pm8.9$    & $304.5\pm8.9$   
             & $313\pm11$       & $-260.8\pm9.3$\\\cline{2-6} 
   & $\ROP$  &        ---       &        ---      
             &        ---       & $1.926\pm0.051$\\\cline{2-6}
$\PSM$&$\UOI$& $1709.5\pm0.4$   & $1527.7\pm0.4$  
             & $1528.1\pm0.5$   & $1525.6\pm0.79$\\ 
\hline\hline
   & $\VOY$  & $4400\pm1700$    & $4500\pm2000$   
             & $4000\pm1800$    & $6000\pm2200$\\\cline{2-6} 
   & $\EH $  & $0.327\pm0.027$  & $0.335\pm0.031$ 
             & $0.343\pm0.032$  & $0.321\pm0.025$\\\cline{2-6} 
$\YUK$&$\VOI$& $1709.7\pm0.5$   & $1527.9\pm0.5$  
             & $1528.0\pm0.6$   & $1527.8\pm0.5$\\
\hline\hline
      & $\SEE$ & $203\pm11$      & $198\pm11$      
               & $198\pm10$      & ---\\\cline{2-6} 
      & $\RO $ & $0.591\pm0.047$ & $0.590\pm0.048$ 
               & $0.590\pm0.048$ & ---\\\cline{2-6} 
$\PAR$& $\YO $ & $1627.6\pm0.6$  & $1446.0\pm0.6$  
               & $1446.1\pm0.6$  & ---\\
\hline\hline
      & $\BETA$ & --- & --- & $1.536\pm0.013$   & ---\\\cline{2-6} 
      & $\RO  $ & --- & --- & $0.4613\pm0.0028$ & ---\\\cline{2-6} 
      & $\UO  $ & --- & --- & $85.47\pm0.69$    & ---\\\cline{2-6} 
$\MOR$& $\UI  $ & --- & --- & $1535.0\pm0.6$    & ---\\
\hline\hline
      & $\EH  $ & --- & --- & --- & $-150.39\pm2.9$  \\\cline{2-6} 
      & $\BETA$ & --- & --- & --- & $2.628\pm0.042$\\\cline{2-6} 
$\CLB$& $\VOI $ & --- & --- & --- & $1609.8\pm2.3$  \\
\hline
\end{tabular}
}
\end{center}\end{table}

\begin{encaptab}{htbp}
\caption[All of the MC vs. Analytic Results Revisited]{
        \footnotesize   MC    vs.     analytic    results for    figures
        \ref{fig:modelae} through \ref{fig:modelde}.}
\label{tb:tfmlbaa}
{
\def\RRR{R\;(fm)}
\def\EMN{E_{min}\;(MeV)}
\def\ALP{\alpha}
\def\BTA{\beta\;(fm^{-1})}
\begin{tabular}{|lr|c|c|c|c|}\hline
Method& &$\RRR$&        $\EMN$    &$\ALP$&$\BTA$\\\hline\hline
Analytic&Eq.~\ref{eq:anaa}
& $0/\infty$ &  1709.61           & 1.75 & 1.37 \\\hline
MC&Fig.~\ref{fig:modelae}
&      0  & $1709.647\pm0.032$ & 1.77 & 1.36 \\\cline{3-6}
& &    5  & $1709.485\pm0.033$ & 1.75 & 1.36 \\\hline\hline
Analytic&Eq.~\ref{eq:anab}
& $0/\infty$ &    1527.07      & 1.74 & 1.37 \\\hline
MC&Fig.~\ref{fig:modelbe}
&      0  & $1527.621\pm0.043$ & 1.73 & 1.38 \\\cline{3-6}
& &    5  & $1527.651\pm0.057$ & 1.70 & 1.38 \\\hline
MC&Fig.~\ref{fig:modelce}
&      5  & $1527.715\pm0.042$ & 1.73 & 1.37 \\\hline
MC&Fig.~\ref{fig:modelde}
&      5  & $1527.703\pm0.043$ & 1.73 & 1.36 \\\hline
\end{tabular}
}
\end{encaptab}

\begin{encaptab}{htbp}
\caption[Summary of the properties of $U(R)$ Revisited]{\footnotesize 
        Summary  of  the properties  of $U(R)$  from parabola and Yukawa
        data  in  table~\ref{tb:tfmlaba}, with $U_\infty$ extracted from
        analytic results in   table~\ref{tb:tfmlba}.  For Morse  data $D
        \approx (85.47\pm0.69) MeV$  and $\mu_{\mbox{\tiny min}} \approx
        (134\pm2) MeV$.\\}
\vspace{-5.5ex}
\label{tb:sumrelaa}
{
\begin{tabular}{|lr||c|c|c|c|c|}\cline{4-7}
\multicolumn{2}{c}{\mbox{}}
           & &Fig.~\ref{fig:modelae}&Fig.~\ref{fig:modelbe}
              &Fig.~\ref{fig:modelce}&Fig.~\ref{fig:modelde}\\
\hline
$D$                     &$(MeV)$& Eq.~\ref{eq:depth} 
    & $82.0\pm0.6$ & $81.1\pm0.6$ & $81.0\pm0.6$ &---\\\hline
$\mu_{\mbox{\tiny min}}$&$(MeV)$& Eq.~\ref{eq:parbnd} 
    & $588\pm33$   & $586\pm34$ & $588\pm31$   &---\\\hline
$m_{\mbox{\tiny ex}}$   &$(MeV)$& Eq.~\ref{eq:exchange} 
    & $603\pm50$   & $589\pm55$ & $575\pm54$   & $615\pm48$\\
\hline
\end{tabular}
}
\end{encaptab}
\clearpage
   It can be immediately seen that there  is a dramatic contrast between
the figures  for   $\PH$ and $\PHH\,$.  For    figures \ref{fig:modelae}
through \ref{fig:modelce}  the wells  are  much deeper and  the  binding
energy    constraints   have    come       down     considerably,    see
table~\ref{tb:sumrelaa}    for  $D$    and  $\mu_{\mbox{\tiny     min}}$
($\le\mu_Q$).  These potentials  are now strong  enough to bind the more
massive quarks, such  as $c$ and $b$.  This  should be of no surprise as
the adiabatic  approximation used requires that  the quarks  be massive.
The     effective  Yukawa      masses,     $m_{\mbox{\tiny   ex}}$    in
table~\ref{tb:sumrelaa}, are roughly the  same order of magnitude as the
$\PH\,$ case, table~\ref{tb:sumrela}, but  with a slightly less dramatic
softening effect. However,  these masses are  too large to explain  pion
exchange.

   An attempt was made to fit the flux-bubble model to a Morse potential
but as before  the fit failed  (see insert in figure~\ref{fig:modelce}).
It is interesting to note however that the pseudo-Morse potential of the
form
\begin{equation}
\tilde U(r)\sim\left\{e^{-2\beta^\alpha(r-r_0)^\alpha}-
2e^{-\beta^\alpha(r-r_0)^\alpha}\right\}\,\tilde\sigma\,r
\end{equation}
({\it  cf}.    equation~\ref{eq:pseudo})    describes  $\bar  U(R)$   in
figures~\ref{fig:modelae} through \ref{fig:modelce} quite well.

   The final figure, figure~\ref{fig:modelde},  of the flux-bubble model
with  moving    colour  is    quite  intriguing.   The     anomalies  in
figure~\ref{fig:modelda}  have  disappeared;  the  light quarks have not
drifted away as an isolated pair to leave a linear potential between the
heavy quarks.  In  fact,  there is  no  more linearity inside  the well.
Na\"{\i}vely, this seems to  suggest that the  problem was with the wave
function and  not $SU_c(2)$.  However this  is not quite the case, since
the   old wave  function gave  an  interior  well  depth  twice as deep.
Therefore, it is  energetically more favourable for  the $Q_2$ system to
dissociate into  two isolated mesons; one  with two light quarks and the
other with two heavy ones.

  An attempt  was also made  to find the  shape  of the interior  of the
well, and  it was found that  it fitted  to a  Coulomb  potential of the
form,
\begin{equation}
V(r)=\frac{a}{r}\,\left(1-\dspexp{-\beta r}\right)+V_\infty
\end{equation}
(see insert  in figure~\ref{fig:modelde}, and table~\ref{tb:tfmlaba} for
fitted  data), with  strength  $a\approx-(150.39\pm2.9)MeV\,fm\,$.   The
term  in the brackets was  included   to mimic the  overlap between  the
charge distributions of  two mesonic systems.   In terms  of $\alpha_s$,
the hyperfine constant for this region of the potential is
\begin{equation}
\alpha_{Q_2}\approx(7.62\pm0.15)\alpha_s\,.
\end{equation}
The interior  part of the  potential is  quite deep, ${\cal O}(270)MeV$,
and bottoms  out   at  ${\cal O}(1255)$  $MeV$,   at   which point   the
$-\alpha_s/r$ term for the     heavy quarks kicks  in ({\it    i.e.}, at
$R=0.1fm$).   The  exterior  part of  the  potential  fits  to  a Yukawa
potential with $m_{\mbox{\tiny  ex}}\approx{\cal O}(600)MeV$, {\it  via}
table~\ref{tb:sumrelaa}.

\section{Discussion}
\label{sec-chpdsc}

   Various aspects   of  model building  for  nuclear  matter  have been
examined in the $Q_2$ system.  The ramifications of these investigations
in regards to nuclear modeling will now be discussed.

   Perhaps  the most  important   result is  the  new variational   wave
function, equation~\ref{eq:psdo}.   This new wave  function made a large
change in the depth of the $Q_2$ potential well.  The depth increased by
a  factor   of $27$, deep   enough  to bind  heavy  quarks:  {\it i.e.},
$m_q\ge{\cal O}(m_c)$.  The  wave function also fulfills the requirement
of handling local flux-bubble  interactions, which becomes apparent when
looking at the before and after pictures (of moving colour) in figures
\ref{fig:modelda}  and \ref{fig:modelde},  respectively.  Therefore,  in
$SU_\ell(2)$ for  a many quark  system this would  suggest the following
$ansatz$:
\begin{equation}
\Psi\sim{\rm Perm}|\PH(r_{p_ip_j})|\,
\prod_{\mbox{\tiny colour}}|\Phi(r_{p_k})|
\end{equation}
where
\begin{equation}
\PH(r_{p_ip_j})=\dspexp{-(\beta r_{p_ip_j})^\alpha}\,,
\end{equation}
${\rm  Perm}|\PH(r_{p_ip_j})|$  is  a totally  symmetric pseudo-hydrogen
wave function and  $|\Phi(r_{p_k})|$  is a totally antisymmetric  Slater
wave function (for  the full three  quark system a similar wave function
would  apply).  This does not  necessarily mean that  this wave function
would  lead  to nuclear binding,   since  binding was  only achieved for
relatively very heavy quarks in the  $Q_2\,$ system.  However, given the
order of magnitude of  increase in the  $Q_2$ well  depth it would  seem
quite plausible that it  might be a  strong  enough effect to produce  a
shallow  well in the nuclear-binding-energy  curve.  A simple test would
be to  consider $SU_\ell(2)$  with  just a  string-flip  potential, with
$\alpha$  fixed, in which case scaling  is restored and  the Monte Carlo
becomes quite straightforward to do.

  The flux-bubble  model   proved  quite  successful at     combining the
colour-Coulomb interactions  with the  flux-tube interactions.  Although
these interactions,  in  general, had very  little affect   on the $Q_2$
system it was useful in demonstrating that extending the flux-tube model
to include local perturbative interactions can be done.  Furthermore, it
was   not  surprising that  this    had very  little   effect  as  it was
hypothesized, in chapter~2, that  the hyperfine interactions should play
the dominant  role.    Therefore, it would   prove   most interesting to
investigate the effect  of adding more  perturbative interactions to the
$Q_2$ system.  With the addition of $SU_c(3)$  this would lead to a more
realistic model  of mesonic molecules which perhaps  could  be tested in
the laboratory.

  When the $SU_c(2)$ flux-bubble model was considered with moving colour
the results were quite   interesting; the $Q_2$ system dissociated  into
one light  and one heavy meson.   However, this  system is not physical.
Perhaps a more useful model would  be to consider the  heavy quarks as a
composite of two light quarks and using $SU_c(3)$ instead.  In this case
it would not be possible  for a flux-tube to form  between the two heavy
quarks causing the system to dissociate. 

  Curiously if the heavy quarks were considered to be a composite of two
light $u$ quarks then
\begin{equation}
\mu\raisebox{-0.75ex}{\tiny$Q\supseteq\{uu\}$}/
\mu_{\mbox{\tiny min}}
\approx{\cal O}(0.6)
\end{equation}
which is getting closer  to binding, but is not  quite sufficient.  This
would seem   to    suggest that  perhaps   slight   nucleon deformations
\cite{kn:CarlsonB} would   help   to  enhance  binding.    Therefore, an
interesting possibility would be to consider a many-body $SU_\ell(2)$ or
$SU_c(3)$ (with moving   colour) model in which   there is an  imbalance
between the quark and anti-quark masses.  In  the case of $SU_c(3)\,$, a
strong enhancement would   be expected given the   depths of the   wells
observed in the $Q_2$ system with $SU_c(2)\,$.

  The distributed minimization  algorithm proved to be quite successful.
The algorithm as it stands  should have very little difficulty  handling
$SU_\ell(2)$ models  in which both $\alpha$ and   $\beta$ are allowed to
vary.   The reason  this can be  said with  any  confidence is that  the
energy-surface topologies of  the many-body $SU_\ell(2)$ models are very
similar to those  in the $Q_2$  system.  From a  practical point of view
the projected time for a network of $m$ computers is
\begin{equation}
\tau\sim{\cal O}(\frac{324}{m})hrs\,,
\end{equation}
for a fixed density parameter, $\rho$, which is just barely
tolerable for $m=8$. If $\alpha$ is fixed then
\begin{equation}
\tau\sim{\cal O}(\frac{162}{m})hrs\,,
\end{equation}
which is more reasonable.

  For $SU_\ell(3)$  the projected times  increase by a factor  of ${\cal
O}(4)$. The simplest  way to circumvent  this problem is to increase the
number  of computers.    With the resources    here, at Carleton, it  is
possible to  go to $m=16$ by  running off of  the 8-node $CPU$ farm with
some slight modifications to the  client-server routine that distributes
the jobs: {\it i.e.}, they have to be modified to use ``unused'' machine
time.

  Neural  networks were also  investigated   as another possibility   of
decreasing  $CPU$  time.  Some   preliminary  work   was done  using   a
backpropagation network  (BPN)  \cite{kn:Freeman}  to  perform flux-tube
sorting.  Within  about   5 minutes of  training,   the $BPN$ found  the
minimum of the string-flip potential to within 20\%.  The $BPN$ was then
compared to the fragmentation procedure (for a 7-quark $q\bar q$ system)
for several  hundred flux-tube configurations  and it obtained  a better
solution about 1\% of the time.
The BPN method sorts in constant time and therefore  is not dependent on
the  density, $\rho$, and has  the potential of  handling  a much larger
number of  quarks.   However, at  its current  stage of development  the
projected  learning rate required to obtain  an accuracy of 1\% is about
two months.  Nonetheless,  the work done   with BPN's thus  far seems to
indicate that they hold a very promising future.

  The program for finding a model of nuclear  matter is now quite clear:
examine the effect of
\begin{itemize}
\item the new wave function on the old $SU_\ell(2)$ model 
      \cite{kn:Watson},
\item $SU_c(3)$ on the $Q_2$ system,
\item adding perturbative interactions to the flux-bubble model on 
      the $Q_2$ system, 
\item the interplay between the aforementioned scenarios,
\item the most relevant of these scenarios on the many-body $SU_\ell(2)$ 
      model,
\item $m_q\not = m_{\bar q}$ on the many-body $SU_\ell(2)$ model,
\end{itemize}
and finally
\begin{itemize}
\item move on to studying $SU_\ell(3)$.
\end{itemize}
All of these investigations, with perhaps the exception of the last, are
now   possible to perform  with the  tools   that were developed in this
thesis.

\section{Conclusions}

   The  $Q_2$ system has proven  to be a very  useful  aid for trying to
sort out  the complexities  of model  building  for nuclear matter.  The
details of the  mechanics, from wave  functions to dynamics to practical
computing   methods, of the flux-bubble model   have now been thoroughly
investigated.  It appears   that the flux-bubble  model may  prove to be
very  successful, not  only  for modeling nuclear  matter  but  also for
modeling mesonic molecules as well.

%
%
\chapter{$L^+L^-$ Production in \Esix}

\label{sec-albert}

\section{Introduction}

In this chapter the production of heavy  leptons pairs, $L^+L^-\,$, {\it
via} superstring inspired \Esix models, at  high energy hadron colliders
will  be   investigated   \cite{kn:BoyceB}.   Here  several   underlying
assumptions will be  made about \Esix  models  in order to restrict  the
computation to a manageable,  but reasonable, calculation.  The specific
model  will be presented followed  by a summary   of the calculation and
then its phenomenological consequences.

{\it  NOTE: Many aspects  of the rank-5 models,  that will be considered
here,  are covered  in the literature.   Unfortunately,  when trying  to
extract  the particular model dependent  information needed for $L^+L^-$
production, it appeared that the existing literature was not consistent.
Therefore, it was felt that in order to avoid any ambiguities, the model
should be carefully reconstructed from the ground up.  When constructing
the model, careful attention was paid to being as consistent as possible
with the literature concerning; factors of two, hypercharge conventions,
signs, ambiguous  notational subtleties, etc.  Much  of  the analysis of
the  model was done  by  using Mathematica~\cite{kn:Wolfram} to generate
the  various  couplings,   mass  matrices,   etc.,   directly from   the
superpotential.  This  enabled  easy comparison with  various literature
sources
\cite{kn:Hewett,kn:EllisI,kn:EllisB,kn:EllisA,kn:BargerA,kn:Gunion}.
Differing  conventions and normalizations   aside, the  most significant
problem arose   with  the charged-Higgs,  equation~\ref{eq:charged}, and
pseudo-scalar-Higgs, equation~\ref{eq:neutral}, mass  terms; a factor of
two  was  missing in front    of the  $\sin\beta\cos\beta$ terms,   {\it
op}. {\it cit}.  For example, in the case of the pseudo-scalar-Higgs the
aforementioned authors disagree  by an overall factor  of two in their
mass-mixing matrices but not in their eigenvalues.
As  a   result, the analysis   of  the  mass  constraints   in the Higgs
sector~\cite{kn:Gunion}  had to be re-evaluated, figures~\ref{fig:mztwo}
through~\ref{fig:mhzerob}.  In        addition,  Appendix~\ref{sec-appc}
contains a summary of the couplings  used for $L^+L^-$ production which,
in general, could not be obtained from the literature.}

\section{A Low Energy \Esix Model}

In chapter 1 a general overview of superstring inspired \Esix models was
given.  In  addition, several  comments  were made about  the type of
model that would be presented.  We shall now expand on these assumptions
to determine their low energy consequences.

 There  are many ways of breaking  \Esix down to SM energies. Invariably
these breaking schemes  lead to SM  phenomenologies which contain  extra
gauge bosons.  Here  a rather simple model was  chosen in which  only an
extra $Z$, the $Z^\prime\,$, is produced
\cite{kn:Hewett,kn:BargerA,kn:EllisA}:
$$\Esix\;\;\longrightarrow\;\;
\SU{3}{c}\otimes\SU{2}{L}\otimes\U{1}{Y}\otimes\U{1}{Y_E}$$ 
\noindent  
({\it cf}.  figure~\ref{fig:Wilson}).  In  general the $Z^\prime\,$  can
mix with the SM $Z$ to produce the mixed states
\begin{eqnarray}
Z_1&=&\;\;\;\cos\phi\,Z+\sin\phi\,Z^\prime\,,\label{eq:zmixa}\\
Z_2&=&   -\sin\phi\,Z+\cos\phi\,Z^\prime\,,\label{eq:zmixb}
\end{eqnarray}
({\it cf}. equation \ref{eq:zzmix}.)

  Recall   that    in  order   to      avoid potential   problems   with
flavour-changing-neutral currents, at the tree level, a basis was chosen
in which the third generation of primed-exotic-sleptons were assigned to
play the role of the Higgses \cite{kn:Hewett,kn:EllisI}:
\begin{equation}
\begin{array}{ccc}
\tLp_3=\Doublet{\tilde\nu^{\prime}_{\tau_\L}}{\tilde \TAUp_\L}\,, &
\tRp_3=
\Doublet{\tilde \tau^{\prime c}_\L}{\tilde\nu^{\prime c}_{\tau_\L}}\,,&
\tilde\nu^{\prime\prime c}_{\tau_\L}\,,\;\;\;\;\;\;\;
\end{array}
\label{eq:slhgs}
\end{equation}
({\it i.e.},   $\tilde\R_3\equiv{\tilde\L}_3^{\mbox{}^{\scriptsize c}}$)
or   by  redefining $\tLp_3\,$, $\tRp_3\,$  and $\tilde\nu^{\prime\prime
c}_{\tau_\L}\,$, in  terms of  the complex-isodoublet fields, $\Phi_1\,$
and   $\Phi_2\,$,   and    the  complex-isoscalar    field,  $\Phi_3\,$,
respectively, equation~\ref{eq:slhgs} becomes
\begin{equation}
\begin{array}{ccc}
\Phi_1=\Doublet{\phi^0_1}{\phi^-_1}\,, &
\Phi_2=\Doublet{\phi^+_2}{\phi^0_2}\,, &
\Phi_3=\phi^0_3\,.
\end{array}
\label{eq:hgs}
\end{equation}
This assignment was accomplished by setting
\begin{equation}
\left.
\begin{array}{l@{\;\;\;\;\;\;\;\;}r}
\lambda_4^{i33}\,=\lambda_4^{3i3}\,=\lambda_4^{33i}\,=0 & i=1,2\;\;\;\\
\lambda_4^{333}=\lambda_4^{3jk}=\lambda_4^{j3k}=\lambda_4^{jk3}\neq0 
& j,k=1,2,3\label{kn:unyuk}
\end{array}
\right\}
\end{equation}
in the superpotential,  equation \ref{eq:supot},  where the $ijk$'s  are
generation indices.  Therefore in order to avoid lepton-number violation
the lepton-numbers of all of the primed and double-primed exotic-leptons
must be zero.

 Further   restrictions were placed  on  the superpotential by requiring
that  the baryon and lepton  numbers of the exotic-quarks, $q^\prime\,$,
of the  {\bf 27}'s  (figure~\ref{fig:esix}),  are the  same as  those of
their non-exotic SM counterparts \cite{kn:Hewett}:
$$B(q^\prime_\L)=\frac{1}{3},\;L(q^\prime_\L)=0\;\;\Rightarrow\;\;
\Lam{6}=\Lam{7}=\Lam{8}=\Lam{9}=\Lam{10}=0\,.$$
Also  $\lambda_{11}$ was set  equal to zero  in order to  avoid any fine
tuning problems with the $\nu^c_{l_\L}$ masses \cite{kn:Hewett}.

  With  all the  aforementioned assumptions about  the Yukawa couplings,
the superpotential now simplifies to
\begin{eqnarray}
W &=& \lambda_1i\Phi_{Q}^T\tau_2\Phi_{\Rp}\Phi_{\UR} 
        +\lambda_2i\Phi_{\Lp}^T\tau_2\Phi_{Q}\Phi_{\DR} 
        +\lambda_3i\Phi_{\Lp}^T\tau_2\Phi_{L}\Phi_{\ER}
        \nonumber\\&& 
        +\lambda_4i\Phi_{\Rp}^T\tau_2\Phi_{\Lp}\Phi_{\NppE}
        +\lambda_5\Phi_{\DpL}\Phi_{\DpR}\Phi_{\NppE}\label{eq:soup}
\end{eqnarray}
where the $\lambda$'s were chosen to be real, plus similar terms for the
other generations and their cross-terms.  The $\Super{A}=\Phi(\psi_A,A)$
are   the  superfields  which   contain a   two-component-spinor  field,
$\psi_A\,$,     and    a   complex-scalar-singlet     field,      $A\,$.
Table~\ref{tb:parpr} summarizes the particle properties of this model.
\begin{encaptab}{htbp}
\caption[Table of 
  $\SU{3}{c}\otimes\SU{2}{L}\otimes\U{1}{Y}\otimes\U{1}{Y_E}$   Particle
  Properties]{\footnotesize     \rule[-1ex]{0ex}{1ex}       Table     of
  $\SU{3}{c}\otimes\SU{2}{L}\otimes\U{1}{Y}\otimes\U{1}{Y_E}$   particle
  properties.}{
\footnotesize
\def\VS{\mbox{\rule[-3.5ex]{0ex}{8ex}}}
\def\IT{\mbox{\tiny$\Doublet{\;\;1/2}{-1/2}$}}
\def\QP{\mbox{\tiny$\Doublet{\;1}{\;0}$}}
\def\QM{\mbox{\tiny$\Doublet{\;\;0}{-1}$}}
\def\QQ{\mbox{\tiny$\Doublet{\;\;2/3}{-1/3}$}}
\def\TH{\mbox{\bf$3$}}
\def\TB{\mbox{\bf$\bar 3$}}
\def\ON{\mbox{\bf$1$}}
\def\CR{\\\hline}
$
\begin{array}{|l|c|c|c|c|c|c|c|c|c|c|c|}\cline{2-12}
\multicolumn{1}{c|}{\mbox{}}
   &Q   &\L  &\UR    &\DR    &\ER   &\NER&\DpL&\DpR  &\Lp &\Rp&\NppE\CR
c  &\TH &\ON &\TB    &\TB    &\ON   &\ON&\TH &\TB    &\ON &\ON &\ON \CR
I_3&\IT &\IT & 0     & 0     & 0    &0  & 0  & 0     &\IT &\IT &0\VS\CR
Y  & 1/3&-1  &-4/3   &\;\;2/3& 2    &0  &-2/3&\;\;2/3&-1  & 1  &0   \CR
Y_E& 2/3&-1/3&\;\;2/3&-1/3   & 2/3  &5/3&-4/3&-1/3   &-1/3&-4/3&5/3 \CR
Q  &\QQ &\QM &-2/3   &\;\;1/3&-1\;\;&0  &-1/3&\;\;1/3&\QM &\QP &0\VS\CR
B  & 1/3& 0  &-1/3   &-1/3   & 0    &0  &1/3 &-1/3   & 0  & 0  &0   \CR
L  & 0  & 1  & 0     & 0     &-1    &-1 & 0  & 0     & 0  & 0  &0   \CR
\end{array}
$}
\label{tb:parpr}
\end{encaptab}

  The   superpotential specifies   all   of the  couplings  between  the
particles  of  the {\bf 27}'s.   According  to appendix B  of  Haber and
Kane~\cite{kn:Haber}, the Yukawa interactions are given by
\begin{equation}
{\cal L}_{\rm Yuk} = -\,\frac{1}{2}\,\left[
\left.\left(\frac{\partial^2W}{\partial A_i\partial A_j}\right)
\right|_{\psi_A's\,=\,0}
\psi_i\psi_j+
\left.\left(\frac{\partial^2W}{\partial A_i\partial A_j}\right)^*
\right|_{\psi_A's\,=\,0}
\bar\psi_i\bar\psi_j
\right]\label{eq:yuk}
\end{equation}
and the scalar interactions are given by
\begin{equation}
V=V_F+V_D+V_{Soft}\,.\label{eq:scalpot}
\end{equation}
In equation~\ref{eq:scalpot},
\begin{eqnarray}
V_F&=&F_i^*F_i\,,\\
V_D&=&\onetwo\,[D^aD^a+(D^\prime)^2]\,,
\end{eqnarray}
with
\begin{eqnarray}
F_i&=&\left.\left(\frac{\partial W}{\partial A_i}\right)
\right|_{\psi_A's\,=\,0}\,,\\
D^a&=&gA^*_iT_{ij}^aA_j\,,\\
D^\prime&=&\mbox{\small$\onetwo$}\, g^\prime Y_i A_i^*A_i\,.
\end{eqnarray}
$g$ represents an SU(N) coupling  constant with generators $T^a\,$,  and
$g^\prime$     represents a U(1)    coupling   constant with hypercharge
$Y$. $V_{Soft}$ is a soft-SUSY-breaking term that was put  in by hand in
order to lift  the  supersymmetric-mass-degeneracy  between $m_{\psi_A}$
and $m_A\,$:
\begin{equation}
V_{Soft}=V_{\tilde q}+V_{\tilde l}+V_{H_{Soft}}\,,
\end{equation}
with
\begin{eqnarray}
V_{\tilde q\;\;\;\;\;\;}\;&\supseteq&\tilde M_Q^2|\tilde Q|^2+
 \tilde M_u^2\tilde u_\R^* \tilde u_\R +
 \tilde M_d^2\tilde d_\R^* \tilde d_\R+
 \tilde M_{d^\prime_\L}^2
 \tilde d^{\prime^*}_\L\tilde d^\prime_\L+
 \tilde M_{d^\prime_\R}^2
 \tilde d^{\prime^*}_\R\tilde d^\prime_\R+\nonumber\\&&
 2\,\real[\lambda_1A_ui\tilde Q^T\tau_2\Phi_2\tilde u_\R^*+
 \lambda_2A_di\Phi_1^T\tau_2\tilde Q\tilde d_\R^*+
 \lambda_5A_{d^\prime}\tilde d^{\prime^*}_\R\tilde d^\prime_\L\Phi_3
 ]\,,\;\;\;\;\;\;\;\;\\
V_{\tilde l\;\;\;\;\;\;\;}\;&\supseteq&\tilde M_L^2|\tilde L|^2+
 \tilde M_e^2\tilde e_\R^* \tilde e_\R +
 \tilde M_{\nu_e}^2\tilde \nu_{e_\R}^* \tilde\nu_{ e_\R}+
 2\lambda_3A_e\,\real[i\Phi_1^T\tau_2\tilde L\tilde e_\R^*]
 \,,\\
V_{H_{Soft}}&=&\mu_1^2|\Phi_1|^2+\mu_2^2|\Phi_2|^2+\mu_3^2|\Phi_3|^2-
 \mbox{\small$\frac{1}{\sqrt{2}}$}\,
 \lambda A(i\Phi_1^T\tau_2\Phi_2\Phi_3+h.c.)\,.\label{eq:softy}
\end{eqnarray}
The coefficients $\tilde M_A^2\,$, $A_A\,$, $\mu_i^2\,$, and $A$ are the
soft-SUSY-breaking parameters, and $\lambda\equiv\lambda_4^{333}\,$.

  The Higgs potential can be extracted directly from equation
\ref{eq:scalpot} and is given by \cite{kn:Hewett,kn:BargerA,kn:Gunion}
\begin{eqnarray}
V_H&=&V_{H_{Soft}}+\lambda^2
(|\Phi_1|^2|\Phi_2|^2+|\Phi_1|^2|\Phi_3|^2+|\Phi_2|^2|\Phi_3|^2)
\nonumber\\&&
+\mbox{\small$\frac{1}{8}$}(g^2+g^{\prime^2})(|\Phi_1|^2-|\Phi_2|^2)^2
+\mbox{\small$\frac{1}{72}$}g^{\prime\prime^2}
(|\Phi_1|^2+4|\Phi_2|^2-5|\Phi_3|^2)^2
\nonumber\\&&
+(\mbox{\small$\frac{1}{2}$}g^2-\lambda^2)|\Phi_1^\dagger\Phi_2|^2\,,
\label{eq:fox}
\end{eqnarray}
where $g\,$,  $g^\prime\,$, and $g^{\prime\prime}$ are the SU(2)$_\L\,$,
U(1)$_Y\,$, and U(1)$_{Y_E}\,$  coupling  constants,  respectively.  
The minimization condition \cite{kn:Hewett}
\begin{equation}
\left.\frac{\partial V_H}{\partial\phi_i^0}\right|_{VEV's}=0\,,
\end{equation}
can   be   used   to  fix  the  $\mu^2_i$     terms  in $V_{H_{Soft}}\,$
[Eq.~(\ref{eq:softy})],  where the  vacuum  expectation values, $VEV$'s,
are given by
\begin{equation}
\langle\Psi\rangle=
\left\{\begin{array}{cl}
{\displaystyle\frac{\nu_i}{\sqrt{2}}}&{\rm if}\;\Psi=\phi^0_i\\
0&{\rm otherwise}
\end{array}\right.\;\;\varepsilon\;\Re\,.
\end{equation}
An   analytical   solution  for the   $\mu_i$'s    can be  found;  using
Mathematica~\cite{kn:Wolfram},
\begin{eqnarray}
\mu^2_i&=&
 \frac{3}{72}\,g^{\prime\prime^2}(v_1^2+4\nu_2^2-5\nu_3^2)Y_{E_i}
 +\,\frac{1}{8}\,(g^2+g^{\prime^2})(\nu_1^2-\nu_2^2)Y_i\nonumber\\&&
 -\sum_{j<k}\tilde\varepsilon_{ijk}\,\left[
 \frac{(v_j^2+v_k^2)}{2}\,\lambda^2-\frac{\nu_j\nu_k}{4\nu_i}
 \,\lambda A\right]
\end{eqnarray}
where    $\tilde\varepsilon_{ijk}=|\varepsilon_{ijk}|\,$.    The kinetic
terms for the scalar fields are given by \cite{kn:Cheng}
\begin{equation}
{\cal L}_{K.E.}\supseteq
|{\cal D_\mu}\Phi_i|^2\stackrel{\mbox{\tiny or}}{=}
|(\partial_\mu-{\textstyle\frac{i}{2}}G_\mu)\Phi_i|^2\,,\label{eq:hke}
\end{equation}
with ({\it cf}. \cite{kn:Hewett})
\begin{eqnarray}
G_\mu&=&(g\tau_3\sin\theta_W+\gp Y\cos\theta_W)A_\mu
+(g\tau_3\cos\theta_W-g^\prime Y\sin\theta_W)Z_\mu\nonumber\\&&
+\mbox{\footnotesize$\sqrt{2}$}g[\tau_+W^-_\mu+\tau_-W^+_\mu]
+g^{\prime\prime}Y_E Z^\prime_\mu\,,\label{eq:hkea}
\end{eqnarray}
where               $\tau_\pm=\onetwo(\tau_1\pm               i\tau_2)$,
$\tau_i|\Phi_3\rangle\equiv0\,$,                                     and
$\gp\approx\gpp$~\cite{kn:Hewett,kn:EllisB}.  The  $\tau_i$'s    are the
Pauli matrices acting in isospin space.

  The $\Phi_i$ fields  have complex components, $\phi^a_i\,$, which were
chosen     to  be      of the    form   \cite{kn:EllisI,kn:Cheng}  ({\it
cf}.~\cite{kn:GunionA})
\begin{equation}
\phi^a_i=\frac{1}{\sqrt{2}}\,(\phi^a_{iR}+i\phi^a_{iI})\,,
\label{eq:pfff}
\end{equation}
where $\phi^a_{iR}/\sqrt{2}\,$ and $\phi^a_{iI}/\sqrt{2}\,$ are the real
and imaginary  parts, respectively.  Therefore, the  $\{\Phi_i\}$ fields
have a total of 10 degrees of freedom: four are ``eaten'' to give masses
to   the   $W^\pm$, $Z$,  and  $Z^\prime$    bosons, and  the  remainder
yield~\cite{kn:Hewett}      two  charged-Higgs  bosons,    $H^\pm$,  one
pseudo-scalar-Higgs    boson, $P^0\,$, and    three scalar-Higgs bosons,
$H^0_{i=1,2,3}\,$.  The mass terms for the  Higgs fields can be obtained
from the second order terms of the expansion  of $V_H(\phi_k)$ about its
minimum \cite{kn:Cheng},
\begin{equation}
V_H(\phi_k)\;\supseteq\;\onetwo\,{\cal M}_{ij}^2\,
(\phi_i-\langle\phi_i\rangle)(\phi_j-\langle\phi_j\rangle)\,,
\end{equation}
where  
\begin{equation}
{\cal M}_{ij}^2=\left.\frac{\partial^2V_H}{\partial\phi_i\partial\phi_j}
\right|_{VEV's}
\end{equation}
is the Higgs-mass-mixing matrix.  Therefore the mass terms for the Higgs
fields are simply
\begin{eqnarray}
{\cal L}_{\cal M}&\supseteq&
-\,(\phi_2^{+^*} \phi_1^-)\,{\cal M}^2_{H^\pm}
\Doublet{\phi_2^+}{\phi_1^{-^*}}
-\frac{1}{2}\,
(\phi_{2I}^0\:\phi_{1I}^0\:\phi_{3I}^0)\,{\cal M}^2_{P^0}
\Triplet{\phi_{2I}^0}{\phi_{1I}^0}{\phi_{3I}^0}\nonumber\\&&
-\frac{1}{2}\,
(\phi_{1R}^0-\nu_1\;\phi_{2R}^0-\nu_2\;\phi_{3R}^0-\nu_3)\,
{\cal M}^2_{H^0_i}\label{kn:simply}
\Triplet{\phi_{1R}^0-\nu_1}{\phi_{2R}^0-\nu_2}{\phi_{3R}^0-\nu_3}\,.
\end{eqnarray}
The mass-mixing matrices are
\begin{eqnarray}
{\cal M}^2_{H^\pm}&=&
\onetwo\left(
\begin{array}{ll}
(\onetwo g^2-\lambda^2)\nu_1^2+\lambda A\,
\mbox{$\displaystyle\frac{\nu_{13}}{\nu_2}$}&
(\onetwo g^2-\lambda^2)\nu_{12}+\lambda A\nu_3\\
(\onetwo g^2-\lambda^2)\nu_{12}+\lambda A\nu_3&
(\onetwo g^2-\lambda^2)\nu_2^2+
\mbox{$\displaystyle\lambda A\,\frac{\nu_{23}}{\nu_1}$}
\end{array}
\right)\,,
\\\nonumber\\
{\cal M}_{P^0}^2&=&\frac{\lambda A\nu_3}{2}
\left(
\begin{array}{ccc}
{\displaystyle\frac{\nu_1}{\nu_2}}&
1&
{\displaystyle\frac{\nu_1}{\nu_3}}\\
1&
{\displaystyle\frac{\nu_2}{\nu_1}}&
{\displaystyle\frac{\nu_2}{\nu_3}}\\
{\displaystyle\frac{\nu_1}{\nu_3}}&
{\displaystyle\frac{\nu_2}{\nu_3}}&
{\displaystyle\frac{\nu_{12}}{\nu_3^2}}
\end{array}
\right)\,,
\\\nonumber\\
{\cal M}_{H^0_i}^2&=&\onetwo
\left(
\begin{array}{lll}
B_1\nu_1^2+\lambda A\,\mbox{$\displaystyle\frac{\nu_{23}}{\nu_1}$}&
B_2\nu_{12}-\lambda A\nu_3&
B_3\nu_{13}-\lambda A\nu_2\\
B_2\nu_{21}-\lambda A\nu_3&
B_4\nu_2^2+\lambda A\,\mbox{$\displaystyle\frac{\nu_{13}}{\nu_2}$}&
B_5\nu_{23}-\lambda A\nu_1\\
B_3\nu_{31}-\lambda A\nu_2&
B_5\nu_{32}-\lambda A\nu_1&
B_6\nu_3^2+\lambda A\,\mbox{$\displaystyle\frac{\nu_{12}}{\nu_3}$}
\end{array}
\right)\label{eq:mhimat}
\end{eqnarray}
where $\nu_{ij}=\nu_i\nu_j\,$. In equation~\ref{eq:mhimat}
\begin{equation}
\mbox{\small$\left.
\begin{array}{@{}lll@{}}
B_1=\txtfrac{1}{2}(g^2+\gps)+\txtfrac{1}{18}\gpps&
B_2=2\lambda^2+\txtfrac{2}{9}\gpps-\txtfrac{1}{2}(g^2+\gps)&
B_3=2\lambda^2-\txtfrac{5}{18}\gpps\\
&B_4=\txtfrac{1}{2}(g^2+\gps)+\txtfrac{8}{9}\gpps&
B_5=2\lambda^2-\txtfrac{10}{9}\gpps\\
&&B_6=\txtfrac{25}{18}\gpps
\end{array}\right\}$}\,.
\end{equation}
The physical  states are obtained by  diagonalizing the  terms in ${\cal
L}_{\cal M}\,$.   The  eigenvectors for  the  charged and  pseudo-scalar
Higgs terms are respectively,
\begin{eqnarray}
H^\pm&=&\;\;\;\;
\cos\beta\,\phi_2^\pm+\sin\beta\,\phi_1^\pm\,,\\
G^\pm&=&-
\sin\beta\,\phi_2^\pm+\cos\beta\,\phi_1^\pm\,,
\end{eqnarray}
and
\begin{eqnarray}
P^0&=&\sqrt{\frac{\lambda A\nu_3}{2m_{P^0}^2}}\,
\left[
\sqrt{\frac{\nu_1}{\nu_2}}\,\phi_{2I}^0+
\sqrt{\frac{\nu_2}{\nu_1}}\,\phi_{1I}^0+
\sqrt{\frac{\nu_{12}}{\nu_3^2}}\,\phi_{3I}^0
\right]\,,\\
G^0_1&=&\frac{\nu_2^2\nu_3}{\sqrt{
(\nu_2^2+\nu_3^2)(\nu_{12}^2+\nu^2\nu_3^2)
}}\,
\left[
\frac{\nu_3}{\nu_2}\,\phi_{2I}^0-
\frac{\nu_1}{\nu_3}\left(1+\frac{\nu_3^2}{\nu_2^2}\right)\phi_{1I}^0+
\phi_{3I}^0
\right]\,,\\
G^0_2&=&\frac{\nu_3}{\sqrt{\nu_2^2+\nu_3^2}}\,
\left[-\,\frac{\nu_2}{\nu_3}\,\phi_{2I}^0+\phi_{3I}^0\right]\,,
\end{eqnarray}
where $\phi^\pm_i=(\phi^\mp_i)^*\,$, $\nu^2 =  \nu_1^2 + \nu_2^2\,$, and
$\tan\beta\equiv\nu_2/\nu_1\,$. Here, the  $H^\pm$ are the charged-Higgs
states with masses
\begin{equation}
m^2_{H^\pm}=\frac{\lambda A\nu_3}{\sin(2\beta)}
+\left(1-2\,\frac{\lambda^2}{g^2}\right)m_W^2\,,
\label{eq:charged}
\end{equation}
the $P^0$ is the pseudo-scalar-Higgs state with mass
\begin{equation}
m_{P^0}^2=\frac{\lambda A\nu_3}{\sin(2\beta)}\,
\left(1+\frac{\nu^2}{4\nu_3^2}\sin^2(2\beta)\right)\,,
\label{eq:neutral}
\end{equation}
and the $G^\pm$ and $G^0_{1,2}$ are the Goldstone-boson states with zero
mass.
The scalar-Higgs   term can be  diagonalized exactly   using Mathematica
\cite{kn:Wolfram}, however the    result   is, in general,   not    very
enlightening.   For our  purposes  it  suffices  to resort  to numerical
techniques.  In the unitary gauge (U-gauge)  the Goldstone modes vanish,
{\it i.e.}, the  $G{\rm 's}=0\,$, and  the fields become physical.  This
allows the change of basis:
\begin{eqnarray}
\phi^\pm_1&=&\sin\beta H^\pm\,,\label{eq:basa}\\
\phi^\pm_2&=&\cos\beta H^\pm\,,\\
\phi^0_{1I}&=&\kappa\nu_{23}\,P^0\,,\\
\phi^0_{2I}&=&\kappa\nu_{13}\,P^0\,,\\
\phi^0_{3I}&=&\kappa\nu_{12}\,P^0\,,\\
\phi^0_{iR}&=&\nu_i+\sum_{j=1}^3U_{ij}H^0_j\,,\label{eq:basb}
\end{eqnarray}
where  $\kappa=1/\sqrt{v_1^2v_2^2+v^2v_3^2}\,$.   The  $U_{ij}$  are the
elements  of the inverse of  the matrix that  was used in the similarity
transformation to diagonalized the  scalar-Higgs-mass term.
With these transformations at hand it is now a straightforward matter to
obtain all of the masses and couplings for the various particles in this
model.

  The mass terms  for the gauge fields can  be found by transforming the
kinetic   terms for the   $\Phi_i$ fields, equation~\ref{eq:hke}, to the
U-gauge basis, equations \ref{eq:basa} through \ref{eq:basb}, yielding:
\begin{equation}
{\cal L}_{K.E.}^{\Phi_i}\supseteq
m_W^2\,W^+_\mu W^{-^\mu}+
\frac{1}{2}\,(Z\;Z^\prime)_\mu\,{\cal M}^2_{Z-Z^\prime}\,
\Doublet{Z}{Z^\prime}^\mu\,.\label{eq:zzmix}
\end{equation}
As a consequence, the $W$ mass is
\begin{equation}
m_W^2=\onefour g^2\nu^2\,,
\end{equation}
and                   the                       $Z-Z^\prime$-mass-mixing
matrix is~\cite{kn:Hewett,kn:EllisB,kn:EllisA,kn:BargerA}
\begin{equation}
{\cal M}^2_{Z-Z^\prime}=
\left(
\begin{array}{cc}
m_Z^2&\delta m^2\\
\delta m^2&m_{Z^\prime}^2
\end{array}
\right)\,,\label{eq:mzzmat}
\end{equation}
with matrix elements:
\begin{eqnarray}
m_{Z^{\;}}^2&=&\onefour(g^2+g^{\prime^2})\,\nu^2\,,\\
m_{Z^\prime}^2&=&\frac{1}{36}\gpps(\nu_1^2+16\nu_2^2+25\nu_3^2)\,,\\
\delta m^2&=&\frac{1}{12}\,\sqrt{g^2+\gps}\,\gpp(4\nu_2^2-\nu_1^2)\,.
\end{eqnarray}
Diagonalization of ${\cal M}^2_{Z-Z^\prime}$ yields the mass eigenstates
given by equations \ref{eq:zmixa} and
\ref{eq:zmixb} with eigenvalues
\begin{eqnarray}
m_{Z_1}^2&=&m_Z^2\cos^2\phi+\delta m^2\sin(2\phi)+
m_{Z^\prime}^2\sin^2\phi\,,\\
m_{Z_2}^2&=&m_Z^2\sin^2\phi-\delta m^2\sin(2\phi)+
m_{Z^\prime}^2\cos^2\phi\,,
\end{eqnarray}
and mixing angle
\begin{equation}
\tan(2\phi)=\frac{2\delta m^2}{m_Z^2-m_{Z^\prime}^2}\,.
\end{equation}
Notice that in  the  large $\nu_3$ limit $\phi\;\rightarrow\;\pi/2$  and
therefore $Z_1\;\rightarrow\;Z$  and $Z_2\;\rightarrow\;Z^\prime\,$.  In
fact,   for the   range   of $VEV$'s   that will    be   considered here
$m_{Z_1}<m_{Z_2}\,$.  Therefore, $Z_1$ will  be  designated the role  of
the observed $Z$ at facilities such as LEP or SLC.

  The mass terms for  the fermions, and  hence the Yukawa couplings, can
be      found     by   evaluating       ${\cal    L}_{Yuk}^{\mbox{}}\,$,
equation~\ref{eq:yuk}, in the U-gauge basis and then using Appendix A of
Haber and  Kane \cite{kn:Haber},  to convert   to four component  spinor
notation.\footnote{{\it   cf}.   equation~\ref{eq:yields}.  For  a  more
explicit example see section  4.2 of Gunion and Haber \cite{kn:GunionA}}
The result is
\begin{equation}
{\cal L}_{\rm Yuk}\supseteq
-\frac{1}{\sqrt{2}}\,\left\{
\lambda_1\nu_2\,\bar u u+
\lambda_2\nu_1\,\bar d d+
\lambda_3\nu_1\,\bar e e+
\lambda_4\nu_3\,\bar e^\prime e^\prime+
\lambda_5\nu_3\,\bar d^\prime d^\prime
\right\}\,.\label{eq:yucky}
\end{equation}
Therefore, the Yukawa couplings for the first generation are given by:
\begin{eqnarray}
\lambda_1&=&\frac{g\,m_u}{\sqrt{2}\,\mw\sin\beta}\,,\label{eq:ycpa}\\
\lambda_2&=&\frac{g\,m_d}{\sqrt{2}\,\mw\cos\beta}\,,\\
\lambda_3&=&\frac{g\,m_e}{\sqrt{2}\,\mw\cos\beta}\,,\\
\lambda_4&=&\frac{\sqrt{2}}{\nu_3}\,m_{e^\prime}\,,\\
\lambda_5&=&\frac{\sqrt{2}}{\nu_3}\,m_{d^\prime}\,,\label{eq:ycpd}
\end{eqnarray}
and similarly for the other generations.

  The sfermion masses are obtained  by evaluating the scalar-interaction
potential, V  (equation \ref{eq:scalpot}),  and then transforming  it to
the U-gauge basis:
\begin{equation}
{\cal L}_{\cal M}\supseteq-V\supseteq
-(\tilde f_L^* \tilde f_R^*){\cal M}^2_{\tilde f}
\Doublet{\tilde f_L}{\tilde f_R}\,,
\end{equation}
with
\begin{equation}
{\cal M}^2_{\tilde f}=
\left(
\begin{array}{cc}
M^2_{LL}&M^2_{LR}\\
M^2_{LR}&M^2_{RR}
\end{array}
\right)\,,
\end{equation}
being the sfermion-mass-mixing  matrix. The mass-mixing  matrix elements
are found to be, using Mathematica~\cite{kn:Wolfram}:
\begin{eqnarray}
M^{(\tilde u)^2}_{LL}&=&
\tilde M_Q^2+m_u^2+\frac{1}{6}\,(3-4\xw)\,m_Z^2\,\cos(2\beta)-
\frac{1}{36}\,\gpps(\nu_1^2+4\nu_2^2-5\nu_3^2)\,,\;\;\;\;\\
M^{(\tilde u)^2}_{RR}&=&
\tilde M_u^2+m_u^2+\frac{2}{3}\,\xw\,m_Z^2\,\cos(2\beta)-
\frac{1}{36}\,\gpps(\nu_1^2+4\nu_2^2-5\nu_3^2)\,,\\
M^{(\tilde u)^2}_{LR}&=&m_u\,(A_u-m_{e^\prime}\cot\beta)\,,
\end{eqnarray}
for the $\tilde u_{L,R}$ squarks;
\begin{eqnarray}
M^{(\tilde d)^2}_{LL}&=&
\tilde M_Q^2+m_d^2-\frac{1}{6}\,(3-2\xw)\,m_Z^2\,\cos(2\beta)-
\frac{1}{36}\,\gpps(\nu_1^2+4\nu_2^2-5\nu_3^2)\,,\;\;\;\;\\
M^{(\tilde d)^2}_{RR}&=&
\tilde M_d^2+m_d^2-\frac{1}{3}\,\xw\,m_Z^2\,\cos(2\beta)+
\frac{1}{72}\,\gpps(\nu_1^2+4\nu_2^2-5\nu_3^2)\,,\\
M^{(\tilde d)^2}_{LR}&=&m_d\,(A_d-m_{e^\prime}\tan\beta)\,,
\end{eqnarray}
for the $\tilde d_{L,R}$ squarks;
\begin{eqnarray}
M^{(\tilde d^\prime)^2}_{LL}&=&
\tilde M_{d^\prime_L}^2+m_{d^\prime}^2+
\frac{1}{6}\,\xw\,m_Z^2\,\cos(2\beta)+
\frac{1}{18}\,\gpps(\nu_1^2+4\nu_2^2-5\nu_3^2)\,,\;\;\;\;
\,\;\;\;\;\;\;\;\\
M^{(\tilde d^\prime)^2}_{RR}&=&
\tilde M_{d^\prime_R}^2+m_{d^\prime}^2-
\frac{1}{3}\,\xw\,m_Z^2\,\cos(2\beta)+
\frac{1}{72}\,\gpps(\nu_1^2+4\nu_2^2-5\nu_3^2)\,,\\
M^{(\tilde d^\prime)^2}_{LR}&=&
m_{d^\prime}\,(A_{d^\prime}-
m_{e^\prime}\,\frac{\nu_{12}}{\nu_3^2})\,,
\end{eqnarray}
for the $\tilde d^\prime_{L,R}$ squarks;
\begin{eqnarray}
M^{(\tilde e)^2}_{LL}&=&
\tilde M_L^2+m_e^2-\frac{1}{2}\,(1-2\xw)\,m_Z^2\,\cos(2\beta)+
\frac{1}{72}\,\gpps(\nu_1^2+4\nu_2^2-5\nu_3^2)\,,\;\;\;\;\\
M^{(\tilde e)^2}_{RR}&=&
\tilde M_e^2+m_e^2-\xw\,m_Z^2\,\cos(2\beta)-
\frac{1}{36}\,\gpps(\nu_1^2+4\nu_2^2-5\nu_3^2)\,,\\
M^{(\tilde e)^2}_{LR}&=&m_e\,(A_e-m_{e^\prime}\tan\beta)\,,
\end{eqnarray}
for the $\tilde e_{L,R}$ sleptons;
\begin{eqnarray}
M^{(\tilde \nu_e)^2}_{LL}&=&
\tilde M_L^2+m_{\nu_e}^2+\frac{1}{2}\,m_Z^2\,\cos(2\beta)+
\frac{1}{72}\,\gpps(\nu_1^2+4\nu_2^2-5\nu_3^2)\,,\;\;\;\;
\,\;\;\;\;\;\;\;\;\;\\
M^{(\tilde \nu_e)^2}_{RR}&=&
\tilde M_{\nu_e}^2-
\frac{5}{72}\,\gpps(\nu_1^2+4\nu_2^2-5\nu_3^2)\,,\\
M^{(\tilde \nu_e)^2}_{LR}&=&0\,,
\end{eqnarray}
for   the  $\tilde \nu_{L,R}$ sleptons,   and  similarly for  the other
generations.  The mass eigenstates are given by
\begin{equation}
\left(\begin{array}{c}
\tilde f_1\\\tilde f_2
\end{array}\right)=
\left(\begin{array}{rr}
\cos\theta_{\tilde f}&\sin\theta_{\tilde f}\\
-\sin\theta_{\tilde f}&\cos\theta_{\tilde f}
\end{array}\right)
\left(\begin{array}{c}
\tilde f_L\\\tilde f_R
\end{array}\right)\label{eq:smix}
\end{equation}
with mass eigenvalues
\begin{eqnarray}
m_{f_1}^2&=&M^2_{LL}\cos^2\theta_{\tilde f}+
M^2_{LR}\sin(2\theta_{\tilde f})+M^2_{RR}\sin^2\theta_{\tilde f}\,,\\
m_{f_2}^2&=&M^2_{LL}\sin^2\theta_{\tilde f}-
M^2_{LR}\sin(2\theta_{\tilde f})+M^2_{RR}\cos^2\theta_{\tilde f}\,,
\end{eqnarray}
and mixing angle
\begin{equation}
\tan(2\theta_{\tilde f})=\frac{M^2_{LR}}{M^2_{LL}-M^2_{RR}}\,.
\end{equation}
Notice that for fairly large $\nu_3$
\begin{equation}
\tan(2\theta_{\tilde f})\;\sim\;
{\cal O}\left(\frac{m_fA_f}{\nu_3^2}\right)\,,
\end{equation}
where  the soft  terms  have been assumed  to  be large and  degenerate.
Therefore,  in  general,  the mixing   is only expected   to  effect the
sfermions that have fairly heavy fermion partners.
   
  The supersymmetric partner,  or spartner, degrees  of freedom  for the
neutral-Higgs   fields  (neutral-Higgsinos),    $\nu_{\tau_L}^\prime\,$,
$\nu_{\tau_L}^{\prime^c}\,$    and   $\nu_{\tau_L}^{\prime \prime^c}\,$,
along with the spartner degrees of freedom for  the neutral gauge fields
(neutral-gauginos),   $\tilde\gamma\,$,    $\tilde   Z\,$,  and  $\tilde
Z^\prime\,$,  mix to  form  a (6\timey6) neutralino, $\tilde  \chi^0\,$,
mass-mixing matrix.\footnote{A detailed  study  of the   $\tilde \chi^0$
mass  spectrum can   be   found  in \cite{kn:thesisa}.}    Similarly the
charged-Higgsinos, $\tau_L^\prime$    and $\tau_L^{\prime^c}\,$, and the
charged-gauginos, $\tilde  W^\pm\,$, form a (2\timey2) chargino, $\tilde
\chi^\pm\,$, mass-mixing matrix.\footnote{The  full  form of  these mass
matrices can  be  found in Ellis,  $et$  $al.$,~\cite{kn:EllisI} and the
details of how to obtain  them can be  found in appendix  B of Haber and
Kane~\cite{kn:Haber}.}  By  virtue  of supersymmetry the  neutralino and
chargino   mass-mixing matrices contain    the  same  Yukawa  and  gauge
couplings as their spartners, modulo soft terms.

  The   real  and imaginary  parts   of  the  sfermion  fields,  $\tilde
\nu_{e_L}^\prime\,$,    $\tilde    \nu_{e_L}^{\prime^c}\,$,      $\tilde
\nu_{e_L}^{\prime\prime^c}\,$, $\tilde   \nu_{\mu_L}^\prime\,$,  $\tilde
\nu_{\mu_L}^{\prime^c}\,$, and  $\tilde \nu_{\mu_L}^{\prime\prime^c}\,$,
yield   two     separate    (6\timey6)-mass-mixing  matrices  for    the
neutral-unHiggses ({\it  cf}. equation \ref{kn:simply} for $\phi^0_{iR}$
and $\phi^0_{iI}$), which contain the Yukawa couplings given in equation
\ref{kn:unyuk}.  In general,  these mass-mixing matrices are expected to
lead to very massive unHiggs states \cite{kn:EllisI}.

  The  spartner   degrees   of  freedom   for   the   neutral-unHiggses,
$\nu_{e_L}^\prime\,$,      $\nu_{e_L}^{\prime^c}\,$,  $\nu_{e_L}^{\prime
\prime^c}\,$,  $\nu_{\mu_L}^\prime\,$,  $\nu_{\mu_L}^{\prime^c}\,$,  and
$\nu_{\mu_L}^{\prime \prime^c}\,$,  form a (6\timey6)-mass-mixing matrix
for  the     neutral-unHiggsinos.  Therefore,    the  neutral-unHiggsino
mass-mixing  matrix contains the   same    Yukawa couplings as     their
neutral-unHiggs partners.
 
  The            sfermion       fields           $\tilde{e}^\prime_L\,$,
$\tilde{e}^{\prime^c}_L\,$,           $\tilde\mu^\prime_L\,$,        and
$\tilde\mu^{\prime^c}_L\,$,  yield two  separate  (2\timey2)-mass-mixing
matrices for  the charged-unHiggses ({\it cf}.  equation \ref{kn:simply}
for $\phi^{\pm}_i$).    These matrices have a  large   number of unknown
parameters  and  quite naturally    acquire a  very   large mass   ({\it
cf}.~\cite{kn:EllisI}).

  Finally,  the  spartner degrees of   freedom for the charged-unHiggses
give diagonalized mass  eigenstates (see equation~\ref{eq:yucky})  which
correspond to the charged heavy leptons.

\section{$L^+L^-$  Production Cross-Section} 
\label{sec-cross}

   Figure~\ref{fig:fusion} shows the Feynman diagrams used for computing
the parton level gluon-gluon fusion to heavy leptons matrix elements.
\begin{encapfig}{htbp}
$$
\begin{array}{lclc}
  a) & \ggqzll  & b) &\ggqhpll  \\ &&& \\
  c) & \ggsqhll & d) &\ggvsqhll
\end{array}
$$
\caption[$gg\rightarrow L^+L^-$ Feynman Diagrams]{\footnotesize
         Feynman  diagrams  for  gluon-gluon  fusion  to  heavy  charged
         leptons.}
\label{fig:fusion}
\end{encapfig}
It was shown \cite{kn:Konig} that the  \Esix matrix element computations
are very   similar  to the corresponding  $MSSM$   calculation  by Cieza
Montalvo, $et$ $al.$,   \cite{kn:Montalvo} and can  be easily  extracted
from their paper.  The matrix elements are as follows:
\begin{enumerate}
\item
For the $Z_{1,2}$ exchange diagram shown in figure~\ref{fig:fusion}.a
\begin{equation}
\hat\sigma^{qZ_{1,2}}_{L^\pm}=
\frac{\alpha^2\alpha_s^2}{128\pi x_W^2}\,
\frac{m_{L}^2}{m_W^4}\,\beta_L
|\sum_{i=1}^2[\tilde C^{L^\pm Z_i}_L-\tilde C^{L^\pm Z_i}_R]\,
\xi^{Z_i}(\hat s)
\sum_q[\tilde C^{qZ_i}_L-\tilde C^{qZ_i}_R](1+2\lambda_qI_q)|^2
\label{eq:hatqz}
\end{equation}
where the left-right Fermion, $f\,$, couplings are given by
\begin{equation}
\left(\begin{array}{c}
   \tilde C^{fZ_1}_{L,R}\\\tilde C^{fZ_2}_{L,R}
\end{array}\right)=
\left(\begin{array}{rr}
   \cos\phi & \sin\phi \\
  -\sin\phi & \cos\phi
\end{array}\right)
\left(\begin{array}{c}
   C^{fZ}_{L,R}\\ C^{fZ^\prime}_{L,R}
\end{array}\right)\,,
\end{equation}
such that
\begin{eqnarray}
C^{fZ}_{L,R}        &=& T_{3_{L,R}}-e_f\xw\,,\label{eq:hatqzc}\\
C^{fZ^\prime}_{L,R} &=& 
\frac{1}{2}\left(\frac{\gpp}{g}\right)
y^\prime_{f_{L,R}}\,\sqrt{1-\xw}\,,\label{eq:hatqzd}
\end{eqnarray}
where  $T_{3_R}  =  -T_{3_L}^c$   is  the   isospin, $y^\prime_{f_R}   =
-y^\prime_{f^c_L}$
is the $Y_E$-hypercharge, and $e_f$ is the electric charge.
\item
For  the    $H^0_{1,2,3}$ and  $P^0$   exchange   diagrams   shown  in
figure~\ref{fig:fusion}.b
\begin{eqnarray}
\hat\sigma^{qH^0_{i}{\rm 's}}_{L^\pm}&=&
\frac{\alpha^2\alpha_s^2}{512\pi x_W^2}\,
\frac{m_{L}^2}{m_W^4}\,\beta_L^3
|\sum_{i=1}^3K^{L^\pm H^0_i}\zeta^{H^0_i}(\hat s)
\sum_qK^{qH^0_i}m_q^2[2+(4\lambda_q-1)I_q]|^2\,,\nonumber\\&&\\
\hat\sigma^{qP^0_{\;}\;\;}_{L^\pm}&=&
\frac{\alpha^2\alpha_s^2}{512\pi x_W^2}\,
\frac{m_{L}^2}{m_W^4}\,\beta_L(K^{L^\pm P^0})^2
|\zeta^{P^0}(\hat s)|^2
|\sum_qK^{qP^0}m_q^2I_q|^2\,,
\end{eqnarray}
where the couplings $K^{fH^0_i}$  and $K^{fP^0}$ are given  by equations
\ref{eq:rata} through \ref{eq:ratb}.
\item
For       the    $H^0_{1,2,3}$     exchange    diagrams       shown   in
figures \ref{fig:fusion}.c and \ref{fig:fusion}.d
\begin{equation}
\hat\sigma^{\tilde qH^0_{i}{\rm 's}}_{L^\pm}=
\frac{\alpha^2\alpha_s^2m_L\beta_L}{512\pi x_W^2(1-\xw)^2}\,|
\sum_{i=1}^3K^{L^\pm H^0_i}\zeta^{H^0_i}(\hat s)
\sum_{\tilde q}\sum_{k=1}^2\tilde K^{\tilde q H^0_i}_k
(1+2\lambda_{\tilde q_k}I_{\tilde q_k})|^2\,,
\end{equation}
where the  sfermion  mass eigenstates, $\tilde  f_{1,2}\,$, couplings to
the $H^0_i$ are given by
\begin{eqnarray}
\tilde K^{\tilde f H^0_i}_1&=&
K^{\tilde f H^0_i}_{LL}\cos^2\theta_{\tilde f}+
K^{\tilde f H^0_i}_{LR}\sin2\theta_{\tilde f}+
K^{\tilde f H^0_i}_{RR}\sin^2\theta_{\tilde f}
\,,\label{eq:kcupa}\\
\tilde K^{\tilde f H^0_i}_2&=&
K^{\tilde f H^0_i}_{LL}\sin^2\theta_{\tilde f}-
K^{\tilde f H^0_i}_{LR}\sin2\theta_{\tilde f}+
K^{\tilde f H^0_i}_{RR}\cos^2\theta_{\tilde f}
\,,\label{eq:kcupb}
\end{eqnarray}
where  $K^{\tilde   f  H^0_i}_{AB}$  ($A,B=L,R$)  are  the corresponding
couplings for  the sfermion  helicity  states, $\tilde f_{L,R}\,$,  with
mixing  angle  $\theta_{\tilde  f}\,$.   The  $K^{\tilde f  H^0_i}_{AB}$
couplings are given by equations \ref{eq:cupd} through
\ref{eq:cupdy}.
\item
For   the    $q(Z_{1,2}$-$P^0)$    interference    terms,    {\it   via}
figures~\ref{fig:fusion}.a and~\ref{fig:fusion}.b,
\begin{eqnarray}
\hat\sigma^{\tilde q(Z_i{\rm 's}-P^0)}_{L^\pm}&=&
\frac{-\alpha^2\alpha_s^2}{128\pi x_W^2}\,\frac{m_L^2}{m_W^4}\,
\beta_L\,K^{L^\pm P^0}\,\real\{\,\zeta^{P^0}(\hat s)
\sum_{i=1}^2\xi^{Z_i}(\hat s)^*\,
[\tilde C^{L^\pm Z_i}_L-\tilde C^{L^\pm Z_i}_R]\nonumber\\&&\times
\sum_qK^{qP^0}m_q^2\,I_q\sum_{q^\prime}\,
[\tilde C^{q^\prime Z_i}_L-\tilde C^{q^\prime Z_i}_R]\,
(1+2\lambda_{q^\prime}I_{q^\prime}^*)\,\}\,.
\end{eqnarray}
\item
For  the   $(\tilde q - q)H^0_{1,2,3}$   interference   terms, {\it via}
figures~\ref{fig:fusion}.b through~\ref{fig:fusion}.d,
\begin{eqnarray}
\hat\sigma^{(\tilde q-q)\tilde H^0_i{\rm 's}}_{L^\pm}&=&
\frac{-\alpha^2\alpha_s^2}{256\pi x_W^2(1-\xw)^2}
\left(\frac{m_L^{\mbox{}}}{m_Z^{\mbox{}}}\right)^2\beta_L^3\,\real\{
\sum_{i=1}^3K^{L_\pm H^0_i}\zeta^{H^0_i}(\hat s)
\;\;\;\;\;\;\;\;\;\;\;\;\nonumber\\&&\times
\sum_qK^{qH^0_i}m_q^2[2+(4\lambda_q-1)I_q]
\sum_{j=1}^3K^{L_\pm H^0_j}\zeta^{H^0_j}(\hat s)^*
\nonumber\\&&\times
\sum_{\tilde q}\sum_{k=1}^2\tilde K^{\tilde qH^0_j}_k
(1+2\lambda_{\tilde q}I_{\tilde q}^*)\}\,.
\end{eqnarray}
\end{enumerate}
In the aforementioned list of cross-section equations:
\begin{eqnarray}
I_p&\equiv& I_p(\lambda_p)=\int_0^1\frac{dx}{x}\ln\left[
1-\frac{(1-x)x}{\lambda_p}\right]\nonumber\\&&\nonumber\\&=&
\left\{\begin{array}{l@{\;\;{\rm if}\;\;}l}
-2\left[\sin^{-1}\left(\frac{1}{2\sqrt{\lambda_p}}\right)\right]^2&
\lambda_p\,>\,\txtfrac{1}{4}\\
\txtfrac{1}{2}\,\ln^2\left(\frac{r_+}{r_-}\right)-
\txtfrac{\pi^2}{2}+i\pi\,\ln\left(\frac{r_+}{r_-}\right)&
\lambda_p\,<\,\txtfrac{1}{4}
\end{array}\right.\,,
\end{eqnarray}
is the loop function \cite{kn:Montalvo}, where
\begin{equation}
\lambda_p=\frac{m_p^2}{\hat s}\,,\\
\end{equation}
and $r_\pm=1\pm\sqrt{1-4\lambda_p}\,$   such  that  $p\,  \varepsilon\,
\{f,\tilde f\}\,$;
\begin{eqnarray}
\;\;\;\;\xi^{Z_i}(\hat s)&=&
\frac{\hat s-m_{Z_i}^2}{\hat s-m_{Z_i}^2
+i\,m_{Z_i}^{\mbox{}}\Gamma_{Z_i}^{\mbox{}}}\,,\\
\zeta^{H^0_i,\,P^0}(\hat s)&=&
\frac{1}{\hat s-m_{H^0_i,P^0}^2
+i\,m_{H^0_i,\,P^0}^{\mbox{}}\Gamma_{H^0_i,\,P^0}^{\mbox{}}}\,,
\end{eqnarray}
are  related to the vector  and  scalar propagators, respectively, where
the  widths,  {\it i.e.},  the   $\Gamma_{V,\phi}$'s, are summarized  in
\S~\ref{sec-wit}, and
\begin{equation}
\beta_L=\sqrt{1-\frac{4m_L^2}{\hat s}}\,,
\end{equation}
is the  lepton velocity in the c.m.  frame.  The details  of the various
components   that have  gone   into this  computation   can be  found in
appendix~\ref{sec-appc}.  Before  the parton  level cross-section can be
used to compute the heavy lepton production rates some assumptions about
the parameters and masses in the model must be made.

\begin{encapfig}{hbtp}
\mbox{\epsfxsize=144mm
	\epsffile{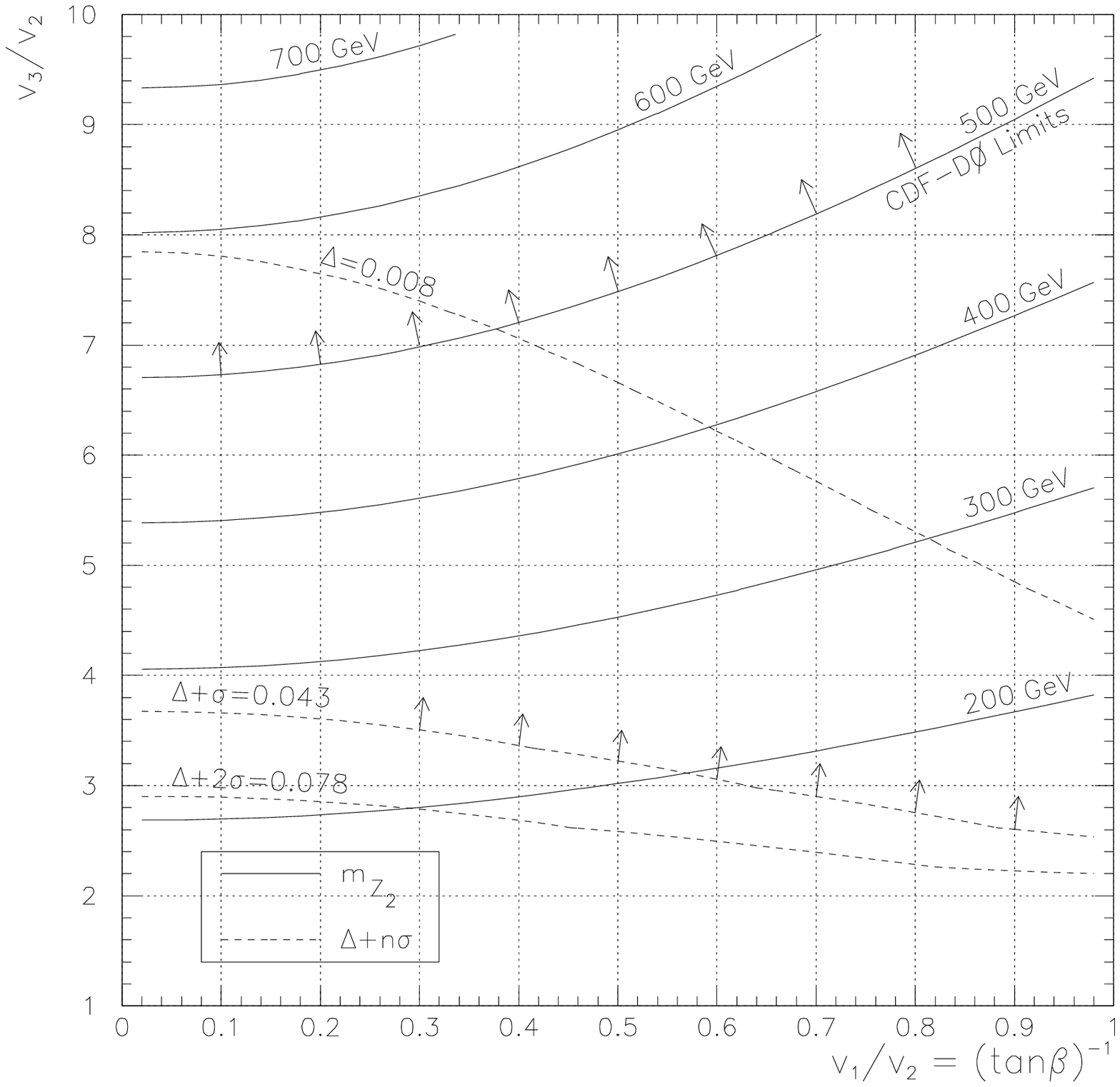}}
\caption[$(v_3/v_2,v_1/v_2)$ Parameter Space Constraints]{\footnotesize   
     Plot   of  $m_{Z_2}^{\mbox{}}$  and $\Delta$   contour  lines  as a
     function of $v_3/v_2$  and $v_1/v_2\,$.  The $\Delta$ contour lines
     are shown at the $0\sigma\,$,   $1\sigma\,$, and $2\sigma$  levels.
     The  arrows  point toward  the  allowed  regions on  the plot ({\it
     cf}.~\cite{kn:EllisA}).   The $m_{Z_2}^{\mbox{}}=500GeV$ line shows
     the $CDF$ and $D0\!\!\!/$  constraints, assuming standard couplings
     \cite{kn:Shochet}.}
\label{fig:mztwo}
\end{encapfig}

     The first thing that has to be constrained are the $VEV\,$'s. It is
reasonable to assume  that $v_1/v_2\approxle1$,  since $m_b<<m_t\,$, for
any reasonable  range  of  Yukawa couplings  \cite{kn:EllisA,kn:Gunion}.
Now the ratios $v_1/v_2$ and $v_3/v_2$  can be constrained by looking at
how the variation in  the $Z_1$ ({\it i.e.},  the ``$Z$'')  mass affects
$\bar\xw$ ($\equiv\sin^2\bar\theta_W^{\mbox{}}$) such that
\cite{kn:EllisA}
\begin{equation}
\sin^2\bar\theta_W^{\mbox{}}\equiv 1-\frac{m_W^2}{m_{Z_1}^2}\;\;<\;\;
\sin^2\theta_W^{\mbox{}}\equiv \left.\frac{\gps}{g^2+\gps}
\right|_{\mu=\mw}\,,
\end{equation}
where $\bar\xw$  and   $\xw$  are  calculated  using the   on-shell  and
$\overline{MS}$ schemes \cite{kn:PDG}, respectively.
$\xw(\mw)$    can     be   found     by  evolving
$\xw(\mz)\approx0.2319\pm0.0005$    \cite{kn:PDG} down   to      $\mw\,$
\cite{kn:Barger},  which  gives $\xw(\mw) \approx 0.233\pm0.035\,$, with
$\alpha^{-1}(\mz)  \approx   127.9\pm0.1$ \cite{kn:PDG},  $\mz   \approx
(91.187\pm0.007)   GeV$ \cite{kn:PDG}, and  $\mw  \approx (80.23\pm0.18)
GeV$           \cite{kn:Shochet}.  Therefore    given  $\bar\xw  \approx
0.2247\pm0.0019$
\cite{kn:PDG} yields
\begin{equation}
\Delta\equiv\xw-\bar\xw\approx0.008\pm0.035\,.
\end{equation}
Figure~\ref{fig:mztwo} shows the $\Delta$ contour line  as a function of
$v_3/v_2$ and $v_1/v_2$,  along with its  $1\sigma$ and  $2\sigma$ level
contour lines.
Also shown  are the $\mztwo$   contour lines.  Taking a $1\sigma$  level
constraint  implies     $v_3/v_2   \approxge     \order{3.5}$      ({\it
cf}. \cite{kn:Hewett,kn:EllisA}) and $\mztwo \approxge  \order{200}GeV$.
Unfortunately  these constraints are  not  that tight due  to the  large
uncertainty in $\alpha(\mz)$.   A stronger  constraint  can be found  by
using   the  $CDF$  and  $D0\!\!\!/$    limits on  the    $\mztwo$  mass
\cite{kn:Shochet}, assuming SM-like   couplings, figure~\ref{fig:mztwo}.
This  constraint is  fairly   reasonable  since $Y_E{\rm    's}\sim{\cal
O}(Y)$'s  ({\it  cf}.   table~\ref{tb:parpr}).  With  these  constraints
$v_3/v_2 \approxge \order{7.5}$ and $\mztwo\approxge \order{500}GeV$.

  Figures \ref{fig:mhzeroa}  through \ref{fig:mhzerob} show $H^0_1$ mass
contour  plots    as   a  function  of   $\mpzero$    and  $\mhpm$   for
$(v_1/v_2,v_3/v_2)=(0.02,6.7)$,     $(v_1/v_2,v_3/v_2)=(0.5,7.7)$,   and
$(v_1/v_2,v_3/v_2)=(0.9,9.1)$, respectively, such that $m_{Z_2}$    lies
roughly around  the $CDF$ and  $D0\!\!\!/$ limits.  These figures  are a
fairly good representation of  the behavior of the $m_{H^0_1}^{\mbox{}}$
contour  lines as   a function of  $v_1/v_2$.   For  fixed $v_1/v_2$ the
contour lines change   very little ({\it i.e.},  $\approxle\order{5}$\%)
for $\order{10}\approxge  v_3/v_2\approxge\order{4.5}\,$.   This  region
corresponds   to  the     $(v_1/v_2,v_3/v_2)$   parameter  space     for
$m_{Z_2}\approxge\order{300}GeV$  depicted  in   figure~\ref{fig:mztwo}.
Further     examination of      the    other   scalar-Higgses      shows
$m_{H^0_3}^{\mbox{}}$ is fairly  insensitive to  variations in $\mpzero$
and $\mhpm$ for  fixed $\mztwo$ and is  slightly sensitive to variations
in $v_3/v_2$, whereas  the behaviour of $m_{H^0_2}^{\mbox{}}$ appears to
be quite sensitive to any variation.  Fortunately for the range of VEV's
considered here ({\it i.e.}, large $v_3$), the only contributions to the
parton level cross-sections turn  out to be  the diagrams  which contain
the  $Z_i$  and $H^0_3$ propagators; the  other   terms are, in general,
suppressed by several orders of  magnitude.\footnote{{\it i.e.}, in  the
large $v_3$   limit  the  couplings $P^0L^+L^-\rightarrow0$, {\it   via}
equations~\ref{eq:cupc}  and~\ref{eq:ratb}, and  $H^0_iL^+L^-\rightarrow
-(\mheavyl/v_3)\,\delta_{3i}\,$, {\it via} equations~\ref{eq:cupb} and
\ref{eq:ratae}, and equation 4.12  of Hewett  and Rizzo~\cite{kn:Hewett}
for  the  $U_{3i}$'s in this  limit,  to $\order{1/v_3}$.} Therefore the
heavy lepton production  cross-section is  insensitive to variations  in
$\mpzero$ and $\mhpm\,$.
Here the $P^0$ mass will be  set to $200GeV$.  The corresponding $H^\pm$
mass was chosen to be $215GeV$ for figure \ref{fig:mhzeroa} and $212GeV$
for figures \ref{fig:mhzeroc}  and \ref{fig:mhzerob}, which  lies within
the allowed  regions on the $m_{H^0_1}^{\mbox{}}$  contour plots.  Based
on   the very limited    experimental  constraints  that  do   exist for
supersymmetric models~\cite{kn:PDG} these appear to be very conservative
choices.   They also lead to  fairly  reasonable values  for the $H^0_i$
masses.

%
%
%
\begin{encapfig}{hbtp}
\mbox{\epsfxsize=144mm
	\epsffile{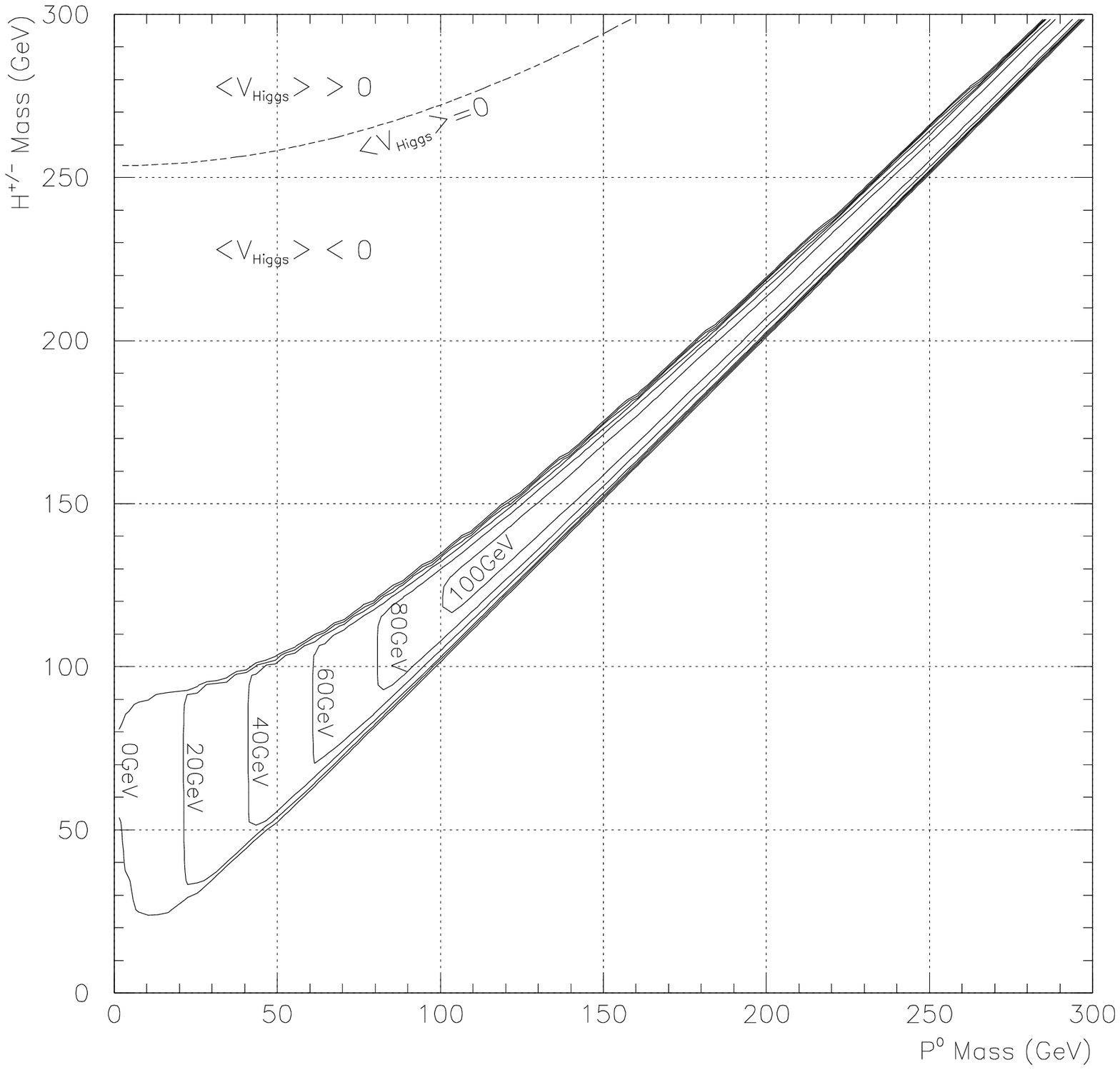}}
\caption[$m_{H^0_1}^{\mbox{}}(\mpzero,\mhpm)$ contour plot for 
         $v_1/v_2=0.02$ and $v_3/v_2=6.7\,$]{\footnotesize A plot of the
         $H^0_1$  mass contour  lines as  a   function of $\mpzero$  and
         $\mhpm$,    for    $v_1/v_2=0.02$        and    $v_3/v_2=6.7\,$
         ($\mztwo\approx496GeV$).    The  dashed   curve  in the   upper
         left-hand corner is a plot  of the zero  of the Higgs potential
         above which it becomes positive.}
\label{fig:mhzeroa}
\end{encapfig}

%
%
%
\begin{encapfig}{hbtp}
\mbox{\epsfxsize=144mm
	\epsffile{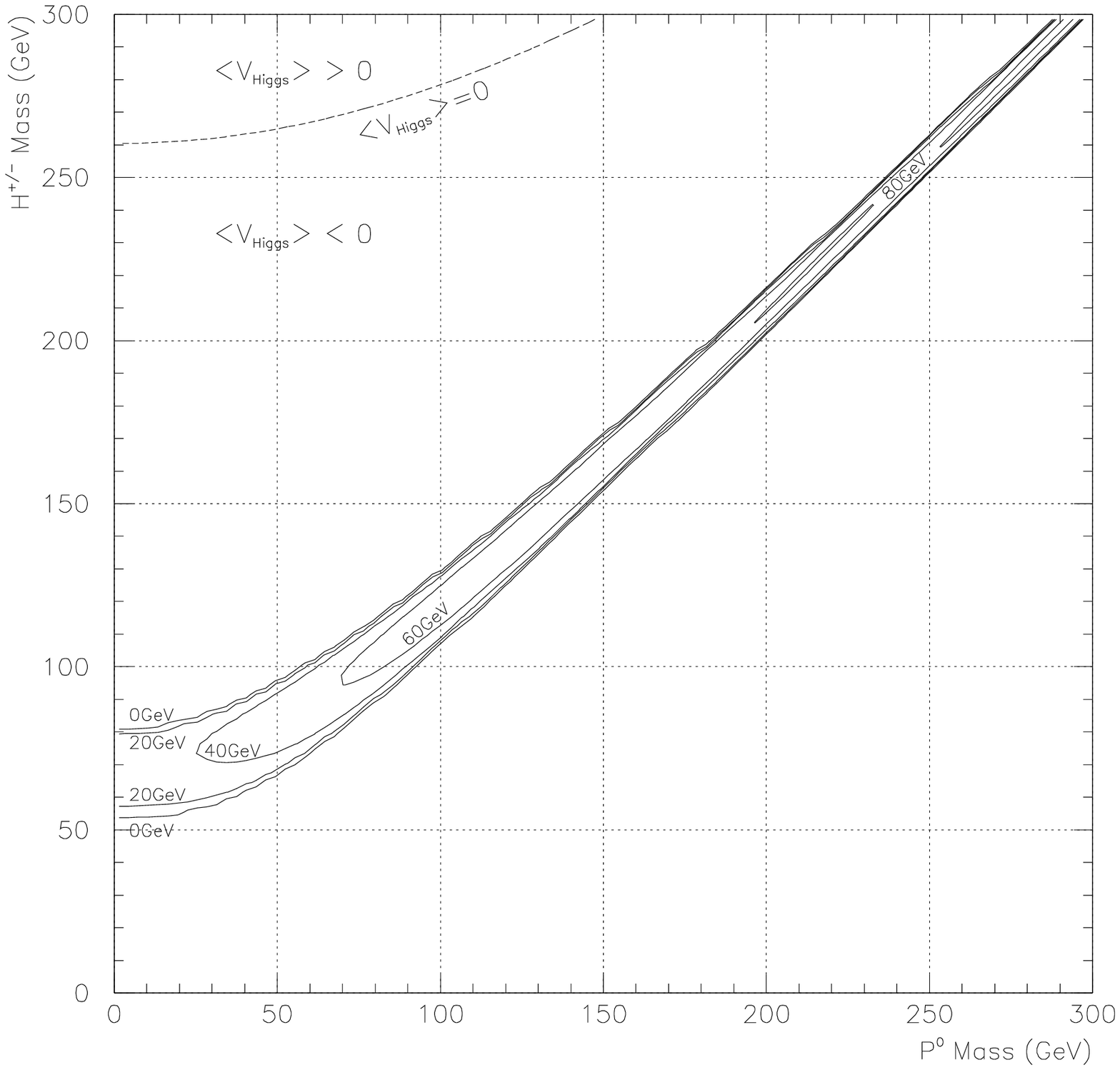}}
\caption[$m_{H^0_1}^{\mbox{}}(\mpzero,\mhpm)$ contour plot for 
         $v_1/v_2=0.02$ and $v_3/v_2=7.7\,$]{\footnotesize A plot of the
         $H^0_1$  mass contour  lines   as a function  of  $\mpzero$ and
         $\mhpm$,         for     $v_1/v_2=0.5$  and     $v_3/v_2=7.7\,$
         ($\mztwo\approx509GeV$).  The    dashed  curve   in   the upper
         left-hand corner is  a plot of the  zero of the Higgs potential
         above which it becomes positive.}
\label{fig:mhzeroc}
\end{encapfig}

%
%
%
\begin{encapfig}{hbtp}
\mbox{\epsfxsize=144mm
	\epsffile{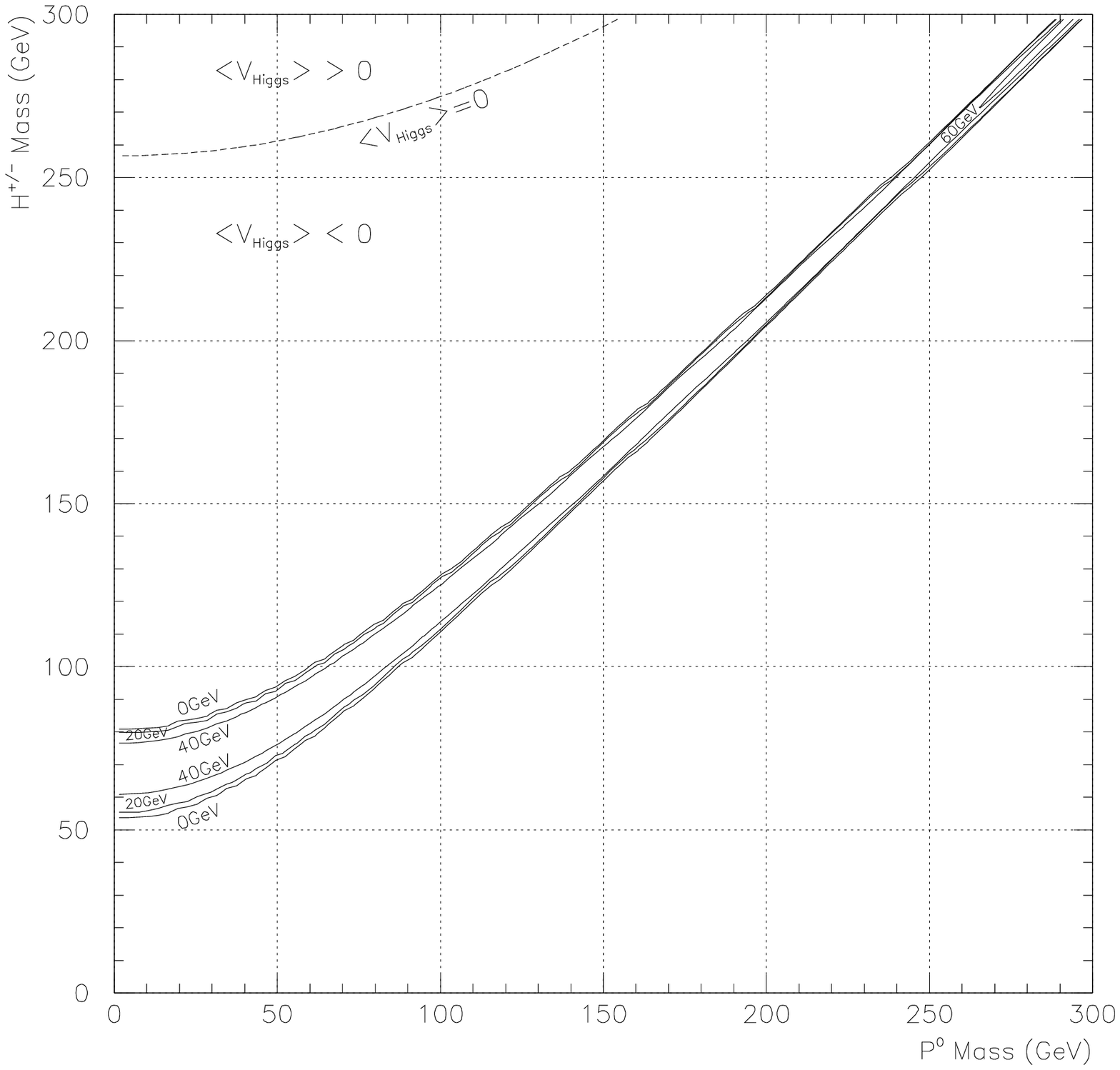}}
\caption[$m_{H^0_1}^{\mbox{}}(\mpzero,\mhpm)$ contour plot for 
         $v_1/v_2=0.9$  and $v_3/v_2=9.1\,$]{\footnotesize A plot of the
         $H^0_1$  mass  contour lines  as  a  function of  $\mpzero$ and
         $\mhpm$,     for   $v_1/v_2=0.9$     and        $v_3/v_2=9.1\,$
         ($\mztwo\approx499GeV$).    The    dashed curve  in  the  upper
         left-hand corner is a plot  of the zero  of the Higgs potential
         above which it becomes positive.}
\label{fig:mhzerob}
\end{encapfig}

   The next parameters that need to be fixed are the soft terms. Exactly
how       these   terms   should behave   at         low  energy is  not
clear~\cite{kn:Konig}.  At   the moment, their   behaviour is very model
dependent  and unless supersymmetric particles  are found this situation
will  most   likely remain so.   Here   the soft terms  will  be treated
parametrically    as  function  of a    single    parameter $\ms\,$.  In
particular, the  soft terms will be   assumed to be  degenerate, $\tilde
M_{f}\approx A_f\approx \ms\,$, with the exception  of the $\lambda$ and
$A$  terms  which were  fixed  by  selecting  the $\mpzero$  and $\mhpm$
masses.  The selection  of   the soft  terms   in this   way,  including
$\lambda$  and $A$ for $m_{P^0,\,H^\pm}^{\mbox{}}\approxle\order{1}TeV$,
typify the generic outcome for the sfermion masses of most SUSY-breaking
models,   {\it    cf}.~\cite{kn:EllisB,kn:Gunion}.  In     these  models
$\order{0.2}TeV\approxle\ms\approxle\order{10}TeV$ \cite{kn:Konig}.  How
low   $\ms$ can be  pushed  depends upon  the  choice of VEV's ($v_3$ in
particular, since it is relatively  large).  For the  VEV's used here it
was  found $\ms\approxge\order{400-450}GeV$.  In  general the  sfermions
with light   spartners have  masses  $\approxge\order{\ms}\,$, which are
roughly degenerate (within  $\order{50}GeV$)  with their mass-eigenstate
partners.  The stops, $\tilde t^\prime_{1,2}\,$, and the exotic squarks,
$\tilde             q^\prime_{1,2}\,$,           have         splittings
$\approxge\order{\ffrac{1}{2}\,m_{t,\,q^\prime}}\,$, for  fermion masses
$\order{200}\approxle  m_{t,\,q^\prime}\approxle\order{600}GeV$, for low
values  of  $\ms\,$.  As  $\ms$  approaches $\order{1}TeV$  all   of the
sfermion mass become degenerate and $\approx\order{\ms}\,$.

  Finally there is  the matter of fixing  the heavy fermion masses.  The
heavy quark masses, $m_{q^\prime}\,$, will be assumed degenerate as will
the   heavy   charged   lepton  masses,   $m_{l^\prime}\,$,   such  that
$m_{q^\prime,\,l^\prime}\approxge\order{0.1}TeV  $~\cite{kn:PDG,kn:MAD}.
The $e^{\prime^\pm}$ will be  designated to play  the role of the  heavy
charged leptons, $L^\pm\,$.

\begin{encapfig}{hbtp}
\mbox{\epsfxsize=144mm
	\epsffile{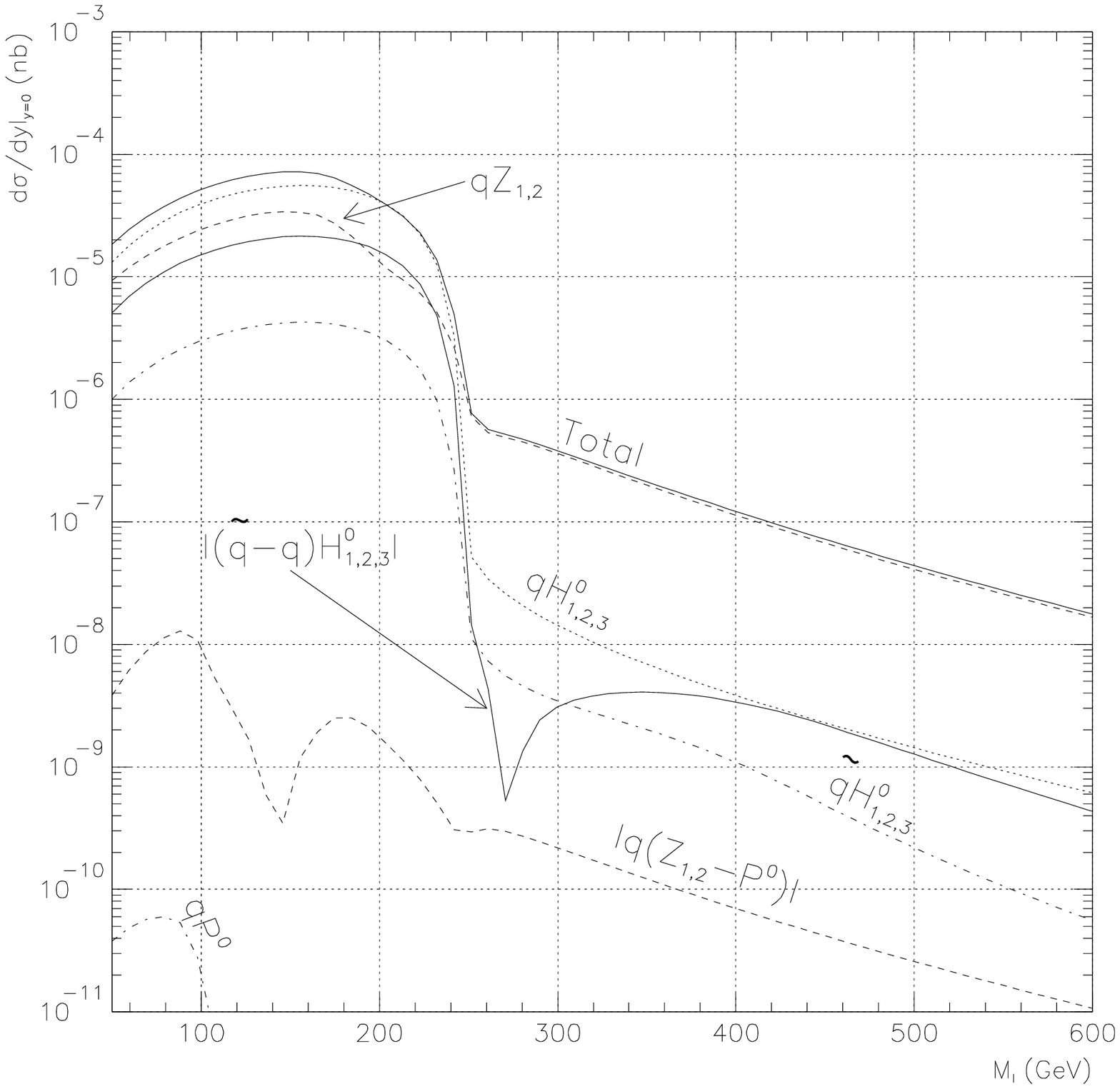}}
\caption[$\frac{d\sigma}{dy}|_{y=0}$ at $LHC$ 
         for     $v_1/v_2=0.02\,$, $v_3/v_2=6.7\,$,  $\ms=400GeV$,   and
         $\mQ=200GeV$]{\footnotesize Rapidity  distribution at $y=0$ for
         heavy charged lepton production   at the $LHC$ ($14TeV$)   as a
         function  of  heavy   lepton   mass,  where   $v_1/v_2=0.02\,$,
         $v_3/v_2=6.7\,$, and $\ms=400GeV$.  The  mass spectrum for  the
         non-SM  particles    involved   in   these   processes     are,
         $\mztwo\approx496GeV$                 ($\gztwo\approx20.9GeV$),
         $\mpzero\approx200GeV$               ($\gpzero\approx16.4GeV$),
         $\mhpm\approx215GeV$,                    $\mhone\approx94.3GeV$
         ($\ghone\approx7.50\times10^{-3}GeV$),    $\mhtwo\approx200GeV$
         ($\ghtwo\approx16.5GeV$),               $\mhthree\approx495GeV$
         ($\ghthree\approx0.230GeV$), $\mQ=200GeV$.}
\label{fig:decaya}
\end{encapfig}

\begin{encapfig}{hbtp}
\mbox{\epsfxsize=144mm
	\epsffile{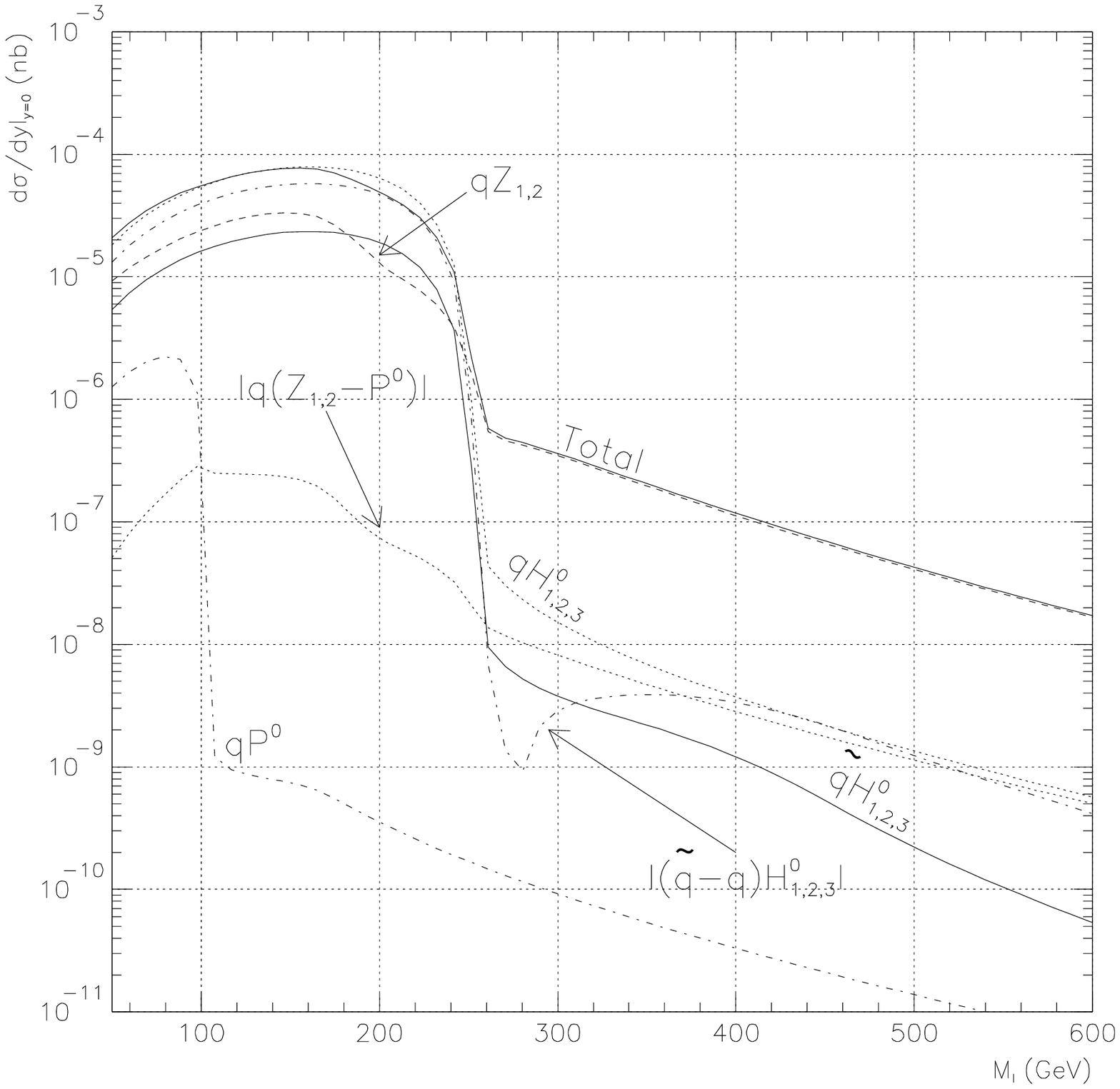}}
\caption[$\frac{d\sigma}{dy}|_{y=0}$ at $LHC$  for $v_1/v_2=0.5\,$,
         $v_3/v_2=7.7\,$,  $\ms=400GeV$, and $\mQ=200GeV$]{\footnotesize
         Rapidity distribution  at  $y=0$   for  heavy   charged  lepton
         production at the $LHC$ ($14TeV$) as a function of heavy lepton
         mass, where $v_1/v_2=0.5\,$, $v_3/v_2=7.7\,$, and $\ms=400GeV$.
         The mass  spectrum for the  non-SM  particles involved in these
         processes are, $\mztwo\approx509GeV$  ($\gztwo\approx21.5GeV$),
         $\mpzero\approx200GeV$  ($\gpzero\approx2.52\times10^{-2}GeV$),
         $\mhpm\approx212GeV$,                    $\mhone\approx75.4GeV$
         ($\ghone\approx3.65\times10^{-3}GeV$),    $\mhtwo\approx212GeV$
         ($\ghtwo\approx7.49\times10^{-2}GeV$),  $\mhthree\approx507GeV$
         ($\ghthree\approx0.198GeV$), $\mQ=200GeV$.}
\label{fig:decayb}
\end{encapfig}

\begin{encapfig}{hbtp}
\mbox{\epsfxsize=144mm
	\epsffile{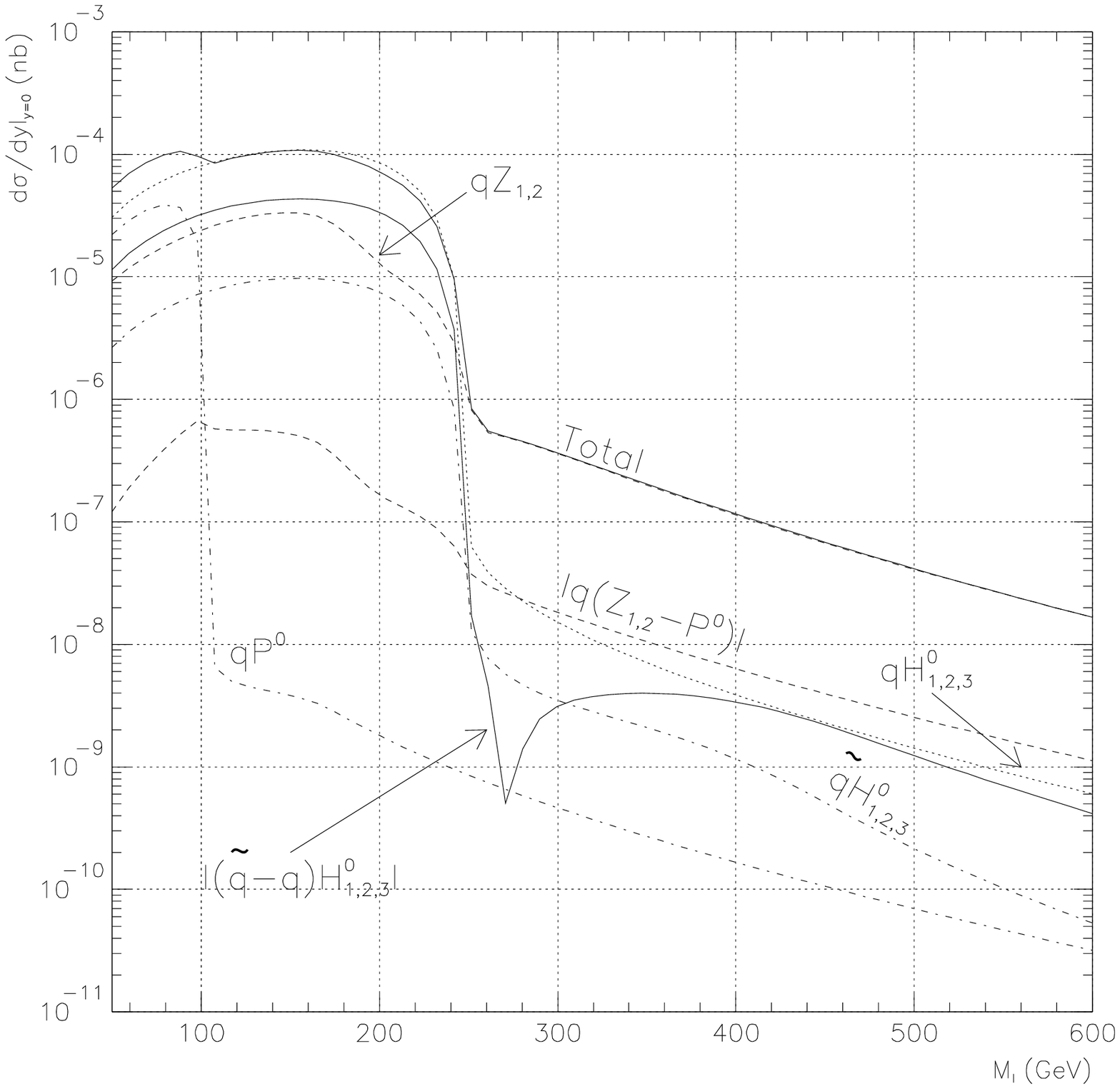}}
\caption[$\frac{d\sigma}{dy}|_{y=0}$  at $LHC$ for 
         $v_1/v_2=0.9\,$,   $v_3/v_2=9.1\,$,      $\ms=400GeV$,      and
         $\mQ=200GeV$]{\footnotesize Rapidity distribution at $y=0$  for
         heavy  charged  lepton production at   the $LHC$ ($14TeV$) as a
         function  of    heavy  lepton  mass,    where  $v_1/v_2=0.9\,$,
         $v_3/v_2=9.1\,$, and $\ms=400GeV$.   The mass spectrum  for the
         non-SM    particles     involved   in  these    processes  are,
         $\mztwo\approx499GeV$                 ($\gztwo\approx20.8GeV$),
         $\mpzero\approx200GeV$  ($\gpzero\approx8.13\times10^{-3}GeV$),
         $\mhpm\approx212GeV$,                    $\mhone\approx52.3GeV$
         ($\ghone\approx1.87\times10^{-3}GeV$),    $\mhtwo\approx216GeV$
         ($\ghtwo\approx1.37\times10^{-2}GeV$),  $\mhthree\approx498GeV$
         ($\ghthree\approx0.130GeV$), $\mQ=200GeV$.}
\label{fig:decayc}
\end{encapfig}

\begin{encapfig}{hbtp}
\mbox{\epsfxsize=144mm
	\epsffile{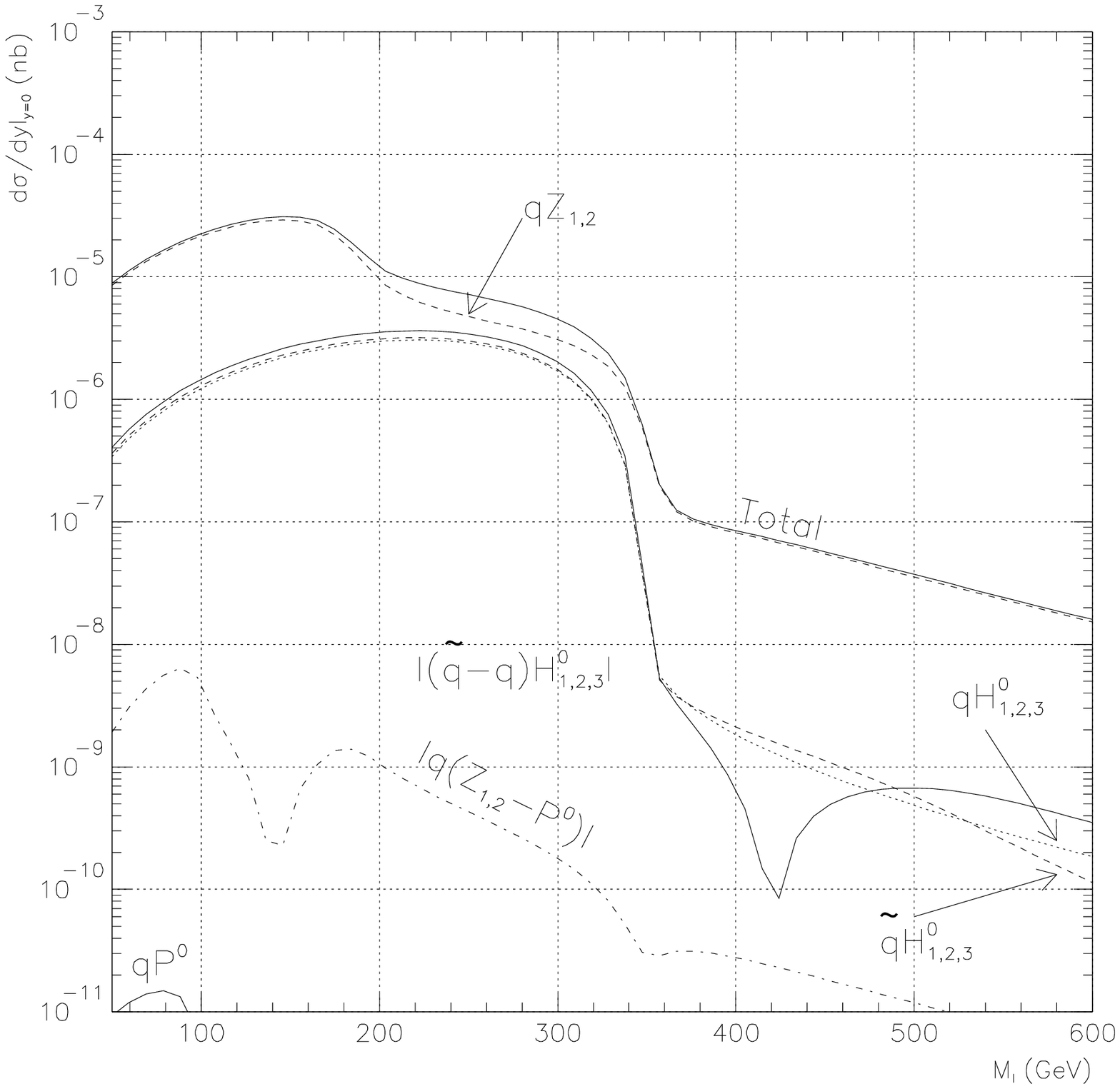}}
\caption[$\frac{d\sigma}{dy}|_{y=0}$  at $LHC$ for $v_1/v_2=0.02\,$, 
         $v_3/v_2=9.5\,$, $\ms=450GeV$, and  $\mQ=200GeV$]{\footnotesize
         Rapidity  distribution  at  $y=0$    for  heavy charged  lepton
         production at the $LHC$ ($14TeV$) as a function of heavy lepton
         mass,   where   $v_1/v_2=0.02\,$,        $v_3/v_2=9.5\,$,   and
         $\ms=450GeV$.   The mass   spectrum for   the non-SM  particles
         involved     in  these   processes  are,  $\mztwo\approx700GeV$
         ($\gztwo\approx31.9GeV$),                $\mpzero\approx200GeV$
         ($\gpzero\approx16.5GeV$),                $\mhpm\approx215GeV$,
         $\mhone\approx94.6GeV$   ($\ghone\approx7.49\times10^{-3}GeV$),
         $\mhtwo\approx200GeV$                 ($\ghtwo\approx16.5GeV$),
         $\mhthree\approx700GeV$             ($\ghthree\approx1.04GeV$),
         $\mQ=200GeV$.}
\label{fig:decayd}
\end{encapfig}

\begin{encapfig}{hbtp}
\mbox{\epsfxsize=144mm
	\epsffile{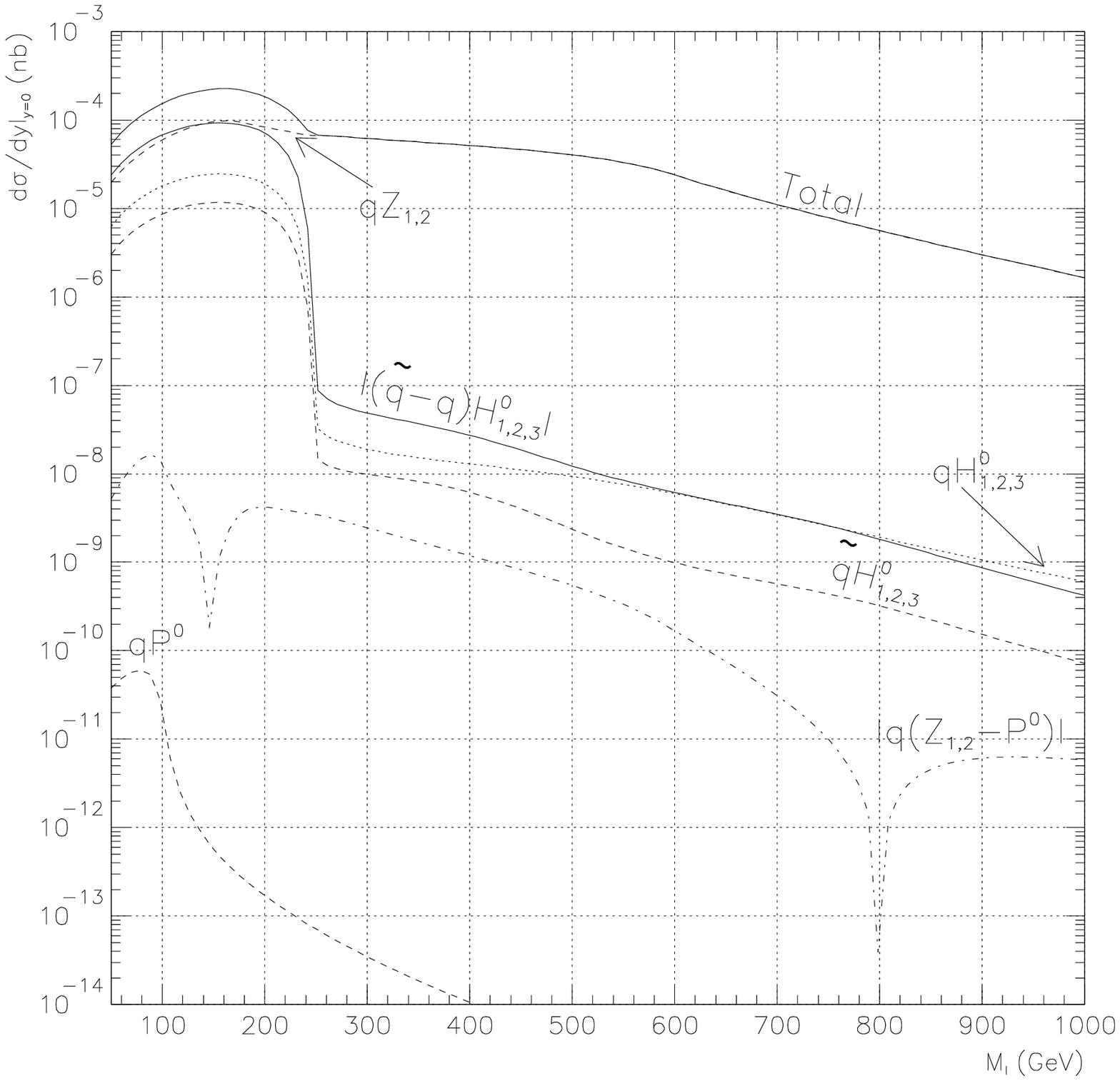}}
\caption[$\frac{d\sigma}{dy}|_{y=0}$  at $LHC$ for 
         $v_1/v_2=0.02\,$,    $v_3/v_2=6.7\,$,       $\ms=400GeV$,   and
         $\mQ=600GeV$]{\footnotesize Rapidity distribution at $y=0$  for
         heavy charged lepton production   at the $LHC$ ($14TeV$) as   a
         function  of     heavy lepton mass,     where $v_1/v_2=0.02\,$,
         $v_3/v_2=6.7\,$, and $\ms=400GeV$.   The  mass spectrum  is for
         the  non-SM   particles  involved   in  these    processes are,
         $\mztwo\approx496GeV$                 ($\gztwo\approx19.4GeV$),
         $\mpzero\approx200GeV$               ($\gpzero\approx16.4GeV$),
         $\mhpm\approx215GeV$,                    $\mhone\approx94.3GeV$
         ($\ghone\approx7.50\times10^{-3}GeV$),    $\mhtwo\approx200GeV$
         ($\ghtwo\approx16.5GeV$),               $\mhthree\approx495GeV$
         ($\ghthree\approx0.138GeV$), $\mQ=600GeV$.}
\label{fig:decaye}
\end{encapfig}

\begin{encapfig}{hbtp}
\mbox{\epsfxsize=144mm
	\epsffile{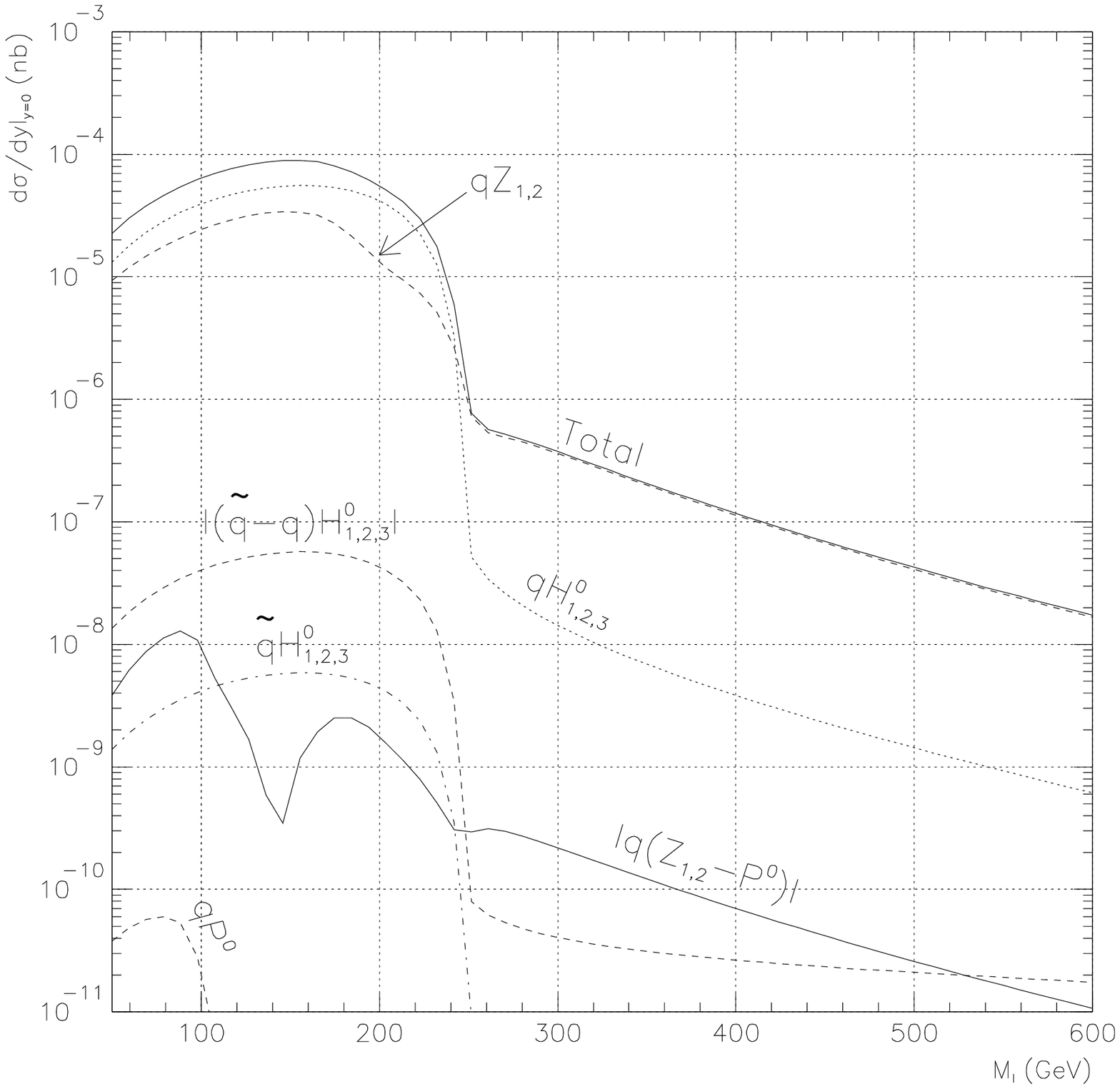}}
\caption[$\frac{d\sigma}{dy}|_{y=0}$  at $LHC$ for 
         $v_1/v_2=0.02\,$,       $v_3/v_2=6.7\,$,    $\ms=1TeV$,     and
         $\mQ=200GeV$]{\footnotesize Rapidity  distribution at $y=0$ for
         heavy charged lepton  production  at the  $LHC$ ($14TeV$)  as a
         function   of  heavy  lepton   mass,   where  $v_1/v_2=0.02\,$,
         $v_3/v_2=6.7\,$, and  $\ms=1TeV$.   The  mass spectrum for  the
         non-SM  particles  involved     in    these  processes     are,
         $\mztwo\approx496GeV$                 ($\gztwo\approx20.9GeV$),
         $\mpzero\approx200GeV$               ($\gpzero\approx16.4GeV$),
         $\mhpm\approx215GeV$,                    $\mhone\approx94.3GeV$
         ($\ghone\approx7.50\times10^{-3}GeV$),    $\mhtwo\approx200GeV$
         ($\ghtwo\approx16.5GeV$),               $\mhthree\approx495GeV$
         ($\ghthree\approx0.230GeV$), $\mQ=200GeV$.}
\label{fig:decayf}
\end{encapfig}

\begin{encapfig}{hbtp}
\mbox{\epsfxsize=144mm
	\epsffile{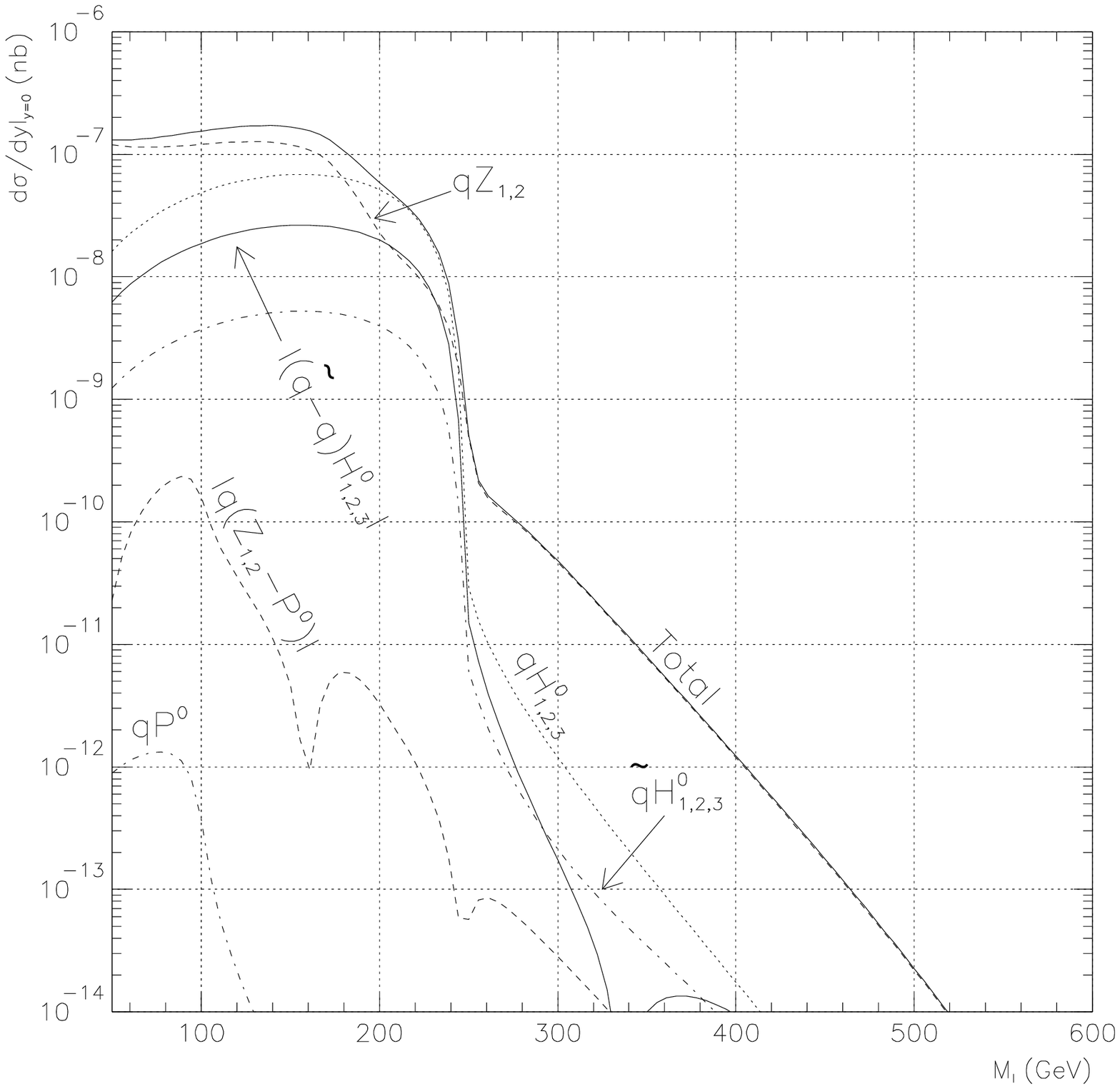}}
\caption[$\frac{d\sigma}{dy}|_{y=0}$  at the 
         $\teva$   for  $v_1/v_2=0.02\,$, $v_3/v_2=6.7\,$, $\ms=400GeV$,
         and  $\mQ=200GeV$]{\footnotesize Rapidity distribution at $y=0$
         for heavy charged  lepton production at the $\teva$  ($1.8TeV$)
         as  a function of  heavy  lepton mass, where  $v_1/v_2=0.02\,$,
         $v_3/v_2=6.7\,$, and $\ms=400GeV$.   The mass spectrum for  the
         non-SM  particles    involved    in   these    processes   are,
         $\mztwo\approx496GeV$                 ($\gztwo\approx20.9GeV$),
         $\mpzero\approx200GeV$               ($\gpzero\approx16.4GeV$),
         $\mhpm\approx215GeV$,                    $\mhone\approx94.2GeV$
         ($\ghone\approx7.50\times10^{-3}GeV$),    $\mhtwo\approx200GeV$
         ($\ghtwo\approx16.5GeV$),               $\mhthree\approx495GeV$
         ($\ghthree\approx0.230GeV$), $\mQ=200GeV$.}
\label{fig:decayg}
\end{encapfig}

    Figures \ref{fig:decaya} through  \ref{fig:decayg} show the rapidity
distribution   at rapidity   y=0  for $p\porbar\rightarrow gg\rightarrow
L^+L^-\,$ as  a  function of $\mheavyl\,$,  for  various scenarios.  The
rapidity, $y$, is defined by
\begin{equation}
y=\onetwo\ln\frac{E+p_z}{E-p_z}
\end{equation}
and is related to the motion of the center  of mass (c.m.)  with respect
to the lab  frame, where $E$  is the c.m.  energy and  $p_z$ is the  net
momentum of    the  incoming    hadrons.  The   rapidity   distribution,
$d\sigma/dy\,$,   is   related  to    the parton   level  cross-section,
$\hat\sigma(gg\rightarrow L^+L^-)$, through
\begin{equation}
\frac{d\sigma}{dy}=\int_{\tau_{min}}^{e^{|y|}}d\tau
G(\sqrt{\tau}e^y,Q^2)G(\sqrt{\tau}e^{-y},Q^2)\hat\sigma(\tau s)\,,
\end{equation}
where   $\tau=\hat s/s\,$,  $\tau_{min}=4m_L^2/s\,$,  $\sqrt{s}$  is the
center-of-mass energy, and $G(x,Q^2)$ is  the gluon structure  function.
The values  of   $\sqrt{s}$  have been    set to   $14TeV$  ($LHC$)  for
figures~\ref{fig:decaya}      through~\ref{fig:decayf},  and    $1.8TeV$
($\teva$)  for   figure~\ref{fig:decayg}.   In  these   figures the $SM$
couplings and masses where extracted  from the PDG~\cite{kn:PDG}, except
for    $m_t\approx174\pm13GeV$~\cite{kn:Shochet}.   For $G(x,Q^2)$   the
leading order Duke and Owens 1.1 ($DO1.1$) \cite{kn:Duke,kn:Owens} gluon
distribution was  used.  The  results  were  compared with the   next to
leading  order $MRSA$~\cite{kn:MRSA}  gluon distribution function, which
yielded a  negligible difference~\cite{kn:MAD}.   Although these results
include squark mixing it was found that  there was no significant change
if mixing  is  not included~\cite{kn:MAD}.  Since  $d\sigma/dy$  is flat
about $y=0$
the  relationship  between  $\frac{d\sigma}{dy}|_{y=0}$  and  the  total
cross-section  is immediate.   Therefore the  total  event rate for  the
$p\porbar\rightarrow gg\rightarrow  L^+L^-$ production  mechanism can be
estimated from $y=0\,$, {\it i.e.},
\begin{equation}
\sigma=\int_{\ln\sqrt{\tau_{min}}}^{-\ln\sqrt{\tau_{min}}}
\frac{d\sigma}{dy}dy\approx -\ln(\tau_{min})\,\sigma\,.
\end{equation}

    Figures       \ref{fig:decaya} through  \ref{fig:decayc}        show
$\frac{d\sigma}{dy}|_{y=0}$ for   different    VEV's  ratios  along  the
$\mztwo\approx\order{500}GeV$  contour line of   figure~\ref{fig:mztwo}.
Notice that as $v_1/v_2$ becomes comparable to $v_3/v_2$ the large $v_3$
limit breaks   down  and  the   generally small  $qP^0$  term  starts to
contribute  (the     $q(Z_{1,2}-P^0)$ contribution also    grows   quite
significantly but  remains a   negligible contribution).   Therefore for
relatively  large   values   of   $v_1/v_2$  variations    in  $\mpzero$
($\approx\mhpm$  up   to at  least  $\order{1}TeV$)  become   important.
However it is more natural  to assume that the intra-generational Yukawa
couplings are of the same order of magnitude and therefore for $v_1/v_2$
to  be small.  For  the rest of  these figures then,  it will be assumed
that $v_1/v_2=0.02\,$.   Figure~\ref{fig:decaya} is  the figure with the
default values.

   Figure~\ref{fig:decayd} shows what happens  when a larger  $Z_2$ mass
of  $\order{700GeV}$  ({\it i.e.},  $v_3/v_2=9.5$)  is  used.  For  this
figure $\ms$ had to be   pushed up slightly   to $450GeV$, in order   to
produce physical squark masses.  The noticeable difference between  this
and  all of the other  figures is that the peak  has broadened.  This is
expected  since the  $Z_2$  can  remain  on-shell  for larger values  of
$\mheavyl\,$.  Notice that the $H^0_3$ resonance cut off seems to follow
the  $Z_2$'s.  More precisely  $\mhthree\approx\mztwo$  for large $v_3$.
This becomes  immediately evident when  taking the large $v_3$ limits of
the $H^0_i$ and   $Z_i$ mass mixing matrices,  equations \ref{eq:mhimat}
and \ref{eq:mzzmat} respectively: {\it i.e.},
\begin{equation}
\lim_{v_3\rightarrow\infty}m_{H^0_3}^2 =
\lim_{v_3\rightarrow\infty}m_{Z_2}^2   =
\frac{25}{36}\,\gpps\,v_3^2         \approx
\frac{25}{9}\,\frac{(v_3/v_2)^2\,\xw}{1+(v_1/v_2)^2}\,m_Z^2\,,
\end{equation}
which  is in fairly  good agreement with  all  of the figures. Also, the
overall production  is slightly suppressed due  to  the smallness of the
gluon distribution function at large momentum fraction.

  Figure~\ref{fig:decaye}  shows what happens  when the heavy quark mass
was pushed up to $600GeV$. The effect is quite dramatic. To see why this
is    so,    notice   the   slight     kink    in   the  curve    around
$\mheavyl\approx600GeV$.  There is also a much  more significant kink in
all   of    the  other graphs   around   $200GeV$,   {\it i.e.},  around
$\mheavyl\approx\mQ$.    Further  examination    of  the  parton   level
cross-section shows that kink occurs when the  heavy quarks in the loops
can no longer be
on shell.

   In figure~\ref{fig:decayf} the scalar mass  was pushed up to  $1TeV$.
Increasing  $\ms$ has  caused  the terms  involving   the squarks to  be
supressed by several orders  of magnitude.  This supression occurs since
there is  not as much  gluon luminosity  available  to allow these heavy
squarks in the loops to be on shell, as there was for the light squarks.
The $qH^0_i$ term now enhances $L^+L^-$ production, below the $\mhthree$
threshold, as the  destructive  interference with $\tilde qH^0_i$  term,
{\it i.e.}, $(\tilde q-q)H^0_i\,$, has been suppressed.

  Finally   figure~\ref{fig:decayg}  shows what  happens  at $\sqrt{s} =
1.8TeV$, the $\teva$.  The  overall topology is the  same as depicted in
figure~\ref{fig:decaya} but the $L^+L^-$ production rate is dramatically
reduced: very   little gluon luminosity  is available   to produce these
heavy particles.

\section{Discussion} 

\begin{encapfig}{hbtp}
\mbox{\epsfxsize=144mm
	\epsffile{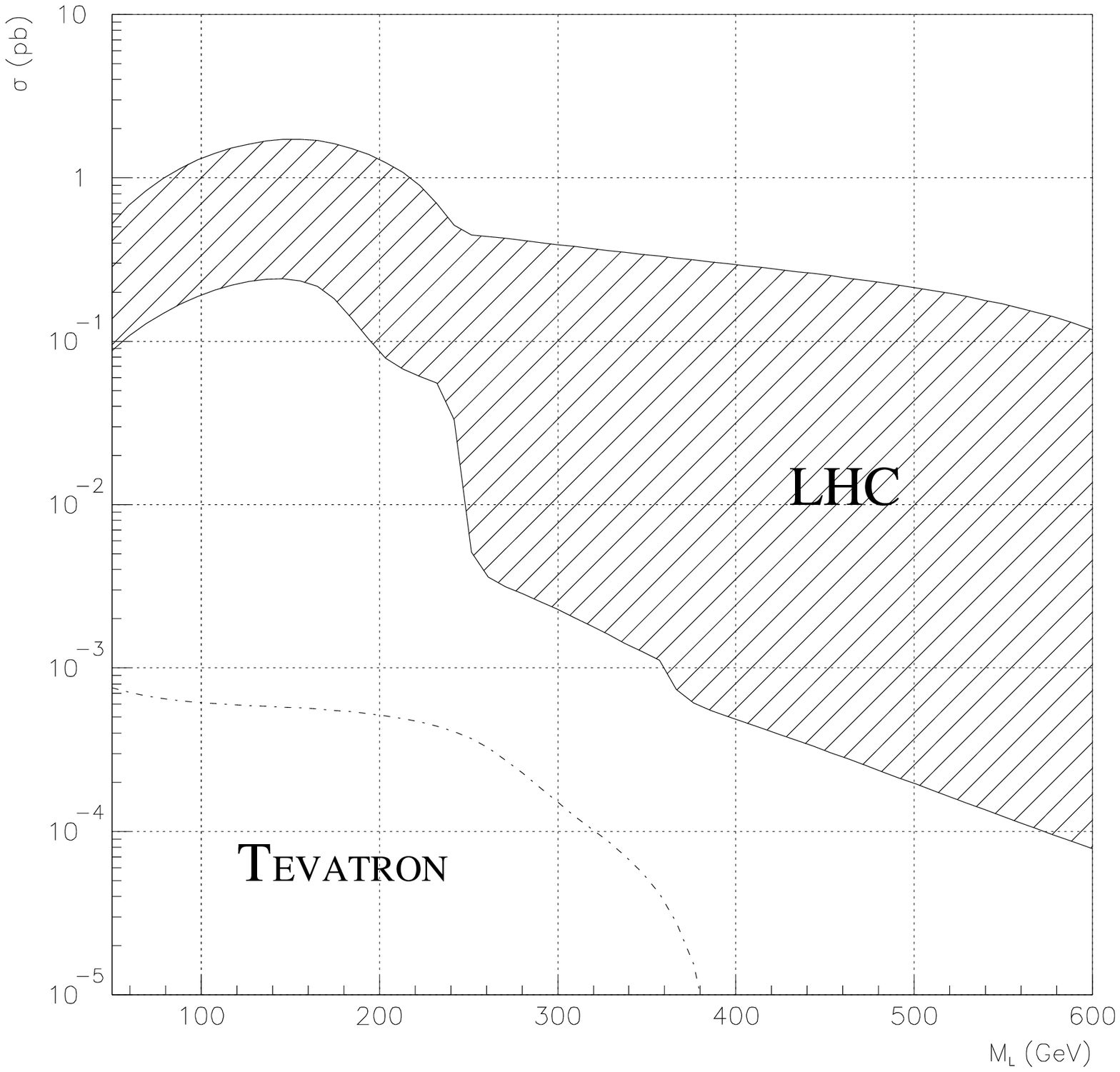}}
\caption[Summary plot of $L^+L^-$     production
              cross-sections]{\footnotesize Summary  plot of results for
              the  total $L^+L^-$ production  cross-section at the $LHC$
              ($\sqrt{s}=14TeV$, ${\cal L}\sim 10^5 pb^{-1}/yr$) and the
              $\teva$  ($\sqrt{s}=1.8TeV$,    ${\cal      L}\sim    10^2
              pb^{-1}/yr$) energies as a  function of $\mheavyl\,$.  The
              hatched region is the results for figures \ref{fig:decaya}
              through \ref{fig:decayf}, and the dash  line is the result
              for figure \ref{fig:decayg}.}
\label{fig:totally}
\end{encapfig}

  Figure~\ref{fig:totally}  gives a  summary  of the total cross-section
for  $L^+L^-$ production for the \Esix  model parameter space studied in
the previous  section.  At  the $\teva$  it is  immediately obvious that
nothing will   be seen, {\it i.e.},  $\approxle\order{0.1}$ $events/yr$.
For  the $LHC$,   over     a  reasonable range  of   parameter    space,
$\order{10^{4\pm1}}$  $events/yr$ are  expected.  The $MSSM$  results of
Cieza   Montalvo,    $et$   $al.$,    ($CM$)~\cite{kn:Montalvo} predicts
$\order{10^{5}}$  $events/yr$, a factor of  at least  $10$ more than our
results~\cite{kn:Konig,kn:MAD}. This is a rather surprising result since
it was expected that the  \Esix event rate would  be enhanced due to the
greater  number  of heavy  particles running around   in the  loop.  The
parameters  in  each  model   were   varied to  try  and  determine  the
difference.  Unfortunately, it turns out that the \Esix parameters space
suppresses   $L^+L^-$  production, since $v_3$   is  fairly large.  This
restriction causes the production to occur  mainly through the $Z_{1,2}$
and    $H^0_3$ terms.  The   $MSSM$ has   two  neutral  Higgses and  one
pseudo-scalar Higgs that are allowed to  contribute to the processes.  A
very simple test of the \Esix model was done by varying $v_3/v_2$, about
$v_1/v_2=0.2$, that showed  for $v_3/v_2=2.8$ and $m_L\approxle100GeV$ a
factor  of 10   increase was  obtained.  However,  this  region of \Esix
parameter space is forbidden, see figure~\ref{fig:mztwo}.

  The $L^\pm\rightarrow\nuheavyl   W^\pm,\,\nuheavyl H^\pm$ decay  modes
are expected to be similar for both models, as these are SM-like decays.
These  modes         depend      upon     the     mass        difference
$\Delta=\mheavyl-\mheavynu\,$.   For $\Delta<  \mw << \mhpm$  the decays
modes will be by virtual $W$'s, $W^*\rightarrow f\bar f\,$, and on shell
for $\Delta>\mw$.  Leptonic decays of the $W$'s offer the possibility of
$L^\pm$ detection by  measuring  $\ell^+\ell^-$ production with  missing
transverse   momentum,    $\missing\,$~\cite{kn:Montalvo}.      The   SM
backgrounds          competing     with   these     processes        are
$pp\rightarrow\tau^+\tau^-,\, W^+W^-,\,  Z^0Z^0$.  Studies  have  shown,
using SM couplings,  that it is  possible  to pull  the $L^+L^-$ signals
from  the  background for $\Delta>\mw$   given  sufficiently large event
rates;     it       is         much       more       difficult       for
$\Delta<\mw$~\cite{kn:Montalvo,kn:BargerB,kn:Hinchliffe}.   Since    the
$MSSM$  event rate is higher than  in \Esix, detection would more likely
indicate   $MSSM$   candidates.    If $\mhpm\approx\order{\mw}\,$,  then
$H^\pm\rightarrow f_i\bar f_j$ dominates,  for naturally large values of
$\tan\beta$.  Since the Higgs likes to couple to massive particles, this
would lead to multiple heavy jet  events which in  general would be very
difficult to pull out of the  background in either  the $MSSM$ or \Esix.
Similar processes are expected to occur for the cases $\mw<\Delta<\mhpm$
and  $\mw\,,\mhpm<\Delta\,$.    For large   enough  $\mhpm$ the sfermion
channels also open, {\it i.e.}, $H^\pm\rightarrow\tilde f_i\tilde f_j^*$
(e.g.  $\tilde  u  \tilde d^*$).   The  sfermions would eventually decay
leaving only the lightest supersymmetric particles ($LSP$'s), which will
escape  undetected along   with   the  $\nuheavyl$'s  leaving  lots   of
$\missing\,$.  In fact, all of the aforementioned processes will lead to
events with   $\missing\,$, as the $\nuheavyl$'s   will pass through the
detector.

  In certain  regions of the $MSSM$  and \Esix model parameter spaces it
may be  possible to distinguished between  the two models.   If $L^+L^-$
event rates are  larger than those predicted by  \Esix, then the  likely
candidate is    the  $MSSM$.  Unlike  the  $MSSM$,  it  is  possible for
$\mhpm<\mw$ in  \Esix~\cite{kn:Hewett} and   therefore if  $H^\pm$'s are
found in  this  mass range the more   likely candidate  would  be \Esix.
Another possible way of telling the models apart is to look for sfermion
production,   {\it i.e.}, $L^\pm\rightarrow  f\tilde f^*\,,\tilde  f\bar
f\,$, which is unique to \Esix, since $L^\pm$ has opposite $R$ parity to
the other $SM$-like fermions in  the {\bf 27}'s (figure~\ref{fig:esix}).
The  sfermion  would eventually   decay to   an  $LSP$  which  is stable
(assuming  $R$  parity conservation),   yielding $jets$+$\missing\,$, in
general.  Whether or  not  it is possible  to distinguish  them from the
$MSSM$ and the $SM$-like backgrounds would  require a much more detailed
study, as  the allowable parameter space  for sfermion masses and Yukawa
couplings  is  quite  large.  Finally,  the    $MSSM$ does have   fairly
stringent unitarity  constraints on  the  heavy lepton and  heavy  quark
masses as  a function of  $\tan\beta$~\cite{kn:Montalvo}, in  particular
$\mheavyl\approxle1200\cos\beta\,$.  Therefore, it should be possible to
eliminate  $(\mheavyl,\tan\beta)$   regions  in   the  $MSSM$   $L^-L^+$
production cross-section plots, as a function of $\mheavyl\,$, such that
only    \Esix  models     are      allowed.  For    example,    assuming
$\mhone\approxge\order{600}GeV$  rules    out        the    $MSSM$   for
$\mheavyl\approxge242GeV$ and   $\tan\beta\approxge5$~\cite{kn:MAD}.  In
the allowed  $MSSM$ region  this gives an  upper  limit on the  $L^+L^-$
production  cross-section of $\order{10}pb$~\cite{kn:MAD}, at the $LHC$.
Also,  in the  $MSSM$   there    are phenomenological constraints     on
$\tan\beta$~\cite{kn:Konig}  which could allow for further restrictions.
A more detailed study of these constraints has not been carried out.

  In closing, it should be pointed out that only a simple model of \Esix
has  been considered.  It is possible  for other \Esix models to produce
results similar to the  model studied here or  to the $MSSM$. Therefore,
in  general, $L^+L^-$ production by   gluon-gluon fusion  should not  be
considered a  definitive means of separating  out  the different models;
several experiments would be required.

\section{Conclusions}

   The   $p\porbar\rightarrow     gg\rightarrow       L^+L^-$ production
cross-section  was  computed for a  simple rank-5  \Esix   model.  For a
fairly conservative survey of  the  various parameters  in the model  we
expect $\order{10^{4\pm1}}$ $events/yr$  at  the  $LHC$; no events   are
expected at the $\teva$.  It was found that the  results were at least a
factor of 10 less than the event rates predicted  for the $MSSM$, due to
the    $CDF$ and  $D0\!\!\!/$  soft limits     ({\it i.e.}, assuming  SM
couplings)  placed  on $\mztwo$~\cite{kn:Shochet}.    These  constraints
resulted in  the  $H^0_{1,2}$ and  $P^0$ contributions   to the $L^+L^-$
production rate being suppressed leaving  only the $H^0_3$ and $Z_{1,2}$
to contribute.  For certain regions   in the $MSSM$ and \Esix  parameter
spaces  it was demonstrated that it   is possible to distinguish between
the two  models, in principle.  However,  it should  be pointed out that
there are  many  candidate \Esix models  which   could yield overlapping
results.

%
%
\chapter{Conclusions}

  The   chapter contains a summary  discussion  of chapters  2 through 4
followed by some closing remarks.

\section{Summary Discussions}
\subsection{String-Flip and Flux-Bubble Models}

  Chapters  2 and 3 of  this  thesis investigated various extensions and
aspects  of string-flip potential  models in an  attempt to extend these
viable models of nuclear matter to obtain nuclear binding.

  In  chapter 2 the  $SU_\ell(3)$   and $SU_h(3)$ three-quark  potential
models were examined  as possible alternatives to  earlier $SU_\ell(2)$,
$SU_h(2)$, and   $SU_h(3)$ (these  models  included   large  multi-quark
structures)   models~\cite{kn:HorowitzI,kn:Watson}  that  may    produce
nuclear  binding.  Similar   results  to  the  $SU_\ell(2)$ models  were
obtained,  but  with no  nuclear binding.    It  was found  that generic
features of all of the aforementioned models were:
\begin{itemize}
\item nucleon gas at low densities with no van der Waals forces
\item nucleon swelling and saturation of nuclear forces with increasing 
density
\item fermi gas of quarks at  higher densities
\end{itemize}
After a detailed analysis it was concluded that  these models would have
to   be extended to  include  perturbative  QCD interactions between the
quarks  in  order   to achieve  nuclear  binding.   Furthermore,  it was
hypothesized that the   colour-hyperfine interactions would prove  to be
the most significant,  as was indicated by some  earlier work  done with
linked-cluster-expansion models~\cite{kn:Nzar}.     Chapter 3 marked the
beginning of these extensions.

  In   chapter  3,  a  new    class  of  potential  models,  called  the
``flux-bubble models,''   were  constructed   that   extended the    old
string-flip potential models to include perturbative interactions.  Some
initial work   was done using  a flux-bubble  model with $SU_\ell(2)$ by
introducing        colour-Coulomb         interactions~\cite{kn:BoyceA}.
Unfortunately, many  problems arose with trying to  construct a new wave
function   which   would take  into   account  the local  nature  of the
perturbative colour-Coulomb  interactions.  Also, the  fact that we were
dealing with  a multi-quark system  which required vast amounts of $CPU$
resources only exacerbated the  problem.  As a  result, a toy model of a
mesonic-molecule ($Q_2$)~\cite{kn:Treurniet}   was     proposed,   which
consisted of two massive quarks ($Q$'s)  bound by an effective potential
produced  by  two light  quarks  ($q$'s),  in order   to sort  out these
difficulties.

   The $Q_2$ system was  analyzed using the old $SU_\ell(2)$ variational
pseudo-hydrogen    wave   function~\cite{kn:Watson},  $\PH\,$,   for the
string-flip and flux-bubble models.  The string-flip models were studied
with and without a  colour-Coulomb term added to the  linear part of the
potential, and the flux-bubble models   were studied using  $SU_\ell(2)$
(with  a colour-Coulomb  extension)  with and  without the colour-charge
being  allowed to flow  from quark to  quark.   In general, it was found
that the old $\PH$ wave function produced a very shallow $Q_2$ potential
well ($\order{3}MeV$ deep), not  enough to bind  the $Q_2$  system.  The
exception was  the  $SU_\ell(2)$ model  in  which the  colour-charge was
allowed to flow.  Here the system dissociated into two mesons, one heavy
and one  light.   Next a   new variational  wave function was  proposed,
coined the  ``pseudo-hydrogen-molecular wave function,''  $\PHH\,$ ({\it
i.e.}, its idea was motivated  by the hydrogen-molecular wave function),
in an attempt to  take into account  the local nature of the flux-bubble
interactions.  The  aforementioned   studies  of  the  string-flip   and
flux-bubble  models  were  repeated using  $\PHH\,$.   Surprisingly, the
$\PHH$ wave function produced a well that was in general $\order{80}MeV$
deep,  deep enough   to bind quarks   with masses $m_q\ge\order{m_c}\,$.
Here the exception to the case, $SU_\ell(2)$ with flowing colour-charge,
did not dissociate.   Although this produced a   state higher in  energy
than with the $\PH$ wave function, it demonstrated  that the $\PHH$ wave
function had the  desired  properties of  localizing the  colour-Coulomb
interactions, as it bound  the  light quarks  to the heavy  quarks which
would have otherwise floated  away.\footnote{It should be re-emphasized,
as stated   earlier in chapter   3, that the   $SU_\ell(2)$ with flowing
colour-charge is not  a physically  viable  model of nuclear  matter.  A
more physical model would use $SU_c(3)$ and assume that heavy quarks are
composites of two light quarks, in which case  it becomes impossible for
the system to dissociate.} Furthermore, the fact that the wells produced
were, in general, much deeper seems to confirm  earlier beliefs that the
old      $SU_\ell(2)$   wave    function  used       to   model  nuclear
matter~\cite{kn:Watson} was a poor choice~\cite{kn:WatsonA}.

  The results of the analysis produced a  more detailed understanding of
intricacies involved in modeling nuclear matter. The situation as it now
stands   consists of    several scenarios    that   were  discussed  in
\S~\ref{sec-chpdsc} which may lead  to nuclear binding. Of  the various
scenarios discussed it is believed that the most likely ones are,
\begin{itemize}
\item the $SU_\ell(2)$ string-flip model using the  many-body wave  
      function, $$\Psi\sim{\rm Perm}|\PH|\,\Phi_{\rm Slater}\,,$$
\item the $SU_c(3)$ flux-bubble model, with flowing colour-charge, 
      using $\Psi$,
\end{itemize}
both for  a simple $q\bar q$ nucleon  model.  These were the models that
produced the most dramatic effects in the $Q_2$  system; the later model
being only  studied for $SU_c(2)$.   Finally,  it should be  pointed out
that this does not necessarily rule  out the earlier hypotheses that the
colour-hyperfine interactions are important, further investigation would
be required.

\subsection{$L^+L^-$ Production}

  In    chapter  4,  the    $p\porbar\rightarrow  gg\rightarrow  L^+L^-$
production cross-section was   computed for a  very  simple ({\it i.e.},
rank-5) superstring inspired \Esix model.  The $L^+L^-$ production rates
were  studied  over  the   \Esix model  parameter  space  for,  $0.02\le
v_1/v_2\le0.9\,$, $6.7\le v_3/v_2\le9.1\,$, $0.4\,TeV\le\ms\le1\,TeV\,$,
$200\,GeV\le\mheavyq\le600\,GeV\,$,  $\mpzero\approx\order{214}\,GeV\,$,
and $\mhpm\approx\order{200}\,GeV\,$, as   a function of  $\mheavyl$ for
$0.05\,TeV\le\mheavyl\le\order{0.6-1}\,TeV\,$.   For this survey of  the
parameter   space, we~\cite{kn:Konig,kn:MAD} expect $\order{10^{4\pm1}}$
$events/yr$ at the  $LHC$  and no  events at  the $\teva$.  It  was also
found that the results were at least a factor  of 10 less than the event
rates predicted for the  $MSSM$, due to  the $CDF$ and $D0\!\!\!/$  soft
limits placed   on $\mztwo$~\cite{kn:Shochet} ({\it  i.e.},  $v_3/v_2$),
requiring $\mztwo\approxge\order{500}\,GeV$.  These soft limits resulted
in the  $H^0_{1,2}$ and   $P^0$   contributions to the   total  $L^+L^-$
production cross-section    to    become suppressed.     It   was   also
demonstrated,  in principle, that for  certain regions in the $MSSM$ and
\Esix parameter spaces it  is  possible to  distinguish between the  two
models.  Finally some discussion  about the possible $L^\pm$ decay modes
ensued, with the general  outcome  resulting in $jets+\missing\,$.   For
these processes it is, in general, expected  to be difficult to pull out
signals from $SM$ backgrounds; further study would be required.

\section{Closing Remarks}

   In closing, the string-flip potential models  have not been ruled out
as candidate models for nuclear matter.  They  correctly predict most of
the bulk   properties of nuclear  matter with  the exception  of nuclear
binding.  It is  expected that the  string-flip potential model with the
inclusion of the newly proposed variational wave function, $\Psi\sim{\rm
Perm}|\PH|\,\Phi_{\rm Slater}\,$, along with perhaps some variant of the
flux-bubble  model, should     produce nuclear  binding.    Finally, for
$L^+L^-$ production  $\order{10^{4\pm1}}$ $events/yr$ are  expect at the
$LHC$. If  heavy leptons  are discovered  it may indicate   the possible
existence of    an  $\Esix$ model.    However,    in general,  different
experiments would be required to truly validate the model.

%
%
\appendix
\chapter{Three-quark K.E.  Computations}
\label{sec-appa}
Given the wave function
\begin{equation}
\Psi(\alpha,\beta,\rho)=\chi(\alpha,\beta)\Phi(\rho)\,,
\end{equation}
where
\begin{equation}
\chi(\alpha,\beta)=
\mathrm{e}^{-
            {\displaystyle \sum_{\{\mathrm{rgb}\}}
            (\beta\xi_{\mathrm{rgb}})^\alpha}
           }
\end{equation}
and
\begin{equation}
\Phi(\rho)=
\prod_{c\,\varepsilon\,\{\mathrm{rgb}\}}|\Phi_{\mathrm{S}_c}(\rho)|
\end{equation}
are the correlation and Fermi parts respectively, the kinetic energy can
be split up as follows:  into a correlation piece $\bar{T}_{\mathrm{C}}=
2\bar{T}_{{\mathrm{C}}_{\mathrm{-s}}}-\overline{F^2_{\mathrm{C}}}\,$,  a
Fermi  piece $\bar{T}_{\mathrm{F}}=2\bar{T}_{{\mathrm{F}}_{\mathrm{-s}}}
-\overline{F^2_{\mathrm{F}}}\,$, and    a mixed  correlation-Fermi piece
$\bar{T}_{\mathrm{CF}}=-2\bar{F}_{\mathrm{CF}}\,$, where
\begin{eqnarray}
\bar{T}_{{\mathrm{C}}_{\mathrm{-s}}}&=&
\frac{-1}{4N_nm_q}\langle\sum_q(\nabla_q^2\ln\chi)\,\rangle\,,\\
\bar{T}_{{\mathrm{F}}_{\mathrm{-s}}}&=&
\frac{-1}{4N_nm_q}\langle\sum_q(\nabla_q^2\ln\Phi)\,\rangle\,,\\
\overline{F^2_{\mathrm{C}}}\;\;&=&
\frac{1}{2N_nm_q}\langle\sum_q(\grad_q\ln\chi)^2\rangle\,,\\
\overline{F^2_{\mathrm{F}}}\;\;&=&
\frac{1}{2N_nm_q}\langle\sum_q(\grad_q\ln\Phi)^2\rangle\,,
\end{eqnarray}
and
\begin{equation}
\bar{F}_{\mathrm{CF}}=\frac{1}{2N_nm_q}\langle\sum_q(\grad_q\ln\chi)
\cdot(\grad_q\ln\Phi)\rangle\,.
\label{eq:mcf}
\end{equation}

The correlation pieces are straightforward to evaluate, and give the
following
\begin{eqnarray}
\grad_\ell\ln\chi
&=&-\,\frac{\alpha}{2}\,\beta^\alpha
   \sum_{\{\mathrm{rgb}\}}\xi_{\mathrm{rgb}}^{\alpha-2}
   \grad_\ell\xi_{\mathrm{rgb}}^2\nonumber\\
&=&-\,\frac{\alpha}{3}\,\beta^\alpha
   \sum_{\{\mathrm{rgb}\}}\xi_{\mathrm{rgb}}^{\alpha-2}
        [
          (\vrrg-\vrbr)\delta_{\mathrm{r}\ell}
         +(\vrgb-\vrrg)\delta_{\mathrm{g}\ell}
         +(\vrbr-\vrgb)\delta_{\mathrm{b}\ell}
        ]\label{eq:chi}\nonumber\\ & &      
\end{eqnarray}
with $\xi_{\mathrm{rgb}}=
\frac{1}{\sqrt{3}}\sqrt{\rrg^2+\rgb^2+\rbr^2}\,$, which yields
\begin{eqnarray}
\sum_\ell(\grad_\ell\ln\chi)^2
&=&-\,\frac{\alpha^2}{9}\,\beta^{2\alpha}
   \sum_{\{\mathrm{rgb}\}}\xi_{\mathrm{rgb}}^{2\alpha-4}
   [(\vrrg-\vrbr)^2+(\vrgb-\vrrg)^2+(\vrbr-\vrgb)^2]\nonumber\\
&=&-\alpha^2\beta^{2\alpha}
   \sum_{\{\mathrm{rgb}\}}\xi_{\mathrm{rgb}}^{2\alpha-2}\,,
\end{eqnarray}
and
\begin{eqnarray}
\sum_\ell(\nabla_\ell^2\ln\chi)
&=&-\,\frac{\alpha}{3}\,\beta^\alpha
   \sum_\ell\sum_{\{\mathrm{rgb}\}}
   [
    \mbox{$\frac{\alpha-2}{2}$}\,\xi_{\mathrm{rgb}}^{\alpha-4}
    (\grad_\ell\xi_{\mathrm{rgb}}^2)
    +\xi_{\mathrm{rgb}}^{\alpha-2}\grad_\ell
   ]\nonumber\\
   &&\cdot
   [
     (\vrrg-\vrbr)\delta_{\mathrm{r}\ell}
    +(\vrgb-\vrrg)\delta_{\mathrm{g}\ell}
    +(\vrbr-\vrgb)\delta_{\mathrm{b}\ell}
   ]\nonumber\\                        
&=&-\,\frac{\alpha}{3}\,\beta^\alpha\sum_{\{\mathrm{rgb}\}}
   \{\mbox{$\frac{\alpha-2}{3}$}\,\xi_{\mathrm{rgb}}^{-2}
   [(\vrrg-\vrbr)^2+(\vrgb-\vrrg)^2+(\vrbr-\vrgb)^2]\nonumber\\ & &
   +18\}\xi_{\mathrm{rgb}}^{\alpha-2}\nonumber\\
&=&-\alpha(\alpha+4)\beta^\alpha\sum_{\{\mathrm{rgb}\}}
   \xi_{\mathrm{rgb}}^{\alpha-2}\,,
\end{eqnarray}
which imply
\begin{equation}
T_{{\mathrm{C}}_{\mathrm{-s}}}=\frac{\alpha(\alpha+4)\beta^\alpha}{4m_q}
\sum_{\{\mathrm{rgb}\}}\xi_{\mathrm{rgb}}^{\alpha-2}\,,
\end{equation}
and
\begin{equation}
F^2_{\mathrm{C}}=\frac{\alpha^2\beta^{2\alpha}}{2m_q}
\sum_{\mathrm{rgb}}\xi_{\mathrm{rgb}}^{2\alpha-2}\,,
\end{equation}
thus giving the desired result, {\it i.e.},
\begin{equation}
T_{\mathrm{C}}=2T_{{\mathrm{C}}_{\mathrm{-s}}}-F_{\mathrm{C}}^2
              =\frac{\alpha\beta^\alpha}{2m_q}
               \sum_{\{\mathrm{rgb}\}}\xi_{\mathrm{rgb}}^{\alpha-2}
               [\alpha(1-(\beta\xi_{\mathrm{rgb}})^\alpha)+4]\;.
\end{equation}

The Fermi pieces are also straight forward once the following identities
are realized \cite{kn:Watson}:
\begin{eqnarray}
\grad_\ell\ln|\Phi_{\mathrm{S}_c}|
&=&\sum_{ij}\bar{\phi}_{ij}\grad_\ell\phi_{ij}\,,\label{eq:phi}\\
   \grad_\ell\bar{\phi}_{ij}
&=&-\sum_{mn}\bar{\phi}_{im}(\grad_\ell\phi_{ij}^T)\bar{\phi}_{nj}\,,\\
   \nabla_\ell^2\ln|\Phi_{\mathrm{S}_c}|
&=&\sum_{ij}[\bar{\phi}_{ij}\nabla_\ell^2\phi_{ij}
   -\sum_{mn}\bar{\phi}_{im}(\grad_\ell\phi_{mn}^T)
   \cdot(\grad_\ell\phi_{ij})\bar{\phi}_{nj}]\,,
\end{eqnarray}
where $\bar{\phi}_{ij}\equiv(\phi^T)^{-1}_{ij}\,$, and
$\phi_{ij}=\sin(\frac{2\pi}{\mathrm{L}}\,\vn_i
\cdot{\vec{\mathrm{r}}}_j+\delta_i)\,$.
These imply
\begin{equation}
\grad_\ell\phi_{ij}=\frac{2\pi}{L}\vn_i\delta_{j\ell}\phi_{ij}^\prime
\label{eq:gphi}
\end{equation}
and
\begin{equation}
\nabla_\ell^2\phi_{ij}=
-\,\frac{4\pi^2}{L^2}\vn_i^2\delta_{j\ell}\phi_{ij}\,,
\end{equation}
where $\phi_{ij}^\prime\equiv\phi_i(\vrj+\frac{L}{4\vn_i^2}\vn_i)\,$.
Therefore
\begin{equation}
\sum_\ell\nabla^2_\ell\ln|\Phi_{\mathrm{S}_c}|
=-\,\frac{4\pi^2}{L^2}
 [
  \sum_{ij}\bar\phi_{ij}\vn_i^2\phi_{ji}^T
  +\sum_{ijk}\bar\phi_{ki}\phi_{ki}^\prime\vn_k
  \cdot\vn_j\phi_{ji}^\prime\bar\phi_{ji}
 ]\,
\end{equation}
and
\begin{equation}
\sum_\ell(\grad_\ell\ln|\Phi_{\mathrm{S}_c}|)^2
=\frac{4\pi^2}{L^2}\sum_{ijk}\bar\phi_{ki}\phi_{ki}^\prime\vn_k
 \cdot\vn_j\phi_{ji}^\prime\bar\phi_{ji}\,,
\end{equation}
imply
\begin{eqnarray}
T_{\mathrm{F}}^{\mathrm{(c)}}
& = &2T_{{\mathrm{C}}_{\mathrm{-s}}}^{\mathrm{(c)}}
     -F_{\mathrm{C}}^{2^{\mathrm{(c)}}}
\;=\;\frac{2\pi^2}{m_qL^2}
     \sum_{ij}\bar\phi_{ij}\vn_i^2\phi_{ji}^T\nonumber\\ 
& = &\frac{2\pi^2}{m_qL^2}\mathrm{Tr}(\phi^T\vec{N}_c^2\bar\phi)
\;=\;\frac{2\pi^2}{m_qL^2}\mathrm{Tr}(\vec{N}_c^2)\,,
\end{eqnarray}
where $(\vec{N}_c^2)_{ij}\equiv\vn^2_i\delta_{ij}\,$. Thus  the  desired
result is obtained by summing over the quark  colour degrees of freedom,
{\it i.e.}, $\sum_cT_{\mathrm{F}}^{\mathrm{(c)}}$, which implies
\begin{equation}
T_{\mathrm{F}}=\frac{2\pi^2}{m_qL^2}\sum_q\vn^2_q\,.
\end{equation}

Finally,  the  mixed   correlation-Fermi  result,   given   by  equation
(\ref{eq:mixed}), is  simply  obtained, via  equation (\ref{eq:mcf}), by
taking the inner product of equations (\ref{eq:chi}) and (\ref{eq:phi}),
summing over $\ell$ and  $q$, and using the  identity given by  equation
(\ref{eq:gphi}).

\chapter{Algorithms}

This appendix contains useful  algorithms that were  used in some of the
Monte Carlo routines in this thesis.

\section{Generating Slater Wave Functions}
\label{sec-gslb}
This section shows how to  generate the Slater  wave function for a given
number of fermions, $n_F$.

The first thing to notice is that equation~(\ref{eq:slaterwave}) can be
rewritten as,
\begin{equation}
\phi_{\vn}(\vr)=\left\{\begin{array}{ll}
  \cos(\frac{2\pi}{\mathrm{L}}\,\vn\cdot{\vr})&
  {\rm if}\;\eta(\vn)\ge 0\\
  \sin(\frac{2\pi}{\mathrm{L}}\,\vn\cdot{\vr})&
  {\rm if}\;\eta(\vn)< 0
\end{array}\right.\,,
\end{equation}
where
\begin{equation}
\eta(\vn) = 3\,\varepsilon({\rm n_1})+
            2\,\varepsilon({\rm n_2})+
            \varepsilon({\rm n_3})\,,
\end{equation}
is the signature function, and
\begin{equation}
\varepsilon(x)=\left\{\begin{array}{rl}
  -1 & {\rm if}\;x<0\\
   0 & {\rm if}\;x=0\\
  +1 & {\rm if}\;x>0
\end{array}\right.\,.
\end{equation}
This definition is much more useful for programming purposes because it
gives a unique wave function for a given lattice vector, $\vn$.

  The  final  step is  to develop  an algorithm  to generate the lattice
vectors: consider a set of lattice vectors,  $\vn$, whose components are
given by
\begin{equation}
n_a=-k,\ldots,k\,,
\end{equation}
where k is an integer. Then the total number of states is $(2k+1)^3$, or
\begin{equation}
k=\frac{1}{2}\,n_F^{1/3}-1\,.
\end{equation}
Therefore the algorithm is as follows:
\begin{itemize}
\item Generate $(2k+1)^3$ vectors $\vn$, with components 
      $n_a=-k,\ldots,k$.
\item Sort the vectors in increasing order of magnitude squared, 
      $\vn^2\,$.
\end{itemize}
The  sorting step makes sure   that the states, $\phi_{\vn}(\vr)\,$, are
increasing in energy: since  ${\rm E}_{\rm Fermi}\sim|\vn|^2\,$.

\section{Distributed Minimization}
\label{sec-distmin}

   The section gives  a detailed outline  of  a distributed minimization
algorithm that was used for the work done in chapter~\ref{chap-fbbl}.

  Consider a $d$-dimensional hypersurface  that is being fitted  locally
to  a  paraboloid  as depicted   in   figure~\ref{fig:distmin}, then the
minimum number of points required for the fitting is
\begin{equation}
p=\frac{1}{2}d(d-1)+2d+1\,,
\end{equation}
and the      distributed   minimization    algorithm,      outlined   in
\S~\ref{sec-mindist}, is as follows:
\begin{enumerate}
\item Submit $\kappa\ge{\cal O}(2p)+\psi$ 
      ``sample'' points, $x_\kappa\,$, s.t. $\psi\sim{\cal O}(m)\,$.
\item Wait for $f(x_\eta)$ points to arrive, 
      s.t. $\eta\sim\kappa-\psi\,$.
\item Push the results onto stack $S$.
\item Fit each point $f(x_k)$, $k=1,\ldots,\eta$, using their
      $N\ge{\cal O}(2p)$   nearest neighbor   points in $S$,
      where $N$ is chosen s.t. $\chi^2_\nu$ is minimized.
\item Fit $S$ in its entirety.
\item Push the critical points, $x_\rho\,$, corresponding to
      parabolic minima onto stack $M$.
\item Purge any redundancy out of $M$.
\item Submit the newly found critical points $x_\rho\varepsilon M\,$
      that survived the purging.
\item If $\rho=0$ and $M$ is not empty choose   a point in $M$ with the
      lowest $\chi^2_\nu$ and refit it to obtain  a new critical point.
\item If this point is a minimum  and is not in  $M$ then push  
      a copy of it onto $M$ and submit  it as $x_{\rho=1}\,$.
\item If this is the first pass submit $(\psi-\rho)\theta(\psi-\rho)$
      ``sample''  points, otherwise if    $\rho=0$ and  there  are 
      $\psi-1$ points pending submit a ``sample'' point.
\item Repeat all of the above steps, except steps 1 \& 5, with
      $\eta=1\,$ until some convergence criterion is meet.
\item Flush all the computers.
\item Return the lowest result,  $(f(x_{min}),x_{min})\,$, off 
      of stack $S$.
\end{enumerate}
Step 1  is  the initialization step  which consists  of  overloading the
system,  of $m$ computers.  In  step 2  of  the initialization phase the
routine waits  for ${\cal O}(2p)$  points to  arrive.  Once these points
have arrived the system is left occupied processing the remaining $\psi$
points. Therefore $\psi$ is chosen by the  user, who decides how many of
the $m$ computers they want to use (keep  loaded).  Although all that is
required for a fitting to a $d$-dimensional paraboloid  is $p$ points it
was found that  ${\cal O}(2p)$ points gave  much better  statistics.  In
steps  2   and 3 all   of   the incoming  Monte Carlo   are  fitted to a
$d$-dimensional paraboloid and stored in a data  base: {\it i.e.}, stack
$S$.  As  this routine passes  through several iterations the data base,
$S$,  builds  and the statistics  becomes  better.  To take advantage of
this as  many points as  possible are  fitted  in the  neighborhood of a
newly arrived data point,   $x_k\,$.  The optimum  number of  points  is
determined by  minimizing $\chi_\nu^2\,$ such  that  there are  at least
${\cal O}(2p)\,$ of them.  Step 5 is  an optional step which attempts to
fit the entire  contents   of $S$.  This   step  is  only  done  in  the
initialization phase of the routine.  Steps 6 through  8 submit only the
critical points that correspond to a parabolic minimum and that have not
already been submitted.   Therefore stack $M$  contains a list of points
(along with other pertinent fitting  information) that have already been
submitted.  If these steps fail to submit any points, steps 9 through 11
take  over.  Steps 9  and 10 attempt to find  a  new parabolic minima to
submit by refitting about the neighborhood of  a point in $M$, using the
data in $S$, that has the lowest $\chi_\nu^2\,$.
%
%
If these steps fail  then step 11 takes  over by submitting more
``sample'' points, as follows.
%
%
If on the first pass no points were  submitted the system becomes doubly
loaded. On future passes no  points are submitted until the  overloading
has vanished, at which point the system is kept  optimally loaded.  Step
12  checks its convergence  criterion to  determine  whether or not  the
routine should continue.  If the convergence  criterion is met, steps 13
and 14 clean up the system and returns the result, which is taken as the
lowest value in $S$.

  The ``sample'' points were   generated by using a  Sobol  quasi-random
sequence \cite{kn:NumRecC} that ensured a uniform sampling of the region
being searched,  such that each point was  only visited once.  Therefore
if all of the  attempts  failed at  fitting parabaloids, to  predict the
next step, the    searching region would  be  continually  sampled: {\it
i.e.}, if becomes effectively a grid search.

   In Step  11 an option  was  used in  which, under the  condition that
after the first pass if $M$ was not empty the  sampling was done locally
about the point that failed to yield a  minimum in step 6.  The locality
of the  sampling was determined by  the  uncertainty in  the position of
critical  point  from its  past parabolic  fit,   in the  previous pass.
Therefore, assuming several passes lead  to execution of this step, more
statistics would be gathered  locally about this  region and the  search
would either converge or wonder  away.  In order  to make sure that this
searching thread did not leave the box, certain boundary conditions were
placed on it: {\it i.e.}, if the point was near the edge of the box only
intersection of its error box with the search region was sampled.

   For the work  presented in this  thesis, a convergence  criterion was
used      in        which   the    search     would    end       if  the
$\chi_\nu^2\approxle\order{1}\,$,   for   the  point   with   the lowest
$\chi_\nu^2$  in  $M$.  Later  work showed   that  minimization  using a
weighted variance  produced  more  satisfactory results.   This   method
reduced the computation time by $\sim1/2\,$.






\chapter{Couplings and Widths for $\hat\sigma(gg\rightarrow L^+L^-)$}
\label{sec-appc}
 
   This appendix gives a summary  of the calculations  that were used to
obtain  the couplings and the  widths  for the $\hat\sigma(gg\rightarrow
L^+L^-)$ matrix elements given in \S~\ref{sec-cross}.

\section{The Couplings}
\label{sec-crscup}
 
   In this section the calculations of the vertex factors used to obtain
the $gg\rightarrow L^+L^-$ matrix elements, given in \S~\ref{sec-cross},
are summarized.

   For  the        $Z_{1,2}$     exchange     diagrams     shown      in
figure~\ref{fig:fusion}.a the following vertex factors were used
\eqnpict{\gaugeff{Z^\mu_{i}}{f}{\bar f}}{
   \frac{-g}{\sqrt{1-x_W}}\,\gamma^\mu\,
   [\tilde C^{fZ_i}_LP_L+\tilde C^{fZ_i}_RP_R]\,,\label{eq:cupa}
}
\noindent
where $i=1,2$,
\begin{eqnarray}
P_L&=&\frac{1}{2}(1-\gamma_5)\,,\\
P_R&=&\frac{1}{2}(1+\gamma_5)\,,
\end{eqnarray}
and $C^{qZ_i}_L$  and  $C^{qZ_i}_R$  were     the couplings  used     in
equation~\ref{eq:hatqz}.  The  gauge-fermion interaction Lagrangian  for
$\SU{2}{L}\otimes\U{1}{Y}\otimes\U{1}{E}$ is given by
\begin{equation}
{\cal L}_{\rm int}\supseteq\,-\,\frac{1}{2}\,
(g_L^{\mbox{}}\hat\tau^a_{ij}L_\mu^a+
g_Y^{\mbox{}}\delta_{ij}\hat Y_{Y_i}Y_\mu
+g_E^{\mbox{}}\hat Y_{E_i}E_\mu)\,\bar\psi_i\bar\sigma^\mu\psi_j\,,
\label{eq:appca}
\end{equation}
where  the  $\psi_i$'s  are two-component  spinors,  see equation~B.2 of
Haber and Kane  (HK)~\cite{kn:Haber}.   Defining $g=g_L\,$, $\gp=g_Y\,$,
$\gpp=g_E\,$, and $\hat Y=\hat  Y_{Y_i}(=2\hat Q-\hat\tau_3)$, and using
the identities
\begin{eqnarray}
aL^3_\mu+bY_\mu&=& (a\cos\theta_W-b\sin\theta_W)Z_\mu
                  +(a\sin\theta_W+b\cos\theta_W)A_\mu\,,\\
E_\mu&=&Z_\mu^\prime\,,
\end{eqnarray}
then equation~\ref{eq:appca} becomes
\begin{equation}
{\cal L}_{\rm int}\supseteq\,-\,\left\{
\frac{g}{\cos\theta_W}(\hat T_3 - \hat Q\,x_W)Z_\mu
+\,\frac{1}{2}\,\gpp\hat Y_{E}Z^\prime_\mu
\right\}_{ij}[\bar\psi_{(f_L)_i}\bar\sigma^\mu\psi_{(f_L)_j}+
\bar\psi_{(f_L^c)_i}\bar\sigma^\mu\psi_{(f_L^c)_j}]\,,
\end{equation}
where           $\hat      T_3=\hat\tau_3/2\,$,     $\tau_i=\sigma_i\,$,
$\tan\theta_W=\gp/g\,$, and $x_W=\sin^2\theta_W\,$. Noting that
\begin{eqnarray}
(\hat T_3 - \hat Q\,x_W)|f^c_L>&=&-(\hat T_3 - \hat Q\,x_W)|f_R>\,,\\
\hat Y_E|f^c_L>&=&-\hat Y_E|f_R>\,,
\end{eqnarray}
yields
\begin{equation}
{\cal L}_{\rm int}\supseteq\,-\,\left\{
\frac{g}{\cos\theta_W}(\hat T_3 - \hat Q\,x_W)Z_\mu
+\,\frac{1}{2}\,\gpp\hat Y_{E}Z^\prime_\mu
\right\}_{ij}[\bar\psi_{(f_L)_i}\bar\sigma^\mu\psi_{(f_L)_j}-
\bar\psi_{(f_R)_i}\bar\sigma^\mu\psi_{(f_R)_j}]\,.
\end{equation}
Using the following identities
\begin{eqnarray}
\;\;\bar\psi_{(f_L)_i}\bar\sigma^\mu\psi_{(f_L)_j}&=&
\bar f_i\gamma^\mu P_Lf_j\,,\\
-\bar\psi_{(f_R)_i}\bar\sigma^\mu\psi_{(f_R)_j}&=&
\bar f_i\gamma^\mu P_Rf_j\,,
\end{eqnarray}
to convert from two-component to four-component spinor notation yields
\begin{equation}
{\cal L}_{\rm int}\supseteq
\frac{-g}{\sqrt{1-x_W}}\sum_{A=L,R}\bar f\left\{
   (T_{3_A}-e_fx_W)\not\!Z
   +\frac{1}{2}\left(\frac{\gpp}{g}\right)
   y^\prime_{f_A}\,\sqrt{1-x_W}\not\!Z^\prime
\right\}P_A\,f\,,
\end{equation}
see     equations~A.28     of     HK  \cite{kn:Haber}.       Then    the
$Z$-$Z^\prime$-vertex factor is
\eqnpict{\gaugeff{Z^\mu,Z^{\prime^\mu}}{f}{\bar f}}{
   \frac{-g}{\sqrt{1-x_W}}\,\gamma^\mu\,
   [C^{fZ,fZ^\prime}_LP_L+ C^{fZ,fZ^\prime}_RP_R]\,,
}
where the $C^{fZ,fZ^\prime}_{L,R}\,$'s are defined by equations
\ref{eq:hatqzc} and \ref{eq:hatqzd}. Using the inverse
of transformations \ref{eq:zmixa} and \ref{eq:zmixb},
\begin{equation}
\left(\begin{array}{l}
   \tilde Z^\prime \\ Z
\end{array}\right)=
\left(\begin{array}{rr}
   \cos\phi & -\sin\phi \\
   \sin\phi &  \cos\phi
\end{array}\right)
\left(\begin{array}{c}
   Z_1\\ Z_2
\end{array}\right)\,,\label{eq:sleep}
\end{equation}
yields the desired result
\begin{equation}
{\cal L}_{\rm int}\supseteq
\frac{-g}{\sqrt{1-x_W}}\sum_{i=1}^2\sum_{A=L,R}\bar f
   \!\not\!Z_i\,\tilde C^{fZ_i}_AP_A\,f\,,\label{eq:zabvrt}
\end{equation}
{\it i.e.}, vertex factor~\ref{eq:cupa}.

   For  the  $H^0_{1,2,3}$ and    $P^0$   exchange  diagrams  shown   in
figure~\ref{fig:fusion}.b the following vertex factors were used
\eqnpict{\scalarff{H^0_i}{f}{\bar f}}{
   -g\,\frac{m_f}{2m_W}\,K^{fH^0_i}\,,\label{eq:cupb}
}
and
\eqnpict{\scalarff{P^0}{f}{\bar f}}{
   ig\,\frac{m_f}{2m_W}\,\gamma_5\,K^{fP^0}\,,\label{eq:cupc}
}
\noindent
respectively, where    $i=1,2,3\,$.  The   $K^{fH^0_i}$   and $K^{fP^0}$
couplings are  obtained    from the Yukawa   interaction   part  of  the
Lagrangian given by equation~\ref{eq:yuk}: noting that $\varepsilon_{ij}
=   (i\tau_3)_{ij}$  and    plugging   W,  equation~\ref{eq:soup},  into
equation~\ref{eq:yuk} yields
\begin{eqnarray}
{\cal L}_{Yuk}&\supseteq&
-\txtfrac{1}{2}\varepsilon_{ij}\{-\lambda_1[
  \Phi_{2_i}^{\mbox{}}(\psi_{Q_j}\psi_{u^c_L}+\psi_{u^c_L}\psi_{Q_j})+
  \Phi_{2_i}^*(\bar\psi_{Q_j}\bar\psi_{u^c_L}+
    \bar\psi_{u^c_L}\bar\psi_{Q_j}
  )
]
\nonumber\\&&+\,
\lambda_2[
  \Phi_{1_i}^{\mbox{}}(\psi_{Q_j}\psi_{d^c_L}+\psi_{d^c_L}\psi_{Q_j})+
  \Phi_{1_i}^*(\bar\psi_{Q_j}\bar\psi_{d^c_L}+
    \bar\psi_{d^c_L}\bar\psi_{Q_j}
  )
]
\nonumber\\&&+\,
\lambda_3[
  \Phi_{1_i}^{\mbox{}}(\psi_{L_j}\psi_{e^c_L}+\psi_{e^c_L}\psi_{L_j})+
  \Phi_{1_i}^*(\bar\psi_{L_j}\bar\psi_{e^c_L}+
    \bar\psi_{e^c_L}\bar\psi_{L_j}
  )
]
\nonumber\\&&+\,
\lambda_4[
  \Phi_{3}^{\mbox{}}(\psi_{R^\prime_i}\psi_{L^\prime_j}+
    \psi_{L^\prime_j}\psi_{R^\prime_i})+
  \Phi_{3}^*(\bar\psi_{R^\prime_i}\bar\psi_{L^\prime_j}+
    \bar\psi_{L^\prime_j}\bar\psi_{R^\prime_i}
  )
]
\nonumber\\&&+\,
\lambda_5[
  \Phi_{3}^{\mbox{}}(\psi_{d^{\prime^c}_L}\psi_{d^\prime_L}+
    \psi_{d^\prime_L}\psi_{d^{\prime^c}_L})+
  \Phi_{3}^*(\bar\psi_{d^{\prime^c}_L}\bar\psi_{d^\prime_L}+
    \bar\psi_{d^\prime_L}\bar\psi_{d^{\prime^c}_L}
  )
]\}\,,\\
&\supseteq&
-\txtfrac{1}{2}\{\lambda_1[
  \phi^0_2(\psi_{u_L}\psi_{u^c_L}+\psi_{u^c_L}\psi_{u_L})+
  \phi^{0^*}_2(\bar\psi_{u_L}\bar\psi_{u^c_L}+
    \bar\psi_{u^c_L}\bar\psi_{u_L}
  )
]
\nonumber\\&&+\,
\lambda_2[
  \phi^0_1(\psi_{d_L}\psi_{d^c_L}+\psi_{d^c_L}\psi_{d_L})+
  \phi^{0^*}_1(\bar\psi_{d_L}\bar\psi_{d^c_L}+
    \bar\psi_{d^c_L}\bar\psi_{d_L}
  )
]
\nonumber\\&&+\,
\lambda_3[
  \phi^0_1(\psi_{e_L}\psi_{e^c_L}+\psi_{e^c_L}\psi_{e_L})+
  \phi^{0^*}_1(\bar\psi_{e_L}\bar\psi_{e^c_L}+
    \bar\psi_{e^c_L}\bar\psi_{e_L}
  )
]
\nonumber\\&&+\,
\lambda_4[
  \phi^0_3(\psi_{e^{\prime^c}_L}\psi_{e^\prime_L}+
    \psi_{e^\prime_L}\psi_{e^{\prime^c}_L})+
  \phi^{0^*}_3(\bar\psi_{e^{\prime^c}_L}\bar\psi_{e^\prime_L}+
    \bar\psi_{e^\prime_L}\bar\psi_{e^{\prime^c}_L}
  )
]
\nonumber\\&&+\,
\lambda_5[
  \phi^0_3(\psi_{d^{\prime^c}_L}\psi_{d^\prime_L}+
    \psi_{d^\prime_L}\psi_{d^{\prime^c}_L})+
  \phi^{0^*}_3(\bar\psi_{d^{\prime^c}_L}\bar\psi_{d^\prime_L}+
    \bar\psi_{d^\prime_L}\bar\psi_{d^{\prime^c}_L}
  )
]\}\,,
\end{eqnarray}
and similarly for the other generations. Defining
\begin{equation}
f=\Doublet{\psi_{f_L}}{\bar\psi_{f^c_L}}
\end{equation}
and using the following identities
\begin{eqnarray}
\psi_{f^c_{1_L}}\psi_{f_{2_L}^{\mbox{}}}=
\psi_{f_{2_L}^{\mbox{}}}\psi_{f^c_{1_L}}
&=&\bar f_1P_Lf_2\,,\\
\bar\psi_{f^c_{1_L}}\bar\psi_{f_{2_L}^{\mbox{}}}=
\bar\psi_{f_{2_L}^{\mbox{}}}\bar\psi_{f^c_{1_L}}
&=&\bar f_2P_Rf_1\,,
\end{eqnarray}
see equations A.24, A.25, and A.28 of HK \cite{kn:Haber}, implies
\begin{eqnarray}
{\cal L}_{Yuk}&\sim&-\lambda_i(
  \phi^0_j\psi_{f^c_L}\psi_{f_L}+
  \phi^{0^*}_j\bar\psi_{f^c_L}\bar\psi_{f_L}
)\,,\\
&=&-\txtfrac{1}{2}\,\lambda_i\,
[\phi^0_j\bar f(1-\gamma_5)f+\phi^{0^*}_j\bar f(1+\gamma_5)f]\,,\\
&=&-\lambda_i\,
[\real(\phi^0_j)\bar ff-i\,\imag(\phi^0_j)\bar f\gamma_5f]\,,\\
&=&-\txtfrac{1}{\sqrt{2}}\,\lambda_i\,
(\phi^0_{jR}\bar ff-i\,\phi^0_{jI}\bar f\gamma_5f)\,.
\end{eqnarray}
Expanding the $\phi^0_i$'s in terms of their physical fields, equations
\ref{eq:basa} through \ref{eq:basb}, yields
\begin{eqnarray}
{\cal L}_{Yuk}&\supseteq&
\underbrace{-\frac{1}{\sqrt{2}}\,\left\{
\lambda_1\nu_2\,\bar u u+
\lambda_2\nu_1\,\bar d d+
\lambda_3\nu_1\,\bar e e+
\lambda_4\nu_3\,\bar e^\prime e^\prime+
\lambda_5\nu_3\,\bar d^\prime d^\prime
\right\}}_{{\bf Equation~\ref{eq:yucky}}}
\nonumber\\&&\nonumber\\&&
-\frac{1}{\sqrt{2}}\sum_{j=1}^3\left\{
\lambda_1U_{2j}\,\bar u u+
U_{1j}(\lambda_2\,\bar d d+
\lambda_3\,\bar e e)+
U_{3j}(\lambda_4\,\bar e^\prime e^\prime+
\lambda_5\,\bar d^\prime d^\prime)
\right\}H^0_j\nonumber\\&&\nonumber\\&&
+\frac{i\kappa}{\sqrt{2}}
\{\lambda_1 v_{13}\,\bar u\gamma_5 u+v_{23}
(\lambda_2\,\bar d\gamma_5 d+
\lambda_3\,\bar e\gamma_5 e)+v_{12}
(\lambda_4\,\bar e^\prime\gamma_5 e^\prime+
\lambda_5\,\bar d^\prime\gamma_5 d^\prime)\}
P^0\,,\label{eq:yields}
\nonumber\\
%
%
\end{eqnarray}
where $\kappa=1/\sqrt{v_1^2v_2^2+v^2v_3^2}\,$.  The couplings can now be
read  directly  and give,   {\it via}   equations~\ref{eq:ycpa}  through
\ref{eq:ycpd},
\begin{eqnarray}
K^{uH^0_i}&=&\frac{1}{\sin\beta}U_{2i}\,,\label{eq:rata}\\
\nonumber\\
K^{dH^0_i}&=&\frac{1}{\cos\beta}U_{1i}\,,\\
\nonumber\\
K^{d^\prime H^0_i}&=&\frac{2\mw}{g\nu_3}U_{3i}\,,\\
\nonumber\\
K^{eH^0_i}&=&\frac{1}{\cos\beta}U_{1i}\,,\\
\nonumber\\
K^{e^\prime H^0_i}&=&\frac{2\mw}{g\nu_3}U_{3i}\,,\label{eq:ratae}
\end{eqnarray}
for the scalar Higgs fields, $H^0_i\,$, and
\begin{eqnarray}
K^{uP^0}&=&\label{eq:ratbb}
\frac{1}{\sin\beta}\,\kappa v_{13}\,,\\
\nonumber\\
K^{dP^0}&=&
\frac{1}{\cos\beta}\,\kappa v_{23}\,,\\
\nonumber\\
K^{d^\prime P^0}&=&
\frac{2 m^{\mbox{}}_W}{gv_3}\,\kappa v_{12}\,,\\
\nonumber\\
K^{eP^0}&=&
\frac{1}{\cos\beta}\,\kappa v_{23}\,,\\
\nonumber\\
K^{e^\prime P^0}&=&
\frac{2 m^{\mbox{}}_W}{gv_3}\,\kappa v_{12}\,,
\label{eq:ratb}
\end{eqnarray}
%
%
for pseudo-scalar Higgs fields, $P^0\,$.

  For the      $H^0_{1,2,3}$   exchange  diagrams shown     in   figures
\ref{fig:fusion}.c and \ref{fig:fusion}.d  the following  vertex factors
were used
\eqnpict{\scalarss{H^0_i}{\tilde f_A}{\tilde f_B^{\mbox{}^*}}}{
  \kappa^{\tilde f H_i^0}_{AB} = 
  -\,\frac{g m_Z}{\sqrt{1-x_W}}\,K^{\tilde f H_i^0}_{AB}
  \,,\label{eq:cupd}
}
\noindent
where $A,B=L,R\,$.  The $\kappa^{\tilde  q H_i^0}_{AB}$ couplings, which
were obtained   directly  by  using  Mathematica  \cite{kn:Wolfram}   to
generate equation~\ref{eq:scalpot} and to extract the couplings from it,
are as follows:
\begin{eqnarray}
%
%
%
\kappa^{\tilde u H_i^0}_{LL}&=&
(\txtfrac{1}{18}\gpps-\txtfrac{1}{4}g^2+\txtfrac{1}{12}\gps)\,
U_{1i}\,\nu_1
+(\txtfrac{2}{9}\gpps+\txtfrac{1}{4}g^2-\txtfrac{1}{12}\gps-\lambda_1^2)\,
U_{2i}\,\nu_2\nonumber\\&&
-\txtfrac{5}{18}\,\gpps U_{3i}\,\nu_3\,,\\
%
%
%
\kappa^{\tilde u H_i^0}_{RR}&=&
(\txtfrac{1}{18}\gpps-\txtfrac{1}{3}\gps)\,
U_{1i}\,\nu_1
+(\txtfrac{2}{9}\gpps+\txtfrac{1}{3}\gps-\lambda_1^2)\,
U_{2i}\,\nu_2\nonumber\\&&
-\txtfrac{5}{18}\gpps U_{3i}\,\nu_3\,,\\
%
%
%
\kappa^{\tilde u H_i^0}_{LR}&=&
\txtfrac{1}{2}\,[
  (U_{3i}\,\nu_1+U_{1i}\,\nu_3)\lambda-\sqrt{2}\,U_{2i}A_u
]\lambda_1\,,\\
%
%
%
\kappa^{\tilde d H_i^0}_{LL}&=&
(\txtfrac{1}{18}\gpps+\txtfrac{1}{4}g^2+\txtfrac{1}{12}\gps-\lambda_2^2)\,
U_{1i}\,\nu_1
+(\txtfrac{2}{9}\gpps-\txtfrac{1}{4}g^2-\txtfrac{1}{12}\gps)\,
U_{2i}\,\nu_2\nonumber\\&&
-\txtfrac{5}{18}\,\gpps U_{3i}\,\nu_3\,,\\
%
%
%
\kappa^{\tilde d H_i^0}_{RR}&=&
-(\txtfrac{1}{36}\gpps-\txtfrac{1}{6}\gps+\lambda_2^2)\,
U_{1i}\,\nu_1
-(\txtfrac{1}{9}\gpps+\txtfrac{1}{6}\gps)\,
U_{2i}\,\nu_2\nonumber\\&&
+\txtfrac{5}{36}\gpps U_{3i}\,\nu_3\,,\\
%
%
%
\kappa^{\tilde d H_i^0}_{LR}&=&
\txtfrac{1}{2}\,[
  (U_{3i}\,\nu_2+U_{2i}\,\nu_3)\lambda-\sqrt{2}\,U_{1i}A_d
]\lambda_2\,,\\
%
%
%
%
\kappa^{\tilde d^\prime H_i^0}_{LL}&=&
-(\txtfrac{1}{9}\gpps+\txtfrac{1}{6}\gps)\,U_{1i}\,\nu_1
-(\txtfrac{4}{9}\gpps-\txtfrac{1}{6}\gps)\,U_{2i}\,\nu_2
\nonumber\\&&
+(\txtfrac{5}{9}\gpps-\lambda_5^2)\,U_{3i}\,\nu_3\,,\\
%
%
%
\kappa^{\tilde d^\prime H_i^0}_{RR}&=&
-(\txtfrac{1}{36}\gpps-\txtfrac{1}{6}\gps)\,U_{1i}\,\nu_1
-(\txtfrac{1}{9}\gpps+\txtfrac{1}{6}\gps)\,U_{2i}\,\nu_2
\nonumber\\&&
+(\txtfrac{5}{36}\gpps-\lambda_5^2)\,U_{3i}\,\nu_3\,,\\
%
%
%
\kappa^{\tilde d^\prime H_i^0}_{LR}&=&
\txtfrac{1}{2}\,[
  (U_{2i}\,\nu_1+U_{1i}\,\nu_2)\lambda-\sqrt{2}\,U_{3i}A_{d^\prime}
]\lambda_5\,,
\end{eqnarray}
for the squark-Higgs couplings, and
\begin{eqnarray}
%
%
%
\kappa^{\tilde e H_i^0}_{LL}\;&=&
-(\txtfrac{1}{18}\gpps-\txtfrac{1}{4}g^2+\txtfrac{1}{4}\gps+\lambda_3^2)\,
U_{1i}\,\nu_1
-(\txtfrac{2}{9}\gpps+\txtfrac{1}{4}g^2-\txtfrac{1}{4}\gps)\,
U_{2i}\,\nu_2\nonumber\\&&
+\txtfrac{5}{36}\,\gpps U_{3i}\,\nu_3\,,\\
%
%
%
\kappa^{\tilde e H_i^0}_{RR}\;&=&
(\txtfrac{1}{18}\gpps+\txtfrac{1}{2}\gps-\lambda_3^2)\,
U_{1i}\,\nu_1
+(\txtfrac{2}{9}\gpps-\txtfrac{1}{2}\gps)\,
U_{2i}\,\nu_2\nonumber\\&&
-\txtfrac{5}{18}\gpps U_{3i}\,\nu_3\,,\\
%
%
%
\kappa^{\tilde e H_i^0}_{LR}\;&=&
\txtfrac{1}{2}\,[
  (U_{3i}\,\nu_2+U_{2i}\,\nu_3)\lambda-\sqrt{2}\,U_{1i}A_e
]\lambda_2\,,\\
%
%
%
\kappa^{\tilde \nu_e H_i^0}_{LL}&=&
-(\txtfrac{1}{36}\gpps+\txtfrac{1}{4}g^2+\txtfrac{1}{4}\gps)\,
U_{1i}\,\nu_1
-(\txtfrac{1}{9}\gpps-\txtfrac{1}{4}g^2-\txtfrac{1}{4}\gps)\,
U_{2i}\,\nu_2\nonumber\\&&
+\txtfrac{5}{36}\,\gpps U_{3i}\,\nu_3\,,\\
%
%
%
\kappa^{\tilde \nu_e H_i^0}_{RR}&=&
\txtfrac{5}{36}\gpps (U_{1i}\,\nu_1+4U_{2i}\,\nu_2-5U_{3i}\,\nu_3)\,,\\
%
%
%
%
\kappa^{\tilde \nu_e H_i^0}_{LR}&=&0\,,\label{eq:cupdy}
\end{eqnarray}
for the slepton-Higgs couplings.   The mass eigenstate couplings $\tilde
K^{\tilde   f H^0_i}_{1,2}\,$,  given   by  equations~\ref{eq:kcupa} and
\ref{eq:kcupb}, were obtained by substituting
\begin{equation}
\left(\begin{array}{c}
\tilde f_L\\\tilde f_R
\end{array}\right)=
\left(\begin{array}{rr}
\cos\theta_{\tilde f}&-\sin\theta_{\tilde f}\\
\sin\theta_{\tilde f}&\cos\theta_{\tilde f}
\end{array}\right)
\left(\begin{array}{c}
\tilde f_1\\\tilde f_2
\end{array}\right)\,,\label{eq:trany}
\end{equation}
which  is just the  inverse  of equation~\ref{eq:smix}, into the  scalar
potential and extracting them {\it via} Mathematica
\cite{kn:Wolfram}.

   The corresponding pseudo-scalar-Higgses couplings
\eqnpict{\scalarss{P^0}{\tilde f_A}{\tilde f_B^{\mbox{}^*}}}{
  i\,\kappa^{\tilde d^\prime P^0}_{AB}=
  -i\,\frac{g m_Z}{\sqrt{1-x_W}}\,K^{\tilde q P^0}_{AB}
  \label{eq:cupe}
}
\noindent
are obtained in a similar fashion as above: {\it i.e.},
\begin{eqnarray}
%
%
%
%
%
\kappa^{\tilde u P^0}_{AB}&=&
-\,\eAB\,\frac{\nu_2}{2\,\sqrt{\nu_{12}^2+\nu^2\nu_3^2}}\,
\left[
(\nu_1^2+\nu_3^2)\lambda+\sqrt{2}\,A_u\nu_{3}\cot\beta
\right]
\lambda_1\,,\\&&\nonumber\\
%
%
%
%
\kappa^{\tilde d P^0}_{AB}&=&
-\,\eAB\,\frac{\nu_1}{2\,\sqrt{\nu_{12}^2+\nu^2\nu_3^2}}\,
\left[
(\nu_2^2+\nu_3^2)\lambda+\sqrt{2}\,A_d\nu_{3}\tan\beta
\right]
\lambda_2\,,\\&&\nonumber\\
%
%
%
%
\kappa^{\tilde d^\prime P^0}_{AB}&=&
-\,\eAB\,\frac{\nu_3}{2\,\sqrt{\nu_{12}^2+\nu^2\nu_3^2}}\,
\left[
\nu^2\lambda+\sqrt{2}\,A_{d^\prime}\frac{\nu_{12}}{\nu_3}
\right]
\lambda_5\,,
\end{eqnarray}
for squark-pseudo-Higgs couplings, and
\begin{eqnarray}
%
%
%
%
\kappa^{\tilde e P^0}_{AB}\;&=&
-\,\eAB\,\frac{\nu_1}{2\,\sqrt{\nu_{12}^2+\nu^2\nu_3^2}}\,
\left[
(\nu_2^2+\nu_3^2)\lambda+\sqrt{2}\,A_e\nu_{3}\tan\beta
\right]
\lambda_3\,,\\
%
%
%
\kappa^{\tilde \nu_e P^0}_{AB}&=&0\,,
\end{eqnarray}
for the slepton-pseudo-Higgs couplings, where
\begin{equation}
\eAB=\left\{
\begin{array}{r@{\;\;\;{\rm if}\;}l}
 1&A=L,\;B=R\\
 0&A=B\\
-1&A=R,\;B=L
\end{array}
\right.\,.
\end{equation}
In general,  these couplings will also  mix to give the  mass eigenstate
couplings  $\tilde  K^{\tilde f  P^0}_{1,2}\,$; which are   defined in a
similar fashion to equations~\ref{eq:kcupa} and \ref{eq:kcupb}.

\section{The Widths}
\label{sec-wit}

  In this section   all of the   tree  level two-body  decay  widths for
$Z_2\,$, $H^0_i\,$, and $P_0$  are computed. The  generic two body decay
formula is given by \cite{kn:Griffiths}
\begin{equation}
\begin{array}{cc}
\raisebox{10mm}{$\displaystyle
  \Gamma=\frac{S|{\cal M}_{ab}|^2}{16\pi m_0}\beta_{ab}\,,
$}&
\mbox{\setlength{\unitlength}{1mm}
  \begin{picture}(10,20)
    \put(20,10){\circle{3}}
    \put(20,12){\vector(0,1){8}}
    \put(20,8){\vector(0,-1){8}}
    \put(23,0){$m_b$}
    \put(23,9){$m_0$}
    \put(23,18){$m_a$}
  \end{picture}
}
\end{array}
\label{eq:width}
\end{equation}
where $m_i\,$, $i=0,a,b\,$, are the masses of the particles, $p_i\,$, in
the decay process $p_0\rightarrow p_ap_b\,$,
\begin{equation}
\beta_{ab}=\sqrt{1-\frac{2(m_a^2+m_b^2)}{m_{0}^2}+
\frac{(m_a^2-m_b^2)^2}{m_{0}^4}}\,,
\end{equation}
such that $\beta_{ab}\equiv\beta_a$ if $a=b\,$, $S$ is a symmetry factor
for the out going particles,  $p_a$ and $p_b\,$,  and ${\cal M}_{ab}$ is
the amplitude for the process.

\subsection{$\Gamma_{Z_2}$}

For the $Z_2$ width the following processes need to be computed:
$$
Z_2\longrightarrow \mbox{\footnotesize $W^+W^-,\,Z_1H^0_i,\,
W^\pm H^\mp,$}\,q_i\bar q_i,\,l_i\bar l_i,\,
\tilde\chi^0_i\bar{\tilde\chi}\mbox{}^0_j,\,
\tilde\chi^+_i\tilde\chi^-_j,\,
\tilde q_i\tilde q_j^*,\,
\tilde l_i\tilde l_j^{\mbox{}^*},\,
\mbox{\footnotesize $H^0_iH^0_j,\,H^+H^-,\,
P^0H^0_i$}\,.
$$

  The $Z_2\rightarrow W^+W^-$ width, which can be  found in Hewett and
Rizzo \cite{kn:Hewett}, is given by
\begin{equation}
\Gamma(\mbox{\footnotesize $Z_2\rightarrow W^+W^-$})=
\frac{g^2\mztwo\,\sin^2\phi}{192\pi(1-x_W)}\,
\left(\frac{\mztwo}{\mz}\right)^4
\beta_W^3
\left[1+
  20\left(\frac{\mw}{\mz}\right)^2+
  12\left(\frac{\mw}{\mz}\right)^4
\right]\,.
\end{equation}

  The $Z_2\rightarrow q_i\bar  q_i,\,  l_i\bar l_i$  vertex factors  are
given by
\eqnpict{
   \varepsilon_\mu(q,\lambda)\!\!\!\!\!
   \overlaystuff{
     \gaugeff{Z_2}{u_f(p)}{\bar v_{\bar f}(p^\prime)}
   }{q}{p}{p^\prime}
}{
   -g\gamma_\mu(v_f-a_f\gamma_5)\,,
}
which  were obtained from  equation~\ref{eq:zabvrt} by  converting to the
$V-A$ basis: {\it i.e.},
\begin{equation}
a\,P_L+b\,P_R=v_f-a_f\gamma_5\,,\label{eq:vabasis}
\end{equation}
where 
\begin{eqnarray}
v_f&=&\txtfrac{1}{2}\,(a+b)\,,\\
a_f&=&\txtfrac{1}{2}\,(a-b)\,,
\end{eqnarray}
which yields
\begin{eqnarray}
v_f&=&\frac{1}{2\sqrt{1-x_W^{\mbox{}}}}\,
(\tilde C^{fZ_2}_L+\tilde C^{fZ_2}_R)\,,\\
a_f&=&\frac{1}{2\sqrt{1-x_W^{\mbox{}}}}\,
(\tilde C^{fZ_2}_L-\tilde C^{fZ_2}_R)\,.\label{eq:vabasisd}
\end{eqnarray}
Therefore
\begin{eqnarray}
\overline{|{\cal M}_{f\bar f}|^2}&=&\txtfrac{1}{3}\,g^2
\sum_\lambda\sum_{spin}
|\bar v(p^\prime)\gamma_\mu(v_f-a_f\gamma_5)u(p)
\varepsilon^\mu(q,\lambda)|^2\,,
\nonumber\\&=&\txtfrac{4}{3}\,g^2
[v_f^2(m^2_{Z_2}+2m_f^2)+a_f^2(m_{Z_2}^2-4m_f^2)]\,.
\end{eqnarray}
Plugging this into equation~\ref{eq:width} gives
\begin{equation}
\Gamma(Z_2\rightarrow f\bar f)=c_f\,\frac{g^2}{12\pi}\,
m_{Z_2}^{\mbox{}}\beta_f^{\mbox{}}
\left[
v_f^2\left(1+\frac{2m_f^2}{m_{Z_2}^2}\right)+
a_f^2\left(1-\frac{4m_f^2}{m_{Z_2}^2}\right)
\right]\,,
\end{equation}
where $c_f$ is a colour factor which is 3 for quarks and 1 for leptons.

  The  $Z_2\rightarrow  \tilde    q_i\tilde  q_j^*,\, \tilde   l_i\tilde
l_j^{\mbox{}^*}$ vertex factors are given by
\eqnpict{
   \varepsilon_\mu(q,\lambda)\!\!\!\!\!
   \overlaystuff{
     \gaugess{Z_2}{\tilde f_i}{\tilde f_j^*}
   }{q}{\!\!p_i}{\!\!p_j^\prime}
}{
   -ig\kappa_{ij}(p_j^\prime-p_i)^\mu\,,
}
where $\tilde f_{k=1,2}$ are sfermion mass eigenstates, and
\begin{equation}
\kappa_{ij}=\left\{\begin{array}{r@{\;\;;\;{\rm if\;}}l}
v_f+a_f\,\cos2\theta_{\tilde f}& i=j=1\\
v_f-a_f\,\cos2\theta_{\tilde f}& i=j=2\\
   -a_f\,\sin2\theta_{\tilde f}& i\not =j
\end{array}\right.\,.
\end{equation}
The vertex factor is obtained by followings steps similar to equations
\ref{eq:appca} through \ref{eq:zabvrt}, 
\begin{eqnarray}
{\cal L}_{\rm int}&\supseteq&-\,\frac{i}{2}\,
(g_L^{\mbox{}}\hat\tau^a_{ij}L_\mu^a+
g_Y^{\mbox{}}\delta_{ij}\hat Y_{Y_i}Y_\mu
+g_E^{\mbox{}}\hat Y_{E_i}E_\mu)\,\tilde f_i^*\lrpartial^\mu\tilde f_j\\
&\supseteq&\frac{-ig}{\sqrt{1-x_W}}\sum_{i=1}^2\sum_{A=L,R}
\tilde C^{fZ_i}_A\,\tilde f_A^*\lrpartial_\mu\tilde f_A\,Z^\mu_i\,,
\end{eqnarray}
followed  by transforming  the sfermions to  their  mass eigenstates  by
using equation~\ref{eq:trany}, 
\begin{eqnarray}
{\cal L}_{\rm int}&\supseteq&\frac{-ig}{\sqrt{1-x_W}}\sum_{i=1}^2\{
[\tilde C^{fZ_i}_L\cos^2\theta_{\tilde f}+
\tilde C^{fZ_i}_R\sin^2\theta_{\tilde f}]
\,\tilde f_1^*\lrpartial_\mu\tilde f_1\nonumber\\&&
+[\tilde C^{fZ_i}_L\sin^2\theta_{\tilde f}+
\tilde C^{fZ_i}_R\cos^2\theta_{\tilde f}]
\,\tilde f_2^*\lrpartial_\mu\tilde f_2\nonumber\\&&
-\,\txtfrac{1}{2}\,[\tilde C^{fZ_i}_L-\tilde C^{fZ_i}_R]
\sin2\theta_{\tilde f}\,(
\tilde f_1^*\lrpartial_\mu\tilde f_2+
\tilde f_2^*\lrpartial_\mu\tilde f_1
)
\}\,Z^\mu_i\,,
\end{eqnarray}
and      then changing   to       the    $V-A$    basis,   {\it     via}
equations~\ref{eq:vabasis} through \ref{eq:vabasisd}, to obtain
\begin{equation}
{\cal L}_{\rm int}\supseteq\,-ig\sum_{i,j,k=1}^2\kappa_{ij}\,
\tilde f_i^*\lrpartial_\mu\tilde f_j\,Z_k^\mu\,.
\end{equation}
Therefore
\begin{eqnarray}
\overline{|{\cal M}_{\tilde f_i^{\mbox{}}\tilde f_j^*}|^2}&=&
\txtfrac{1}{3}\,g^2\sum_\lambda
|\varepsilon_\mu(q,\lambda)\kappa_{ij}(p_j^\prime-p_i)^\mu|^2\,,
\nonumber\\
&=&
\txtfrac{1}{3}\,g^2\,m_{Z_2}^2\kappa_{ij}^2
\beta_{\tilde f_i\tilde f_j}^2\,.
\end{eqnarray}
Plugging this into equation~\ref{eq:width} gives
\begin{equation}
\Gamma(Z_2\rightarrow \tilde f_i^{\mbox{}}\tilde f_j^*)=
c_f\frac{g^2\,m_{Z_2}}{48\pi}\,\kappa_{ij}^2\,
\beta_{\tilde f_i\tilde f_j}^2\,.
\end{equation}
   
  For the range of VEV's that will be considered here ({\it i.e.}, large
$v_3$  in   particular)   the  $    Z_2\rightarrow   \mbox{\footnotesize
$Z_1H^0_i,\,  W^\pm H^\mp,\,H^0_iH^0_j,\,H^+H^-,\,P^0H^0_i$}\, $  widths
can be approximated by
\begin{equation}
\Gamma(Z_2\rightarrow V+S)\approx\frac{17g^2\xw}{864\pi(1-\xw)}
\,\mztwo\,,
\end{equation}
where      the   $H^0_iH^0_j$   are    kinematically      forbidden   or
suppressed~\cite{kn:Hewett}.

  The      $Z_2\rightarrow  \tilde\chi^0_i\bar{\tilde\chi}\mbox{}^0_j,\,
\tilde\chi^+_i\tilde\chi^-_j$ widths  are  quite complicated to compute,
due to  the complex nature of the  mass matrices, and can  contribute as
much   as 10-20\%    to  the   total  width,   neglecting  phase   space
suppression~\cite{kn:Hewett}. Here  its  contribution  will be  taken as
15\%; this proved to have no noticeable impact on $L^+L^-$ production.

\subsection{$\Gamma_{H^0_i}$}
For the $H^0_i$ widths the following processes need to be computed:
\begin{eqnarray}
H^0_i&\longrightarrow&\mbox{\footnotesize $Z_jZ_k,\,W^+W^-,$}\,
q_j\bar q_j,\,l_j\bar l_j,\,
\tilde\chi^0_j\bar{\tilde\chi}\mbox{}^0_k,\,
\tilde\chi^+_j\tilde\chi^-_k,\,
\tilde q_j\tilde q_k^*,\,
\tilde l_j\tilde l_k^{\mbox{}^*},\,
\mbox{\footnotesize $H^0_jH^0_k,\,H^+H^-,\,P^0P^0$}
\,.\nonumber
\end{eqnarray}

  The $H^0_i\rightarrow Z_jZ_k,\,W^+W^-$ vertex factors are given by
\eqnpict{
   \overlaystuff{
     \scalarvv{H^0_i}{V_a^\mu}{V_b^{\nu^*}}
   }{q}{\!p_a^{\mbox{}}}{\!p_b^{\mbox{}}}
}{
   iC^{H^0_i}_{V_aV_b}\,g_{\mu\nu}\,,
}
where
\begin{eqnarray}
%
%
%
%
%
C^{H^0_i}_{Z_1Z_1}&=&C^{H^0_i}_{ZZ}\,\cos^2\phi-
2\,C^{H^0_i}_{Z Z^\prime}\sin2\phi+
C^{H^0_i}_{Z^\prime Z^\prime}\,\sin^2\phi\,,\\
%
%
%
%
%
%
%
C^{H^0_i}_{Z_1Z_2}&=&
C^{H^0_i}_{Z Z^\prime}\cos2\phi
-(C^{H^0_i}_{Z\,Z\,}-C^{H^0_i}_{Z^\prime Z^\prime})\sin2\phi\,,\\
%
%
%
%
%
C^{H^0_i}_{Z_2Z_2}&=&C^{H^0_i}_{ZZ}\,\sin^2\phi+
2\,C^{H^0_i}_{Z Z^\prime}\sin2\phi+
C^{H^0_i}_{Z^\prime Z^\prime}\,\cos^2\phi\,,
\end{eqnarray}
for the $Z_i$'s, with
\begin{eqnarray}
%
%
%
C^{H^0_i}_{Z\,Z\,}&=&\frac{1}{4}\,(g\cos\tw+\gp\sin\tw)^2\,
(U_{1i} v_1 + U_{2i} v_2)\,,\\
C^{H^0_i}_{Z Z^\prime}&=&\frac{\gpp}{6}\,
(g\cos\tw+\gp\sin\tw)\,(U_{1i} v_1 - 4 U_{2i} v_2)\,,\\
%
%
%
C^{H^0_i}_{Z^\prime Z^\prime}&=&\frac{\gpps}{36}\,
(U_{1i} v_1 + 16 U_{2i} v_2 + 25 U_{3i} v_3)\,,
\end{eqnarray}
and
%
%
\begin{equation}
C^{H^0_i}_{W^+W^-}=\frac{g^2}{2}\,(U_{1i}v_1+U_{2i}v_2)\,,
\end{equation}
for  the $W$'s.   The    vertex factors,  $C^{H^0_i}_{V_aV_b}\,$,   were
obtained,  {\it    via}   Mathematica~\cite{kn:Wolfram},  by    plugging
equation~\ref{eq:sleep}          and     equations         \ref{eq:basa}
through~\ref{eq:basb}   into the  kinetic   terms for  the  scalar-Higgs
fields, equation~\ref{eq:hke}. Therefore
\begin{eqnarray}
\overline{|{\cal M}_{ab}^i|^2}&=&
\sum_{\lambda_a\lambda_b}|\varepsilon_\mu(p_a,\lambda_a)
C^{H^0_i}_{V_aV_b}\varepsilon_\nu^*(p_b,\lambda_b)g^{\mu\nu}|^2
\nonumber\\&=&
\frac{m^4_{H^0_i}\,
C^{H^0_i}_{V_aV_b^*}\raisebox{2.5ex}{\scriptsize2}}{4(m_am_b)^2}
\left[1-\frac{2(m_a^2+m_b^2)}{m_{H^0_i}^2}+
\frac{(m_a^2+m_b^2)^2+8(m_am_b)^2}{m_{H^0_i}^4}
\right]\,,\;\;\;\;\;
\end{eqnarray}
which yields, {\it via} equation~\ref{eq:width},
\begin{equation}
\Gamma(H^0_i\rightarrow V_aV_b)=
\frac{S\,C^{H^0_i}_{V_aV_b}\raisebox{2.5ex}{\scriptsize2}
m^3_{H^0_i}\,\beta_{ab}}{64\pi(m_am_b)^2}\,
\left[1-\frac{2(m_a^2+m_b^2)}{m_{H^0_i}^2}+
\frac{(m_a^2+m_b^2)^2+8(m_am_b)^2}{m_{H^0_i}^4}
\right]\,,
\end{equation}
where  $S=\ffrac{1}{2}\,$ for identical $Z_i$'s,  otherwise $S=1$. 

  The $H^0_i\rightarrow q_i\bar q_i,\,l_i\bar l_i$ decay width is
\begin{equation}
\Gamma(H^0_i\rightarrow f\bar f)=
\frac{c_fg^2}{32\pi}\left(\frac{m_f}{\mw}\right)^2K^{fH^0_i}
\raisebox{2.5ex}{\scriptsize2}
\beta_{H^0_i}^3\,m_{H^0_i}^{\mbox{}}\,,
\end{equation}
{\it via} equation~\ref{eq:width} with amplitude
\begin{equation}
\overline{|{\cal M}_{f\bar f}|^2}=
\frac{g^2}{2}\left(\frac{m_f}{\mw}\right)^2K^{fH^0_i}
\raisebox{2.5ex}{\scriptsize2}\,
[m_{H^0_i}^2-4m_f^2]\,,
\end{equation}
where the $K^{fH^0_i}$ couplings are defined by equation~\ref{eq:cupb}.

  For the scalar processes $H^0_i\rightarrow\tilde  \phi \phi^*$ the
vertex factor is
\eqnpict{
   \overlaystuff{
     \scalarss{H^0_i}{\phi_a}{\phi_b^*}
   }{q}{\!p_a^{\mbox{}}}{\!p_b^{\mbox{}}}
}{
   C^{H^0_i}_{\phi_a\phi_b}\,,
}
which yields the following decay width
\begin{equation}
\Gamma(H^0_i\rightarrow \phi_a\phi_b^*)=
\frac{c_f}{16\pi m_{H^0_i}^{\mbox{}}}\,
|C^{H^0_i}_{\phi_a\phi_b}|^2\beta_{ab}\,,
\label{eq:widdy}
\end{equation}
{\it via} equation~\ref{eq:width}, with amplitude
\begin{equation}
\overline{|{\cal M}_{\phi_a\phi_b}|^2}=
|C^{H^0_i}_{\phi_a\phi_b}|^2\,.
\end{equation}
For  $H^0_i\rightarrow\tilde     q_j\tilde   q_k^*,\,\tilde  l_j\tilde
l_k^{\scriptsize\mbox{}^*}\,$ the vertex factors are
\begin{equation}
C^{H^0_i}_{\tilde f_j\tilde f_k^*}=
\frac{g\,m_Z}{\sqrt{1-\xw}}\,
K^{\tilde f H_i^0}_{jk},
\end{equation}
where    the  $K^{\tilde  f   H_i^0}_{jk}$   couplings   are  given   by
equations~\ref{eq:rata} through \ref{eq:ratae}.  For   $H^0_i\rightarrow
H^0_jH^0_k,\,H^+H^-,\,P^0P^0$ the vertex factors are:
\begin{eqnarray}
C^{H^0_2}_{H^0_1H^0_1}&=&
\frac{1}{2}\,\lambda A\,
(U_{12} U_{21} U_{31} + U_{11} U_{22} U_{31} + U_{11} U_{21} U_{32})
\nonumber\\&&
+\left\{U_{12} \left[\frac{-1}{24}(\gpps+9g^2+9\gps) U_{11}^2
-\left(\frac{1}{18}\gpps-\frac{1}{8}g^2-\frac{1}{8}\gpps\right) U_{21}^2
\right.\right.
\nonumber\\&&
\left.+\frac{5}{72}\gpps U_{31}^2-\frac{1}{2}\lambda^2
(U_{21}^2  + U_{31}^2)\right]
+ U_{11} \left[
\left(\frac{-1}{18}\gpps+\frac{1}{8}g^2+\frac{1}{8}\gps\right)  U_{21}
U_{22}
\right.
\nonumber\\&&
+\left.\left.\frac{5}{36}\gpps  U_{31} U_{32}
-\lambda^2(U_{21} U_{22} + U_{31} U_{32})\right]\right\}v_1
\nonumber\\&&
+\left\{U_{22}\left[
\left(\frac{-1}{18}\gpps+\frac{1}{8}g^2+\frac{1}{8}\gpps\right) U_{11}^2
-\frac{1}{24}(16\gpps+9g^2+9\gps) U_{21}^2\right.\right.
\nonumber\\&&
+\left.\frac{5}{18}\gpps U_{31}^2
-\frac{1}{2}\lambda^2(U_{11}^2  + U_{31}^2 )\right]
+U_{21} \left[
\left(\frac{-1}{9}\gpps+\frac{1}{4}g^2+\frac{1}{4}\gps\right)U_{11}U_{12}
\right.
\nonumber\\&&
+\left.\left.\frac{5}{9}\gpps U_{31} U_{32}
-\lambda^2 (U_{11} U_{12} + U_{31} U_{32})\right]\right\}v_2
\nonumber\\&&
+\left\{U_{31}\left[
\frac{5}{36}\gpps(U_{11}U_{12}+4U_{21}U_{22})
-\lambda^2(U_{11}U_{12}+U_{21} U_{22})\right]\right.
\nonumber\\&&
+\left.U_{32}\left[\frac{5}{72}\gpps(U_{11}^2 +4U_{21}^2- 15 U_{31}^2 )
-\frac{1}{2} (U_{11}^2   + U_{21}^2  )\right]\right\}v_3
\,,\\
C^{H^0_3}_{H^0_1H^0_1}&=&
\frac{1}{2}\,\lambda A\,
(U_{13} U_{21} U_{31} + U_{11} U_{23} U_{31} + U_{11} U_{21} U_{33})
\nonumber\\&&
+\left\{U_{13}\left[\frac{-1}{24}(\gpps+9g^2+9\gps) U_{11}^2
-\left(\frac{1}{18}\gpps-\frac{1}{8}g^2-\frac{1}{8}\gpps\right) U_{21}^2
\right.\right.
\nonumber\\&&
\left.+\frac{5}{72}\gpps U_{31}^2-\frac{1}{2}\lambda^2
(U_{21}^2  + U_{31}^2)\right]
+U_{11} \left[
\left(\frac{-1}{9}\gpps+\frac{1}{4}g^2+\frac{1}{4}\gps\right)  U_{21}
U_{23}
\right.
\nonumber\\&&
+\left.\left.\frac{5}{36}\gpps  U_{31} U_{33} 
-\lambda^2(U_{21} U_{23} + U_{31} U_{33})\right]\right\}v_1
\nonumber\\&&
+\left\{U_{23}\left[
\left(\frac{-1}{18}\gpps+\frac{1}{8}g^2+\frac{1}{8}\gpps\right) U_{11}^2
-\frac{1}{24}(16\gpps+9g^2+9\gps) U_{21}^2\right.\right.
\nonumber\\&&
+\left.\frac{5}{18}\gpps U_{31}^2
-\frac{1}{2}\lambda^2(U_{11}^2  + U_{31}^2 )\right]
+U_{21} \left[
\left(\frac{-1}{9}\gpps+\frac{1}{4}g^2+\frac{1}{4}\gps\right)U_{11}U_{13}
\right.
\nonumber\\&&
+\left.\left.\frac{5}{9}\gpps U_{31}U_{33}
-\lambda^2 (U_{11} U_{13} + U_{31} U_{33})\right]\right\}v_2
\nonumber\\&&
+\left\{U_{31}\left[
\frac{5}{36}\gpps(U_{11}U_{13}+4U_{21}U_{23})
-\lambda^2(U_{11}U_{13}+U_{21} U_{23})\right]\right.
\nonumber\\&&
+\left.U_{33}\left[\frac{5}{72}\gpps(U_{11}^2 +4U_{21}^2- 15 U_{31}^2 )
-\frac{1}{2} (U_{11}^2   + U_{21}^2  )\right]\right\}v_3
\,,\\
C^{H^0_3}_{H^0_1H^0_2}&=&
\frac{1}{2}\,\lambda A\,
(U_{13} U_{22} U_{31} + U_{12} U_{23} U_{31} + U_{13} U_{21} U_{32} 
\nonumber\\&& 
+ U_{11} U_{23} U_{32} + U_{12} U_{21} U_{33} + U_{11} U_{22} U_{33}) 
\nonumber\\&& 
-\left\{\left(\frac{1}{9}\gpps-\frac{1}{4}g^2-\frac{1}{4}\gpps\right)
(U_{12} U_{21} + U_{11} U_{22}) U_{23} \right.
\nonumber\\&& 
+U_{13}\left[\frac{1}{12}(\gpps+9g^2+9\gps)U_{11} U_{12} 
+\left(\frac{1}{9}\gpps-\frac{1}{4}g^2-\frac{1}{4}\gpps\right)
U_{21} U_{22} \right.
\nonumber\\&& 
\left.
-\frac{5}{36}\,\gpps U_{31} U_{32}
+\lambda^2 (U_{21} U_{22} + U_{31} U_{32})\right]
\nonumber\\&& 
-\left.\frac{5}{36}\,\gpps (U_{12} U_{31} + U_{11} U_{32}) 
U_{33}\right\}v_1
\nonumber\\&& 
-\lambda^2
(U_{12} U_{21} U_{23} + U_{11} U_{22} U_{23} + U_{12} U_{31} U_{33} 
+ U_{11} U_{32} U_{33})v_1
\nonumber\\&& 
-\left\{\left(\frac{1}{9}\gpps-\frac{1}{4}g^2-\frac{1}{4}\gpps\right)
 U_{13} (U_{12} U_{21} + U_{11} U_{22}) \right.
\nonumber\\&& 
+U_{23}\left[
\left(\frac{1}{9}\gpps-\frac{1}{4}g^2-\frac{1}{4}\gpps\right)
U_{11} U_{12}
+\left(\frac{4}{3}\gpps+\frac{1}{3}g^2+\frac{1}{3}\gpps\right) 
U_{21} U_{22}
\right.
\nonumber\\&& 
-\left.\frac{5}{9}\,\gpps U_{31} U_{32}
+\lambda^2(U_{11} U_{12} + U_{31} U_{32})\right]
\nonumber\\&& 
-\left.\frac{5}{9}\,\gpps 
(U_{22} U_{31} + U_{21} U_{32}) U_{33}\right\}v_2
\nonumber\\&& 
-\lambda^2
(U_{12} U_{13} U_{21} + U_{11} U_{13} U_{22} + 
U_{22} U_{31} U_{33} + U_{21} U_{32} U_{33})v_2
\nonumber\\&& 
+\left\{
\left[\frac{5}{36}\,\gpps(U_{12} U_{13} + 4 U_{22} U_{23})
-\lambda^2(U_{12} U_{13} + U_{22} U_{23})\right] U_{31}
\right.
\nonumber\\&& 
+\left[\frac{5}{36}\,\gpps(U_{11} U_{13} + 4 U_{21} U_{23})
-\lambda^2(U_{11} U_{13} + U_{21} U_{23})\right] U_{32}
\nonumber\\&& 
+\left[\frac{5}{36}\,\gpps(U_{11} U_{12} + 
4 U_{21} U_{22} - 15 U_{31} U_{32})\right.
\nonumber\\&& 
\left.\left.-\lambda^2(U_{11} U_{12} + U_{21} U_{22})
\rule[-1.5ex]{0ex}{3ex}\right] U_{33}
\right\}v_3\,,\\
C^{H^0_3}_{H^0_2H^0_2}&=&
\frac{1}{2}\,\lambda A\,
(U_{13} U_{22} U_{32} + U_{12} U_{23} U_{32} + U_{12} U_{22} U_{33})
\nonumber\\&&
+\left\{U_{13}\left[\frac{-1}{24}(\gpps+9g^2+9\gps) U_{12}^2
-\left(\frac{1}{18}\gpps-\frac{1}{8}g^2-\frac{1}{8}\gpps\right) U_{22}^2
\right.\right.
\nonumber\\&&
\left.+\frac{5}{72}\gpps U_{32}^2-\frac{1}{2}\lambda^2
(U_{22}^2  + U_{32}^2)\right]
+U_{12} \left[
\left(\frac{-1}{9}\gpps+\frac{1}{4}g^2+\frac{1}{4}\gps\right)  U_{22}
U_{23}
\right.
\nonumber\\&&
+\left.\left.\frac{5}{36}\gpps  U_{32} U_{33} 
-\lambda^2(U_{22} U_{23} + U_{32} U_{33})\right]\right\}v_1
\nonumber\\&&
+\left\{U_{23}\left[
\left(\frac{-1}{18}\gpps+\frac{1}{8}g^2+\frac{1}{8}\gpps\right) U_{12}^2
-\frac{1}{24}(16\gpps+9g^2+9\gps) U_{22}^2\right.\right.
\nonumber\\&&
+\left.\frac{5}{18}\gpps U_{32}^2
-\frac{1}{2}\lambda^2(U_{12}^2  + U_{32}^2 )\right]
+U_{22} \left[
\left(\frac{-1}{9}\gpps+\frac{1}{4}g^2+\frac{1}{4}\gps\right)U_{12}U_{13}
\right.
\nonumber\\&&
+\left.\left.\frac{5}{9}\gpps U_{32}U_{33}
-\lambda^2 (U_{12} U_{13} + U_{32} U_{33})\right]\right\}v_2
\nonumber\\&&
+\left\{U_{32}\left[
\frac{5}{36}\gpps(U_{12}U_{13}+4U_{22}U_{23})
-\lambda^2(U_{12}U_{13}+U_{22} U_{23})\right]\right.
\nonumber\\&&
+\left.U_{33}\left[\frac{5}{72}\gpps(U_{12}^2 +4U_{22}^2- 15 U_{32}^2 )
-\frac{1}{2} (U_{12}^2   + U_{22}^2  )\right]\right\}v_3
\,,
\end{eqnarray}
for the neutral-scalar-Higgses;
\begin{eqnarray}
%
%
%
%
%
C^{H^0_i}_{H^+H^-}&=&\frac{-1}{4\,(1+\cot^2\beta)}\left\{
v_1(3g^2+\gps-4\lambda^2)(U_{1i}+U_{2i}\cot\beta)
\right.
\nonumber\\&&
+\left(g^2-\gps+\frac{4}{9}\gpps\right)
(v_1U_{1i}\cot^2\beta+v_2U_{2i})
\nonumber\\&&
+\frac{1}{9}\gpps[ v_1(v_1U_{1i}+16U_{2i}\cot\beta)-5v_3
(1+4\cot^2\beta)U_{3i}]
\nonumber\\&&
+4\,(\lambda^2v_3(1+\cot^2\beta)+\lambda \cot\beta)U_{3i}
\,,
\end{eqnarray}
for the charged-scalar-Higgses;
\begin{eqnarray}
C^{H^0_i}_{P^0P^0}&=&\frac{v_3^2}{2(v_1^2v_2^2+v^2v_3^2)}\left\{
\left[m_Z^2\cos2\beta-
\frac{1}{9}\left(\frac{\gpp}{g}\right)^2m_W^2-\lambda^2
\frac{v_1^2v_2^2}{v_3^2}\right]
(v_1U_{1i}+v_2U_{2i})\right.
\nonumber\\&&
-\lambda(v_1^3U_{1i}+v_2^3U_{2i})
-\lambda^2v_3(v_1^2+v_2^2)U_{3i}
-\lambda A\frac{v_1v_2}{v_3}
(v_1U_{1i}+v_2U_{2i}+v_3U_{3i})
\nonumber\\&&\left.
+\frac{5}{36}\gpps\left[\frac{v_1^2v_2^2}{v_3^2}
(v_1U_{1i}+4v_2U_{2i}-5v_3U_{3i})
+v_3(4v_1^2+v_2^2)U_{3i}\right]\right\}\,,
\end{eqnarray}
for the pseudo-scalar-Higgses,   which  were all extracted,    {\it via}
Mathematica~\cite{kn:Wolfram},    by    plugging equations \ref{eq:basa}
through  \ref{eq:basb}   for the  physical Higgs  fields  into the Higgs
potential, equation~\ref{eq:fox}.

  The $H^0_i\rightarrow
\tilde\chi^0_j\bar{\tilde\chi}\mbox{}^0_k,\,
\tilde\chi^+_j\tilde\chi^-_k$ decay processes
are quite complicated to compute.  Here a  simple approximation was made
in which for  $\mpzero\approxle\order{500}GeV$  its contribution to  the
width was 15\%,  otherwise 50\% \cite{kn:Gunion,kn:MAD}.   This addition
had    a   negligible    affect     on   $L^+L^-$  production,     since
$\mpzero=200GeV$.

\subsection{$\Gamma_{P^0}$}

For the $P^0$ width the following processes need to be computed:
\begin{eqnarray}
P^0&\longrightarrow&Z_iZ_j,\,W^\pm H^\mp,\,
q_i\bar q_i,\,l_i\bar l_i,\,
\tilde\chi^0_i\bar{\tilde\chi}\mbox{}^0_j,\,
\tilde\chi^+_i\tilde\chi^-_j,\,
\tilde q_i\tilde q_j^*,\,
\tilde l_i\tilde l_j^{\mbox{}^*}
\,.\nonumber
\end{eqnarray}

  For  $P^0\rightarrow Z_iZ_j,\,W^\pm H^\mp$ the  widths  are zero since
here     $\mpzero<\mztwo$        and    $\mpzero\approx\mhpm\,$:     see
figures~\ref{fig:mztwo}  through  \ref{fig:mhzerob}    and    discussion
therein.

  The $P^0\rightarrow q_i\bar q_i,\,l_i\bar l_i$ decay widths are
\begin{equation}
\Gamma(P^0\rightarrow f\bar f)=
\frac{c_fg^2}{32\pi}\left(\frac{m_f}{\mw}\right)^2K^{fP^0}
\raisebox{2.5ex}{\scriptsize2}
\beta_{P^0}^{\mbox{}}m_{P^0}^{\mbox{}}\,,
\end{equation}
{\it via} equation~\ref{eq:width}, with amplitudes
\begin{equation}
\overline{|{\cal M}_{f\bar f}|^2}=
\frac{g^2}{2}\left(\frac{m_f}{\mw}\right)^2K^{fP^0}
\raisebox{2.5ex}{\scriptsize2}\,m_{P^0}^2\,,
\end{equation}
where the $K^{fP^0}$ couplings defined by equation~\ref{eq:cupc}.

  The  $P^0\rightarrow\tilde     q_j\tilde   q_k^*,\,\tilde  l_j\tilde
l_k^{\mbox{}^*}\,$ decay widths are
\begin{equation}
\Gamma(P^0\rightarrow \tilde f_j\tilde f_k^*)=
\frac{c_fg^2\,m_Z^2}{16\pi(1-\xw)m_{P^0}^{\mbox{}}}
K^{\tilde f P^0}_{jk}\raisebox{2.5ex}{\scriptsize2}
\beta_{\tilde f_j\tilde f_k}\,,
\end{equation}
{\it via} equation~\ref{eq:widdy}, with vertex factors
\begin{equation}
C^{P^0}_{\tilde f_j\tilde f_k^*}=
\frac{g\,m_Z}{\sqrt{1-\xw}}\,
K^{\tilde f P^0}_{jk}\,,
\end{equation}
where    the   $K^{\tilde  f  P^0}_{jk}$    couplings    are  given   by
equations~\ref{eq:ratbb} through \ref{eq:ratb}.

   In  this work    $\mpzero$ was  fixed  at   $200GeV$.  At  this  mass
$P^0\rightarrow \tilde   \chi^0_i           \bar{\tilde\chi} \mbox{}^0_j
,\,\tilde\chi^+_i\tilde\chi^-_j$ decays are suppressed~\cite{kn:Gunion}.

%
%
%
\bibliographystyle{unsrt}
\bibliography{refs}

\end{document}